\documentclass[twoside,12pt]{article}
\usepackage{epsfig}
\usepackage{rotating}
\usepackage{axodraw}
\usepackage[dvips]{color}

\newcommand{\be}{\begin{equation}}
\newcommand{\ee}{\end{equation}}
\newcommand{\bea}{\begin{eqnarray}}
\newcommand{\eea}{\end{eqnarray}}

\newcommand{\lsim}{\raisebox{-0.13cm}{~\shortstack{$<$ \\[-0.07cm] $\sim$}}~}

\begin{document}

\title{
\vspace*{-2cm}
\begin{flushright}
{\small PSI--PR--16--18}
\end{flushright}
\vspace{1cm} Higgs Boson Production and Decay at Hadron Colliders}
\author{Michael Spira \\ \\
Paul Scherrer Institut, CH--5232 Villigen PSI, Switzerland}
\maketitle
\begin{abstract} 
Higgs physics at hadron colliders as the LHC is reviewed within the
Standard Model (SM) and its minimal supersymmetric extension (MSSM) by
summarizing the present state-of-the-art of theoretical predictions for
the production cross sections and decay rates.
\end{abstract}

\tableofcontents

\section{Introduction}
\subsection{\it Organization of the Review}
In this work we will review all Higgs decay widths and branching ratios
as well as all relevant Higgs boson production cross sections at the LHC
within the SM and MSSM. Previous reviews can be found in
Refs.~\cite{review}. This work is a substantial update of
Ref.~\cite{higgs97}.

The paper is organized as follows. This Section 1 will provide an
introduction to the SM and MSSM Higgs sectors. In Section 2 we will
review the decay rates of the SM and the MSSM Higgs particles providing
explicit analytical results of the individual partial decay widths where
possible. Section 3 will discuss the production cross sections for the
Higgs bosons of the SM and MSSM and the impact of the numerical results
on the profile of the Higgs bosons at the LHC. A summary will be given
in Section 4.

\subsection{\it Standard Model \label{sec:smintro}}
The discovery of a resonance at the LHC \cite{discovery} that is
compatible with the Standard-Model (SM) Higgs boson \cite{couplings}
marked a milestone in particle physics. The existence of the Higgs boson
is inherently related to the mechanism of spontaneous symmetry breaking
\cite{hi64} while preserving the full gauge symmetry and the
renormalizability of the SM \cite{smren}. Its mass, the last missing
parameter of the SM, has been measured to be ($125.09 \pm 0.24$) GeV
\cite{couplings}. The existence of the Higgs boson allows the SM
particles to be weakly interacting up to high-energy scales without
violating the unitarity bounds of scattering amplitudes
\cite{unitarity}. This, however, is only possible for particular
Higgs-boson couplings to all other particles so that with the knowledge
of the Higgs-boson mass all its properties are uniquely fixed. The
massive gauge bosons and fermions acquire mass through their interaction
with the Higgs field that develops a finite vacuum expectation value in
its ground state thus hiding the still unbroken electroweak gauge
symmetry.  The minimal model requires the introduction of one weak
isospin doublet of the Higgs field and leads after spontaneous symmetry
breaking to the existence of one scalar Higgs boson, while non-minimal
Higgs sectors predict in general more than one Higgs boson.

Within the SM with one Higgs doublet the Higgs mass is constrained by
consistency conditions as the absence of a Landau pole for the Higgs
self-coupling up to high energy scales and the stability of the
electroweak ground state \cite{triviality}. If the SM is required to
fulfill these conditions for energy scales up to the scale of Grand
Unified theories (GUTs) of $\sim 10^{16}$ GeV the Higgs mass is
constrained between 129 GeV and about 190 GeV \cite{hmassbounds}. The
last condition of vacuum stability can be relaxed by demanding the
life-time of the ground state to be larger than the age of our universe
\cite{metastability}. This reduces the lower bound on the Higgs mass to
about 110 GeV \cite{hmassbounds}. The measured value of the Higgs mass
indicates that our universe is unstable with a large lifetime far beyond
the age of the universe.

If the SM is extended to the GUT scale, radiative corrections to the
Higgs-boson mass tend to push the Higgs mass towards the GUT scale, if
it couples to particles of that mass order. In order to obtain the Higgs
mass at the electroweak scale the counter term has to be fine-tuned to
cancel these large corrections thus establishing a very unnatural
situation that requires a solution. This is known as the hierarchy
problem \cite{hierarchy}. Possible solutions are the introduction of
supersymmetry (SUSY) \cite{susy,susyrev}, a collective symmetry between SM
particles and heavier partners as in Little Higgs theories
\cite{littlehiggs} or an effective reduction of the Planck and GUT
scales as in extra-dimension models at the TeV scale \cite{addrs}.

SM Higgs bosons are dominantly produced via the loop-induced
gluon-fusion process $gg\to H$ with top quarks providing the leading
loop contribution \cite{glufus}. Other production processes as
vector-boson fusion $qq\to qqH$ \cite{vvh}, Higgs-strahlung $q\bar q\to
H+W/Z$ \cite{vhv} and Higgs radiation off top quarks $gg,q\bar q\to
t\bar t H$ \cite{htt} are suppressed by more than one order of
magnitude. While the dominant gluon-fusion process allows for the
detection of the Higgs boson in the rare decay modes into $\gamma
\gamma$, four charged leptons and $W$-boson pairs, other decay modes are
only detectable in the subleading production modes as e.g.~the main
Higgs-boson decay into bottom quarks in strongly boosted Higgs-strahlung
\cite{vhbb} or the Higgs-boson decay into $\tau$-leptons in the
vector-boson-fusion process \cite{vbftautau}.

\subsection{\it Minimal Supersymmetric Standard Model \label{sec:mssmintro}}
Supersymmetric extensions of the SM are motivated by providing
a possible solution to the hierarchy problem if the supersymmetric
particles appear at the few-TeV scale \cite{hierarchysusy}.
Supersymmetry connects fermionic and bosonic degrees of freedom and thus
links internal and external symmetries \cite{susy}. It is the last
possible symmetry-type of $S$-matrix theories, i.e.~the last possible
extension of the Poincar\'e algebra. The minimal supersymmetric extension
of the SM (MSSM) predicts the Weinberg angle in striking agreement with
experimental measurements if embedded in a GUT \cite{sw2sgut}.
Moreover, it allows to generate electroweak symmetry breaking
radiatively \cite{radewsb} and yields a candidate for Dark Matter if
$R$-parity is conserved \cite{susydm} which renders the lightest
supersymmetric particle stable. Finally it increases the proton lifetime
beyond experimental bounds in the context of supersymmetric GUTs
\cite{sw2sgut,susyprotondecay}.

The MSSM requires the introduction of two isospin doublets of Higgs
fields in order to maintain the analyticity of the superpotential and
the anomaly-freedom with respect to the gauge symmetries
\cite{susyprotondecay, twohdoublets}. Moreover, two Higgs doublets are
needed for the generation of the up- and down-type fermion masses.  The
Higgs sector is a Two-Higgs-Doublet model (2HDM) of type II. The
mass eigenstates consist of a light ($h$) and heavy ($H$) scalar, a
pseudoscalar ($A$) and two charged ($H^\pm$) states. The
self-interactions of the Higgs fields are entirely fixed in terms of the
electroweak gauge couplings so that the self-couplings are constrained
to small values. This leads to an upper bound on the light scalar Higgs
mass that has to be smaller than the $Z$-boson mass $M_Z$ at leading
order (LO). This is, however, broken by radiative corrections, which are
dominated by top-quark-induced contributions \cite{mssmrad}. The
parameter $\mbox{tg$\beta$}$, defined as the ratio of the two vacuum
expectation values of the scalar Higgs fields, will in general be
assumed to be in the range $1 < \mbox{tg$\beta$} < m_t/m_b$, where
$m_t(m_b)$ denotes the top (bottom) mass, consistent with the assumption
that the MSSM is the low-energy limit of a supergravity model
\cite{zwirner} and to avoid non-perturbative phenomena.

If the soft SUSY-breaking parameters do not contain any complex phases
the input parameters of the MSSM Higgs sector at LO are generally chosen
to be the mass $M_A$ of the pseudoscalar Higgs boson and
$\mbox{tg$\beta$}$.  All other masses and the mixing angle $\alpha$
between the scalar CP-even Higgs states can be derived from these basic
parameters (and the top mass and SUSY parameters, which enter through
radiative corrections).  The radiative corrections can be approximated
by the parameter $\epsilon$, which grows with the fourth power of the
top quark mass and logarithmically with the stop masses
$m_{\tilde{t}_{1,2}}$, supplemented by terms originating from soft
SUSY-breaking parameters, i.e.~the trilinear coupling $A_t$, the
higgsino mass $\mu$ and the third-generation squark mass $M_S$ with
$X_t=A_t-\mu/\mbox{tg$\beta$}$,
\begin{equation} 
\epsilon = \frac{3G_F}{\sqrt{2}\pi^2}
\frac{\overline{m}_t^4(M_S)}{\sin^2\beta} \left\{ \log\left(
\frac{m_{\tilde{t}_1} m_{\tilde{t}_2}}{\overline{m}_t^2(M_S)} \right) +
\frac{X_t^2}{M_S^2} \left( 1 - \frac{X_t^2}{12 M_S^2} \right) \right\}
\label{eq:epsusy}
\end{equation}
where $G_F$ denotes the Fermi constant and $\overline{m}_t(M_S)$ the
$\overline{\rm MS}$ top mass at the scale $M_S$.  These corrections are
positive and reach a maximum (related to $X_t$) for $|X_t| = \sqrt{6}
M_S$. They increase the squared mass of the light neutral Higgs boson $h$ to
\begin{equation}
M^2_h \le M_Z^2 \cos^2 2\beta + \epsilon \sin^2\beta
\label{eq:hbound}
\end{equation}
In this approximation, the upper bound on $M_h$ is shifted from the tree
level value $M_Z$ up to $\sim$ 145 GeV.  The mass of
the lightest scalar state $h$ is given by
\begin{eqnarray}
M^2_h & = & \frac{1}{2} \left[ M_A^2 + M_Z^2 + \epsilon \right.
\nonumber \\
& & \left. - \sqrt{(M_A^2+M_Z^2+\epsilon)^2
-4 M_A^2M_Z^2 \cos^2 2\beta
-4\epsilon (M_A^2 \sin^2\beta + M_Z^2 \cos^2\beta)} \right]
\label{eq:hmass}
\end{eqnarray}
The masses of the heavy neutral and charged Higgs bosons are determined
by the sum rules (valid for this approximation),
\begin{eqnarray}
M_H^2 & = & M_A^2 + M_Z^2 - M_h^2 + \epsilon \nonumber \\
M_{H^\pm}^2 & = & M_A^2 + M_W^2
\end{eqnarray}
The effective mixing parameter $\alpha$ between the CP-even scalar Higgs
states can be derived as
\begin{equation}
\mbox{tg} 2 \alpha = \mbox{tg} 2\beta \frac{M_A^2 + M_Z^2}{M_A^2 - M_Z^2 +
\epsilon/\cos 2\beta}
\end{equation}
The radiative corrections to the MSSM Higgs sector have been calculated
up to the two-loop level in the effective potential approximation and in
the diagrammatic approach \cite{mssmrad}. The leading corrections are
also known at the three-loop level within the effective potential
approach \cite{mssmrad3}. The corrections beyond next-to-leading order
(NLO) are dominated by the QCD corrections to the top-quark-induced
contributions. They decrease the upper bound on the light scalar Higgs
mass $M_h$ by about 10 GeV to $\sim 135$ GeV, while leaving a residual
uncertainty of $\sim 3$ GeV on the light scalar Higgs mass. In the
context of increasing lower mass bounds for the supersymmetric
particles, a resummation of large logarithms related to the
SUSY-particle masses has been performed for the calculation of the MSSM
Higgs-boson masses \cite{mssmradres}. For large SUSY-particle masses
this resummation may lead to effects of the order of 5 GeV (or larger
for small values of $\mbox{tg}\beta$) on the light scalar Higgs mass.

For on-shell external Higgs bosons there is an additional relevant
effect emerging from finite external Higgs momenta leading to an
additional mixing between different types of Higgs bosons. This can be
treated by the introduction of an external mixing matrix as discussed in
Ref.~\cite{mssmzmat} that ensures that the external Higgs bosons are
defined as proper on-shell states. The numerical effects are sizeable
for certain MSSM scenarios and have to be taken into account in all
processes with external MSSM Higgs bosons.

The calculation of the radiative corrections has also been extended to
the effective trilinear and quartic self-interactions of the MSSM Higgs
bosons that are consistently defined at vanishing external momenta. The
NLO corrections at ${\cal O}(\alpha_t)$ and ${\cal O}(\alpha_b)$ are
available since a long time \cite{lambdanlo}, where $\alpha_t = G_F
m_t^2/(\sqrt{2}\sin^2\beta)$ and $\alpha_b = G_F
m_b^2/(\sqrt{2}\cos^2\beta)$. In the leading approximation in terms of
the parameter $\epsilon$ the trilinear couplings of the neutral Higgs
bosons are given up to ${\cal O}(\alpha_t)$ by \cite{lambdanlo}
\begin{eqnarray}
\lambda_{hhh} & = & 3\cos(2\alpha) \sin(\beta+\alpha)
+ \frac{3 \epsilon}{M_Z^2} \frac{\cos^3\alpha}{\sin \beta} \nonumber \\
\lambda_{Hhh} & = & 2 \sin(2\alpha) \sin(\beta+\alpha) 
                - \cos(2\alpha) \cos(\beta+\alpha)
+ \frac{3 \epsilon}{M_Z^2} \frac{\sin\alpha\cos^2\alpha}{\sin \beta}
\nonumber \\
\lambda_{HHh} & = & - 2 \sin(2\alpha) \cos(\beta+\alpha) 
                - \cos(2\alpha) \sin(\beta+\alpha)
+ \frac{3 \epsilon}{M_Z^2} \frac{\sin^2\alpha\cos\alpha}{\sin \beta}
\nonumber \\
\lambda_{HHH} & = & 3\cos(2\alpha) \cos(\beta+\alpha)
+ \frac{3 \epsilon}{M_Z^2} \frac{\sin^3\alpha}{\sin \beta} \nonumber \\
\lambda_{hAA} & = & \cos(2\beta) \sin(\beta+\alpha)
+ \frac{\epsilon}{M_Z^2} \frac{\cos\alpha\cos^2\beta}{\sin \beta}
\nonumber \\
\lambda_{HAA} & = & - \cos(2\beta) \cos(\beta+\alpha)
+ \frac{\epsilon}{M_Z^2} \frac{\sin\alpha\cos^2\beta}{\sin \beta}
\end{eqnarray}
which have been normalized to $\sqrt{\sqrt{2} G_F} M_Z^2$. Similar but more
involved expressions can be obtained for the quartic Higgs
self-couplings. Some time ago the NLO corrections to the Higgs
self-couplings have been extended by the next-to-next-to-leading order
(NNLO) corrections of ${\cal O}(\alpha_t\alpha_s)$ \cite{lambdannlo},
where $\alpha_s$ denotes the strong coupling constant. The radiative
corrections are large in general, while the NNLO part is of moderate
size but reduces the theoretical uncertainties significantly to the
few per-cent level.

The couplings of the neutral Higgs bosons to fermions and gauge
bosons depend on the angles $\alpha$ and $\beta$. Normalized to the SM
Higgs couplings, they are listed in Table~\ref{tb:hcoup}.  The
pseudoscalar particle $A$ does not couple to gauge bosons at tree level,
and its couplings to down (up)-type fermions are (inversely)
proportional to $\mbox{tg$\beta$}$.
\begin{table}[hbt]
\renewcommand{\arraystretch}{1.5}
\begin{center} 
\begin{tabular}{|lc||ccc|} \hline
\multicolumn{2}{|c||}{$\Phi$} & $g^\Phi_u$ & $g^\Phi_d$ &  $g^\Phi_V$ \\
\hline \hline
SM~ & $H$ & 1 & 1 & 1 \\ \hline 
MSSM~ & $h$ & $\cos\alpha/\sin\beta$ & $-\sin\alpha/\cos\beta$ &
$\sin(\beta-\alpha)$ \\
& $H$ & $\sin\alpha/\sin\beta$ & $\cos\alpha/\cos\beta$ &
$\cos(\beta-\alpha)$ \\
& $A$ & $ 1/\mbox{tg}\beta$ & $\mbox{tg}\beta$ & 0 \\ \hline
\end{tabular}
\renewcommand{\arraystretch}{1.2}
\caption{\label{tb:hcoup}
\it MSSM Higgs couplings to up- and down-type fermions and gauge bosons
($V=W,Z$) relative to the corresponding SM couplings.}
\end{center}
\end{table}
For large values of $\mbox{tg$\beta$}$ the Yukawa couplings to (up)
down-type quarks are (suppressed) enhanced and vice versa apart from the
regions where the light (heavy) scalar is close to its upper (lower)
mass bound, where their coupling factors approach the SM-coupling
strengths (up to sign differences for the heavy scalar). The couplings
of the light scalar Higgs particle approach the SM values for large
pseudoscalar masses, i.e.~in the decoupling regime. Thus it will be
difficult to distinguish the light scalar MSSM Higgs boson from the SM
Higgs particle in the region where all Higgs particles except the light
scalar one are very heavy.

\begin{figure}[hbt]
\begin{center}
\setlength{\unitlength}{1pt}
\begin{picture}(100,100)(50,0)
\DashLine(0,50)(50,50){5}
\ArrowLine(75,75)(100,100)
\ArrowLine(100,0)(75,25)
\Line(75,75)(75,25)
\Gluon(75,25)(75,75){-3}{4}
\DashLine(50,50)(75,75){5}
\DashLine(75,25)(50,50){5}
\put(5,55){$\phi$}
\put(50,65){$\tilde b$}
\put(80,48){$\tilde g$}
\put(105,98){$b$}
\put(105,-2){$\bar b$}
\end{picture}
\begin{picture}(100,100)(-30,0)
\DashLine(0,50)(50,50){5}
\ArrowLine(75,75)(100,100)
\ArrowLine(100,0)(75,25)
\Line(75,75)(75,25)
\Photon(75,25)(75,75){-3}{4}
\DashLine(50,50)(75,75){5}
\DashLine(75,25)(50,50){5}
\put(5,55){$\phi$}
\put(50,65){$\tilde t$}
\put(80,48){$\tilde \chi^\pm$}
\put(105,98){$b$}
\put(105,-2){$\bar b$}
\end{picture}
\setlength{\unitlength}{1pt}
\caption{\label{fg:deltab} \it Feynman diagrams contributing to the
bottom Yukawa couplings at ${\cal O}(\alpha_s)$ (left) and ${\cal
O}(\lambda_t^2)$ (right).}
\end{center}
\end{figure}
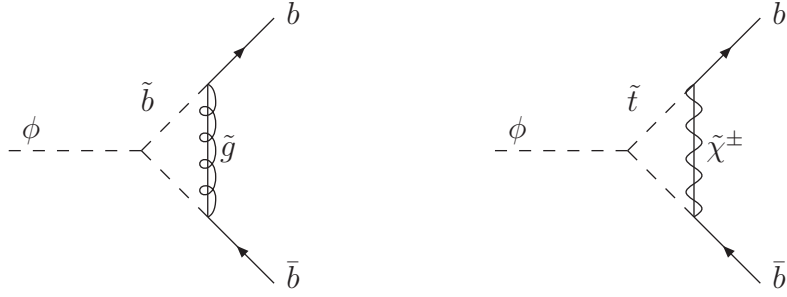

In addition to these coupling factors the bottom Yukawa coupling in
particular is strongly modified by radiative corrections due to
sbottom-gluino exchange in the genuine SUSY--QCD part and by
stop-chargino exchange in the leading SUSY-electroweak part, see
Fig.~\ref{fg:deltab}.
The leading contribution behaves as ${\cal O}(\mu \, m_{\tilde
g}\,\mbox{tg}\beta/M_{SUSY}^2\,\alpha_s/\pi)$ (with the gluino mass
$m_{\tilde g}$) and can thus be very large for large values of
$\mbox{tg}\beta$ \cite{deltab}. In order to improve the perturbative
results these contributions have to be resummed up to all orders. This
can be done in the current-eigenstate basis involving the two Higgs
doublets $\phi_1$ and $\phi_2$ that couple to down- and up-type quarks,
respectively. The leading correction to the bottom Yukawa coupling
emerges from a coupling of the 'wrong' doublet $\phi_2$ to the bottom
quarks and can be discussed in terms of the effective Lagrangian
\begin{eqnarray}
{\cal L}_{eff} & = & -\lambda_b \overline{b_R} \left[ \phi_1^0 +
\frac{\Delta_b}{\mbox{tg}\beta} \phi_2^{0*} \right] b_L + h.c. \nonumber \\
& = & -m_b \bar b \left[1+i\gamma_5 \frac{G^0}{v}\right] b
-\frac{m_b/v}{1+\Delta_b} \bar b \left[ g_b^h \left(
1-\frac{\Delta_b}{\mbox{tg}\alpha~\mbox{tg}\beta}\right) h \right.
\nonumber \\
& & \hspace*{2cm} \left. + g_b^H \left( 1+\Delta_b
\frac{\mbox{tg}\alpha}{\mbox{tg}\beta}\right) H
- g_b^A \left(1-\frac{\Delta_b}{\mbox{tg}^2\beta} \right) i \gamma_5 A
  \right] b
\label{eq:leffdeltab}
\end{eqnarray}
in the low-energy limit where the relations between current and mass
eigenstates of the neutral Higgs components
\begin{eqnarray}
\phi_1^0 & = & \frac{1}{\sqrt{2}}\left[ v_1 + H\cos\alpha - h\sin\alpha
+ iA\sin\beta - iG^0\cos\beta \right] \nonumber \\
\phi_2^0 & = & \frac{1}{\sqrt{2}}\left[ v_2 + H\sin\alpha + h\cos\alpha
+ iA\cos\beta + iG^0\sin\beta \right]
\end{eqnarray}
have been used and $G^0$ denotes the neutral would-be Goldstone
component. The indices $L,R$ denote the chiralities of the bottom
states, $\lambda_b$ the bottom Yukawa coupling of the MSSM Lagrangian and
$v\approx 246$ GeV the SM vacuum expectation value. The leading
contributions to the correction $\Delta_b$ at NLO are given by
\cite{deltab}
\begin{eqnarray}
\Delta_b & = & \Delta_b^{QCD} + \Delta_b^{elw,t} + \Delta_b^{elw,1} +
\Delta_b^{elw,2} \nonumber
\\
\Delta_b^{QCD} & = & \frac{C_F}{2}~\frac{\alpha_s}{\pi}~m_{\tilde
g}~\mu~\mbox{tg}\beta~ I(m^2_{\tilde{b}_1},m^2_{\tilde{b}_2},m^2_{\tilde
g}) \nonumber \\
\Delta_b^{elw,t} & = &
\frac{\lambda_t^2}{(4\pi)^2}~A_t~\mu~\mbox{tg}\beta~
I(m_{\tilde{t}_1}^2,m_{\tilde{t}_2}^2,\mu^2) \nonumber \\
\Delta_b^{elw,1} & = &
-\frac{\alpha_1}{12\pi}~M_1~\mu~\mbox{tg}\beta~\left\{
\frac{1}{3}I(m_{\tilde b_1}^2,m_{\tilde b_2}^2,M_1^2)
+\left( \frac{c_b^2}{2}+s_b^2\right)I(m_{\tilde b_1}^2,M_1^2,\mu^2)
\right. \nonumber \\
& & \left. \hspace*{4cm} +\left( \frac{s_b^2}{2}+c_b^2\right)
I(m_{\tilde b_2}^2,M_1^2,\mu^2)
\right\} \nonumber \\
\Delta_b^{elw,2} & = &
-\frac{\alpha_2}{4\pi}~M_2~\mu~\mbox{tg}\beta~\left\{
c_t^2 I(m_{\tilde t_1}^2,M_2^2,\mu^2)
+ s_t^2 I(m_{\tilde t_2}^2,M_2^2,\mu^2)
\right. \nonumber \\
& & \left. \hspace*{4cm} +\frac{c_b^2}{2} I(m_{\tilde
b_1}^2,M_2^2,\mu^2)
+\frac{s_b^2}{2} I(m_{\tilde b_2}^2,M_2^2,\mu^2)
\right\}
\label{eq:deltab}
\end{eqnarray}
where $C_F=4/3$ and $\lambda_t$ denotes the top Yukawa coupling and
$\alpha_1 = {g'}^2/4\pi$, $\alpha_2 = {g}^2/4\pi$ the electroweak gauge
couplings. The masses $m_{{\tilde b}_{1,2}}$ and $m_{{\tilde t}_{1,2}}$
are the sbottom and stop masses, $\mu$ the higgsino mass parameter and
$M_{1,2}$ the soft SUSY-breaking bino and wino mass parameters.  The
variables $s/c_{t,b} = \sin/\cos \theta_{t,b}$ are related to the
stop/sbottom mixing angles $\theta_{t,b}$. The function $I$ is
generically defined as
\begin{equation}
I(a,b,c) = \frac{\displaystyle ab\log\frac{a}{b} + bc\log\frac{b}{c}
+ ca\log\frac{c}{a}}{(a-b)(b-c)(a-c)}
\end{equation}
The effective Lagrangian of Eq.~(\ref{eq:leffdeltab}) can be
parametrized in compact form as
\begin{equation}
{\cal L}_{eff} = -\frac{m_b}{v}\ \bar{b}\ \big[\ \tilde g^h_b\ h +
\tilde g^H_b\ H - \tilde g^A_b\ i\gamma_5\ A\ ]\ b
\end{equation}
with the effective (resummed) couplings
\begin{eqnarray}
\tilde g^h_b & = & \frac{g^h_b}{1+\Delta_b}\left[ 1 -
\frac{\Delta_b}{\mbox{tg}\alpha\mbox{tg}\beta}  \right] \nonumber \\
\tilde g^H_b & = & \frac{g^H_b}{1+\Delta_b}\left[ 1 + \Delta_b
\frac{\mbox{tg}\alpha}{\mbox{tg}\beta} \right] \nonumber \\
\tilde g^A_b & = & \frac{g^A_b}{1+\Delta_b}\left[ 1 -
\frac{\Delta_b}{\mbox{tg}^2\beta} \right]
\label{eq:rescoup}
\end{eqnarray}
It should be noted that in the decoupling limit of large pseudoscalar
mass $M_A$ the mixing angle factor $\mbox{tg}\alpha$ approaches
$-1/\mbox{tg}\beta$ such that $\tilde g^h_b \to  g^h_b \to 1$ so that
the light scalar Higgs boson still becomes SM-like. By means of power
counting it can be shown that the Lagrangian of
Eq.~(\ref{eq:leffdeltab}) is valid up to all orders in
$\mu\,\mbox{tg}\beta$ and thus $\Delta_b$
\cite{deltabres0,deltabres,deltabnnlo} and potentially sizeable
$A_b$-contributions can be absorbed in a refined definition of
$\Delta_b$ \cite{deltabres}. The rewriting in terms of mass
eigenstates describes the proper resummation of these terms. Two-loop
QCD corrections to the dominant $\Delta_b^{QCD}$ and $\Delta_b^{elw,t}$
contributions have been calculated. They modify the size by a moderate
amount of about 10\% and reduce the scale dependence considerably to the
level of a few per-cent \cite{deltabnnlo, deltabnnlo2} and thus yield a
reliable prediction of the effective bottom Yukawa couplings. The
couplings to all down-type fermions including the charged leptons can be
dressed by the corresponding leading $\Delta_f$ terms analogously. They
play a role for the strange and $\tau$ Yukawa couplings. While the
$\Delta_s$ contributions can be obtained from $\Delta_b$ by
straightforward replacements of the corresponding masses, the
corrections for $\tau$ leptons are given explicitly by \cite{deltab}
\begin{eqnarray}
\Delta_\tau & = & \Delta_\tau^{elw,1} + \Delta_\tau^{elw,2} \nonumber \\
\Delta_\tau^{elw,1} & = &
\frac{\alpha_1}{4\pi}~M_1~\mu~\mbox{tg}\beta~\left\{
I(m_{\tilde \tau_1}^2,m_{\tilde \tau_2}^2,M_1^2)
+\left( \frac{c_\tau^2}{2}-s_\tau^2\right)I(m_{\tilde
\tau_1}^2,M_1^2,\mu^2)
\right. \nonumber \\
& & \left. \hspace*{3cm} +\left( \frac{s_\tau^2}{2}-c_\tau^2\right)
I(m_{\tilde \tau_2}^2,M_1^2,\mu^2)
\right\} \nonumber \\
\Delta_\tau^{elw,2} & = &
-\frac{\alpha_2}{4\pi}~M_2~\mu~\mbox{tg}\beta~\left\{
I(m_{\tilde \nu_\tau}^2,M_2^2,\mu^2)
+ \frac{c_\tau^2}{2} I(m_{\tilde \tau_1}^2,M_2^2,\mu^2)
\right. \nonumber \\
& & \left. \hspace*{4cm} + \frac{s_\tau^2}{2}
I(m_{\tilde \tau_2}^2,M_2^2,\mu^2) \right\} \;,
\end{eqnarray}
where $s/c_\tau = \sin/\cos \theta_\tau$ is related to the $\tilde \tau$
mixing angle $\theta_\tau$ and $m_{\tilde{\tau}_{1,2}}, 
m_{\tilde{\nu}_\tau}$ denote the stau and tau sneutrino masses,
respectively.

\section{Higgs-Boson Decays}
Higgs-boson couplings to SM particles grow with their corresponding
masses so that for the SM Higgs boson its couplings to the intermediate
gauge bosons $W,Z$ and to the top and bottom quarks as well as
$\tau$-leptons are the most relevant ones for decay and production
processes.

\subsection{\it Fermionic Higgs boson decays}
\subsubsection{\it Standard Model}
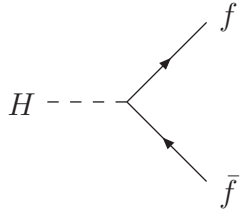
\begin{figure}[hbt]
\begin{center}
\setlength{\unitlength}{1pt}
\begin{picture}(350,100)(-150,200)


\ArrowLine(30,250)(60,280)
\ArrowLine(60,220)(30,250)
\DashLine(0,250)(30,250){5}
\put(-15,246){$H$}
\put(65,280){$f$}
\put(65,213){$\bar f$}
\end{picture}
\setlength{\unitlength}{1pt}
\caption[ ]{\label{fg:hffdia} \it Feynman diagram describing $H\to
f\bar f$ at lowest order.}
\end{center}
\end{figure}
The decay widths of the SM Higgs boson into fermion pairs mediated by
the fermion Yukawa couplings (see Fig.~\ref{fg:hffdia}) is given, for
$M_H^2\gg m_f^2$, by \cite{prohiggs,hll}
\begin{equation}
\Gamma [H\to f\bar f] = \frac{N_c G_F M_H } {4\sqrt{2}\pi}
\ m_f^2\ [1 + \delta_{\rm QCD}+\delta_t + \delta_{mixed}]
\left(1+\delta_{elw} \right)
\label{eq:hqq}
\end{equation}
with $N_c=1(3)$ for leptons (quarks), the Higgs-boson mass $M_H$ and the
fermion mass $m_f$. In the leptonic case we use the lepton pole mass,
while for quarks we use the ${\overline{\rm MS}}$ quark mass
$\overline{m}_Q(M_H)$\footnote{Running-mass effects with respect to QED
corrections are not taken into account.}. The QCD corrections for the
decays into quark pairs can be expressed as \cite{drees,h2bb,chet}
\begin{eqnarray}
\delta_{\rm QCD} & = & 5.67 \frac{\alpha_s (M_H)}{\pi} + (35.94 - 1.36
N_F) \left( \frac{\alpha_s (M_H)}{\pi} \right)^2
+ (164.14 - 25.77 N_F + 0.259 N_F^2) \left( \frac{\alpha_s(M_H)}{\pi}
\right)^3 \nonumber \\
& & +(39.34-220.9 N_F+9.685 N_F^2-0.0205 N_F^3) \left(
\frac{\alpha_s(M_H)}{\pi} \right)^4 \nonumber \\
\delta_t & = & \left(\frac{\alpha_s (M_H)}{\pi}\right)^2 \left[ 1.57 -
\frac{2}{3} \log \frac{M_H^2}{M_t^2} + \frac{1}{9} \log^2
\frac{\overline{m}_Q^2 (M_H)}{M_H^2} \right]
\label{eq:hqqdelta}
\end{eqnarray}
with the number $N_F$ of contributing quark flavours. The running quark
mass and the QCD coupling are defined at the scale of the Higgs mass,
absorbing in this way large mass logarithms. The quark mass effects
beyond the Yukawa coupling can be neglected in general, except for heavy
quark decays in the threshold region\footnote{The inclusion of the full
quark mass effects is motivated by a proper treatment of off-shell Higgs
splittings into quark pairs and by determining all subleading quark-mass
effects up to NLO. Moreover, the heavy scalar MSSM Higgs boson can
acquire a mass close to the location of the top-antitop threshold.}. The
QCD corrections in this case are given, in terms of the quark {\it pole}
mass $M_Q$, by \cite{drees}
\begin{eqnarray}
\Gamma [H \to Q\bar{Q}\,]= \frac{3 G_F M_H}{4 \sqrt{2} \pi} \,
\, M_Q^2 \, \beta^3 \, \left[ 1 +\frac{4}{3} \frac{\alpha_s}{\pi}
\delta^H \right]
\label{eq:qcdmass}
\end{eqnarray}
where $\beta = (1-4M_Q^2/M_H^2)^{1/2}$ denotes the velocity of the heavy
quarks $Q$. The NLO QCD correction factor reads \cite{drees}
\begin{eqnarray}
\delta^H = \frac{1}{\beta}A(\beta) + \frac{1}{16\beta^3}(3+34\beta^2-
13 \beta^4)\log \frac{1+\beta}{1-\beta} +\frac{3}{8\beta^2}(7 \beta^2-1)
\label{eq:dqcdmass}
\end{eqnarray}
with the function
\begin{eqnarray}
A(\beta) &= & (1+\beta^2) \left[ 4 {\rm Li}_2 \left( \frac{1-\beta}{1+\beta}
\right) +2 {\rm Li}_2 \left( -\frac{1-\beta}{1+\beta} \right) -3 \log
\frac{1+\beta}{1-\beta} \log \frac{2}{1+\beta} \right. \nonumber \\
& & \left. -2 \log \frac{1+\beta}{1-\beta} \log \beta \right] -
3 \beta \log \frac{4}{1-\beta^2} -4 \beta \log \beta
\label{eq:abeta}
\end{eqnarray}
(Li$_2$ denotes the Spence function, Li$_2(x)= -\int_0^x dy y^{-1} \log(1-y)$.)

Electroweak corrections to heavy quark and lepton decays are well under control
\cite{hqvelw,hqqelw}. In the intermediate mass range they can be
approximated by
\cite{hqqelwapp}
\begin{equation}
\delta_{\rm elw} = \frac{3}{2} \frac{\alpha}{\pi}e_f^2 \left(\frac{3}{2} -
\log \frac{M_H^2}{M_f^2} \right) + \frac{G_F}{8 \pi^2 \sqrt{2}}
\left\{ k_f M_t^2 + M_W^2 \left[ - 5 + \frac{3}{s_W^2} \log c_W^2 \right]
- M_Z^2 \frac{6 v_f^2 - a_f^2}{2} \right\}
\end{equation}
with $v_f = 2I_{3f} - 4e_f s_W^2$ and $a_f = 2I_{3f}$. $I_{3f}$ denotes
the third component of the electroweak isospin, $e_f$ the electric
charge of the fermion $f$ and $s_W = \sin\theta_W$ the sine of the
Weinberg angle; $\alpha$ denotes the QED coupling, $M_t$ the top quark
mass and $M_{W,Z}$ the $W,Z$ boson masses.  The large logarithm $\log
M_H^2/M_f^2$ can be absorbed in the running fermion mass analogous to
the QCD corrections.  The coefficient $k_f$ is equal to 7 for decays
into leptons and light quarks; for $b$ quarks it is reduced to 1 due to
additional contributions involving top quarks inside the vertex
corrections. Within this approximation the two- and three-loop QCD
corrections to the $k_f$ terms have been computed by means of low-energy
theorems \cite{hbb,hbb3}.  The results imply the simple replacements at
NNLO
\begin{eqnarray}
k_f & \to &
k_f\times\left\{1-\frac{1}{7}\left(\frac{3}{2}+\zeta_2\right)
\frac{\alpha_s(M_t)}{\pi} \right\} \hspace{1cm}\mbox{for $f\neq
b$}\nonumber \\
k_b & \to &
k_b\times\left\{1-4\left(1+\zeta_2\right)\frac{\alpha_s(M_t)}{\pi}
\right\}
\end{eqnarray}
with $\zeta_2 = \pi^2/6$.
The three-loop QCD corrections to the $k_f$ term can be found in
\cite{hbb3}.  Recently, the full mixed QCD-electroweak corrections
$\delta_{mixed}$ have been determined \cite{hbbqcdelw}. These turn out
to be in the per-mille range if included in the factorized expression
for QCD and electroweak corrections as shown in Eq.~(\ref{eq:hqq}). The
electroweak corrections in total are small, of ${\cal O}(5\%)$, in the
intermediate mass range. The residual theoretical uncertainties in the
intermediate mass range have been estimated as 0.2\% for the QCD part
and 0.5\% for electroweak corrections including mixed contributions in
the factorized approach \cite{yr4}.

\subsubsection{\it Minimal supersymmetric extension}
At lowest order the leptonic decay width of neutral MSSM Higgs
boson\footnote{In the following we denote the different types of neutral
Higgs particles collectively by $\Phi = h,H,A$.} decays is given by
\cite{prohiggs,hll}
\begin{equation}
\Gamma [\Phi \to l^+ l^-] = \frac{G_F M_\Phi }{4\sqrt{2}\pi}
(g_l^\Phi)^2 m_l^2 \beta^p
\end{equation}
where $g_l^\Phi$ denotes the corresponding MSSM coupling factor, presented in
Table \ref{tb:hcoup}, $\beta = (1-4m_l^2/M_\Phi^2)^{1/2}$ the velocity
of the final-state leptons and $p=3\, (1)$ the exponent for scalar
(pseudoscalar) Higgs particles. The $\tau$ pair decays play a
significant role, with a branching ratio of up to about 10\%.  Muon
decays can develop branching ratios of a few $10^{-4}$. All other
leptonic decay modes are phenomenologically irrelevant.

The analogous leptonic decay width of the charged Higgs boson reads
\begin{equation}
\Gamma [H^{+} \to  \nu{\overline{l}}] = \frac{G_F
M_{H^\pm}}{4\sqrt{2}\pi}
m_l^2 \mbox{tg}^2 \beta \left(1-\frac{m_l^2}{M_{H^\pm}^2} \right)^3
\label{eq:hcnl}
\end{equation}
The decay mode into $\tau^+ \nu_\tau$ reaches branching ratios of more
than 90\% below the $t\bar b$ threshold and the muonic one ranges at a few
$10^{-4}$.  All other leptonic decay channels of the charged Higgs
bosons are unimportant.

For large Higgs masses [$M_\Phi^2 \gg M_Q^2$] the QCD-corrected decay
widths of the MSSM Higgs particles into quarks can be obtained from
evaluating the analogous diagrams as presented in Fig.~\ref{fg:hffdia},
where the SM Higgs particle $H$ has to be substituted by the
corresponding MSSM Higgs boson $\Phi$ \cite{drees,chet}:
\begin{equation}
\Gamma [\Phi \, \to \, Q{\overline{Q}}] =
\frac{3G_F M_\Phi }{4\sqrt{2}\pi} \overline{m}_Q^2(M_\Phi) (g_Q^\Phi)^2
\left[ 1 + \delta_{\rm QCD} + \delta_t^\Phi \right]
\end{equation}
Neglecting regular quark mass effects, the QCD corrections $\delta_{\rm
QCD}$ are presented in Eq.~(\ref{eq:hqqdelta}) and the top quark induced
contributions read as \cite{chet} 
\begin{eqnarray}
\delta_t^{h/H} & =& \frac{g_t^{h/H}}{g_Q^{h/H}}~\left(\frac{\alpha_s
(M_{h/H})}{\pi}
\right)^2 \left[ 1.57 - \frac{2}{3} \log \frac{M_{h/H}^2}{M_t^2}
+ \frac{1}{9} \log^2 \frac{\overline{m}_Q^2
(M_{h/H})}{M_{h/H}^2}\right]\nonumber \\
\delta_t^A & = & \frac{g_t^A}{g_Q^A}~\left(\frac{\alpha_s (M_A)}{\pi}
\right)^2
\left[ 3.83 - \log \frac{M_A^2}{M_t^2} + \frac{1}{6} \log^2
\frac{\overline{m}_Q^2 (M_A)}{M_A^2} \right] \nonumber
\end{eqnarray}
Analogous to the SM case the large logarithmic contributions
of the QCD corrections are absorbed in the running $\overline{\rm MS}$
quark mass $\overline{m}_Q(M_\Phi)$ at the scale of the corresponding
Higgs mass $M_\Phi$.

The quark decay width of the charged Higgs boson reads, in the large
Higgs mass regime $M_{H^\pm} \gg M_U + M_D$, as \cite{hud1,hud}
\begin{equation}
\Gamma [\,H^{+} \to \, U{\overline{D}}\,] =
\frac{3 G_F M_{H^\pm}}{4\sqrt{2}\pi} \, \left| V_{UD} \right|^2 \,
\left[ \overline{m}_U^2(M_{H^\pm}) (g_U^{A})^2 +
\overline{m}_D^2(M_{H^\pm})
 (g_D^{A})^2 \right] (1 + \delta_{\rm QCD})
\label{eq:hcud}
\end{equation}
(Eq.~(\ref{eq:hcud}) is valid if either the first or the second term is
dominant.) where $U(D)$ denote heavy up(down)-type quarks. The relative
couplings $g_{Q}^A$ have been collected in Table \ref{tb:hcoup} and the
coefficient $V_{UD}$ denotes the CKM matrix element of the transition of
$D$ to $U$ quarks. The QCD correction factor $\delta_{\rm QCD}$ is given
in Eq.~(\ref{eq:hqqdelta}), where large logarithmic terms are again
absorbed in the running $\overline{\rm MS}$ masses $\overline{m}_{U,D}
(M_{H^\pm})$ at the scale of the charged Higgs mass $M_{H^\pm}$.  In the
threshold regions mass effects play a significant role. The partial
decay widths of the neutral Higgs bosons $\Phi=h,H$ and $A$ into heavy
quark pairs, in terms of the quark {\it pole} mass $M_Q$, can be cast
into the form \cite{drees}
\begin{eqnarray}
\Gamma [\Phi \to Q\bar{Q}]= \frac{3 G_F M_\Phi}{4 \sqrt{2} \pi} \,
(g_Q^\Phi)^2 \, M_Q^2 \, \beta^p \, \left[ 1 +\frac{4}{3}
\frac{\alpha_s}{\pi}
\delta^{\Phi} \right]
\label{eq:phiqqmass}
\end{eqnarray}
where $\beta = (1-4M_Q^2/M_\Phi^2)^{1/2}$ denotes the velocity of the
final-state quarks and $p=3\, (1)$ the exponent for scalar
(pseudoscalar) Higgs bosons. The different powers in $\beta$ are related
to the property that the $Q\bar Q$ pairs are produced in a ${\cal
P}$-wave (${\cal S}$-wave) for (pseudo)scalar Higgs decays. To
next-to-leading order, the QCD correction factor is given by
Eq.~(\ref{eq:dqcdmass}) for the scalar Higgs particles $h,H$, while for
the CP-odd Higgs boson $A$ it reads as \cite{drees}
\begin{eqnarray}
\delta^{A} = \frac{1}{\beta}A(\beta) + \frac{1}{16\beta}(19+2\beta^2+
3 \beta^4)\log \frac{1+\beta}{1-\beta} +\frac{3}{8}(7 -\beta^2)
\end{eqnarray}
with the function $A(\beta)$ defined in Eq.~(\ref{eq:abeta}). The
QCD corrections in the $t\bar t$ threshold region are moderate, apart
from a Coulomb singularity, which is regularized by taking into account
the finite top quark decay width.

The partial decay width of the charged Higgs particles into heavy quarks
can be expressed as \cite{hud}
\begin{eqnarray}
\Gamma [H^+\rightarrow U\bar{D}\,] &=& \frac{3 G_F
M_{H^\pm}}{4\sqrt{2}\pi}
|V_{UD}|^2 \, \lambda^{1/2} \, \left\{ (1-\mu_U -\mu_D) \left[
\frac{M_U^2}
{ {\rm tg}^2
\beta } \left( 1+ \frac{4}{3} \frac{\alpha_s}{\pi} \delta_{UD}^+ \right)
\right. \right. \label{eq:h+udmass} \\
&& \left. \left. +M_D^2 {\rm tg}^2 \beta \left( 1+ \frac{4}{3}
\frac{\alpha_s}
{\pi} \delta_{DU}^+ \right) \right]
-4M_UM_D \sqrt{\mu_U \mu_D} \left( 1+ \frac{4}{3}
\frac{\alpha_s}{\pi} \delta_{UD}^- \right) \right\} \nonumber
\end{eqnarray}
where $\mu_i=M_i^2/M_{H^\pm}^2$, and
$\lambda=(1-\mu_U-\mu_D)^2-4\mu_U\mu_D$ denotes the usual two-body
phase-space function.  The quark masses $M_{U,D}$ are the {\it pole}
masses. The QCD factors $\delta_{ij}^\pm~~(i,j=U,D)$ are given by
\cite{hud}
\begin{eqnarray}
\delta_{ij}^{+} &=&  \frac{9}{4} + \frac{ 3-2\mu_i+2\mu_j}{4} \log
\frac{\mu_i}{\mu_j} + \frac{ (\frac{3}{2}-\mu_i-\mu_j) \lambda+5 \mu_i
\mu_j}{2 \lambda^{1/2} (1-\mu_i -\mu_j)} \log x_i x_j  +B_{ij} \nonumber \\
\delta_{ij}^{-} &=&  3 + \frac{ \mu_j-\mu_i}{2} \log \frac{\mu_i}{\mu_j}
+ \frac{ \lambda +2(1-\mu_i-\mu_j)} { 2 \lambda^{1/2} } \log x_i x_j
+B_{ij}
\end{eqnarray}
with the scaling variables $x_i= 2\mu_i/[1-\mu_i-\mu_j+\lambda^{1/2}]$
and the
generic function
\begin{eqnarray*}
B_{ij} &=& \frac{1-\mu_i-\mu_j} { \lambda^{1/2} } \left[ 4{\rm
Li_{2}}(x_i
x_j)- 2{\rm Li_{2}}(-x_i) -2{\rm Li_{2}}(-x_j) +2 \log x_i x_j \log
(1-x_ix_j)
\right. \nonumber \\
&& \left. \hspace*{2.2cm} - \log x_i \log (1+x_i) - \log x_j \log
(1+x_j)
\right] \nonumber \\
&& - 4 \left[ \log (1-x_i x_j)+ \frac{x_i x_j}{1-x_i x_j} \log x_i x_j
\right]\nonumber
\\ && +\frac{ \lambda^{1/2}+\mu_i-\mu_j } {\lambda^{1/2} } \left[
\log (1+x_i) -\frac{x_i}{1+x_i} \log x_i \, \right] \nonumber \\
&& +\frac{\lambda^{1/2}-\mu_i+\mu_j} { \lambda^{1/2} } \left[
\log (1+x_j) -\frac{x_j}{1+x_j} \log x_j \right]
\end{eqnarray*}
The transition from the threshold region, involving mass effects, to the
renormalization-group-improved large Higgs mass regime is provided by a
smooth linear interpolation analogous to the SM case in all heavy quark
decay modes in the program {\sc Hdecay} \cite{hdecay}.

The full SUSY--electroweak and SUSY--QCD corrections to the fermionic
decay modes have been computed \cite{deltabres0, deltabres, dabel}. They
turn out to be moderate for small values of $\mbox{tg}\beta$, of the
order of about 10\%.  Only for large values of $\mbox{tg$\beta$} > 10$
do the gluino corrections induce large corrections due to the $\Delta_f$
terms of down-type fermions as discussed in the introduction. These
effects are in particular relevant for the MSSM Higgs boson decays into
bottom quarks. The final NLO decay width into bottom quarks can be
written as
\begin{equation}
\Gamma [\Phi \, \to \, b{\overline{b}}] =
\frac{3G_F M_\Phi }{4\sqrt{2}\pi} \overline{m}_b^2(M_\Phi)
\tilde{g}_b^\Phi
\left[ 1 + \delta_{\rm QCD} + \delta_t^\Phi \right] \left[
\tilde{g}_b^\Phi + g_b^\Phi \delta_{SQCD}^{rem} \right]
\end{equation}
where the effective couplings $\tilde{g}_b^\Phi$ are given in
Eq.~(\ref{eq:rescoup}) and thus resum the leading $\Delta_b$ terms. The
contribution $\delta_{SQCD}^{rem}$ denotes the remainder of the full
SUSY--QCD corrections after absorbing the $\Delta_b$ terms in the
effective Yukawa couplings. This remainder turns out to be small for all
MSSM Higgs bosons even for large Higgs masses \cite{deltabres}.

Below the $t\bar t$ threshold, heavy neutral Higgs boson decays into
off-shell top quarks are sizeable, thus modifying the profile of these
Higgs particles significantly in this region. The dominant
below-threshold contributions can be obtained from the SM expression
for the scalar Higgs bosons \cite{1OFF}
\begin{equation}
\frac{d\Gamma}{dx_1 dx_2} (\phi\to tt^*\to Wtb) = (g_t^\phi)^2
\frac{3G_F^2}{32\pi^3} M_t^2 M_\phi^3 \frac{\Gamma^\phi_0}{y_1^2 + \gamma_t
\kappa_t}
\label{eq:mssmhttdalitz}
\end{equation}
with the reduced energies $x_{1,2}=2E_{t,b}/M_\phi~(\phi=h,H,A)$, the
scaling variables $y_{1,2} = 1-x_{1,2}$, $\kappa_i = M_i^2/M_\phi^2$ and
the reduced decay widths of the virtual particles
$\gamma_i=\Gamma_i^2/M_\phi^2$. The squared amplitudes can be written as
\cite{1OFF}
\begin{eqnarray}
\Gamma^H_0 & = & y_1^2(1-y_1-y_2+\kappa_W-5\kappa_t) +
2\kappa_W(y_1y_2-\kappa_W -2\kappa_ty_1+4\kappa_t\kappa_W) \nonumber \\
& & -\kappa_ty_1y_2+\kappa_t(1-4\kappa_t)(2y_1+\kappa_W+\kappa_t)
\nonumber \\
\Gamma^A_0 & = & y_1^2(1-y_1-y_2+\kappa_W-\kappa_t) +
2\kappa_W(y_1y_2-\kappa_W)
-\kappa_t(y_1y_2-2y_1-\kappa_W-\kappa_t)
\end{eqnarray} 
The differential decays width of Eq.~(\ref{eq:mssmhttdalitz})
has to be integrated over the $x_1, x_2$ region, bounded by
\begin{equation}
\left| \frac{2(1-x_1-x_2+\kappa_t+\kappa_b-\kappa_W) + x_1x_2}
{\sqrt{x_1^2-4\kappa_t} \sqrt{x_2^2-4\kappa_b}} \right| \leq 1
\label{eq:dalitzbound}
\end{equation}
In these formulae $W$- and charged-Higgs-boson exchange contributions
are neglected, because they are suppressed with respect to the off-shell
top quark contribution to $Wtb$ final states. However, for the sake of
completeness they are included in {\sc Hdecay} \cite{hdecay}. Their
explicit expressions can be found in \cite{1OFF}. The transition from
below to above the threshold is provided by a smooth cubic
interpolation. Below-threshold decays yield a $t\bar t$ branching ratio
at the per-cent level already for heavy scalar and pseudoscalar Higgs
masses $M_{H,A}\sim 300$ GeV.

Below the $t\bar b$ threshold off-shell charged-Higgs decays $H^+\to
t^*\bar b \to b\bar b W^+$ are important. For $M_{H^\pm} < M_t + M_b -
\Gamma_t$, where $\Gamma_t$ denotes the top width, their expression can
be cast into the form \cite{1OFF}
\begin{eqnarray}
\Gamma (H^+\to t^*\bar b \to Wb\bar b) & = &
\frac{3G_F^2M_t^4}{64\pi^3\mbox{tg}^2\beta} M_{H^\pm}
\left\{\frac{\kappa_W^2}{\kappa_t^3}(4\kappa_W\kappa_t + 3\kappa_t -
4\kappa_W)
\log \frac{\kappa_W(\kappa_t-1)}{\kappa_t-\kappa_W} \right.  \nonumber
\\
& & +(3\kappa_t^2 - 4\kappa_t - 3\kappa_W^2 + 1)
\log\frac{\kappa_t-1}{\kappa_t-\kappa_W} - \frac{5}{2} \label{eq:pttoff}
\\
& & \left. +\frac{1-\kappa_W}{\kappa_t^2} (3\kappa_t^3 -
\kappa_t\kappa_W
- 2\kappa_t \kappa_W^2 + 4\kappa_W^2) +
  \kappa_W\left(4-\frac{3}{2}\kappa_W
\right) \right\} \nonumber
\end{eqnarray}
with the scaling variables $\kappa_i = M_i^2/M_{H^\pm}^2~~(i=t,W)$. The
bottom mass has been neglected in Eq.~(\ref{eq:pttoff}), but it is taken
into account in {\sc Hdecay} by performing a numerical integration of
the corresponding Dalitz plot density, given in \cite{1OFF}.  The
off-shell branching ratio can reach the per-cent level for charged Higgs
masses above about 100~GeV for small $\mbox{tg$\beta$}$, which is
significantly below the $t\bar b$ threshold $M_{H^\pm}\sim 180$ GeV.

\subsection{\it Higgs boson decays into intermediate gauge bosons}
\subsubsection{\it Standard Model}
\begin{figure}[hbt]
\begin{center}
\setlength{\unitlength}{1pt}
\begin{picture}(200,100)(-50,0)

\DashLine(0,50)(50,50){5}
\Photon(50,50)(100,80){3}{6}
\Photon(50,50)(100,20){-3}{6}
\put(-15,46){$H$}
\put(105,18){$W,Z$}
\put(105,78){$W,Z$}

\end{picture}  \\
\setlength{\unitlength}{1pt}
\caption{\label{fg:hvvdia} \it Diagram contributing to $H\to VV$
[$V=W,Z$].}
\end{center}
\end{figure}
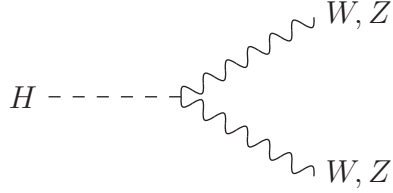
\noindent
Above the $WW$ and $ZZ$ decay thresholds, the partial decay widths into
on-shell pairs of massive gauge bosons ($V = W,Z$) at lowest order (see
Fig.~\ref{fg:hvvdia}) are given by \cite{unitarity}\footnote{Even though
the SM Higgs mass is too small these expressions play a role for
off-shell Higgs bosons splitting into on-shell $W$- or $Z$-boson pairs.}
\begin{equation}
\Gamma(H\to VV) = \delta_V \frac{G_F M_H^3}{16\sqrt{2}\pi} \beta
(1-4x+12x^2)
\end{equation}
with the abbreviations $x = M_V^2/M_H^2$, $\beta = \sqrt{1-4x}$ and
$\delta_V = 2\,(1)$ for $V = W\, (Z)$.

The NLO electroweak corrections are small and amount to less than about
5\% in the intermediate mass range \cite{hqvelw,hvvelw}.  Furthermore
the QCD corrections to the leading top mass corrections of ${\cal O}(G_F
M_t^2)$ have been calculated up to N$^3$LO. At NNLO they rescale the $WW,
ZZ$ decay widths by \cite{let,hzzmt}
\begin{eqnarray}
\Gamma(H\to ZZ) & = & \Gamma_{LO}(H\to ZZ) \left\{1 - x_t \left[ 5 -
(15-2\zeta_2)\frac{\alpha_s}{\pi} \right] \right\} \\
\Gamma(H\to WW) & = & \Gamma_{LO}(H\to WW) \left\{1 - x_t \left[ 5 -
(9-2\zeta_2)\frac{\alpha_s}{\pi} \right] \right\}
\label{eq:hvvxt}
\end{eqnarray}
with $x_t = G_F M_t^2 / (8\sqrt{2}\pi^2)$. The three-loop corrections can be
found in \cite{hvv3}.

Below threshold the decays into off-shell gauge particles are important
\cite{hvvp,2OFF}. For Higgs masses slightly larger than the
corresponding gauge boson mass the decay widths into pairs of off-shell
gauge bosons can be cast into the form \cite{2OFF}
\begin{equation}
\Gamma(H\to V^*V^*) = \frac{1}{\pi^2}\int_0^{M_H^2} \frac{dQ_1^2 M_V
\Gamma_V}
{(Q_1^2 - M_V^2)^2 + M_V^2 \Gamma_V^2}
\int_0^{(M_H-Q_1)^2} \frac{dQ_2^2 M_V \Gamma_V} {(Q_2^2 - M_V^2)^2 +
M_V^2
\Gamma_V^2} \Gamma_0
\label{eq:hvvoff}
\end{equation}
with $Q_1^2, Q_2^2$ being the squared invariant masses of the virtual gauge
bosons, $M_V$ and $\Gamma_V$ their masses and total decay widths; $\Gamma_0$
is given by
\begin{equation}
\Gamma_0 = \delta_V \frac{G_F M_H^3}{16\sqrt{2}\pi}
\sqrt{\lambda(Q_1^2,Q_2^2;M_H^2)} \left[ \lambda(Q_1^2,Q_2^2;M_H^2) +
12\frac{Q_1^2Q_2^2}{M_H^4} \right]
\label{eq:h2vv}
\end{equation}
with the phase-space factor
\begin{equation}
\lambda(x,y;z) = \left(1-\frac{x}{z}-\frac{y}{z}
\right)^2-4\frac{xy}{z^2}.
\label{eq:kallen}
\end{equation}
The branching ratios of double off-shell decays reach the per cent level
for Higgs masses above about 100 (110) GeV for off-shell $W(Z)$ boson
pairs.  Off-shell effects modify the partial decay widths by ${\cal
O}(10\%)$ above the $WW$ and $ZZ$ thresholds.  During the last years the
complete electroweak corrections have been obtained for the full Higgs boson
decays into four fermions that include the resonant and non-resonant
contributions to Higgs boson decays into intermediate gauge bosons $W,Z$
\cite{prophecy4f}. In Eq.~(\ref{eq:h2vv}) they can be approximated by
the improved Born approximation \cite{prophecy4f}, \\[0.5cm]
$\underline{H\to ZZ\to 4f:}$
\begin{eqnarray*}
\Gamma_0 & \to & \Gamma_0 \times \Re e\left\{1+
\frac{G_F \mu_t^2}{8\sqrt{2}\pi^2}
\Biggl[1-\frac{6 \cos\theta_W}{\sin\theta_W}
\left(\frac{e_{f_1}}{g^{\sigma_1}_{Z f_1 f_1}}
+\frac{e_{f_3}}{g^{\sigma_3}_{Z f_3 f_3}}\right)
+\tau_{HZZ}\left(\frac{M_H^2}{\mu_t^2}\right)
\right] \nonumber \\
&&\quad\qquad{}
+\frac{G_F M_H^2}{8\sqrt{2}\pi^2}
\left(\frac{5\pi^2}{6}-3\sqrt{3}\pi+\frac{19}{2} \right)
+62.0308(86) \left(\frac{G_F M_H^2}{16\pi^2\sqrt{2}}\right)^2 \nonumber \\
&&\quad\qquad{}
+\delta_{Z\to f_1\bar f_2}^{QCD}
+\delta_{Z\to f_3\bar f_4}^{QCD}
+c_{HZZ}
\Biggr\}, \nonumber
\end{eqnarray*}
$\underline{H\to WW\to 4f:}$
\begin{eqnarray}
\Gamma_0 & \to & \Gamma_0 \times \Re e\Biggl\{1+
\frac{G_F\mu_t^2}{8\pi^2\sqrt{2}}
\left[-5 +\tau_{HWW}\left(\frac{M_H^2}{\mu_t^2}\right)
\right] \nonumber \\
&&\quad\qquad{}
+\frac{G_F M_H^2}{8\sqrt{2}\pi^2}
\left(\frac{5\pi^2}{6}-3\sqrt{3}\pi+\frac{19}{2} \right)
+62.0308(86) \left(\frac{G_F M_H^2}{16\pi^2\sqrt{2}}\right)^2 \nonumber \\
&&\quad\qquad{}
+g(\bar\beta)
\delta_{Coul}\left(M_H^2,Q_1^2,Q_2^2\right)
+\delta_{W\to f_1\bar f_2}^{QCD}
+\delta_{W\to f_3\bar f_4}^{QCD}
+c_{HWW}
\Biggr\}
\end{eqnarray}
with the electric charges $e_f$ of the fermions and the fermionic
couplings to the $Z$ boson
\begin{equation}
g_{Zff}^+ = -\frac{\sin\theta_W}{\cos\theta_W} e_f, \qquad
g_{Zff}^- = -\frac{\sin\theta_W}{\cos\theta_W} e_f +
\frac{I_{3f}}{\cos\theta_W \sin\theta_W}
\end{equation}
and the auxiliary functions
\begin{eqnarray}
\tau_{HZZ}\left(\frac{M_H^2}{\mu_t^2}\right) &=&
20+6\beta_{t}^2+3\beta_{t}(\beta_{t}^2+1)\log x_{t}
+3(1-\beta_{t}^2)\log^2 x_{t},
\nonumber \\
\tau_{HWW}\left(\frac{M_H^2}{\mu_t^2}\right) &=&
8+12\beta_{t}^2+3\beta_{t}(3\beta_{t}^2-1)\log x_{t}
+\frac{3}{2}(1-\beta_{t}^2)^2\log^2 x_{t},
\end{eqnarray}
where
\begin{equation}
\beta_{t} = \sqrt{1-\frac{4\mu_t^2}{M_H^2}}, \qquad
x_{t} = \frac{\beta_{t}-1}{\beta_{t}+1}
\end{equation}
and $\mu_t$ denotes the complex top mass, given by $\mu_t^2 = m_t^2 - i
m_t \Gamma_t$ with the top width $\Gamma_t$. The factor $\delta_{Coul}$
describes the Coulomb singularity,
\begin{eqnarray}
\delta_{Coul}(s,Q_1^2,Q_2^2) &=& \frac{\alpha(0)}{\bar\beta}
\Im m\left\{\log\left(\frac{\beta-\bar\beta+\Delta_M}
        {\beta+\bar\beta+\Delta_M}\right)\right\} \nonumber \\
\bar\beta &=&
\frac{\sqrt{s^2+Q_1^4+Q_2^4-2sQ_1^2-2sQ_2^2-2Q_1^2Q_2^2}}{s} \nonumber \\
\beta &=&  \sqrt{1-\frac{4\mu_W^2}{s}}, \qquad
\Delta_M = \frac{|Q_1^2-Q_2^2|}{s},
\end{eqnarray}
where $\alpha(0)$ denotes the fine-structure constant in the Thomson
limit and the complex $W$ mass $\mu_W^2 = M_W^2 - iM_W\Gamma_W$ is given in
terms of the $W$ pole mass $M_W$ and the $W$ decay width $\Gamma_W$. The
function $g(\bar\beta)$ reads
\begin{equation}
g(\bar\beta) = \left(1-\bar\beta^2\right)^2
\end{equation}
Finally the QCD correction factors to the gauge-boson decays are given
by
\begin{equation}
\delta_{V\to l_i\bar l_j}^{QCD} = 0, \qquad
\delta_{V\to q_i\bar q_j}^{QCD} =
\frac{\alpha_s}{\pi}
\end{equation}
and the continuum contributions can be fitted as
\begin{equation}
c_{HZZ} = 3\%, \qquad
c_{HWW} = 4\%
\end{equation}
They constitute auxiliary corrections beyond the previous terms to
compensate the remaining offset between the exact corrections and the
improved Born approximation \cite{prophecy4f}. In general there are
additional non-resonant LO contributions beyond those of
Eq.~(\ref{eq:hvvoff}). However, they turn out to range at the per-cent
level \cite{prophecy4f}. Recently the full electroweak NLO expressions
to the leptonic Higgs decays $H\to 4\ell$ have been matched to parton
showers \cite{hto4l}, indicating that there are a few exclusive
observables where multiple photon effects may be relevant. The
theoretical uncertainties due to missing corrections beyond NLO have
been estimated as 0.5\% in the intermediate mass range for the inclusive
decay rate into four fermions \cite{brthupu, yr3}.

\subsubsection{\it Minimal supersymmetric extension}
The dominant part of the partial widths of the scalar MSSM Higgs bosons
into $W$ and $Z$ boson pairs can be obtained from the SM Higgs decay
widths by rescaling with the corresponding MSSM couplings $g_V^{h/H}$
as listed in Table \ref{tb:hcoup}:
\begin{equation}
\Gamma(h/H \to V^{(*)}V^{(*)}) = (g^{h/H}_V)^2 \Gamma(H_{SM} \to
V^{(*)}V^{(*)})
\end{equation}
Their branching ratios are strongly reduced by reduced MSSM couplings
and additional open leading decay modes, and thus do not play a dominant
role as in the SM case.  Nevertheless the $WW,ZZ$ branching ratios can
reach values of ${\cal O}(10\%)$ for the heavy scalar Higgs boson $H$
for small $\mbox{tg$\beta$}$. Off-shell $WW,ZZ$ decays can pick up
several per-cent of the light scalar Higgs decays at the upper end of
its mass range. The pseudoscalar Higgs particle does not couple to $W$
and $Z$ bosons at tree level. The SUSY-electroweak corrections to scalar
Higgs boson decays into off-shell $WW$ and $ZZ$ pairs have been computed
\cite{h24fselw}. While being similar to the SM-Higgs case they can be
enhanced to the 10\%-level in particular MSSM scenarios.

\subsection{\it Higgs boson decays into gluons}
\subsubsection{\it Standard Model}
\begin{figure}[hbt]
\begin{center}
\setlength{\unitlength}{1pt}
\begin{picture}(180,100)(0,0)

\Gluon(100,20)(150,20){-3}{5}
\Gluon(100,80)(150,80){3}{5}
\ArrowLine(100,20)(100,80)
\ArrowLine(100,80)(50,50)
\ArrowLine(50,50)(100,20)
\DashLine(0,50)(50,50){5}
\put(-15,46){$H$}
\put(105,46){$t,b$}
\put(155,18){$g$}
\put(155,78){$g$}

\end{picture}  \\
\setlength{\unitlength}{1pt}
\caption{\label{fg:hgglodia} \it Diagrams contributing to $H\to gg$
at lowest order.}
\end{center}
\end{figure}
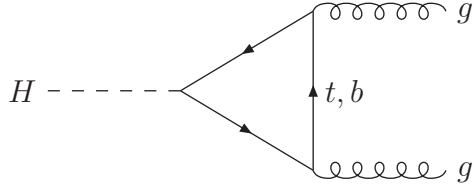
\noindent
The decay of the Higgs boson into gluons is mediated by heavy quark
loops in the SM, with top quarks providing the dominant
contribution, see Fig.~\ref{fg:hgglodia}. The partial decay width
\cite{prohiggs,higgsqcd,hgg0,hgg} is at lowest order given by
\begin{eqnarray}
\Gamma_{LO}\, [H\to gg] = \frac{G_{F}\, \alpha_{s}^{2}\,M_{H}^{3}}
{36\,\sqrt{2}\,\pi^{3}} \left| \sum_{Q} A_Q^H(\tau_Q) \right|^2
\label{eq:hgglo}
\end{eqnarray}
with the form factor
\begin{eqnarray}
A_Q^H (\tau) & = & \frac{3}{2} \tau \left[ 1+(1-\tau) f(\tau)
\right] \nonumber \\
f(\tau) & = & \left\{ \begin{array}{ll}
\displaystyle \arcsin^2 \frac{1}{\sqrt{\tau}} & \tau \ge 1 \\
\displaystyle - \frac{1}{4} \left[ \log \frac{1+\sqrt{1-\tau}}
{1-\sqrt{1-\tau}} - i\pi \right]^2 & \tau < 1
\end{array} \right.
\label{eq:ftau}
\end{eqnarray}
The parameter $\tau_Q= 4M_Q^2/M_H^2$ is defined by the pole mass $M_Q$
of the heavy loop quark $Q$.  For large quark masses the form factor
approaches unity. The QCD corrections are known up to NLO including the
full quark mass dependence \cite{higgsqcd} and up to N$^3$LO in the
limit of heavy top quarks \cite{hgg, hgg3, hgg4}. For a Higgs mass
$M_{H} \sim 125$ GeV they can be expressed as
\begin{eqnarray}
\Gamma\,[H\to gg] & = & \Gamma_{LO}\, \left\{1+(23.75-1.167 N_F +
\Delta_m)\frac{\alpha_{s}(M_H)}{\pi} \right. \nonumber \\
& + & \left(370.20- 47.19 N_F + 0.902 N_F^2
+ (2.375 + 0.667 N_F)
\log\frac{M_H^2}{m_t^2}\right) \left( \frac{\alpha_{s}(M_H)}{\pi}
\right)^2 \nonumber \\
& + & \left. \left( 4533.46 - 1062.82 N_F + 52.62 N_F^2 - 0.5378 N_F^2
+ (66.66 + 14.60 N_F \phantom{\log\frac{M_H^2}{m_t^2}} \right. \right.
\nonumber \\
&& \left. \left. - 0.6887 N_F^2) \log\frac{M_H^2}{m_t^2}
+ (6.53 + 1.44 N_F - 0.111 N_F^2) \log^2\frac{M_H^2}{m_t^2} \right)
\left( \frac{\alpha_{s}(M_H)}{\pi} \right)^3 \right\} \nonumber \\
& \approx & \Gamma_{LO}\, \left\{1+0.67+0.20+0.02 \right\}
\label{eq:hgg}
\end{eqnarray}
for $N_{F}=5$ light quark flavours. The term $\Delta_m\approx 0.76$
denotes the NLO quark-mass effects from the top, bottom and charm quarks
\cite{higgsqcd}. The radiative corrections turn out to be very large:
the decay width is increased by about 90\% in the intermediate mass
range. The approximation of the heavy top limit is valid for the partial
gluonic decay width within about 5--10\% for the whole Higgs mass range
up to 1 TeV, while it is valid at the per-cent level in the intermediate
mass range. The reason for the suppressed quark mass dependence of the
relative QCD corrections is the dominance of soft and collinear gluon
contributions, which do not resolve the Higgs coupling to gluons and
thus lead to a simple rescaling factor. The three-loop \cite{hgg3} and
four-loop \cite{hgg4} QCD corrections to the gluonic decay width have
been evaluated in the limit of a heavy top quark. They contribute a
further amount of ${\cal O}(20\%)$ relative to the lowest order result
and thus increase the full NLO expression by ${\cal O}(10\%)$. The
reduced size of these corrections signals a significant stabilization of
the perturbative result and thus a reliable theoretical prediction.

The QCD corrections in the heavy quark limit can also be obtained by
means of a low-energy theorem \cite{prohiggs,let0}. The starting
point is that, for vanishing Higgs momentum $p_H\to 0$, the entire
interaction of the Higgs particle with $W,Z$ bosons and fermions can be
generated by the substitution
\begin{equation}
M_i \to M_i \times \left[ 1 + \frac{H}{v} \right] \hspace*{1cm}
(i=f,W,Z)
\label{eq:shift}
\end{equation}
where the Higgs field $H$ acts as a constant complex number. At higher
orders this substitution has to be expressed in terms of bare or
$\overline{\rm MS}$-subtracted parameters \cite{let,higgsqcd}. In most
of the practical cases the external Higgs particles are treated
on-shell, so that $p_H^2 = M_H^2$ and the mathematical limit of
vanishing Higgs momentum coincides with the limit of small Higgs masses
or large loop-particle masses.  In order to calculate the effective
Higgs couplings to gluons in the heavy-quark limit one starts from the
heavy quark $Q$ contribution to the kinetic gluon operator \cite{leff4}
that is related to the decoupling relation of the strong coupling
constant \cite{leff3,decoup},
\begin{eqnarray}
{\cal L}_{g} & = & -\frac{1}{4\zeta_{\alpha_s}} \hat G^{a\mu\nu} \hat
G^a_{\mu\nu} \nonumber \\[0.5cm]
\alpha_s^{(N_F)}(\mu_R^2) & = &
\zeta_{\alpha_s}~\alpha_s^{(N_F+1)}(\mu_R^2)
\label{eq:leffgg}
\end{eqnarray}
where $N_F=5$ is the number of light quark flavours. The gluonic field
strength tensor $\hat G^a_{\mu\nu}$ is defined in the $(N_F+1)$-flavour
theory, while the superscript of the strong coupling $\alpha_s$ defines
the number of active flavours taken into account. The inverse of the
decoupling factor $\zeta_{\alpha_s}$ develops a perturbative expansion
that contains a logarithmic top mass dependence, i.e.~it can be
expressed in terms of the logarithm $L_t =
\log(\mu_R^2/\overline{m_t}^2(\mu_R^2))$, where
$\overline{m_t}^2(\mu_R^2)$ denotes the $\overline{\rm MS}$ top mass at
the renormalization scale $\mu_R$. The substitution of
Eq.~(\ref{eq:shift}) implies the shift
\begin{equation}
L_t \to \bar L_t = L_t - 2\log\left(1 + \frac{H}{v} \right)
\end{equation}
Keeping only Higgs-field dependent terms and transforming the gluonic
field strength operator and the strong coupling constant by their
$N_F$-flavour expressions one arrives at the effective Lagrangian in the
heavy top-quark limit \cite{leff4},
\begin{eqnarray}
{\cal L}_{eff} & = & \frac{\alpha_s}{12\pi} \left\{ (1 + \delta) \log
\left(1+\frac{H}{v}\right) - \frac{\eta}{2} \log^2
\left(1+\frac{H}{v}\right) \right. \nonumber \\
& & \hspace*{2cm} \left. + \frac{\rho}{3} \log^3
\left(1+\frac{H}{v}\right)
- \frac{\sigma}{4} \log^4\left(1+\frac{H}{v}\right)
\right\} G^{a\mu\nu} G^a_{\mu\nu}
\label{eq:leff}
\end{eqnarray}
with the QCD corrections up to N$^4$LO
\begin{eqnarray}
\delta & = & \delta_1 \frac{\alpha_s}{\pi} + \delta_2 \left(
\frac{\alpha_s}{\pi} \right)^2 + \delta_3 \left(
\frac{\alpha_s}{\pi} \right)^3 + \delta_4 \left(
\frac{\alpha_s}{\pi} \right)^4 + {\cal O}(\alpha_s^5) \nonumber \\
\eta & = & \eta_2 \left( \frac{\alpha_s}{\pi} \right)^2 + \eta_3 \left(
\frac{\alpha_s}{\pi} \right)^3 + \eta_4 \left(
\frac{\alpha_s}{\pi} \right)^4 + {\cal O}(\alpha_s^5) \nonumber \\
\rho & = & \rho_3 \left( \frac{\alpha_s}{\pi} \right)^3 + \rho_4 \left(
\frac{\alpha_s}{\pi} \right)^4 + {\cal O}(\alpha_s^5) \nonumber \\
\sigma & = & \sigma_4 \left( \frac{\alpha_s}{\pi} \right)^4
+ {\cal O}(\alpha_s^5)
\end{eqnarray}
The perturbative coefficients are explicitly given by
\begin{eqnarray}
\delta_1 & = & \frac{11}{4} \qquad \qquad \qquad \qquad \qquad \qquad
\delta_2 = \frac{2777}{288} + \frac{19}{16} L_t + N_F
\left(\frac{L_t}{3}-\frac{67}{96} \right) \nonumber \\
\delta_3 & = & \frac{897943}{9216} \zeta_3 - \frac{2892659}{41472} +
\frac{209}{64} L_t^2 + \frac{1733}{288} L_t \nonumber \\
& + & N_F \left(\frac{40291}{20736} - \frac{110779}{13824} \zeta_3 +
\frac{23}{32} L_t^2 + \frac{55}{54} L_t \right) + N_F^2
\left(-\frac{L_t^2}{18} + \frac{77}{1728} L_t - \frac{6865}{31104}
\right) \nonumber \\
\delta_4 & = & 
-\frac{121}{1440} N_F \log^{5} 2 +\frac{3751}{2880}
\log^{5} 2+\frac{685}{41472} N_F^2 \log^{4} 2 +\frac{11679301}{
1741824} N_F \log^{4} 2 \nonumber \\
& - & \frac{93970579}{870912}
\log^{4} 2+\frac{121}{144} N_F \zeta_2 \log^{3} 2-\frac{3751}{288}
\zeta_2 \log^{3} 2-\frac{685}{6912} N_F^{2} \zeta_2 \log^{2 } 2
\nonumber \\
& - & \frac{11679301}{290304} N_F \zeta_2 \log^{2} 2 +
\frac{93970579}{145152} \zeta_2 \log^{2} 2 +\frac{2057}{192} N_F \zeta_4
\log 2 - \frac{63767}{384} \zeta_4 \log 2 \nonumber \\
& + & \frac{685}{1728} N_F^{2} a_4
+\frac{11679301}{72576} N_F a_4 -\frac{93970579}{36288}
a_4+\frac{121}{12} N_F a_5 -\frac{3751}{24} a_5+\frac{211}{1728} N_F^{3}
\zeta_3 \nonumber \\
& - & \frac{270407}{1492992} N_F^{3}+\frac{4091305}{331776} N_F^{2}
\zeta_3-\frac{576757}{ 55296} N_F^{2} \zeta_4-\frac{115}{384} N_F^{2}
\zeta_5-\frac{48073}{27648} N_F^{2} \nonumber \\
& - & \frac{151369}{725760} N_F
X_0-\frac{12171659669}{38707200} N_F \zeta_3+\frac{608462731}{ 11612160}
N_F \zeta_4+\frac{313489}{6912} N_F
\zeta_5 \nonumber \\
& + & \frac{76094378783}{522547200} N_F+\frac{4692439}{1451520}
X_0+\frac{28121193841}{19353600} \zeta_3+\frac{4674213853}{ 2903040}
\zeta_4-\frac{807193}{1728} \zeta_5 \nonumber \\
& - & \frac{854201072999}{522547200}
+ \left(\frac{481}{5184}
N_F^{3}+\frac{28297}{9216} N_F^{2} \zeta_3-\frac{21139}{3456}
N_F^{2}-\frac{32257}{288} N_F \zeta_3 \right. \nonumber \\
& & \left. + \frac{5160073}{41472} N_F+
\frac{9364157}{12288}
\zeta_3-\frac{49187545}{55296}\right) L_t
+ \left(-\frac{77}{6912}
N_F^{3}-\frac{1267}{13824} N_F^{2}+\frac{4139}{2304} 
N_F \right. \nonumber \\
& & \left. +\frac{8401}{384}\right) L_t^2 
+ \left(\frac{1}{108} N_F^{3}-\frac{157}{576} N_F^{2}+\frac{275}{192}
N_F+\frac{2299}{256} \right) L_t^3
\end{eqnarray}
and
\begin{eqnarray}
\eta_2 & = & \frac{35}{24} + \frac{2}{3} N_F \nonumber \\
\eta_3 & = & \frac{1333}{432} + \frac{589}{48} L_t + N_F \left(
\frac{1081}{432} + \frac{191}{72} L_t \right) + N_F^2 \left(
\frac{77}{864}
- \frac{2}{9} L_t \right) \nonumber \\
\eta_4 & = & 
\frac{481}{2592} N_F^{3}+N_F^{2} \left( \frac{28297}{4608}
\zeta_3-\frac{373637}{31104} \right)
+N_F \left(\frac{429965}{1728}-\frac{2985893}{13824} \zeta_3 \right)
\nonumber \\
& + & \frac{26296585}{18432} \zeta_3-
\frac{143976701}{82944}
+ \left(-\frac{77}{1728} N_F^{3}-\frac{1421}{3456}
N_F^{2}+\frac{9073}{1728} N_F+\frac{45059}{576} \right) L_t \nonumber \\
& + & \left(\frac{N_F^{3}}{18} -\frac{455}{288} N_F^{2}+\frac{63}{8}
N_F+\frac{6479}{128}\right) L_t^2
\nonumber \\
\rho_3 & = & \frac{1697}{144} + \frac{175}{72} N_F - \frac{2}{9} N_F^2
\nonumber \\
\rho_4 & = & \frac{130201}{1728} + \frac{18259}{192} L_t + N_F\left(
\frac{5855}{1728} + \frac{2077}{144} L_t \right) - N_F^2\left(
\frac{175}{384} + \frac{439}{144} L_t \right) \nonumber \\
& + & N_F^3 \left( \frac{L_t}{9} - \frac{77}{1728} \right)
\nonumber \\
\sigma_4 & = & \frac{51383}{864} + \frac{317}{36} N_F - \frac{47}{24}
N_F^2 + \frac{2}{27}N_F^3
\end{eqnarray}
where $G^a_{\mu\nu}$ denotes the gluon field strength tensor and
$\alpha_s$ the strong coupling constant with $N_F=5$ active flavours.
The constants used in these expressions are defined as $a_n=Li_n(1/2)$
($n=4,5$) and \cite{x0}
\begin{equation}
X_0= \frac{873}{2}\zeta_5 - 384~a_5 + \frac{16}{5} \log^5 2
- 32~\zeta_2 \log^3 2 - 318~\zeta_4 \log 2 = 1.8088795462\ldots
\end{equation}
This effective Lagrangian describes the multi-Higgs couplings to gluons
after integrating out the heavy top quark.

\begin{figure}[hbt]
\begin{center}
\setlength{\unitlength}{1pt}
\begin{picture}(500,100)(-15,0)

\DashLine(0,50)(50,50){5}
\Gluon(50,50)(75,65){3}{3}
\Gluon(75,65)(100,80){3}{3}
\Gluon(50,50)(75,35){-3}{3}
\Gluon(75,35)(100,20){-3}{3}
\Gluon(75,65)(75,35){3}{3}
\GCirc(50,50){10}{0.5}
\put(-15,46){$H$}
\put(105,18){$g$}
\put(105,78){$g$}

\DashLine(140,50)(190,50){5}
\GlueArc(215,50)(25,0,180){3}{8}
\GlueArc(215,50)(25,180,360){3}{8}
\Gluon(240,50)(290,80){3}{5}
\Gluon(240,50)(290,20){3}{5}
\GCirc(190,50){10}{0.5}
\put(125,46){$H$}
\put(295,18){$g$}
\put(295,78){$g$}

\DashLine(330,50)(380,50){5}
\Gluon(380,50)(405,65){3}{3}
\Gluon(405,65)(430,80){3}{2}
\Gluon(405,65)(430,50){-3}{2}
\Gluon(380,50)(430,20){-3}{5}
\GCirc(380,50){10}{0.5}
\put(315,46){$H$}
\put(435,18){$g$}
\put(435,78){$g$}
\put(435,48){$g$}

\end{picture}  \\
\setlength{\unitlength}{1pt}
\caption{\label{fg:hgglimdia} \it Typical effective diagrams
contributing to the NLO QCD corrections to $H\to gg$ in the heavy quark
limit.}
\end{center}
\end{figure}
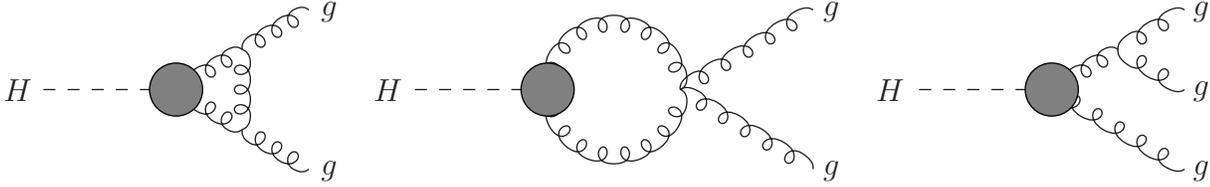

For the full calculation of the heavy quark limit given in
Eq.~(\ref{eq:hgg}) the effective $Hgg(g)$ couplings have to be inserted
into the blobs of the effective diagrams shown in
Fig.~\ref{fg:hgglimdia} for the NLO contribution. After evaluating these
effective massless one-loop contributions the result coincides with the
explicit calculation of the two-loop corrections in the heavy quark
limit of Eq.~(\ref{eq:hgg}).  This calculation has been extended to
N$^3$LO analogously \cite{hgg3, hgg4}.

Using the discussed low-energy theorem, the electroweak corrections of
${\cal O}(G_F M_t^2)$ to the gluonic decay width, which are mediated by
virtual top quarks, can be obtained from the ${\cal O}(G_F M_t^2)$
contribution to the kinetic gluon operator.  The final result leads to a
simple rescaling of the lowest order decay width \cite{hbb3,abdelgambino}
\begin{equation}
\Gamma(H\to gg) = \Gamma_{LO}(H\to gg) \left[ 1+
\frac{G_F M_t^2}{8\sqrt{2}\pi^2} \right]
\label{eq:hggelw}
\end{equation}
These corrections enhance the gluonic decay width by about 0.3\% and are
thus negligible. These ${\cal O}(G_F M_t^2)$ corrections are part of the
full electroweak corrections that have been determined first
analytically for the partial light-fermion contributions \cite{hggelwf}
and finally numerically for the full electroweak corrections
\cite{hggelw}. The electroweak corrections increase the partial Higgs
decay width into gluons by about 5\%. The residual theoretical
uncertainties due to missing electroweak corrections beyond NLO have
been estimated as 1\% and due to missing higher-order QCD corrections as
3\% so that the total uncertainty ranges at about 3\% \cite{yr4,
brthupu}.

\subsubsection{\it Minimal supersymmetric extension}
\label{sc:hgg}
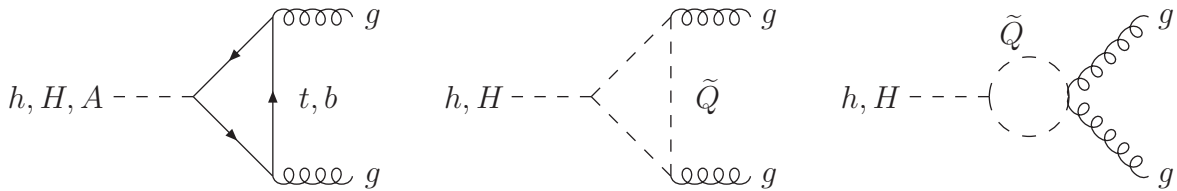
\begin{figure}[hbt]
\begin{center}
\setlength{\unitlength}{1pt}
\begin{picture}(400,100)(-20,0)

\Gluon(60,20)(90,20){-3}{4}
\Gluon(60,80)(90,80){3}{4}
\ArrowLine(60,20)(60,80)
\ArrowLine(60,80)(30,50)
\ArrowLine(30,50)(60,20)
\DashLine(0,50)(30,50){5}
\put(-40,46){$h,H,A$}
\put(70,46){$t,b$}
\put(95,18){$g$}
\put(95,78){$g$}

\Gluon(210,20)(240,20){-3}{4}
\Gluon(210,80)(240,80){3}{4}
\DashLine(210,80)(210,20){5}
\DashLine(210,20)(180,50){5}
\DashLine(180,50)(210,80){5}
\DashLine(150,50)(180,50){5}
\put(125,46){$h,H$}
\put(220,46){$\widetilde{Q}$}
\put(245,18){$g$}
\put(245,78){$g$}

\DashLine(300,50)(330,50){5}
\DashCArc(345,50)(15,0,180){5}
\DashCArc(345,50)(15,180,360){5}
\Gluon(360,50)(390,80){3}{5}
\Gluon(360,50)(390,20){-3}{5}
\put(275,46){$h,H$}
\put(335,70){$\widetilde{Q}$}
\put(395,18){$g$}
\put(395,78){$g$}

\end{picture}  \\
\setlength{\unitlength}{1pt}
\caption{\label{fg:mssmhgglodia} \it Typical diagrams contributing to
$\Phi \to gg$ at lowest order.}
\end{center}
\end{figure}
\noindent 
Since the $b$-quark couplings to the Higgs bosons may be strongly
enhanced for large $\mbox{tg$\beta$}$ and the $t$-quark couplings
suppressed in the MSSM, $b$ loops can contribute significantly to the
$gg$ coupling so that the approximation $M_{Q}^{2} \gg M_{H}^{2}$ can in
general no longer be applied. The leading order width for $h,H \to gg$
is generated by quark and squark loops, the latter contributing
significantly for squark masses below about 400~GeV \cite{SQCD}. The
contributing diagrams are depicted in Fig.~\ref{fg:mssmhgglodia}. The
partial decay widths are given by \cite{higgsqcd,SQCD,hunter}
\begin{eqnarray}
\Gamma_{LO} (h/H\rightarrow gg) & = & \frac{G_F \alpha_s^2M_{h/H}^3}
{36\sqrt{2}\pi^3} \left| \sum_Q g_Q^{h/H} A^{h/H}_Q(\tau_Q) +
\sum_{\widetilde{Q}} g_{\widetilde{Q}}^{h/H} A_{\widetilde{Q}}^{h/H}
(\tau_{\widetilde{Q}}) \right|^2 \label{eq:mssmhgg} \\
A_Q^{h/H} (\tau) & = & \frac{3}{2}\tau [1+(1-\tau) f(\tau)] \nonumber \\
A_{\widetilde Q}^{h/H} (\tau) & = & -\frac{3}{4}\tau [1-\tau f(\tau)]
\end{eqnarray}
with $\tau_i = 4M_i^2/M_{h/H}^2~~(i=Q, \widetilde{Q})$.  The function
$f(\tau)$ is defined in Eq.~(\ref{eq:ftau}) and the MSSM couplings
$g^{h/H}_Q$ can be found in Table \ref{tb:hcoup}. The squark couplings
$g^{h/H}_{\widetilde{Q}}$ are summarized in Table \ref{tb:hsqcoup}.  The
amplitudes approach constant values in the limit of large loop particle
masses:
\begin{eqnarray*}
A_Q^{h/H} (\tau) & \to & 1 \hspace*{1cm} \mbox{for $M_{h/H}^2 \ll
4M_Q^2$} \\
A_{\widetilde{Q}}^{h/H} (\tau) & \to & \frac{1}{4} \hspace*{1cm}
\mbox{for
$M_{h/H}^2 \ll 4M_{\widetilde{Q}}^2$}
\end{eqnarray*}
The QCD corrections to the quark loops can be adopted from the SM Higgs
case as given in Eq.~(\ref{eq:hgg}) with the Higgs mass $M_H$ replaced
by the corresponding MSSM Higgs mass $M_{h/H}$. The QCD corrections to
the squark loops are known up to NLO including the full mass dependence
\cite{gghsqnlo}. Up to NLO the partial decay width is given by
\begin{eqnarray}
\Gamma [\,h/H\,\to \, gg] &=&
\Gamma_{LO}\,\left\{1+E \frac{\alpha_{s}(M_{h/H})}{\pi}\ \right\}
\label{eq:mssmhggqcd} \\
E & \to &\frac{95}{4}-\frac{7}{6}N_{F} + \frac{7}{2} \Re e
\left\{ \frac{\sum_{\widetilde{Q}} g_{\widetilde{Q}}^{h/H}
A_{\widetilde{Q}}^{h/H} (\tau_{\widetilde{Q}})} {\sum_Q g_Q^{h/H}
A^{h/H}_Q(\tau_Q)} \right\} \hspace*{1cm} \mbox{for $M_{h/H}^2 \ll 4
M_{Q,\widetilde{Q}}^2$} \nonumber
\end{eqnarray}
Analogous to the quark contributions the heavy squark loop correction
can be obtained by means of the extension of the previously described
low-energy theorem to scalar squark particles \cite{SQCD}. The effective
NLO Lagrangian for the squark part is given by \cite{SQCD,
gghsqnlo}\footnote{Taking into account the squark four-point couplings
for degenerate squark masses, too, the coefficient 9/2 is changed to
29/6 \cite{SQCD, gghsqnlo, hggleffsq}. The proper decoupling of gluino
contributions in the heavy-mass limit modifies this coefficient to 37/6
for mass-degenerate stop states \cite{gghdec}.}
\begin{equation}
{\cal L}_{eff} = \frac{\alpha_s}{48\pi} G^{a\mu\nu} G^a_{\mu\nu}
\frac{H}{v} \left[ 1 + \frac{9}{2} \frac{\alpha_s}{\pi} \right]
\label{eq:leffsq}
\end{equation}
where only non-SUSY-QCD corrections are taken into account,
i.e.~diagrams with gluino exchange and squark four-point couplings are
omitted. Thus the only difference to the quark loops in the heavy loop
mass limit arises in the virtual corrections. This leads to the
additional last term of Eq.~(\ref{eq:mssmhggqcd}).

\begin{table}[hbt]
\renewcommand{\arraystretch}{2.0}
\begin{center}
\begin{tabular}{|lc||cc|} \hline
\multicolumn{2}{|c||}{$\Phi$} & $H^\pm$ & $\tilde \chi^\pm_i$ \\
\hline \hline
SM~ & $H$ & 0 & 0 \\ \hline
MSSM~ & $h$ & $\frac{M_W^2}{M_{H^\pm}^2} \left[
\sin(\beta-\alpha)+\frac{\cos
2\beta \sin (\beta+\alpha)}{2\cos^2 \theta_W} \right]$
& $2\frac{M_W}{M_{\tilde \chi^\pm_i}}(S_{ii}\cos\alpha -
Q_{ii}\sin\alpha)$ \\
& $H$ & $\frac{M_W^2}{M_{H^\pm}^2} \left[ \cos(\beta-\alpha)-\frac{\cos
2\beta \cos (\beta+\alpha)}{2\cos^2 \theta_W} \right]$
& $2\frac{M_W}{M_{\tilde \chi^\pm_i}}(S_{ii}\sin\alpha +
Q_{ii}\cos\alpha)$ \\
& $A$ & 0 & $2\frac{M_W}{M_{\tilde \chi^\pm_i}}(-S_{ii}\cos\beta -
Q_{ii}
\sin\beta)$ \\[0.2cm] \hline
\end{tabular} \\[0.3cm]

\begin{tabular}{|lc||c|} \hline
\multicolumn{2}{|c||}{$\Phi$} & $\tilde f_{L,R}$ \\
\hline \hline
SM~ & $H$ & 0 \\ \hline
MSSM~ & $h$ & $\frac{M_f^2}{M_{\tilde f}^2} g_f^h \mp
\frac{M_Z^2}{M_{\tilde f}^2} (I_3^f - e_f \sin^2\theta_W) \sin
(\alpha+\beta)$ \\
& $H$ & $\frac{M_f^2}{M_{\tilde f}^2} g_f^H \pm
\frac{M_Z^2}{M_{\tilde f}^2} (I_3^f - e_f \sin^2\theta_W) \cos
(\alpha+\beta)$ \\ 
& $A$ & 0 \\ \hline
\end{tabular}
\renewcommand{\arraystretch}{1.2} 
\caption{\label{tb:hsqcoup} \it MSSM Higgs couplings to charged Higgs
bosons, charginos and sfermions relative to the corresponding SM Yukawa
couplings.  $Q_{ii}$ and $S_{ii}$ $(i=1,2)$ are related to the mixing
angles between the charginos $\tilde \chi^\pm_1$ and $\tilde
\chi^\pm_2$, see Refs.~\cite{hunter,mssmbase}. The indices $L,R$ for the
sfermion states define the weak charges according to their fermionic
superpartners.}
\end{center}
\end{table}

For the pseudoscalar Higgs decays only quark loops are contributing, and
we find \cite{higgsqcd}
\begin{eqnarray}
\Gamma_{LO}\,[A\to gg]&=&\frac{G_{F}\,
\alpha_{s}^{2}\,M_{A}^{3}}{16\,\sqrt{2}\,
\pi^{3}} \left| \sum_{Q} g_Q^A A_Q^A(\tau_Q) \right|^2
\label{eq:mssmagg} \\
A_Q^A (\tau) & = & \tau f(\tau) \nonumber \\
\end{eqnarray}
with $\tau_Q = 4M_Q^2/M_A^2$.  The MSSM couplings $g_Q^A$ can be found
in Table \ref{tb:hcoup}.  For large quark masses the quark amplitude
approaches unity.  In order to get a consistent result for the QCD
corrections, the pseudoscalar $\gamma_5$ coupling has been regularized
in the 't~Hooft--Veltman scheme \cite{thoovel} or its extension by Larin
\cite{larin}, which requires an additional finite renormalization of the
$AQ\bar Q$ vertex \cite{higgsqcd,gghsusy}. An alternative is provided by
the scheme of Ref.~\cite{kreimer} that avoids the additional counter
terms by giving up the cyclicity of the traces involving Clifford
matrices and thus requires a fixed reading point in all diagrams. The
heavy quark limit $M_A^2 \ll 4M_Q^2$ provides a reasonable approximation
in the MSSM parameter range where this decay mode is significant. At the
threshold $M_A=2M_t$, the QCD corrections develop a Coulomb singularity,
which is regularized by including the finite top decay width
\cite{agagathresh}. The pseudoscalar decay width has been calculated up
to NNLO in QCD within the heavy quark limit \cite{aggnnlo}, while the
NLO corrections are known exactly \cite{higgsqcd},
\begin{eqnarray}
\Gamma\,[A\to gg] & = & \Gamma_{LO}\, \left\{1+(24.25-1.167 N_F +
\Delta_m)\frac{\alpha_{s}(M_A)}{\pi} \right. \nonumber \\
& + & \left. \left(392.22- 48.58 N_F + 0.888 N_F^2 + N_F
\log\frac{M_A^2}{m_t^2}\right) \left( \frac{\alpha_{s}(M_A)}{\pi}
\right)^2 \right\} \nonumber \\
& \approx & \Gamma_{LO}\, \left\{1+0.66+0.20 \right\}
\label{eq:agg}
\end{eqnarray}
where the numbers of the last line are for a pseudoscalar mass $M_A=200$
GeV.  The top, bottom and charm mass effects amount to $\Delta_m\approx
1.3$ for $\mbox{tg}\beta = 1$. For larger values of $\mbox{tg}\beta$
quark mass effects are larger due to the increase of the bottom
contribution.  The heavy-quark limit can also be obtained by means of a
low-energy theorem. The starting point is the Adler--Bell--Jackiw (ABJ)
anomaly in the divergence of the axial vector current \cite{ABJ},
\begin{equation}
\partial^\mu j_\mu^5 = 2M_Q \bar Q i\gamma_5 Q + \frac{\alpha_s}{4\pi}
G^{a\mu\nu} \widetilde{G}^a_{\mu\nu}
\label{eq:abj}
\end{equation}
with $\widetilde{G}^a_{\mu\nu} = \frac{1}{2}
\epsilon_{\mu\nu\alpha\beta} G^{a\alpha\beta}$ denoting the dual field
strength tensor. Since, according to the Sutherland--Veltman paradox
\cite{suthvel}, the matrix element $\langle gg| \partial^\mu j_\mu^5 | 0
\rangle$ vanishes for zero momentum transfer, the matrix element
$\langle gg| M_Q \bar Q i\gamma_5 Q | 0 \rangle$ of the Higgs source can
be related to the ABJ anomaly in Eq.~(\ref{eq:abj}). Thanks to the
Adler--Bardeen theorem \cite{adlerbardeen}, the ABJ anomaly is not
modified by radiative corrections at vanishing momentum transfer, so
that the effective Lagrangian
\begin{equation}
{\cal L}_{eff} =
-g_Q^A\frac{\alpha_s}{8\pi}G^{a\mu\nu}\widetilde{G}^a_{\mu\nu}
\frac{A}{v}
\end{equation}
is valid to all orders of perturbation theory\footnote{This has been
confirmed explicitly up to NNLO in Ref.~\cite{aggnnlo}, if the QCD
corrections are expanded in terms of the strong coupling $\alpha_s$ with
only five light active flavours.}. In order to calculate the full QCD
corrections to the $gg$ decay width, this effective coupling has to be
inserted in the effective diagrams analogous to those of
Fig.~\ref{fg:hgglimdia}. The final result agrees with the explicit
expansion of the two-loop diagrams in terms of the heavy quark mass at
NLO.

The electroweak corrections to the pseudoscalar Higgs decays into gluons
have been calculated in the limit of heavy top quarks and $W$ bosons
some time ago within the 2HDM, i.e.~involving the non-SUSY part of the
MSSM. The corrections modify the partial decay width by \cite{aggelw}
\begin{equation}
\Gamma(A\to gg) = \Gamma_{LO} \times\left\{1-x_t \left( 7 +
\frac{10}{\mbox{tg}^2\beta} \right) \right\}
\end{equation}
with $x_t$ defined after Eq.~(\ref{eq:hvvxt}). The correction ranges at the
per-cent level.

For large values of $\mbox{tg}\beta$ the bottom-quark loops will take
over the dominant role. In the limit of small quark masses the form
factors develop logarithmic structures,
\begin{eqnarray}
A_Q^{h/H} (\tau) & \to & -\frac{3}{2} \frac{m_Q^2}{M_{h/H}^2}
\left[\log\frac{M_{h/H}^2}{m_Q^2}
- i\pi \right]^2 \nonumber \\
A_Q^A (\tau) & \to & -\frac{m_Q^2}{M_A^2} \left[\log\frac{M_A^2}{m_Q^2}
- i\pi \right]^2
\end{eqnarray}
so that the quark contributions to the partial decay widths of the
scalar and pseudoscalar Higgs bosons approach each other, i.e.~the
chiral symmetry is restored [the relative factor 3/2 is compensated by
the different global coefficients of the partial decay widths of
Eqs.~(\ref{eq:mssmhgg}, \ref{eq:mssmagg})]. The same is also true for
the NLO QCD corrections \cite{higgsqcd}. The finite parts of the virtual
corrections approach common logarithmic expressions,
\begin{eqnarray}
\Gamma\,[\Phi\to gg] & = & \Gamma_{LO}\, \left\{1+ (E^\Phi_{virt} +
E^\Phi_{real}) \frac{\alpha_{s}(M_\Phi)}{\pi} \right\} \nonumber \\
E^\Phi_{virt} (\tau_Q) & \to & \frac{C_A-C_F}{12}
\left[\log\frac{M_H^2}{m_Q^2} - i\pi \right]^2 - C_F
\left[\log\frac{M_H^2}{m_Q^2} - i\pi \right] \hspace*{1cm} \mbox{for
$M_\Phi^2 \gg 4M_Q^2$}
\label{eq:hgglog}
\end{eqnarray}
with $C_A=3$ if the quark mass is defined to be the pole mass. The
Abelian part of the large logarithms is related to the Sudakov form
factor \cite{sudakov} at the virtual $\Phi b\bar b$ vertex and have been
resummed down to the subleading logarithmic level \cite{hgagares}. The
non-Abelian part of these logarithms, however, has not been resummed so
far. The real corrections determined by the coefficient $E^\Phi_{real}$
are of subleading non-Abelian logarithmic order.

The genuine SUSY--QCD corrections have been calculated in the limit of
heavy SUSY-particle masses by means of a heavy mass expansion for the
top-stop-gluino contributions \cite{hggtsqcd, hggtsqcd1, hggtbsqcd,
hggtbsqcd1} and a mixed heavy mass and large momentum expansion for the
bottom-sbottom-gluino corrections \cite{hggtbsqcd, hggtbsqcd1,
hggbsqcd}.  A large part of the latter corrections in the bottom case
can be absorbed by the effective bottom Yukawa couplings of
Eq.~(\ref{eq:rescoup}). However, there is a sizeable remainder. This
feature is confirmed by the full calculation of the SUSY--QCD
corrections by means of a numerical integration of the corresponding
two-loop diagrams \cite{hggsqcd0,hggsqcd}. The agreement with the
approximate calculations is reasonable.
\begin{figure}[ht]
\begin{center}
\begin{picture}(150,240)(0,0)
\put(-130,-140.0){\includegraphics{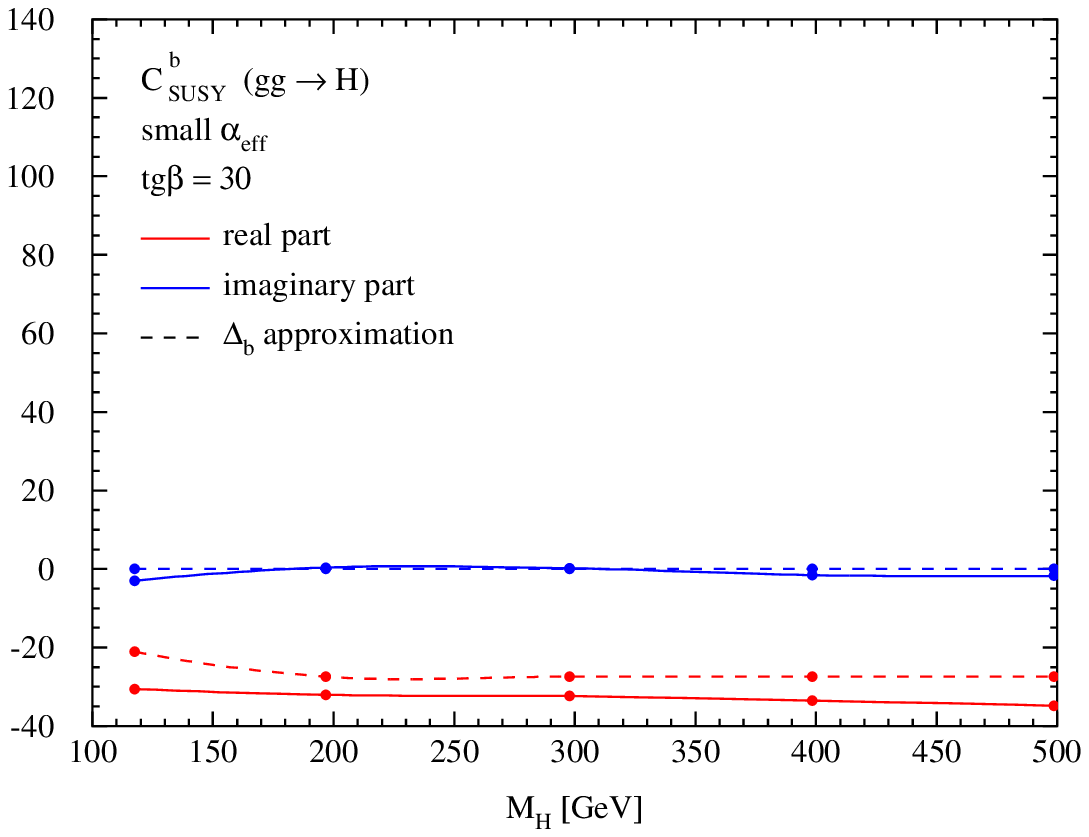}}
\put(110.0,210.0){\small Preliminary}
\end{picture}
\caption{\it The genuine SUSY--QCD corrections normalized to the LO bottom
quark form factor. Real part: red, imaginary part: blue, compared to the
$\Delta_b$ approximation (dashed lines). $A_b$ has been renormalized in
the $\overline{\mbox{MS}}$ scheme. From Ref.~\cite{hggsqcd}.}
\label{fig:totalnlo}
\end{center}
\end{figure}
The numerical results of the full calculation are presented in
Fig.~\ref{fig:totalnlo} for the modified small $\alpha_{eff}$ scenario
\cite{alphaeff}, defined by the following choices of MSSM parameters
[$m_t=172.6$~GeV],
\begin{equation}
\begin{array}{llllll}
M_{\tilde{Q}} &=& 800\;\mbox{GeV} & \qquad \tan\beta &=& 30 \\
M_{\tilde{g}} &=& 1000\;\mbox{GeV} & \qquad \mu &=& 2\;\mbox{TeV} \\
M_2 &=& 500\;\mbox{GeV} & \qquad A_b = A_t &=& -1.133\;\mbox{TeV} \;.
\end{array}
\end{equation}
In this scenario the squark masses amount to
\begin{equation}
\begin{array}{llllll}
m_{\tilde{t}_1} &=& 679\;\mbox{GeV} & \qquad m_{\tilde{t}_2} &=&
935\;\mbox{GeV} \\
m_{\tilde{b}_1} &=& 601\;\mbox{GeV} & \qquad m_{\tilde{b}_2} &=&
961\;\mbox{GeV}
\end{array}
\end{equation}
Fig.~\ref{fig:totalnlo} displays the genuine SUSY--QCD corrections
normalized to the  LO bottom quark form factor, i.e.\ $A_b^{h/H}
(\tau_b) \to A_b^{h/H} (\tau_b) (1+C^b_{SUSY} \frac{\alpha_s}{\pi})$,
where the LO form factor has been defined in terms of the bottom Yukawa
coupling $g_b^{h/H}$ {\it without} $\Delta_b$-resummation.  The
corrections can be sizeable, but can be described reasonably well with
the usual $\Delta_b$ approximation, if $A_b$ is renormalized in the
$\overline{\mbox{MS}}$ scheme, however, with a sizeable remainder that
determines the differences between the full and dashed curves of
Fig.~\ref{fig:totalnlo}.

\subsection{\it Higgs boson decays into photons}
\subsubsection{\it Standard Model}
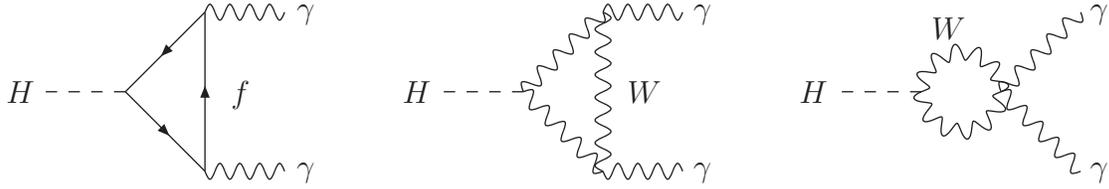
\begin{figure}[hbt]
\begin{center}
\setlength{\unitlength}{1pt}
\begin{picture}(400,100)(0,0)

\Photon(60,20)(90,20){-3}{4}
\Photon(60,80)(90,80){3}{4}
\ArrowLine(60,20)(60,80)
\ArrowLine(60,80)(30,50)
\ArrowLine(30,50)(60,20)
\DashLine(0,50)(30,50){5}
\put(-15,46){$H$}
\put(70,46){$f$}
\put(95,18){$\gamma$}
\put(95,78){$\gamma$}

\Photon(210,20)(240,20){-3}{4}
\Photon(210,80)(240,80){3}{4}
\Photon(210,80)(210,20){3}{7}
\Photon(210,20)(180,50){3}{5}
\Photon(180,50)(210,80){3}{5}
\DashLine(150,50)(180,50){5}
\put(135,46){$H$}
\put(220,46){$W$}
\put(245,18){$\gamma$}
\put(245,78){$\gamma$}

\DashLine(300,50)(330,50){5}
\PhotonArc(345,50)(15,0,180){3}{6}
\PhotonArc(345,50)(15,180,360){3}{6}
\Photon(360,50)(390,80){3}{5}
\Photon(360,50)(390,20){3}{5}
\put(285,46){$H$}
\put(335,70){$W$}
\put(395,18){$\gamma$}
\put(395,78){$\gamma$}

\end{picture}  \\
\setlength{\unitlength}{1pt}
\caption{\label{fg:hgagalodia} \it Typical diagrams contributing to
$H\to \gamma \gamma$ at lowest order.}
\end{center}
\end{figure}
\noindent
The decay of the Higgs boson into photons is mediated by $W$ and heavy
fermion loops in the SM, see Fig.~\ref{fg:hgagalodia}; at
lowest order the partial decay width is given by \cite{let0}
\begin{eqnarray}
\Gamma\, [H\to \gamma\gamma] = \frac{G_{F}\, \alpha^{2}\,M_{H}^{3}}
{128\,\sqrt{2}\,\pi^{3}} \left| \sum_{f} N_{cf} e_f^2 A_f^H(\tau_f) +
A^H_W(\tau_W) \right|^2
\label{eq:hgaga}
\end{eqnarray}
with the form factors
\begin{eqnarray*}
A_f^H (\tau) & = & 2 \tau \left[ 1+(1-\tau) f(\tau) \right] \\
A_W^H (\tau) & = & - \left[ 2+3\tau+3\tau (2-\tau) f(\tau) \right]
\end{eqnarray*}
and the function $f(\tau)$ defined in Eq.~(\ref{eq:ftau}).  The
parameters $\tau_i= 4M_i^2/M_H^2~~(i=f,W)$ are defined in terms of the
corresponding of the heavy loop-particle masses. For large loop masses
the form factors approach constant values:
\begin{eqnarray}
A_f^H & \to & \frac{4}{3} \hspace*{1cm} \mbox{for $M_H^2 \ll 4 M_Q^2$}
\nonumber \\
A_W^H & \to & -7 \hspace*{0.75cm} \mbox{for $M_H^2 \ll 4 M_W^2$}
\end{eqnarray}
The $W$ loop is dominant in the intermediate Higgs mass range, and the
fermion loops interfere destructively. Only far above the thresholds,
for Higgs masses $M_H\sim 600$ GeV, does the top quark loop become
competitive, nearly cancelling the $W$ loop contribution. This feature
is still relevant for off-shell Higgs contributions.

In the past the full two-loop QCD corrections to the quark loops have
been calculated \cite{higgsqcd,hgaga,hgaga1}. The QCD corrections simply
rescale the lowest order quark amplitude by a factor that only depends
on the ratios of the Higgs and quark masses
\begin{eqnarray}
A_Q^H(\tau_Q) & \to & A_Q^H(\tau_Q) \times \left[1+ C_H(\tau_Q)
\frac{\alpha_s}{\pi} \right] \nonumber \\
C_H(\tau_Q) & \to & -1 \hspace*{1cm} \mbox{for $M_H^2\ll 4M_Q^2$}
\label{eq:hgagaqcd}
\end{eqnarray}
According to the low-energy theorem discussed before, the QCD
corrections in the heavy quark limit can be obtained from the effective
Lagrangian \cite{prohiggs,let,higgsqcd,let0}
\begin{eqnarray}
{\cal L}_{eff} = e_t^2 \frac{\alpha}{2\pi} \left\{ (1 + \delta) \log
\left(1+\frac{H}{v}\right) - \frac{\eta}{2} \log^2
\left(1+\frac{H}{v}\right) + \frac{\rho}{3} \log^3
\left(1+\frac{H}{v}\right)
\right\} F^{\mu\nu} F_{\mu\nu}
\label{eq:leffhgaga}
\end{eqnarray}
with the QCD corrections up to N$^3$LO
\begin{eqnarray}
\delta & = & \delta_1 \frac{\alpha_s}{\pi} + \delta_2 \left(
\frac{\alpha_s}{\pi} \right)^2 + \delta_3 \left(
\frac{\alpha_s}{\pi} \right)^3 + {\cal O}(\alpha_s^4) \nonumber \\
\eta & = & \eta_2 \left( \frac{\alpha_s}{\pi} \right)^2 + \eta_3 \left(
\frac{\alpha_s}{\pi} \right)^3 + {\cal O}(\alpha_s^4) \nonumber \\
\rho & = & \rho_3 \left( \frac{\alpha_s}{\pi} \right)^3 +
{\cal O}(\alpha_s^4)
\end{eqnarray}
The perturbative coefficients are explicitly given by
\begin{eqnarray}
\delta_1 & = & -1 \qquad \qquad \qquad \qquad \qquad
\delta_2 = -\frac{8}{9} - \frac{31}{12} L_t + N_F
\left(\frac{L_t}{6}-\frac{1}{18}\right) + \frac{\sum_{i=1}^{N_F}
e_{q_i}^2}{e_t^2} \left( -\frac{11}{72} + \frac{L_t}{6}\right) \nonumber \\
\delta_3 & = & \frac{7835}{288} \zeta_3 - \frac{95339}{2592} -
\frac{961}{144} L_t^2 - \frac{541}{108} L_t
+ N_F \left(\frac{4693}{1296} - \frac{125}{144} \zeta_3 +
\frac{31}{36} L_t^2 + \frac{101}{216} L_t \right) \nonumber \\
& + & N_F^2 \left(-\frac{L_t^2}{36} + \frac{L_t}{54} - \frac{19}{324}
\right)
+ \frac{\sum_{i=1}^{N_F} e_{q_i}^2}{e_t^2} \left[\frac{53}{108} -
\frac{55}{54} \zeta_3 + \frac{11}{54} L_t + \frac{29}{72} L_t^2 - N_F
\left( \frac{449}{3888} + \frac{L_t^2}{36} \right) \right] \nonumber \\
& + & \frac{\left(\sum_{i=1}^{N_F+1} e_{q_i}\right)^2 -
\left(\sum_{i=1}^{N_F} e_{q_i}\right)^2}{e_t^2} \left( \frac{55}{216} -
\frac{5}{9} \zeta_3 \right) \nonumber \\
\eta_2 & = & \frac{31}{24} - \frac{N_F}{6} - \frac{\sum_{i=1}^{N_F}
e_{q_i}^2}{e_t^2} \frac{1}{6} \nonumber \\
\eta_3 & = & \frac{541}{108} - \frac{101}{216} N_F - \frac{N_F^2}{54} +
\left( \frac{961}{72} - \frac{31}{18} N_F + \frac{N_F^2}{18} \right) L_t
+ \frac{\sum_{i=1}^{N_F}e_{q_i}^2}{e_t^2} \left( -\frac{11}{54} -
\frac{29}{36} L_t + \frac{N_F}{18} L_t \right) \nonumber \\
\rho_3 & = & -\frac{961}{108} + \frac{31}{27} N_F - \frac{N_F^2}{27} +
\frac{\sum_{i=1}^{N_F}e_{q_i}^2}{e_t^2} \left( \frac{29}{54} -
\frac{N_F}{27} \right)
\end{eqnarray}
(where $F_{\mu\nu}$ denotes the photon field strength tensor and
$\alpha_s$ the strong coupling constant with $N_F=5$ active flavours)
that can be derived from the top-quark contribution to the kinetic
photon operator analogous to the QCD case. The results for $\delta_2$
and $\delta_3$ agree with previous determinations \cite{leff3,hgaga3}.
Partial results are also known up to N$^4$LO \cite{hgaga4}.

In order to improve the perturbative behaviour of the quark loop
contributions they should be expressed preferably in terms of the
running quark masses $m_Q(M_H/2)$, which are normalized to the {\it
pole} masses $M_Q$ via
\begin{equation}
m_Q(\mu_Q=M_Q) = M_Q
\end{equation}
with their scale identified with $\mu_Q=M_H/2$ within the photonic decay
mode. These definitions imply a proper definition of the $Q\bar Q$
thresholds $M_H = 2 M_Q$, without artificial displacements due to finite
shifts between the {\it pole} and running quark masses, as is the case
for the running $\overline{\rm MS}$ masses. The residual QCD corrections
are small in the intermediate Higgs-mass range increasing the photonic
Higgs decay width by about 2\%.  Recently the full three-loop QCD corrections
to the photonic Higgs decay have been calculated in the heavy top-quark
limit \cite{hgaga3f}. They lead to a further contribution of a few per mille.

The two-loop electroweak corrections have been evaluated in
Ref.~\cite{hgagaelw} for the $W$-boson and top-quark induced
contributions. They decrease the partial photonic decay width by about
2\% thus nearly cancelling the NLO QCD corrections. Both types of
corrections only become large for Higgs masses above the $t\bar
t$-threshold, i.e.~for $M_H > 2m_t$.

\subsubsection{\it Minimal supersymmetric extension}
\begin{figure}[hbt]
\begin{center}
\setlength{\unitlength}{1pt}
\begin{picture}(400,100)(-20,0)

\Photon(60,20)(90,20){-3}{4}
\Photon(60,80)(90,80){3}{4}
\ArrowLine(60,20)(60,80)
\ArrowLine(60,80)(30,50)
\ArrowLine(30,50)(60,20)
\DashLine(0,50)(30,50){5}
\put(-40,46){$h,H,A$}
\put(70,46){$f,\tilde \chi^\pm$}
\put(95,18){$\gamma$}
\put(95,78){$\gamma$}

\Photon(210,20)(240,20){-3}{4}
\Photon(210,80)(240,80){3}{4}
\DashLine(210,80)(210,20){5}
\DashLine(210,20)(180,50){5}
\DashLine(180,50)(210,80){5}
\DashLine(150,50)(180,50){5}
\put(125,46){$h,H$}
\put(215,46){$W, H^\pm, \widetilde{f}$}
\put(245,18){$\gamma$}
\put(245,78){$\gamma$}

\DashLine(300,50)(330,50){5}
\DashCArc(345,50)(15,0,180){4}
\DashCArc(345,50)(15,180,360){4}
\Photon(360,50)(390,80){3}{5}
\Photon(360,50)(390,20){3}{5}
\put(275,46){$h,H$}
\put(325,70){$W, H^\pm, \widetilde{f}$}
\put(395,18){$\gamma$}
\put(395,78){$\gamma$} 

\end{picture}  \\
\setlength{\unitlength}{1pt}
\caption{\label{fg:mssmhgagalodia} \it Typical diagrams contributing
to $\Phi \to \gamma \gamma$ at lowest order.}
\end{center}
\end{figure}
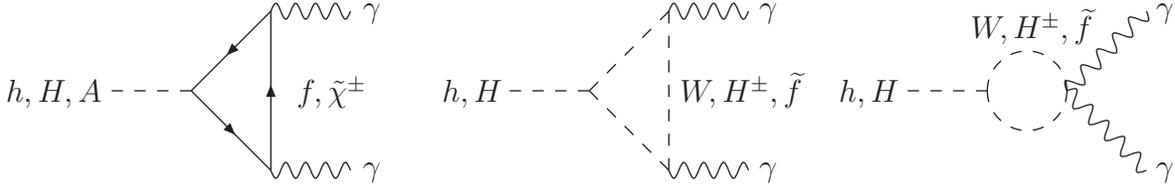
\noindent
The decays of the scalar Higgs bosons into photons are mediated by $W$ and
heavy fermion loops as in the SM and, in addition, by
charged Higgs, sfermion and chargino loops, see
Fig.~\ref{fg:mssmhgagalodia}. The partial decay widths
\cite{higgsqcd,hunter} can be expressed as
\begin{eqnarray}
\!\!\!\!\!\!
\Gamma [h/H\to \gamma\gamma] & = & \frac{G_{F} \alpha^{2}M_{h/H}^{3}}
{128\sqrt{2}\pi^{3}} \left| \sum_{f} N_{cf} e_f^2 g_f^{h/H}
A_f^{h/H}(\tau_f) + g^{h/H}_W A^{h/H}_W(\tau_W)
\right. \nonumber \\
& + & \left. g_{H^\pm}^{h/H} A_{H^\pm}^{h/H}(\tau_{H^\pm})
+ \sum_{\tilde \chi^\pm} g_{\tilde \chi^\pm}^{h/H}
A_{\tilde \chi^\pm}^{h/H} (\tau_{\tilde \chi^\pm}) +
\sum_{\tilde f}N_{cf}e_{\tilde f}^2 g_{\tilde f}^{h/H} A_{\tilde
f}^{h/H} (\tau_{\tilde f}) \right|^2
\label{eq:mssmhgaga}
\end{eqnarray}
involving the form factors
\begin{eqnarray*}
A_{f,\tilde \chi^\pm}^{h/H} (\tau) & = & 2 \tau \left[ 1+(1-\tau)
f(\tau)
\right] \\ \\
A_{H^\pm,\tilde f}^{h/H} (\tau) & = & - \tau \left[1-\tau f(\tau)
\right] \\ \\
A_W^{h/H} (\tau) & = & -\left[ 2+3\tau+3\tau (2-\tau) f(\tau) \right]
\end{eqnarray*}
and the function $f(\tau)$ as defined in Eq.~(\ref{eq:ftau}). For large
loop-particle masses the form factors approach constant values,
\begin{eqnarray*}
A_{f,\tilde \chi^\pm}^{h/H} (\tau) & \to & \frac{4}{3} \hspace*{1cm}
\mbox{for $M_{h/H}^2 \ll 4M_{f,\tilde \chi^\pm}^2$} \nonumber \\
A_{H^\pm,\widetilde{f}}^{h/H} (\tau) & \to & \frac{1}{3} \hspace*{1cm}
\mbox{for $M_{h/H}^2 \ll 4M_{H^\pm,\widetilde{f}}^2$} \nonumber \\
A_W^{h/H} (\tau) & \to & - 7 \hspace*{1cm} \mbox{for $M_{h/H}^2 \ll
4M_W^2$}
\end{eqnarray*}
Sfermion loops start to be sizeable for sfermion masses
$M_{\widetilde{f}} \lsim 300$ GeV, while for larger sfermion masses they are
negligible in general.

The photonic decay mode of the pseudoscalar Higgs boson is generated by
heavy charged-fermion and chargino loops, see
Fig.~\ref{fg:mssmhgagalodia}. The partial decay width reads
\cite{higgsqcd,hunter}
\begin{equation}
\Gamma(A \rightarrow\gamma\gamma)=
\frac{G_F\alpha^2M_A^3}{32\sqrt{2}\pi^3}
\left| \sum_f N_{cf} e_f^2 g_f^A A_f^A (\tau_f)
+ \sum_{\tilde \chi^\pm} g_{\tilde \chi\pm}^A A_{\tilde \chi^\pm}^A
(\tau_{\tilde \chi^\pm}) \right|^2
\label{eq:mssmagaga}
\end{equation}
with the amplitudes
\begin{eqnarray}
A_{f, \tilde \chi^\pm}^A (\tau) & = & \tau f(\tau)
\end{eqnarray}
For large loop-particle masses the pseudoscalar amplitudes approach
unity. The parameters $\tau_i= 4M_i^2/M_\Phi^2~~(i=f,W,H^\pm,\tilde
\chi^\pm, \tilde f)$ are defined in terms of the corresponding heavy
loop-particle masses and the MSSM couplings $g^\phi_{f,W,H^\pm,\tilde
\chi^\pm, \tilde f}$ are summarized in Tables \ref{tb:hcoup} and
\ref{tb:hsqcoup}.

The QCD corrections to the quark and squark loop contributions have been
calculated. They are known for finite quark, squark and Higgs masses
\cite{higgsqcd,gghsqnlo,hgaga,hgaga1}. The QCD corrections rescale the lowest
order quark amplitudes \cite{higgsqcd,gghsqnlo,hgaga,hgaga1,hgagasq},
\begin{eqnarray}
A_Q^\Phi (\tau_Q) & \to & A_Q^\Phi (\tau_Q) \left[ 1+C_\Phi(\tau_Q)
\frac{\alpha_s}{\pi} \right] \\
C_{h/H} (\tau_Q) & \to & -1 \hspace*{1cm} \mbox{for $M_{h/H}^2 \ll
4M_Q^2$}
\nonumber \\
C_A (\tau_Q) & \to & 0 \hspace*{1.33cm} \mbox{for $M_A^2 \ll 4M_Q^2$}
\nonumber \\
\nonumber \\
A_{\widetilde{Q}}^{h/H}(\tau_{\widetilde{Q}}) & \to &
A_{\widetilde{Q}}^{h/H}
(\tau_{\widetilde{Q}}) \left[
1+\widetilde{C}_{h/H}(\tau_{\widetilde{Q}})
\frac{\alpha_s}{\pi} \right] \\
\widetilde{C}_{h/H} (\tau_{\widetilde{Q}}) & \to & 3
\hspace*{1cm}
\mbox{for $M_{h/H}^2 \ll 4M_{\widetilde{Q}}^2$}
\nonumber
\end{eqnarray}
The QCD corrections to the $\gamma\gamma$ decay width are defined in
terms of the running quark masses in the same way as the SM photonic
decay width. The QCD radiative corrections are moderate in the
intermediate mass range \cite{higgsqcd,hgaga,hgaga1}, where this decay mode
will be important.  Owing to the narrow-width approximation of the
virtual quarks, the QCD corrections to the pseudoscalar (scalar) decay
width exhibit a Coulomb singularity at the $t\bar t$ ($\tilde Q
\bar{\tilde Q}$) threshold, which is regularized by taking into account
the finite top quark and squarks decay widths \cite{agagathresh}.

The QCD corrections to the quark loops in the heavy quark limit can be
obtained by means of the low-energy theorems for scalar as well as
pseudoscalar Higgs particles, which have been discussed before. The
result for the scalar Higgs bosons agrees with the SM result of
Eq.~(\ref{eq:hgagaqcd}), and the QCD corrections to the pseudoscalar
decay mode vanish in this limit due to the Adler--Bardeen theorem
\cite{adlerbardeen}. In complete analogy to the gluonic decay mode, the
effective Lagrangian can be derived from the ABJ anomaly \cite{ABJ} and
is given to all orders of perturbation theory by \cite{higgsqcd}
\begin{equation}
{\cal L}_{eff} = -g_Q^A e_Q^2 \frac{3\alpha}{4\pi}F^{\mu\nu}
\widetilde{F}_{\mu\nu} \frac{A}{v}
\end{equation}
Since there are no effective diagrams generated by light particle
interactions that contribute to the photonic decay width at
next-to-leading order, the QCD corrections to the pseudoscalar decay
width vanish, in agreement with the explicit expansion of the massive
two-loop result. Starting at NNLO there are light quark-loop induced
contributions involving the effective $Agg$-vertex in the heavy-quark
limit.

Completely analogous the QCD corrections to the squark loops for the
scalar Higgs particles in the heavy squark limit can be obtained by the
extension of the scalar low-energy theorem to the scalar squarks. Their
coupling to photons at NLO can be described by the effective Lagrangian
\cite{hgagasq}\footnote{Taking into account the squark four-point
couplings for degenerate squark masses, too, the coefficient 3 is
changed to 10/3 \cite{SQCD, gghsqnlo, hggleffsq}.}
\begin{equation}
{\cal L}_{eff} = g_{\widetilde{Q}}^H e_{\widetilde{Q}}^2
\frac{\alpha}{8\pi}
F^{\mu\nu} F_{\mu\nu}
\frac{H}{v} \left[ 1 + 3 \frac{\alpha_s}{\pi} \right]
\end{equation}
where only non-SUSY-QCD corrections are taken into account,
i.e.~diagrams with gluino exchange or the squark four-point couplings
have been omitted. The genuine SUSY--QCD corrections can be deduced from
the analogous gluon-gluon case as described and mentioned in
Refs.~\cite{hggtsqcd1, hggtbsqcd1, hggbsqcd}. Their numerical impact,
however, has not been analysed so far.

The electroweak corrections to the pseudoscalar Higgs decays into photons
have been calculated in the limit of heavy top quarks and $W$ bosons
some time ago within the 2HDM, i.e.~involving the non-SUSY part of the
MSSM. The corrections modify the partial decay width by \cite{aggelw}
\begin{equation}
\Gamma(A\to \gamma\gamma) = \Gamma_{LO} \times\left\{1-x_t \left( 4 +
\frac{7}{\mbox{tg}^2\beta} \right) \right\}
\end{equation}
with $x_t$ defined after Eq.~(\ref{eq:hvvxt}). The correction ranges at the
per-cent level.

For large values of $\mbox{tg}\beta$ the bottom-quark loops will take
over the dominant role. In the limit of small quark masses the form
factors develop logarithmic structures,
\begin{eqnarray}
A_Q^{h/H} (\tau) & \to & -2 \frac{m_Q^2}{M_{h/H}^2}
\left[\log\frac{M_{h/H}^2}{m_Q^2}
- i\pi \right]^2 \nonumber \\
A_Q^A (\tau) & \to & -\frac{m_Q^2}{M_A^2} \left[\log\frac{M_A^2}{m_Q^2}
- i\pi \right]^2
\end{eqnarray}
so that the quark contributions to the partial decay widths of the
scalar and pseudoscalar Higgs bosons approach each other, i.e.~the
chiral symmetry is restored [the relative factor 2 is compensated by the
different global coefficients of the partial decay widths of
Eqs.~(\ref{eq:mssmhgaga}, \ref{eq:mssmagaga})]. The same is also true
for the NLO QCD corrections,
\begin{equation}
C_\phi (\tau_Q) \to C_F \left\{ -\frac{1}{24}
\left[\log\frac{M_\phi^2}{m_Q^2} - i\pi \right]^2 - \frac{1}{2}
\left[\log\frac{M_\phi^2}{m_Q^2} - i\pi \right] + \frac{3}{2}
\log\frac{\mu_R^2}{m_Q^2}\right\} \hspace*{1cm} \mbox{for $M_\phi^2 \gg
4M_Q^2$}
\end{equation}
where $\mu_R$ denotes the renormalization scale of the running quark
mass. The large logarithms are related to the Sudakov form factor
\cite{sudakov} at the virtual $\Phi b\bar b$ vertex and have been
resummed down to the subleading logarithmic level \cite{hgagares}.

\subsection{\it Higgs boson decays into photon and $Z$ boson and Dalitz decays}
\subsubsection{\it Standard Model}
\paragraph{On-shell $Z$ boson.}
\begin{figure}[hbt]
\begin{center}
\setlength{\unitlength}{1pt}
\begin{picture}(400,100)(0,0)

\Photon(60,20)(90,20){-3}{4}
\Photon(60,80)(90,80){3}{4}
\ArrowLine(60,20)(60,80)
\ArrowLine(60,80)(30,50)
\ArrowLine(30,50)(60,20)
\DashLine(0,50)(30,50){5}
\put(-15,46){$H$}
\put(70,46){$f$}
\put(95,18){$\gamma$}
\put(95,78){$Z$}

\Photon(210,20)(240,20){-3}{4}
\Photon(210,80)(240,80){3}{4}
\Photon(210,80)(210,20){3}{7}
\Photon(210,20)(180,50){3}{5}
\Photon(180,50)(210,80){3}{5}
\DashLine(150,50)(180,50){5}
\put(135,46){$H$}
\put(220,46){$W$}
\put(245,18){$\gamma$}
\put(245,78){$Z$}

\DashLine(300,50)(330,50){5}
\PhotonArc(345,50)(15,0,180){3}{6}
\PhotonArc(345,50)(15,180,360){3}{6}
\Photon(360,50)(390,80){3}{5}
\Photon(360,50)(390,20){3}{5}
\put(285,46){$H$}
\put(335,70){$W$}
\put(395,18){$\gamma$}
\put(395,78){$Z$}

\end{picture}  \\
\setlength{\unitlength}{1pt}
\caption{\label{fg:hzgalodia} \it Typical diagrams contributing to
$H\to Z \gamma$ at lowest order.}
\end{center}
\end{figure}
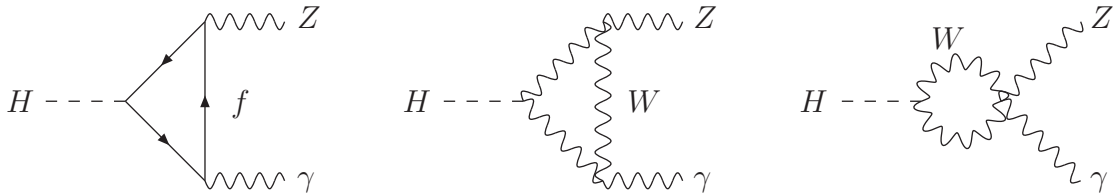
\noindent
The decay of the Higgs boson into a photon and a $Z$ boson proceeds via
$W$ and heavy fermion loops, see Fig.~\ref{fg:hzgalodia}. The partial
decay width can be obtained at leading order as \cite{hunter,hzga}
\begin{eqnarray}
\Gamma\, [H\to Z\gamma] = \frac{G^2_{F}M_W^2\, \alpha\,M_{H}^{3}}
{64\,\pi^{4}} \left( 1-\frac{M_Z^2}{M_H^2} \right)^3 \left| \sum_{f}
A_f^H(\tau_f,\lambda_f) + A^H_W(\tau_W,\lambda_W) \right|^2
\label{eq:hzga}
\end{eqnarray}
with the $W$ and fermion form factors
\begin{eqnarray}
A_f^H (\tau,\lambda) & = & 2 N_{cf} \frac{e_f (I_{3f} -
2e_f\sin^2\theta_W )}
{\cos\theta_W} \left[I_1(\tau,\lambda) - I_2(\tau,\lambda)
\right] \nonumber \\
A_W^H (\tau,\lambda) & = & \cos \theta_W \left\{ 4(3-\tan^2\theta_W)
I_2(\tau,\lambda) \right. \nonumber \\
& & \left. + \left[ \left(1+\frac{2}{\tau}\right) \tan^2\theta_W
- \left(5+\frac{2}{\tau} \right) \right] I_1(\tau,\lambda) \right\}
\label{eq:hzgaform}
\end{eqnarray}
The functions $I_1,I_2$ are defined as
\begin{eqnarray*}
I_1(\tau,\lambda) & = & \frac{\tau\lambda}{2(\tau-\lambda)} +
\frac{\tau^2\lambda^2}{2(\tau-\lambda)^2} \left[ f(\tau) - f(\lambda)
\right] + \frac{\tau^2\lambda}{(\tau-\lambda)^2} \left[ g(\tau) -
g(\lambda) \right] \\
I_2(\tau,\lambda) & = & - \frac{\tau\lambda}{2(\tau-\lambda)}\left[
f(\tau)
- f(\lambda) \right]
\end{eqnarray*}
where the function $g(\tau)$ is given by
\begin{equation}
g(\tau) = \left\{ \begin{array}{ll}
\displaystyle \sqrt{\tau-1} \arcsin \frac{1}{\sqrt{\tau}} & \tau \ge 1
\\
\displaystyle \frac{\sqrt{1-\tau}}{2} \left[ \log \frac{1+\sqrt{1-\tau}}
{1-\sqrt{1-\tau}} - i\pi \right] & \tau < 1
\end{array} \right.
\label{eq:gtau}
\end{equation}
and $f(\tau)$ is defined in Eq.~(\ref{eq:ftau}).  The parameters
$\tau_i= 4M_i^2/M_H^2$ and $\lambda_i= 4M_i^2/M_Z^2~~(i=f,W)$ are
defined in terms of the corresponding masses of the heavy loop
particles.  Due to charge conjugation invariance, only the vectorial $Z$
coupling contributes to the fermion loop so that problems with the axial
$\gamma_5$ coupling do not arise.  The $W$ loop provides the dominant
contribution in the intermediate Higgs mass range, and the heavy fermion
loops interfere destructively.

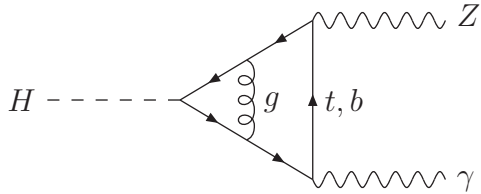
\begin{figure}[hbt]
\begin{center}
\setlength{\unitlength}{1pt}
\begin{picture}(200,100)(0,0)

\DashLine(0,50)(50,50){5}
\ArrowLine(100,20)(100,80)
\ArrowLine(100,80)(75,65)
\ArrowLine(75,65)(50,50)
\ArrowLine(50,50)(75,35)
\ArrowLine(75,35)(100,20)
\Gluon(75,65)(75,35){3}{3}
\Photon(100,20)(150,20){-3}{6}
\Photon(100,80)(150,80){3}{6}
\put(-15,46){$H$}
\put(105,46){$t,b$}
\put(82,48){$g$}
\put(155,18){$\gamma$}
\put(155,78){$Z$}

\end{picture}  \\
\setlength{\unitlength}{1pt}
\caption{\label{fg:hzgaqcddia} \it Typical diagram contributing to
the QCD corrections to $H\to Z \gamma$.}
\end{center}
\end{figure}
The two-loop QCD corrections to the top quark loops have been calculated
numerically \cite{hzgaqcd} and analytically very recently
\cite{hzgaqcd2}. They are generated by virtual gluon exchange inside the
quark triangle (see Fig.~\ref{fg:hzgaqcddia}).  Due to charge
conjugation invariance and color conservation, radiation of a single
gluon is not possible. Hence the QCD corrections can simply be expressed
as a rescaling of the lowest order amplitude by a factor that only
depends on the ratios $\tau_i$ and $\lambda_i~~(i=f,W)$, defined above:
\begin{eqnarray}
A_Q^H(\tau_Q,\lambda_Q) & \to & A_Q^H(\tau_Q,\lambda_Q) \times \left[1+
D_H(\tau_Q,\lambda_Q) \frac{\alpha_s}{\pi} \right] \nonumber \\
D_H(\tau_Q,\lambda_Q) & \to & -1 \hspace*{1cm} \mbox{for $M_Z^2 \ll
M_H^2\ll 4M_Q^2$}
\label{eq:hzgaqcd}
\end{eqnarray}
In the limit $M_Z\to 0$ the quark amplitude approaches the corresponding
form factor of the photonic decay mode ({\it modulo} couplings), which
has been discussed before.  Hence the QCD correction in the heavy quark
limit for small $Z$ masses has to coincide with the heavy quark limit of
the photonic decay mode of Eq.~(\ref{eq:hgagaqcd}). The QCD corrections
for finite Higgs, $Z$ and quark masses are presented in \cite{hzgaqcd,
hzgaqcd2} as a function of the Higgs mass. They amount to less than
0.3\% in the intermediate mass range, where this decay mode is relevant,
and can thus be neglected. More important is the inclusion of off-shell
effects of the final-state $Z$ boson due to its finite width. This
requires the analysis of Dalitz decays $H\to\gamma f\bar f$.

\paragraph{Dalitz decays.}
\begin{figure}[hbt]
\setlength{\unitlength}{1pt}
\begin{picture}(140,70)(-200,0)
\DashLine(20,50)(50,50){5}
\ArrowLine(50,50)(75,65)
\ArrowLine(75,65)(100,80)
\ArrowLine(100,20)(50,50)
\Photon(75,65)(100,50){-3}{3}
\put(5,46){$H$}
\put(105,78){$f$}
\put(105,18){$\bar f$}
\put(105,48){$\gamma$}
\end{picture} \\[0.5cm]
\begin{picture}(140,100)(0,0)
\DashLine(20,50)(50,50){5}
\Photon(50,50)(100,80){-3}{6}
\Photon(100,80)(100,20){-3}{6}
\Photon(100,20)(50,50){-3}{6}
\Photon(100,20)(150,20){-3}{6}
\Photon(100,80)(125,80){-3}{3}
\ArrowLine(125,80)(150,100)
\ArrowLine(150,60)(125,80)
\put(5,46){$H$}
\put(55,65){$W$}
\put(155,98){$f$}
\put(155,58){$\bar f$}
\put(155,18){$\gamma$}
\put(100,90){$\gamma^\ast,Z^\ast$}
\end{picture}
\begin{picture}(140,100)(-30,0)
\DashLine(20,50)(50,50){5}
\PhotonArc(75,50)(25,0,180){3}{9}
\PhotonArc(75,50)(25,180,360){3}{9}
\Photon(100,50)(125,65){3}{3}
\Photon(100,50)(150,20){3}{5}
\ArrowLine(125,65)(150,80)
\ArrowLine(150,50)(125,65)
\put(5,46){$H$}
\put(65,80){$W$}
\put(98,70){$\gamma^\ast,Z^\ast$}
\put(155,78){$f$}
\put(155,48){$\bar f$}
\put(155,18){$\gamma$}
\end{picture}
\begin{picture}(140,100)(-60,0)
\DashLine(20,50)(50,50){5}
\ArrowLine(50,50)(100,80)
\ArrowLine(100,80)(100,20)
\ArrowLine(100,20)(50,50)
\Photon(100,20)(150,20){-3}{6}
\Photon(100,80)(125,80){-3}{3}
\ArrowLine(125,80)(150,100)
\ArrowLine(150,60)(125,80)
\put(5,46){$H$}
\put(55,75){$t,b,\ldots$}
\put(155,98){$f$}
\put(155,58){$\bar f$}
\put(155,18){$\gamma$}
\put(100,90){$\gamma^\ast,Z^\ast$}
\end{picture} \\[0.5cm]
\begin{picture}(140,100)(0,0)
\DashLine(20,50)(50,50){5}
\Photon(50,50)(100,80){-3}{6}
\Photon(100,50)(100,20){-3}{3}
\Photon(100,20)(50,50){-3}{6}
\Photon(100,20)(150,20){-3}{6}
\ArrowLine(100,80)(150,80)
\ArrowLine(100,50)(100,80)
\ArrowLine(150,50)(100,50)
\put(5,46){$H$}
\put(55,65){$W$}
\put(155,78){$f$}
\put(155,48){$\bar f$}
\put(155,18){$\gamma$}
\end{picture}
\begin{picture}(140,100)(-30,0)
\DashLine(20,50)(50,50){5}
\Photon(50,50)(100,80){-3}{6}
\Photon(100,20)(50,50){-3}{6}
\ArrowLine(100,80)(150,80)
\ArrowLine(100,50)(100,80)
\ArrowLine(100,20)(100,50)
\ArrowLine(150,20)(100,20)
\Photon(100,50)(150,50){-3}{6}
\put(5,46){$H$}
\put(55,65){$Z$}
\put(155,78){$f$}
\put(155,18){$\bar f$}
\put(155,48){$\gamma$}
\end{picture}
\begin{picture}(140,100)(-60,0)
\DashLine(20,50)(50,50){5}
\Photon(50,50)(100,80){-3}{6}
\Photon(100,20)(50,50){-3}{6}
\ArrowLine(100,80)(125,80)
\ArrowLine(125,80)(150,60)
\ArrowLine(100,20)(100,80)
\ArrowLine(150,20)(100,20)
\Photon(125,80)(150,100){-3}{3}
\put(5,46){$H$}
\put(55,75){$W,Z$}
\put(155,58){$f$}
\put(155,18){$\bar f$}
\put(155,98){$\gamma$}
\end{picture}
\setlength{\unitlength}{1pt}
\caption{\label{fg:dalitzdia} \it Generic diagrams contributing to
the Dalitz decays $H\to \gamma f\bar f$.}
\end{figure}
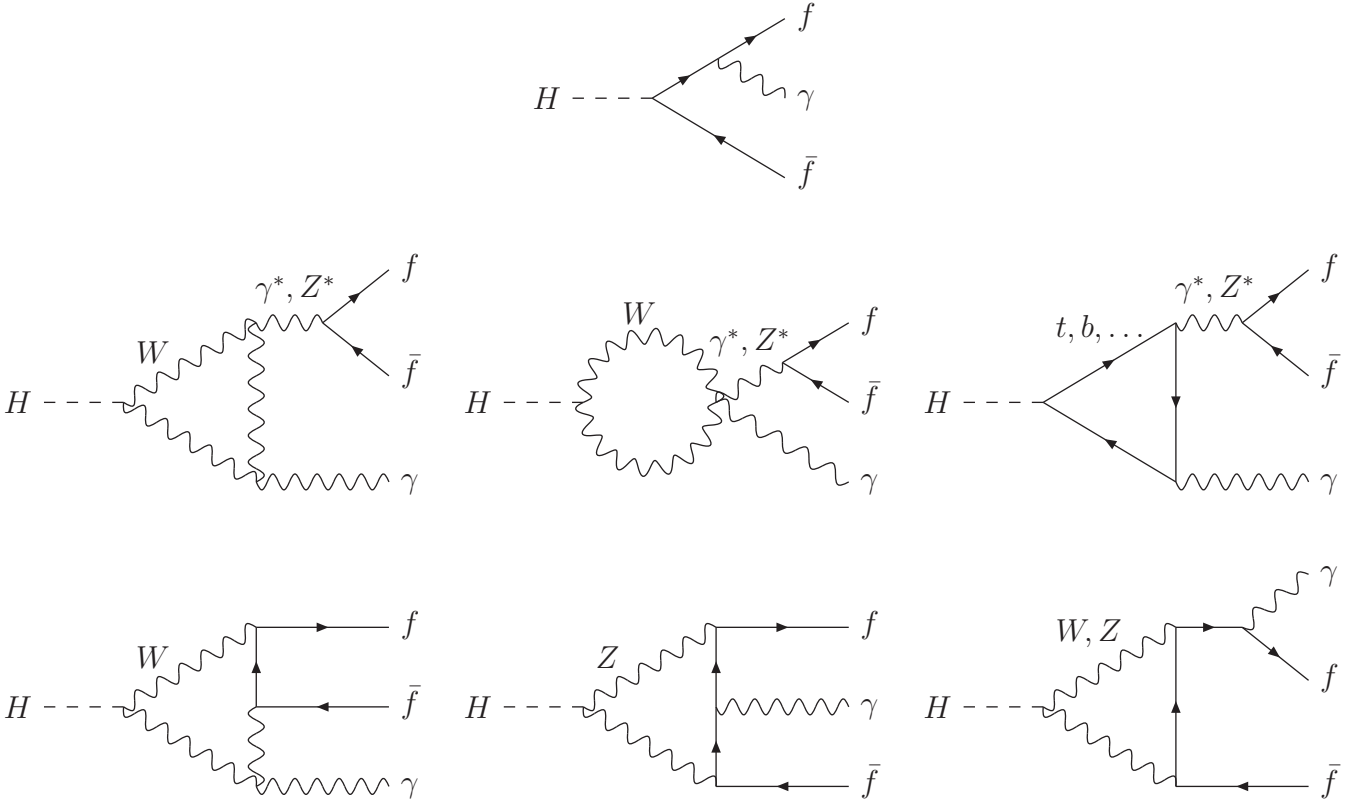
Dalitz decays of the Higgs boson \cite{dalitz1,dalitz} are described by
the diagrams of Fig.~\ref{fg:dalitzdia}. The first diagram corresponds
to the direct contributions, since these are mediated by the Yukawa
coupling of the fermion $f$. The second line shows the triangle diagrams
which include the resonant $H\to \gamma Z\to \gamma f\bar f$
contribution as well as photon conversion $H\to\gamma\gamma^\ast\to
\gamma f\bar f$. The leading contributions to the triangles emerge from
$W$, top and bottom loops. The last line exhibits box contributions and
final-state photon radiation which, however, are numerically suppressed
compared to the triangle diagrams, but are required for a
gauge-invariant result.

\begin{figure}[hbtp]

\begin{picture}(400,500)(0,0)
\put(400,450){$H\to \gamma e^+e^-$}
\put(400,275){$H\to \gamma \mu^+\mu^-$}
\put(400,100){$H\to \gamma \tau^+\tau^-$}
\put(100,350){\epsfxsize=10cm \epsfbox{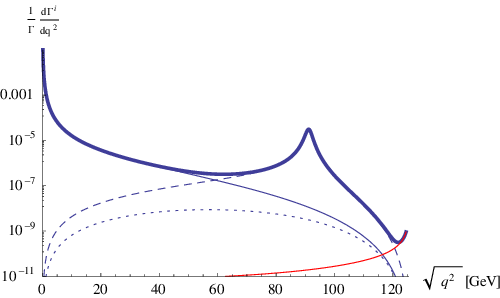}}
\put(100,175){\epsfxsize=10cm \epsfbox{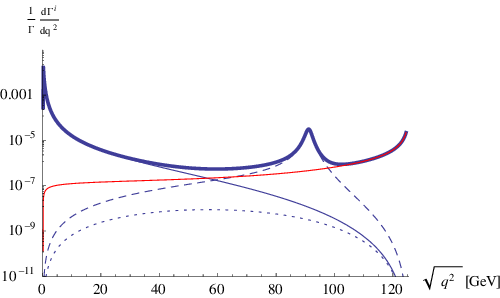}}
\put(100,00){\epsfxsize=10cm \epsfbox{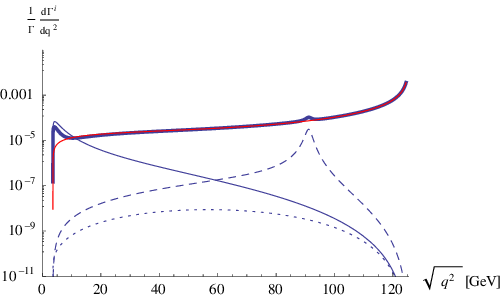}}
\end{picture}

\caption{\label{fg:dalitz} \it The invariant mass distributions in
$\sqrt{q^2} = M_{\ell^+\ell^-}$ of the Dalitz decays $H\rightarrow
\gamma + e^+ e^-/\mu^+\mu^-/\tau^+\tau^-$ normalized to
$\Gamma(H\rightarrow \gamma\gamma)$ with a cut $E_\gamma > 1$ GeV on the
photon energy. The red lines show the contribution of the tree diagrams,
the thin solid lines denote the contribution of the photon conversion
$H\to \gamma\gamma^*\to \gamma\ell^+\ell^-$, and the dashed line the
contribution from the $Z^*$ exchange diagrams, while the thick lines
present the total contributions. The dotted lines denote the
contribution from the box diagrams (in 't Hooft--Feynman gauge). From
Ref.~\cite{dalitz}.}
\end{figure}
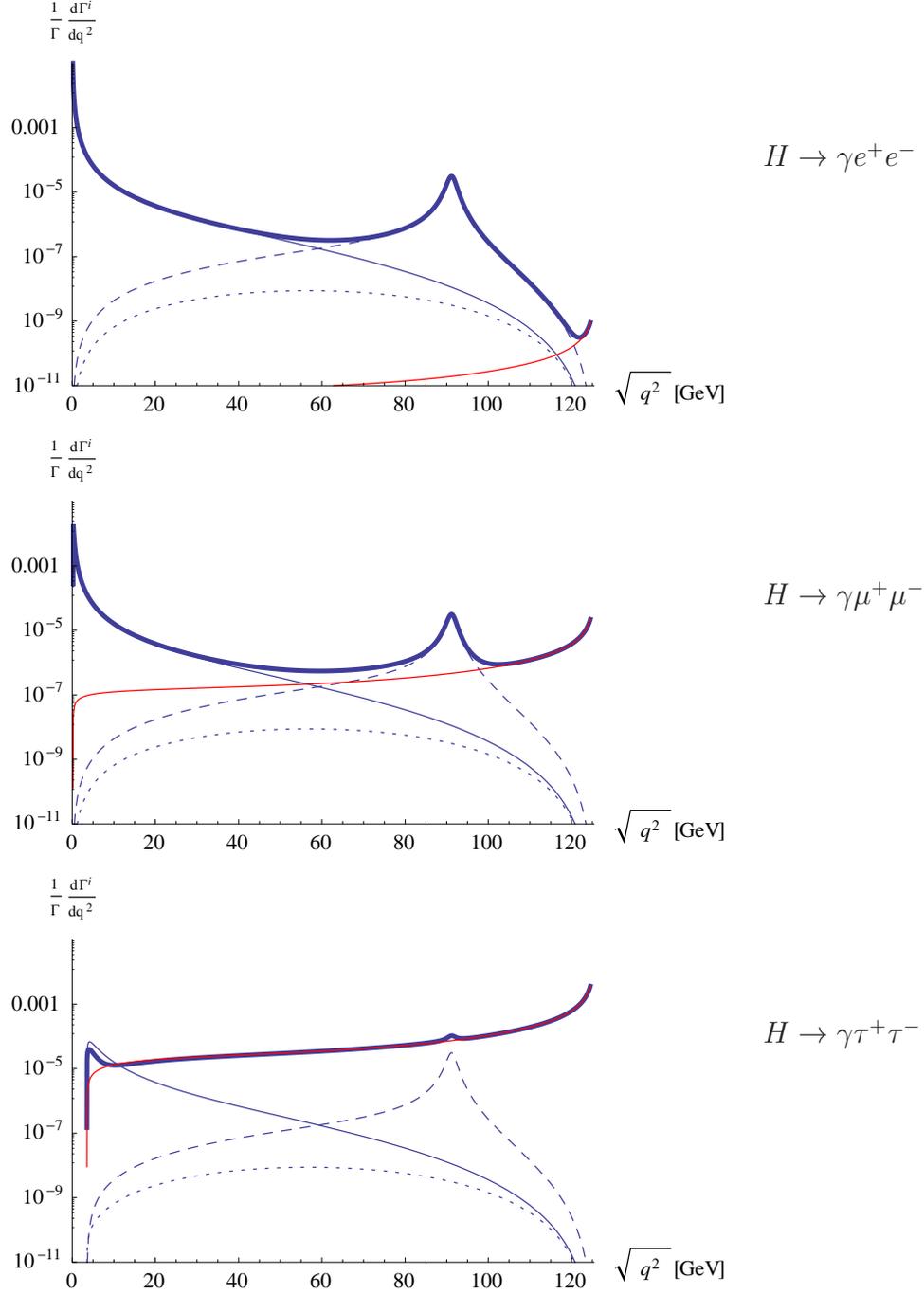
Fig.~\ref{fg:dalitz} shows the invariant mass distributions for $e^+e^-,
\mu^+\mu^-$ and $\tau^+\tau^-$ pairs including the individual
contributions of the different types of diagrams normalized to the
partial decay width into photons. A lower cut of $E_\gamma> 1$ GeV has
been imposed on the photon energy to avoid the infrared singularity of
photon radiation. For small invariant masses the photon conversion
$H\to\gamma\gamma^\ast\to \gamma \ell^+\ell^-$ provides the dominant
contribution, while for invariant masses around the $Z$-boson mass the
$Z$-boson contribution $H\to \gamma Z^\ast\to \gamma f\bar f$ takes over the
dominant role. At the end-point $q^2\lsim M_H^2$ of the spectrum the
direct contribution determines the distributions.  This rises with
growing Yukawa coupling, i.e.~it is largest for $H\to \gamma
\tau^+\tau^-$ (where it dominates in the whole $q^2$-range). For a clean
separation of the $H\to\gamma\gamma$, $H\to\gamma\gamma^\ast\to \gamma
\ell^+\ell^-$, $H\to Z\gamma$ and $H\to \ell^+\ell^-$ contributions
appropriate cuts have to be implemented for the Dalitz decays. The
low-$q^2$ part has to be attributed to $H\to\gamma\gamma$, the
$q^2$-part around $M_Z^2$ to $H\to Z\gamma$ and the end-point region
close to $M_H^2$ to the QED corrections to $H\to\ell^+\ell^-$.

\subsubsection{\it Minimal supersymmetric extension}
\paragraph{On-shell $Z$ boson.}
\begin{figure}[hbt]
\begin{center}
\setlength{\unitlength}{1pt}
\begin{picture}(400,100)(0,0)

\Photon(60,20)(90,20){-3}{4}
\Photon(60,80)(90,80){3}{4}
\ArrowLine(60,20)(60,80)
\ArrowLine(60,80)(30,50)
\ArrowLine(30,50)(60,20)
\DashLine(0,50)(30,50){5}
\put(-40,46){$h,H,A$}
\put(65,46){$f,\tilde \chi^\pm$}
\put(95,78){$Z$}
\put(95,18){$\gamma$}

\Photon(210,20)(240,20){-3}{4}
\Photon(210,80)(240,80){3}{4}
\DashLine(210,80)(210,20){5}
\DashLine(210,20)(180,50){5}
\DashLine(180,50)(210,80){5}
\DashLine(150,50)(180,50){5}
\put(125,46){$h,H$}
\put(220,46){$W, H^\pm, \widetilde{f}$}
\put(245,78){$Z$}
\put(245,18){$\gamma$}

\DashLine(300,50)(330,50){5}
\DashCArc(345,50)(15,0,180){4}
\DashCArc(345,50)(15,180,360){4}
\Photon(360,50)(390,80){3}{5}
\Photon(360,50)(390,20){3}{5}
\put(275,46){$h,H$}
\put(325,70){$W, H^\pm, \widetilde{f}$}
\put(395,78){$Z$}
\put(395,18){$\gamma$}

\end{picture}  \\
\setlength{\unitlength}{1pt}
\caption{\label{fg:mssmhzgalodia} \it Typical diagrams contributing
to $\Phi \to Z \gamma$ at lowest order.}
\end{center}
\end{figure}
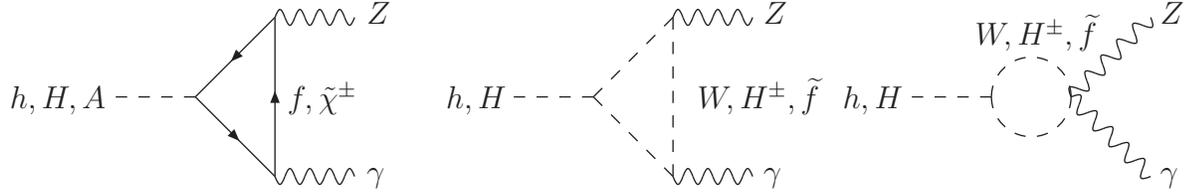
\noindent
The decays of the scalar Higgs bosons into a $Z$ boson and photon are
mediated by $W$ and heavy fermion loops as in the Standard Model and, in
addition, by charged Higgs, sfermion and chargino loops as shown in
Fig.~\ref{fg:mssmhzgalodia}. The partial decay widths are given by
\cite{higgsqcd,hzgamssm}
\begin{eqnarray}
\Gamma\, [h/H\to Z\gamma] & = & \frac{G^2_{F}M_W^2\,
\alpha\,M_{h/H}^{3}}
{64\,\pi^{4}} \left(1-\frac{M_Z^2}{M_{h/H}^2} \right)^3
\left|\sum_{f}g_f^{h/H}
A_f^{h/H}(\tau_f,\lambda_f)
\right. \nonumber \\
& & \left. + g^{h/H}_W A^{h/H}_W(\tau_W,\lambda_W)
+ g_{H^\pm}^{h/H} A_{H^\pm}^{h/H}(\tau_{H^\pm},\lambda_{H^\pm})
\right. \nonumber \\
& & \left. + \sum_{\tilde \chi^\pm_i, \tilde \chi^\mp_j}
g_{\tilde \chi^\pm_i \tilde \chi^\mp_j}^{h/H}
g_{\tilde \chi^\mp_i \tilde \chi^\pm_j}^Z
A_{\tilde \chi^\pm_i \tilde \chi^\mp_j}^{h/H}
+ \sum_{\tilde f_i, \tilde f_j} g_{\tilde f_i \tilde f_j}^{h/H}
g_{\tilde f_i \tilde f_j}^Z A_{\tilde f_i \tilde f_j}^{h/H} \right|^2
\end{eqnarray}
with the form factors $A_f^{h/H},A_W^{h/H}$ of Eq.~(\ref{eq:hzgaform}), and
\begin{equation}
A_{H^\pm}^{h/H} (\tau,\lambda) = \frac{\cos 2\theta_W}{\cos\theta_W}
I_1(\tau,\lambda)
\end{equation}
where the function $I_1(\tau,\lambda)$ is defined after
Eq.~(\ref{eq:hzgaform}).

The $Z \gamma$ decay mode of the pseudoscalar Higgs boson is generated
by heavy charged-fermion and chargino loops, see
Fig.~\ref{fg:mssmhzgalodia}. The partial decay width reads \cite{hzgamssm}
\begin{equation}
\Gamma(A \rightarrow Z\gamma)=
\frac{G^2_F M_W^2 \alpha M_A^3}{64\pi^4} \left(1-\frac{M_Z^2}{M_A^2}
\right)^3
\left| \sum_f g_f^A A_f^A (\tau_f,\lambda_f)
+ \sum_{\tilde \chi^\pm_i, \tilde \chi^\mp_j}
g_{\tilde \chi^\pm_i \tilde \chi^\mp_j}^A
g_{\tilde \chi^\mp_i \tilde \chi^\pm_j}^Z
A_{\tilde \chi^\pm_i \tilde \chi^\pm_j}^A \right|^2
\end{equation}
with the fermion amplitudes
\begin{equation}
A_f^A (\tau,\lambda) = 2 N_{cf} \frac{e_f (I_{3f} - 2e_f\sin^2\theta_W
)}
{\cos\theta_W}~I_2(\tau,\lambda)
\end{equation}
The contributions of charginos and sfermions involve mixing terms. Their
analytical expressions can be found in \cite{hzgamssm}.  For large
loop-particle masses and small $Z$-boson mass, the form factors approach
the photonic amplitudes {\it modulo} couplings.  The parameters $\tau_i=
4M_i^2/M_\Phi^2, \lambda_i= 4M_i^2/M_Z^2~~(i=f,W,H^\pm, \tilde
\chi^\pm,\tilde f)$ are defined by the corresponding heavy loop-particle
masses and the non-mixing MSSM couplings
$g^\phi_{f,W,H^\pm,\tilde\chi^\pm,\tilde f}$ are summarized in Tables
\ref{tb:hcoup} and \ref{tb:hsqcoup}, while the mixing and $Z$ boson
couplings $g_i^Z$ can be found in \cite{hunter}. The branching ratios of
the $Z\gamma$ decay modes range at a level of up to a few $10^{-4}$ in
the intermediate mass ranges of the Higgs bosons and are thus
phenomenologically less important in the MSSM. Dalitz decays have not
been studied within the MSSM framework. They involve additional
contributions from charged Higgs bosons, charginos and sfermions in the
loop contributions including the box diagrams.

\subsection{\it Supersymmetric Higgs boson decays into Higgs particles}
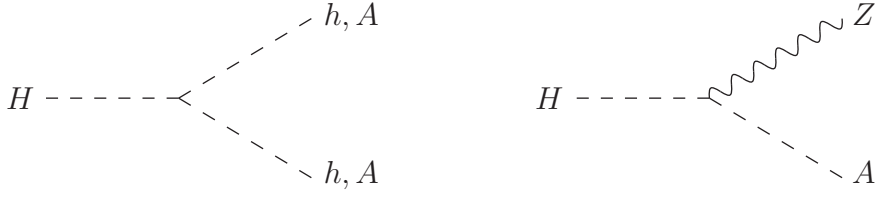
\begin{figure}[hbt]
\begin{center}
\setlength{\unitlength}{1pt}
\begin{picture}(350,100)(0,0)

\DashLine(0,50)(50,50){5}
\DashLine(50,50)(100,80){5}
\DashLine(50,50)(100,20){5}
\put(-15,46){$H$}
\put(105,18){$h,A$}
\put(105,78){$h,A$}

\DashLine(200,50)(250,50){5}
\Photon(250,50)(300,80){3}{6}
\DashLine(250,50)(300,20){5}
\put(185,46){$H$}
\put(305,18){$A$}
\put(305,78){$Z$}

\end{picture}  \\
\setlength{\unitlength}{1pt}
\caption{\label{fg:haadia} \it Typical diagrams contributing to Higgs
decays with Higgs bosons in the final state.}
\end{center}
\end{figure}
The heavy scalar Higgs boson can decay into pairs of light scalar as
well as pseudoscalar Higgs bosons, see Fig.~\ref{fg:haadia}. At LO the
partial decay widths are given by \cite{hunter}
\begin{eqnarray}
\Gamma(H \to hh) & = & \lambda_{Hhh}^2 \frac{G_F M_Z^4}{16\sqrt{2}\pi
M_H} \sqrt{1-4\frac{M_h^2}{M_H^2}} \\
\Gamma(H \to AA) & = & \lambda_{HAA}^2 \frac{G_F M_Z^4}{16\sqrt{2}\pi
M_H} \sqrt{1-4\frac{M_A^2}{M_H^2}}
\end{eqnarray}
The self-couplings $\lambda_{Hhh}$ and $\lambda_{HAA}$ can be derived
from the effective Higgs potential \cite{lambdanlo, lambdannlo}.  The
decay mode into pseudoscalar particles is restricted to small regions of
the MSSM parameter space, where the pseudoscalar mass $M_A$ is small.
The decay into light scalar bosons is dominant for small
$\mbox{tg$\beta$}$ below the $t\bar t$ threshold. The SUSY-electroweak
corrections to these decays have been calculated \cite{h2hh}. They
modify the partial decay widths by a moderate amount in general
depending on the MSSM scenario if the radiatively improved trilinear
Higgs couplings and masses are used, absorbing in this way the bulk of
the corrections in terms of effective parameters. Only in parameter
regions where the LO decay widths are suppressed the radiative
corrections can be larger, but there these decay modes are
phenomenologically irrelevant.

The contributions of final states containing off-shell scalar or
pseudoscalar Higgs bosons may be significant and are thus included in
{\sc Hdecay} \cite{hdecay} and {\sc FeynHiggs} \cite{feynhiggs}. Their
expressions read as \cite{1OFF}
\begin{eqnarray}
\Gamma(H\to \phi\phi^*) & = & \lambda_{H\phi\phi}^2
g_{\phi bb}^2 m_b^2 \frac{3G_F^2M_Z^4}{32\pi^3M_H}
\left\{ (\kappa_\phi -1) \left(2-\frac{1}{2}\log
\kappa_\phi \right) \right. \nonumber \\
& & \left. + \frac{1-5\kappa_\phi}{\sqrt{4\kappa_\phi-1}} \left( \arctan
\frac{2\kappa_\phi-1}{\sqrt{4\kappa_\phi-1}} - \arctan
\frac{1}{\sqrt{4\kappa_\phi-1}} \right) \right\}
\end{eqnarray}
where $\kappa_\phi = M_\phi^2 / M_H^2$. They slightly extend the
regions, where the $hh,AA$ decay modes of the heavy scalar Higgs boson
$H$ are sizeable. Analogous to the decays into $W$ and $Z$ boson pairs
double off-shell decays play a role in some regions of the MSSM
parameter space below the corresponding thresholds. Neglecting the
final-state fermion masses they can be cast into the form
\cite{nmssmcalc}
\begin{equation}
\Gamma(H\to \phi^*\phi^*) = \frac{1}{\pi^2}\int_0^{M_H^2} \frac{dQ_1^2~
M_\phi \Gamma_\phi} {(Q_1^2 - M_\phi^2)^2 + M_\phi^2 \Gamma_\phi^2}
\int_0^{(M_H-Q_1)^2} \frac{dQ_2^2~M_\phi \Gamma_\phi} {(Q_2^2 -
M_\phi^2)^2 + M_\phi^2 \Gamma_\phi^2}~\Gamma_0
\label{eq:hhhoff}
\end{equation}
with $Q_1^2, Q_2^2$ being the squared invariant masses of the virtual Higgs
bosons, $M_\phi$ and $\Gamma_\phi$ their masses and total decay widths;
$\Gamma_0$ is given by \cite{nmssmcalc}
\begin{equation}
\Gamma_0 = \lambda^2_{H\phi\phi}~\frac{G_F M_Z^4}{16\sqrt{2}\pi M_H}
\sqrt{\lambda(Q_1^2,Q_2^2;M_H^2)} \frac{Q_1^2Q_2^2}{M_\phi^4}
\label{eq:h2hh}
\end{equation}
with the phase-space factor $\lambda(x,y;z)$ of Eq.~(\ref{eq:kallen}).

Moreover, Higgs bosons can decay into a gauge and a Higgs boson, see
Fig.~\ref{fg:haadia}. The various partial widths can be expressed as
\begin{eqnarray}
\Gamma (H\to AZ) & = & \lambda_{HAZ}^2 \frac{G_F M_H^3}{8\sqrt{2}\pi}
\lambda^{3/2}(M_A^2,M_Z^2;M_H^2) \\
\Gamma (H\to H^\pm W^\mp) & = & \lambda_{HH^+W}^2 \frac{G_F
M_H^3}{8\sqrt{2}\pi} \lambda^{3/2}(M_{H^\pm}^2,M_W^2;M_H^2) \\
\Gamma (A\to hZ) & = & \lambda_{hAZ}^2 \frac{G_F M_A^3}{8\sqrt{2}\pi}
\lambda^{3/2}(M_h^2,M_Z^2;M_A^2) \\
\Gamma (H^+\to h W^+) & = & \lambda_{hH^+W}^2 \frac{G_F
M_{H^\pm}^3}{8\sqrt{2}\pi} \lambda^{3/2}(M_h^2,M_W^2;M_{H^\pm}^2)
\end{eqnarray}
where the couplings $\lambda_{ijk}$ can be determined from the effective
Higgs potential \cite{lambdanlo, lambdannlo}.  The branching ratios of
these decay modes may be sizeable in specific regions of the MSSM
parameter space. The electroweak corrections within the 2HDM framework
have been calculated \cite{h2hv}.  They are of moderate size if the
radiatively corrected Higgs couplings and masses are used at tree-level.
The genuine SUSY-electroweak corrections are only known for the
effective parameters $\lambda_{ijk}$ and the Higgs masses.

Below-threshold decays into a Higgs particle and an off-shell gauge
boson turn out to be very important for the heavy Higgs bosons of the
MSSM. The individual contributions read as \cite{1OFF, nmssmcalc}
\begin{eqnarray}
\Gamma (H\to AZ^*) & = & \lambda_{HAZ}^2 \delta'_Z
\frac{9G_F^2M_Z^4M_H}{8\pi^3} G_{AZ} \\
\Gamma (H\to H^\pm W^{\mp*}) & = & \lambda_{HH^\pm W}^2
\frac{9G_F^2M_W^4M_H} {16\pi^3} G_{H^\pm W} \\
\Gamma (A\to hZ^*) & = & \lambda_{hAZ}^2 \delta'_Z
\frac{9G_F^2M_Z^4M_A}{8\pi^3} G_{hZ} \\
\Gamma (H^+\to h W^{+*}) & = & \lambda_{hH^\pm W}^2
\frac{9G_F^2M_W^4M_{H^\pm}} {16\pi^3} G_{h W} \\
\Gamma (H^+\to A W^{+*}) & = & \frac{9G_F^2M_W^4M_{H^\pm}}{16\pi^3}
G_{A W}
\end{eqnarray}
with the generic functions $G_{ij}$
\begin{eqnarray}
G_{ij} & = & \frac{1}{4} \left\{
2(-1+\kappa_j-\kappa_i)\sqrt{\lambda_{ij}} \left[ \frac{\pi}{2} +
\arctan \left(\frac{\kappa_j (1-\kappa_j+\kappa_i) -
\lambda_{ij}}{(1-\kappa_i) \sqrt{\lambda_{ij}}} \right) \right] \right. \\
& & \left. + (\lambda_{ij}-2\kappa_i) \log \kappa_i + \frac{1}{3}
(1-\kappa_i) \left[ 5(1+\kappa_i) - 4\kappa_j + \frac{2}{\kappa_j}
\lambda_{ij} \right] \right\}
\end{eqnarray}
using the parameters
\begin{equation}
\lambda_{ij} = -1+2\kappa_i+2\kappa_j-(\kappa_i-\kappa_j)^2,
\hspace{2cm}
\kappa_i = \frac{M_i^2}{M_\phi^2}
\end{equation}
with $\delta'_Z = 7/12 - 10\sin^2\theta_W/9 + 40 \sin^4\theta_W/27$.
Off-shell $hZ^*$ decays are important for the pseudoscalar Higgs boson
for masses above about 130 GeV for small $\mbox{tg$\beta$}$ \cite{1OFF}.
The decay modes $H^\pm \to hW^*, AW^*$ can reach branching ratios of several
tens of per cent and lead to a significant reduction of the dominant
branching ratio into $\tau\nu$ final states to a level of 60--70\% for
small $\mbox{tg$\beta$}$ \cite{1OFF}. This analysis can be extended to
the double off-shell cases analogous to the previous decay modes into
Higgs-boson pairs. If final-state fermion masses are neglected the
corresponding expressions are given by
\begin{equation}
\Gamma(\Phi\to \phi^*V^*) = \frac{1}{\pi^2}\int_0^{M_\Phi^2} \frac{dQ_1^2~
M_\phi \Gamma_\phi} {(Q_1^2 - M_\phi^2)^2 + M_\phi^2 \Gamma_\phi^2}
\int_0^{(M_\Phi-Q_1)^2} \frac{dQ_2^2~M_V \Gamma_V} {(Q_2^2 -
M_V^2)^2 + M_V^2 \Gamma_V^2}~\Gamma_0
\label{eq:hhvoff}
\end{equation}
with $Q_1^2, Q_2^2$ being the squared invariant masses of the virtual
Higgs and gauge bosons, $M_\phi, M_V$ and $\Gamma_\phi, \Gamma_V$ their
masses and total decay widths; $\Gamma_0$ is given by
\begin{equation}
\Gamma_0 = \lambda^2_{\Phi\phi V}~\frac{G_F
M_\Phi^3}{8\sqrt{2}\pi}~\lambda^{3/2}(Q_1^2,Q_2^2;M_\Phi^2)
\label{eq:h2hv}
\end{equation}
with the phase-space factor $\lambda(x,y;z)$ of Eq.~(\ref{eq:kallen}).

\subsection{\it Supersymmetric Higgs boson decays}
\paragraph{Sfermion masses and couplings.}
The scalar partners $\tilde f_{L,R}$ of the left- and right-handed
fermion components mix with each other. The mass eigenstates $\tilde
f_{1,2}$ of the sfermions $\tilde f$ are related to the current
eigenstates $\tilde f_{L,R}$ by mixing angles $\theta_f$,
\begin{eqnarray}
\tilde f_1 & = & \tilde f_L \cos\theta_f + \tilde f_R \sin \theta_f
\nonumber \\
\tilde f_2 & = & -\tilde f_L\sin\theta_f + \tilde f_R \cos \theta_f
\label{eq:sfmix}
\end{eqnarray}
which are proportional to the masses of the related fermions. Thus
mixing effects are only important for the third-generation sfermions
$\tilde t, \tilde b, \tilde \tau$, the mass matrix of which is given by
\cite{mssmbase}
\begin{equation}
{\cal M}_{\tilde f} = \left[ \begin{array}{cc}
\tilde{M}_{\tilde f_L}^2 + M_f^2 & M_f (A_f-\mu r_f) \\
M_f (A_f-\mu r_f) & \tilde{M}_{\tilde f_R}^2 + M_f^2
\end{array} \right]
\end{equation}
with the parameters $r_b = r_\tau = 1/r_t = \mbox{tg$\beta$}$. The
parameters $A_f$ denote the trilinear Yukawa mixing parameters of the soft
supersymmetry breaking part of the Lagrangian and $\tilde{M}_{\tilde
f_{L/R}}$ contain the soft SUSY-breaking squark-mass parameters
$M_{\tilde f_{L/R}}$ and $D$-terms,
\begin{eqnarray}
\tilde{M}^2_{\tilde f_{L/R}} & = & M^2_{\tilde
f_{L/R}} + D_{\tilde f_{L/R}} \nonumber \\
D_{\tilde f_L} & = & M_Z^2 (I^f_{3L} - e_f \sin^2\theta_W) \cos 2\beta
\nonumber \\
D_{\tilde f_R} & = & M_Z^2 e_f \sin^2\theta_W \cos 2\beta
\label{eq:dterms_b}
\end{eqnarray}
The mixing angles acquire the explicit form
\begin{equation}
\sin 2\theta_f = \frac{2M_f (A_f-\mu r_f)}{M_{\tilde f_1}^2 - M_{\tilde
f_2}^2} ~~~,~~~
\cos 2\theta_f = \frac{\tilde M_{\tilde f_L}^2 - \tilde
M_{\tilde f_R}^2}{M_{\tilde f_1}^2 - M_{\tilde f_2}^2}
\end{equation}
and the masses of the squark eigenstates are given by
\begin{equation}
M_{\tilde f_{1,2}}^2 = M_f^2 + \frac{1}{2}\left[ \tilde M_{\tilde f_L}^2 +
\tilde M_{\tilde f_R}^2 \mp \sqrt{(\tilde M_{\tilde f_L}^2 - \tilde
M_{\tilde f_R}^2)^2 + 4M_f^2 (A_f - \mu r_f)^2} \right]
\end{equation}
The neutral Higgs couplings to sfermions can be derived as \cite{DSUSY}
\begin{eqnarray}
g_{\tilde f_L \tilde f_L}^\Phi & = & M_f^2 g_1^\Phi + M_Z^2 (I_{3f}
- e_f\sin^2\theta_W) g_2^\Phi \nonumber \\
g_{\tilde f_R \tilde f_R}^\Phi & = & M_f^2 g_1^\Phi + M_Z^2
e_f\sin^2\theta_W
g_2^\Phi \nonumber \\
g_{\tilde f_L \tilde f_R}^\Phi & = & -\frac{M_f}{2} (\mu g_3^\Phi
- A_f g_4^\Phi)
\label{eq:hsfcouprl}
\end{eqnarray}
with the couplings $g_i^\Phi$ listed in Table \ref{tb:hsfcoup}.
\begin{table}[hbt]
\renewcommand{\arraystretch}{1.5}
\begin{center}
\begin{tabular}{|l|c||c|c|c|c|} \hline
$\tilde f$ & $\Phi$ & $g^\Phi_1$ & $g^\Phi_2$ & $g^\Phi_3$ & $g^\Phi_4$
\\
\hline \hline
& $h$ & $\cos\alpha/\sin\beta$ & $-\sin(\alpha+\beta)$ &
$-\sin\alpha/\sin\beta$ & $\cos\alpha/\sin\beta$ \\
$\tilde u$ & $H$ & $\sin\alpha/\sin\beta$ & $\cos(\alpha+\beta)$ &
$\cos\alpha/\sin\beta$ & $\sin\alpha/\sin\beta$ \\
& $A$ & 0 & 0 & 1 & $-1/\mbox{tg$\beta$}$ \\ \hline
& $h$ & $-\sin\alpha/\cos\beta$ & $-\sin(\alpha+\beta)$ &
$\cos\alpha/\cos\beta$ & $-\sin\alpha/\cos\beta$ \\
$\tilde d$ & $H$ & $\cos\alpha/\cos\beta$ & $\cos(\alpha+\beta)$ &
$\sin\alpha/\cos\beta$ & $\cos\alpha/\cos\beta$ \\
& $A$ & 0 & 0 & 1 & $-\mbox{tg$\beta$}$ \\ \hline
\end{tabular}
\renewcommand{\arraystretch}{1.2}
\caption{\label{tb:hsfcoup}
\it Coefficients of the neutral MSSM Higgs couplings to sfermion pairs.}
\end{center}
\end{table}
The charged Higgs couplings to sfermion pairs \cite{DSUSY} can be
expressed as ($\alpha,\beta = L,R$)
\begin{equation}
g_{\tilde u_\alpha \tilde d_\beta}^{H^\pm} = -\frac{1}{\sqrt{2}}
[g_1^{\alpha
\beta} + M_W^2 g_2^{\alpha\beta} ]
\label{eq:chsfcouprl}
\end{equation}
with the coefficients $g_{1,2}^{\alpha\beta}$ summarized in Table
\ref{tb:chsfcoup}.
\begin{table}[hbt]
\renewcommand{\arraystretch}{1.5}
\begin{center}
\begin{tabular}{|l||c|c|c|c|} \hline
$i$ & $g^{LL}_i$ & $g^{RR}_i$ & $g^{LR}_i$ & $g^{RL}_i$\\ \hline \hline
1 & $M_u^2/\mbox{tg$\beta$}+M_d^2 \mbox{tg$\beta$}$ & $M_u M_d
(\mbox{tg$\beta$} + 1/\mbox{tg$\beta$})$ & $M_d (\mu+A_d\mbox{tg$\beta$})$
&
$M_u (\mu+A_u/\mbox{tg$\beta$})$ \\
2 & $-\sin 2\beta$ & 0 & 0 & 0 \\ \hline
\end{tabular} 
\renewcommand{\arraystretch}{1.2}
\caption{\label{tb:chsfcoup}
\it Coefficients of the charged MSSM Higgs couplings to sfermion pairs.}
\end{center}
\end{table}

In the past SUSY--QCD corrections have been computed to the squark
masses \cite{h2sqcorr, hsqsqqcd}. They are significant in several MSSM
parameter regions. The basic outline of these radiative corrections can
be discussed starting from the $\overline{\rm MS}$ parameters in the
squark mass matrix at LO,
\begin{equation}
{\cal M}_{\tilde q} = \left[ \begin{array}{cc}
\tilde{\overline{M}}_{\tilde q_L}^2(Q_0) + \hat m_q^2(Q_0) & \hat
m_q(Q_0)
[\bar{A}_q(Q_0)-\mu r_q] \\ \hat m_q(Q_0) [\bar{A}_q(Q_0)-\mu r_q]
& \tilde{\overline{M}}_{\tilde q_R}^2(Q_0) + \hat m_q^2(Q_0)
\end{array} \right]
\end{equation}
where $\bar{A}_q(Q_0)$ denotes the running trilinear $\overline{\rm MS}$
coupling and $\tilde{\overline{M}}_{\tilde q_{L/R}}^2(Q_0)$ the left-
and right-handed soft SUSY-breaking squark $\overline{\rm MS}$ mass
parameters $\overline{M}_{\tilde q_{L/R}}(Q_0)$ at the scale $Q_0$ plus
the corresponding $D$-terms [see Eq.~(\ref{eq:dterms_b})].  In order to
accommodate the large $\Delta_b$ terms in the sbottom mass matrix the
effective bottom-mass parameter is defined as
\begin{equation}
\hat m_b(Q) = \frac{\overline{m}_b(Q)}{1+\Delta_b}
\end{equation}
where $\overline{m}_b(Q)$ represents the usual $\overline{\rm MS}$ bottom
mass at the scale $Q$. For the stop mass matrix $\hat m_t(Q)$ is just
identified with the $\overline{\rm MS}$ top mass, $\hat m_t(Q) =
\overline{m}_t(Q)$.

At NLO the masses of the stop/sbottom eigenstates are given by
\begin{eqnarray}
m_{\tilde q_{1/2}}^2 & = & \hat m_q^2(Q_0) + \frac{1}{2}\left[
\tilde{\overline{M}}_{\tilde q_L}^2(Q_0) + \tilde{\overline{M}}_{\tilde
q_R}^2(Q_0) \right. \nonumber \\
& & \left. \hspace*{1.5cm} \mp \sqrt{[\tilde{\overline{M}}_{\tilde
q_L}^2(Q_0) - \tilde{\overline{M}}_{\tilde q_R}^2(Q_0)]^2 + 4\hat
m_q^2(Q_0) [\bar{A}_q(Q_0) - \mu r_q]^2} \right] + \Delta m_{\tilde
q_{1/2}}^2 \nonumber \\
\Delta m_{\tilde q_{1/2}}^2 & = & \Sigma_{11/22}(m_{\tilde q_{1/2}}^2) +
\delta \hat m_{\tilde q_{1/2}}^2
\label{eq:sqmass}
\end{eqnarray}
where $\Sigma_{11/22}$ denote the diagonal parts of the stop/sbottom
self-energies and $\delta \hat m_{\tilde q_{1/2}}^2$ the mass counter
terms. At LO the mixing angles are defined by
\begin{equation}
\sin 2\tilde\theta_q = \frac{2\hat m_q(Q_0) [\bar{A}_q(Q_0)-\mu r_q]}{
m_{\tilde q_1}^2 - m_{\tilde q_2}^2} \quad , \quad \cos 2\tilde\theta_q
= \frac{\tilde M_{\tilde q_L}^2(Q_0) - \tilde M_{\tilde q_R}^2(Q_0)}
{m_{\tilde q_1}^2 - m_{\tilde q_2}^2}
\label{eq:sqmix}
\end{equation}
The radiative corrections to the diagonal matrix elements are
compensated by shifts in the soft mass parameters $\overline{M}_{\tilde
q_{L/R}}(Q_0)$,
\begin{equation}
\tilde M_{\tilde q_{L/R}}^2(Q_0) = \tilde{\overline{M}}^2_{\tilde
q_{L/R}}(Q_0) + \Delta \overline{M}_{\tilde q_{L/R}}^2
\label{eq:mssmshift}
\end{equation}
in order to arrive at tree-level-like expressions at NLO for the
stop/sbottom masses. The shifted SUSY-mass parameters are determined
from the sum rules
\begin{eqnarray}
\tilde M^2_{\tilde q_L}(Q_0) & = &
m^2_{\tilde q_1} \cos^2 \tilde\theta_q + m^2_{\tilde q_2} \sin^2
\tilde\theta_q - \hat m_q^2(Q_0) \nonumber \\ 
\tilde M^2_{\tilde q_R}(Q_0) & = &
m^2_{\tilde q_1} \sin^2 \tilde\theta_q + m^2_{\tilde q_2} \cos^2
\tilde\theta_q - \hat m_q^2(Q_0)
\label{eq:mlr}
\end{eqnarray}
The tree-level definition of the mixing angle $\tilde\theta_q$ in
Eq.~(\ref{eq:sqmix}) corresponds to the following counter term at NLO,
\begin{eqnarray}
\delta\tilde\theta_q & = & \frac{{\rm tg}~2\tilde\theta_q}{2} \left\{
\frac{\delta \hat m_q}{\hat m_q(Q_0)} + \frac{\delta \bar A_q}{\bar
A_q(Q_0)-\mu r_q} - \frac{\delta m_{\tilde q_1}^2 - \delta m_{\tilde
q_2}^2}{m_{\tilde q_1}^2 - m_{\tilde q_2}^2} \right\}
\label{eq:sqmixct1}
\end{eqnarray}
\begin{eqnarray}
\delta m_{\tilde q_{1/2}}^2 & = & -\Sigma_{11/22}(m_{\tilde q_{1/2}}^2)
\label{eq:sqmassct}
\end{eqnarray}
with the $\overline{\rm MS}$ counter term $\delta \bar A_q$, while
$\delta \hat m_q$ includes the finite $\Delta_b$ contributions
in addition.  However, in order to avoid artificial singularities in
physical observables for stop/sbottom masses $m_{\tilde q_{1,2}}$ close
to each other, in {\sc Hdecay} the mixing angle of the squark fields has
been renormalized via the anti-Hermitian counter term
\cite{h2sqcorr,hsqimp},
\begin{eqnarray}
\delta \theta_q & = & \frac{1}{2} \frac{\Re e\Sigma_{12}(m_{\tilde
q_1}^2) + \Re e\Sigma_{12}(m_{\tilde q_2}^2)}{m_{\tilde q_2}^2
- m_{\tilde q_1}^2}
\label{eq:sqmixct}
\end{eqnarray}
where $\Sigma_{12}$ denotes the off-diagonal part of the stop/sbottom
self-energy describing transitions from the first to the second mass
eigenstate or {\it vice versa}.  This implies a finite shift $\Delta
\tilde\theta_q$ to the mixing angle $\tilde\theta_q$ of
Eq.~(\ref{eq:sqmix}),
\begin{equation}
\theta_q = \tilde\theta_q + \Delta \tilde\theta_q \qquad , \qquad
\Delta \tilde\theta_q = \delta \tilde\theta_q - \delta \theta_q
\label{eq:sqmix1}
\end{equation}
which modifies the relations of Eq.~(\ref{eq:mlr}) by replacing $\tilde
\theta_q = \theta_q-\Delta \tilde \theta_q$.

\paragraph{Decays into sleptons and squarks.}
The sfermionic decay widths of the MSSM Higgs bosons $H_k$ ($k =
1,2,3,4$ corresponds to $H,h,A,H^\pm$ and $i,j = 1,2$) can be written as
\cite{DSUSY}
\begin{equation}
\Gamma(H_k\to\tilde f_i\overline{\tilde f}_j)=\frac{N_c G_F}{2\sqrt{2}\pi
M_{H_k}} \sqrt{\lambda_{\tilde f_i \tilde f_j, H_k}}(g^{H_k}_{\tilde f_i
\tilde f_j})^2
\end{equation}
where $N_c=3(1)$ for squarks (sleptons).  The physical MSSM couplings
$g^{H_k}_{\tilde f_i \tilde f_j}$ can be obtained from the couplings
shown in Eqs.~(\ref{eq:hsfcouprl}) and (\ref{eq:chsfcouprl}) by means of
the mixing relations in Eq.~(\ref{eq:sfmix}). The symbol $\lambda_{ij,k}
= \lambda(M_i^2,M_j^2;M_k^2)$ denotes the usual two-body phase-space
factor of Eq.~(\ref{eq:kallen}).  In the limit of massless fermions,
which is a valid approximation for the first two generations, the
pseudoscalar Higgs boson $A$ does not decay into sfermions due to the
suppression of sfermion mixing by the fermion mass. In the decoupling
regime, where the Higgs masses $M_{H,H^\pm}$ are large, the decay widths
of the heavy scalar and charged Higgs particles into sfermions are
proportional to \cite{DSUSY}
\begin{equation}
\Gamma(H,H^\pm\to\tilde f\overline{\tilde f})\propto
\frac{G_F M_W^4}{M_{H,H^\pm}} \sin^2 2\beta
\end{equation}
They are only important for small $\mbox{tg$\beta$} \sim 1$. However,
they are suppressed by an inverse power of the large Higgs masses,
rendering unimportant the sfermion decays of the first two generations.

Decay widths into third-generation sfermions ($\tilde t,\tilde b,\tilde
\tau$) can be much larger, thanks to the significantly larger fermion
masses.  For instance, in the asymptotic regime the heavy scalar Higgs
decay into stop pairs of the same helicity is proportional to
\cite{DSUSY}
\begin{equation}
\Gamma(H\to\tilde t\overline{\tilde t})\propto
\frac{G_F M_t^4}{M_H\mbox{tg$^2\beta$}}
\end{equation}
which will be enhanced by large coefficients compared to the
first/second-generation squarks for small $\mbox{tg$\beta$}$. At large
$\mbox{tg$\beta$}$ sbottom decays will be significant. Moreover, for
large Higgs masses the decay widths of heavy neutral CP-even and CP-odd
Higgs particles into stop pairs of different helicity will be
proportional to \cite{DSUSY}
\begin{equation}
\Gamma (H,A\to\tilde t\overline{\tilde t})\propto\frac{G_F
M_t^2}{M_{H,A}} \left[\mu + \frac{A_t}{\mbox{tg$\beta$}} \right]^2
\end{equation}
and hence will be of the same order of magnitude as standard fermion and
chargino/neutralino decay widths. In summary, if third-generation
sfermion decays are kinematically allowed, they have to be taken into
account.

Some time ago the SUSY-QCD corrections to the stop and sbottom decays of
the MSSM Higgs bosons have been calculated \cite{h2sqcorr, hsqsqqcd,
hsqimp}. They reach about 20--40\%, but are larger close to the
threshold regions due to the Coulomb singularities. It is relevant to
resum the large $\Delta_b$ terms as e.g.~discussed in the previous
paragraph in order to absorb the bulk of the higher-order corrections in
effective bottom and sbottom parameters. These corrections are included
in {\sc Hdecay} \cite{hdecay}, {\sc FeynHiggs} \cite{feynhiggs} and {\sc
Hfold} \cite{hfold}. The SUSY-electroweak corrections are known, too,
and modify the partial decay widths by less than about 10\%
\cite{h2sqelw}. They are included in {\sc FeynHiggs} and {\sc Hfold}.

\paragraph{Chargino/neutralino masses and couplings.}
The chargino/neutralino masses and couplings to the MSSM Higgs bosons
are derived from the Higgs mass parameter $\mu$ and the SU(2) gaugino
mass parameter $M_2$.  The mass matrix of the charginos reads as
\cite{mssmbase}
\begin{equation}
{\cal M}_{\chi^\pm} = \left[ \begin{array}{cc}
M_2 & \sqrt{2} M_W \sin\beta \\
\sqrt{2} M_W \cos\beta & \mu
\end{array} \right]
\end{equation}
It can be diagonalized by two mixing matrices $U,V$, yielding the
masses of the physical $\chi^\pm_{1,2}$ states:
\begin{eqnarray}
M_{\chi^\pm_{1,2}} & = & \frac{1}{\sqrt{2}} \left\{ M_2^2 + \mu^2 +
2M_W^2 \right. \nonumber \\
& & \left. \mp \sqrt{(M_2^2-\mu^2)^2 + 4M_W^4\cos^2 2\beta + 4M_W^2
(M_2^2 + \mu^2 + 2M_2\mu\sin 2\beta)} \right\}^{1/2}
\end{eqnarray}
If either $\mu$ or $M_2$ is large, one chargino corresponds to a pure
gaugino state and the other to a pure higgsino state. The Higgs
couplings to charginos \cite{DSUSY,DSUSY1} can be expressed as ($k =
1,2,3,4$ corresponds to $H,h,A,H^\pm$)
\begin{equation}
H_k \to \chi_i^+ \chi_j^-: \hspace*{1cm} F_{ijk} = \frac{1}{\sqrt{2}}
[e_k V_{i1}U_{j2} - d_k V_{i2}U_{j1}]
\label{eq:hccoup}
\end{equation}
where the coefficients $e_k$ and $d_k$ are determined as
\begin{equation}
\begin{array}{lllclllclll}
e_1 & = & \cos\alpha & , & e_2 & = & \sin\alpha & , & e_3 & =
&-\sin\beta \\
d_1 & = &-\sin\alpha & , & d_2 & = & \cos\alpha & , & d_3 & = &
\cos\beta
\end{array}
\label{eq:ekdk}
\end{equation}
The mass matrix of the four neutralinos depends in addition on the U(1)
gaugino mass parameter $M_1$, which is constrained by SUGRA models to be
$M_1 = \frac{5}{3} \tan\theta_W M_2$. In the bino-wino-higgsino basis,
it acquires the form \cite{mssmbase}
\begin{equation}
{\cal M}_{\chi^0} = \left[ \begin{array}{cccc}
M_1 & 0   & -M_Z \sin\theta_W \cos\beta &  M_Z \sin\theta_W \sin\beta \\
0   & M_2 &  M_Z \cos\theta_W \cos\beta & -M_Z \cos\theta_W \sin\beta \\
-M_Z \sin\theta_W \cos\beta &  M_Z \cos\theta_W \cos\beta & 0    & -\mu
\\
 M_Z \sin\theta_W \sin\beta & -M_Z \cos\theta_W \sin\beta & -\mu & 0
\\
\end{array} \right] 
\end{equation}
which can be diagonalized by a single mixing matrix $N$. The final
results are too involved to be presented here. They can be found in
\cite{DSUSY}. If either $\mu$ or $M_2$ is large, two neutralinos are
pure gaugino states and the other two pure higgsino states.  The Higgs
couplings to neutralino pairs \cite{DSUSY,DSUSY1} can be written as ($k
= 1,2,3$ corresponds to $H,h,A$)
\begin{equation}
H_k \to \chi_i^0 \chi_j^0: \hspace*{1cm} F_{ijk} = \frac{1}{2} (N_{j2} -
\tan\theta_W N_{j1})(e_k N_{i3} + d_k N_{i4}) + (i\leftrightarrow j)
\label{eq:hncoup}
\end{equation}
with the coefficients $e_k, d_k$ defined in Eq.~(\ref{eq:ekdk}). 

The charged Higgs couplings to chargino--neutralino pairs are fixed to
be \cite{DSUSY} 
\begin{eqnarray}
H^\pm \to \chi_i^\pm \chi_j^0: \hspace*{1cm} F_{ij4} & = & \cos\beta
\left[ V_{i1}N_{j4} + \frac{1}{\sqrt{2}} V_{i2} (N_{j2} + \tan\theta_W
N_{j1}) \right]
\nonumber \\
F_{ji4} & = & \sin\beta \left[ U_{i1}N_{j3} - \frac{1}{\sqrt{2}} U_{i2}
(N_{j2} + \tan\theta_W N_{j1}) \right]
\label{eq:chcncoup}
\end{eqnarray}
Analogous to the previous Higgs decays into sfermions the gaugino masses
and couplings can be loop-corrected by NLO electroweak effects via
shifts of the SUSY parameters involved in the derivation of these masses
and couplings \cite{h2gaugau}. The electroweak corrections to the
gaugino sector are of moderate size.

\paragraph{Decays into charginos and neutralinos.}
The decay widths of the MSSM Higgs particles $H_k$ ($k = 1,2,3,4$
corresponds to $H,h,A,H^\pm$) into neutralino and chargino pairs can be
cast into the form \cite{DSUSY,DSUSY1}
\begin{eqnarray}
\Gamma (H_k \to \chi_i \chi_j) & = & \frac{G_F M_W^2}{2\sqrt{2}\pi}
\frac{M_{H_k} \sqrt{\lambda_{ij,k}}}{1+\delta_{ij}} \left[
(F_{ijk}^2 + F_{jik}^2) \left( 1 - \frac{M_{\chi_i}^2}{M_{H_k}^2} -
\frac{M_{\chi_j}^2}{M_{H_k}^2} \right) \right. \nonumber \\
& & \left. -4\eta_k \epsilon_i \epsilon_j F_{ijk} F_{jik}
\frac{M_{\chi_i} M_{\chi_j}}{M_{H_k}^2} \right]
\end{eqnarray}
where $\eta_{1,2,4}=+1,~\eta_3 = -1$ and $\delta_{ij}=0$ unless the
final state consists of two identical (Majorana) neutralinos, in which
case $\delta_{ii}=1$; $\epsilon_i = \pm 1$ stands for the sign of the
$i$'th eigenvalue of the neutralino mass matrix, which can be positive
or negative. For charginos these parameters are always equal to unity.
The symbols $\lambda_{ij,k} = \lambda(M_{\chi_i}^2, M_{\chi_j}^2;
M_{\chi_k}^2)$ denote the usual two-body phase-space functions of
Eq.~(\ref{eq:kallen}).

If chargino/neutralino decays are kinematically allowed, which may be
the case for the heavy MSSM Higgs particles $H,A,H^\pm$, their branching
ratios can reach large values below the corresponding top quark
thresholds.  They can thus jeopardize the Higgs search at the LHC due to
the invisibility of a significant fraction of these decay modes
\cite{DSUSY}.  Even above the corresponding top quark thresholds the
chargino/neutralino branching ratios can be sizeable. For large Higgs
masses they can reach common values of a couple of 10\%.  In the
asymptotic regime $M_{H_k} \gg M_\chi$, the total sum of decay widths
into charginos and neutralinos acquires the simple form
\cite{DSUSY,DSUSY1}
\begin{equation}
\Gamma\left(H_k \to \sum_{i,j}\chi_i\chi_j\right) =
\frac{3G_F M_W^2}{4\sqrt{2}\pi}M_{H_k}
\left(1+\frac{1}{3} \tan^2\theta_W \right)
\end{equation}
for all three Higgs bosons $H,A,H^\pm$, which is independent of any MSSM
parameter ($\mbox{tg$\beta$},\mu,A_{t,b},M_2$). Normalized to the total
width, which is dominated by $t\bar t, b\bar b$ $(t\bar b)$ decay modes
for the neutral (charged) Higgs particles the branching ratio of
chargino/neutralino decays will exceed a level of about 20\% even for
small and large $\mbox{tg$\beta$}$. In some part of the MSSM parameter
space, invisible light scalar Higgs boson decays into the lightest
neutralino $h\to \chi_1^0 \chi_1^0$ will be possible with a relevant
branching ratio \cite{DSUSY,DSUSY1}.

The full SUSY-electroweak corrections to the Higgs boson decays into
gaugino pairs have been calculated \cite{h2gaugau}. Apart from parameter
regions where the LO decay width is suppressed they are of moderate
size, i.e.~not larger than about 20\%. It should be noted that a
consistent renormalization of the gaugino sector is mandatory for
reliable predictions of these partial decay widths.

\subsection{\it Branching ratios and total decay width}
\label{sc:br}
\subsubsection{\it Standard Model}
The determination of the branching ratios of Higgs-boson decays requires
the inclusion of the aforementioned higher-order corrections and a
sophisticated estimate of the theoretical and parametric uncertainties.
The first analyses of this type have been performed some time ago
\cite{brthupu0} and recently \cite{brthupu, brthupu1}. The parametric
uncertainties are dominated by the uncertainties in the top, bottom and
charm quark masses as well as the strong coupling $\alpha_s$. We have
used the $\overline{\rm MS}$ masses for the bottom and charm quark
\cite{yr4},
\begin{equation}
\overline{m}_b (\overline{m}_b) = (4.18 \pm 0.03)~{\rm GeV}, \qquad
\overline{m}_c (3~{\rm GeV}) = (0.986 \pm 0.026)~{\rm GeV}
\end{equation}
and the top quark pole mass
\begin{equation}
M_t = (172.5 \pm 1)~{\rm GeV}
\end{equation}
according to the conventions of the LHC Higgs Cross Section WG (HXSWG)
\cite{yr4}. The $\overline{\rm MS}$ bottom and charm masses are evolved
from the input scale to the scale of the decay process with 4-loop
accuracy in QCD. If needed the pole mass of the bottom quark is obtained
from the 3-loop conversion between the $\overline{\rm MS}$ and pole
mass \cite{msbarpole},
\begin{eqnarray}
M_b & = & \overline{m_b}(\mu_R^2) \left\{ 1 + \left(\frac{4}{3} + \log
\frac{\mu_R^2}{M_b^2} \right) \frac{\alpha_s(\mu_R^2)}{\pi}  +
\left[ 16.110 + 8.847 \log \frac{\mu_R^2}{M_b^2} + 1.791 \log^2
\frac{\mu_R^2}{M_b^2} \right. \right. \nonumber \\
& & \left. \left. - N_F \left( 1.041 + 0.361 \log
\frac{\mu_R^2}{M_b^2} + 0.0833 \log^2 \frac{\mu_R^2}{M_b^2} \right)
+ \frac{4}{3} \sum_{1\leq i\leq N_F} \Delta \left(
\frac{M_i}{M_b} \right) \right]
\left(\frac{\alpha_s(\mu_R^2)}{\pi} \right)^2 \right. \nonumber \\
& & + \left[ 239.297 + 129.420 \log \frac{\mu_R^2}{M_b^2} + 32.105 \log^2
\frac{\mu_R^2}{M_b^2} + 3.683 \log^3 \frac{\mu_R^2}{M_b^2} - N_F \left(
29.701 \phantom{\log \frac{\mu_R^2}{M_b^2}} \right. \right.  \nonumber \\
& & \left. \left. \left. + 15.371 \log \frac{\mu_R^2}{M_b^2}
+ 3.053 \log^2 \frac{\mu_R^2}{M_b^2} + 0.370 \log^3 \frac{\mu_R^2}{M_b^2}
\right) + N_F^2 \left( 0.653 + 0.320 \log \frac{\mu_R^2}{M_b^2} \right.
\right. \right. \nonumber \\
& & \left. \left. \left. 
+ 0.0602 \log^2 \frac{\mu_R^2}{M_b^2} + 0.00926 \log^3 \frac{\mu_R^2}{M_b^2}
\right) \right] \left(\frac{\alpha_s(\mu_R^2)}{\pi} \right)^3 \right\}
+ {\cal O}(\alpha_s^4)
\end{eqnarray}
where $N_F = 4$ for the bottom quark and the strong coupling $\alpha_s$
evolves with 5 active flavours. The mass-dependent term involving
the light flavours can (for $0\leq x \leq 1$) be approximated by
\begin{equation}
\Delta(x) = \frac{\pi^2}{8}~x - 0.579~x^2 + 0.230~x^3
\end{equation}
The charm pole mass $M_c$ is then determined by the renormalon-free
relation \cite{mcmbrel}
\begin{equation}
M_c = M_b - 3.41~{\rm GeV}
\end{equation}
from the bottom pole mass $M_b$. The strong coupling $\alpha_s$ is fixed
by the input value at the $Z$-boson mass scale
\begin{equation}
\alpha_s(M_Z) = 0.118 \pm 0.0015
\end{equation}
The lepton as well as $W,Z$-boson masses and widths are chosen according
to their PDG values \cite{pdg}
\begin{eqnarray}
m_e    & = & (0.510998928\pm 0.000000011)~{\rm MeV} \nonumber \\
m_\mu  & = & (105.6583715\pm 0.0000035)~{\rm MeV} \nonumber \\
m_\tau & = & (1776.82 \pm 0.16)~{\rm MeV} \nonumber \\
M_W & =& (80.385\pm 0.015)~{\rm GeV} \qquad\qquad\qquad\quad \Gamma_W =
(2.085\pm 0.042)~{\rm GeV} \nonumber \\
M_Z & = & (91.1876\pm 0.0021)~{\rm GeV} \qquad\qquad\qquad\; \Gamma_Z =
(2.4952\pm 0.0023)~{\rm GeV}
\end{eqnarray}
The complex pole masses of the $W$ and $Z$ boson can be obtained from
the relations
\begin{equation}
M_V^{\rm pole} - i\Gamma_V^{\rm pole} 
= \frac{M_V-i\Gamma_V}{\sqrt{1+\Gamma^2_V/M^2_V}} \qquad
(V = W,Z)
\end{equation}
These are used for the predictions of the Higgs boson decay widths into
four fermions according to Ref.~\cite{prophecy4f}. Moreover, the Fermi
constant has been chosen as \cite{pdg}
\begin{equation}
G_F=1.166 378 7(6)\cdot 10^{-5}~{\rm GeV}^{-2}
\end{equation}
The total parametric uncertainty for each branching ratio has been
derived from a quadratic sum of the individual impacts of the input
parameters on the decay modes.

\begin{figure}[hbt]
\vspace*{0.5cm}

\hspace*{3.0cm}
\epsfxsize=10cm \epsfbox{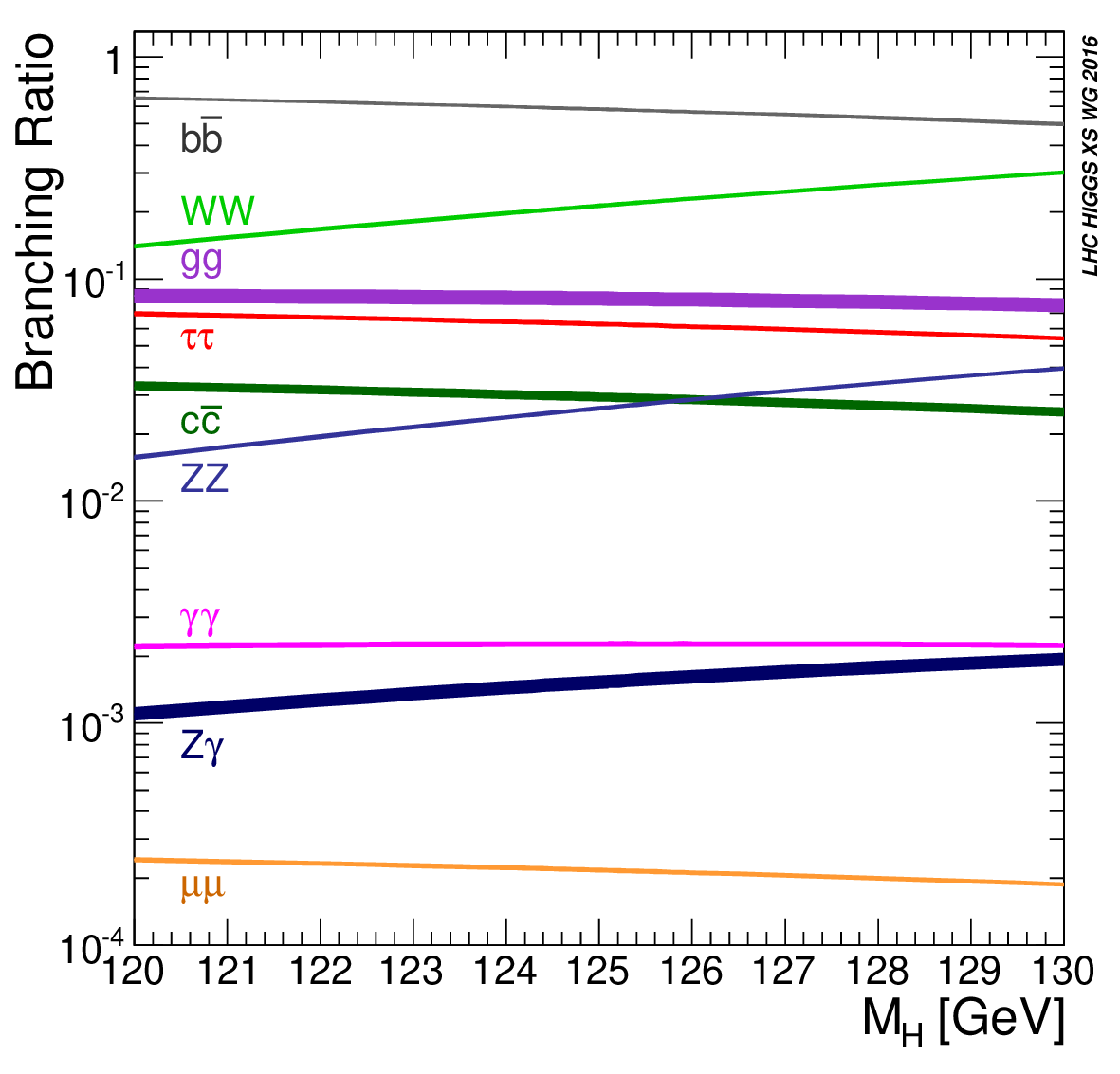}
\vspace*{-0.3cm}

\caption{\label{fg:br} \it Higgs branching ratios and their
uncertainties for the mass range around $125$ GeV. From Ref.~\cite{yr4}
(page 24).}
\end{figure}

The theoretical uncertainties from missing higher orders in the
perturbative expansion are summarized in Table \ref{tab:uncertainty} as
discussed before for the individual partial decay processes. In order to
be conservative the total parametric uncertainties are added linearly to
the theoretical uncertainties. The final result for the branching ratios
is shown in Fig.~\ref{fg:br} for the leading Higgs decay modes with
branching ratio larger than $10^{-4}$ within the Higgs-mass range
between 120 and 130 GeV. They have been obtained with {\sc Prophecy4f}
\cite{prophecy4f} for the decays $H\to WW,ZZ$ and {\sc Hdecay}
\cite{hdecay} for the other decay modes. The bands represent
the total uncertainties of the individual branching ratios. For a Higgs
mass $M_H=125$ GeV the total uncertainty of the leading decay mode $H\to
b\bar b$ amounts to less than 2\%, since the bulk of it cancels out
within the branching ratio. The uncertainty of $\Gamma(H\to b\bar b)$,
however, generates a significant increase of the uncertainties for the
subleading decay modes. The total uncertainties of $BR(H\to WW/ZZ)$ and
$BR(H\to \tau^+\tau^-/\mu^+\mu^-)$ amount to $\sim 2\%$, while the
uncertainties of $BR(H\to gg)$ and $BR(H\to c\bar c)$ range at $\sim
6-7\%$, of $BR(H\to\gamma\gamma)$ at $\sim 3\%$ and of $BR(H\to
Z\gamma)$ at $\sim 7\%$. The total decay width of $\sim 4.1$ MeV can be
predicted with $\sim 2\%$ total uncertainty. These results constitute a
basic ingredient of the corresponding LHC analyses.
\begin{table}[hbt]
\renewcommand{\arraystretch}{1.3}
\setlength{\tabcolsep}{1.5ex}%
\begin{center}
\begin{tabular}{llll}
\hline
{\rm Partial~width} & {\rm QCD} & {\rm electroweak} & {\rm total} \\
\hline
$H \to b\bar b/c\bar c$ & $\sim 0.2\%$ & $\sim 0.5\%$ & $\sim \pm 0.5 \%$ \\
$H\to \tau^+ \tau^-/\mu^+\mu^-$ & & $\sim0.5\%$ & $\sim \pm 0.5 \%$ \\
$H \to gg$ & ${\sim 3\%}$ & $\sim 1\%$ & $\sim\pm 3.2\%$ \\
$H \to \gamma \gamma$ & ${<1\%}$ & $<1\%$ & $\sim\pm 1\%$ \\
$H \to Z \gamma$  & ${<1\%}$ & $\sim 5\%$ & $\sim\pm 5\%$ \\
$H \to WW/ZZ\to 4 f$ & $<0.5\%$ &  $\sim 0.5\%$ & $\sim\pm 0.5\%$ \\
\hline
\end{tabular}
\caption{\it \label{tab:uncertainty} Estimated theoretical uncertainties
from missing higher orders in the intermediate Higgs-mass range. From
Ref.~\cite{yr4} (page 22).}
\end{center}
\renewcommand{\arraystretch}{1.0}
\end{table}

\subsubsection{\it Minimal supersymmetric extension}
\begin{figure}[hbtp]
\begin{center}
\begin{picture}(150,640)(0,0)
\put(-270,370.0){\includegraphics{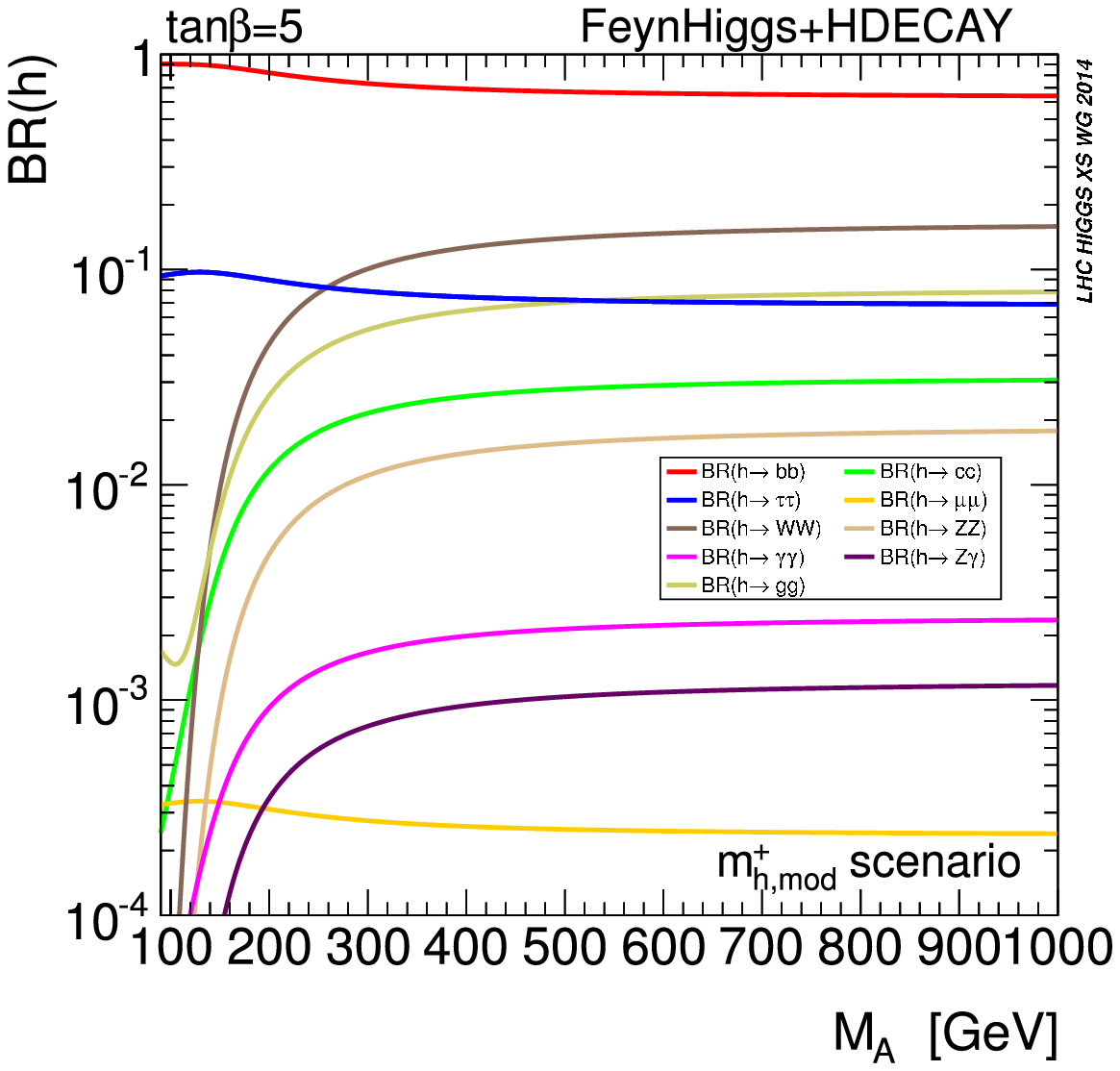}}
\put(-20,370.0){\includegraphics{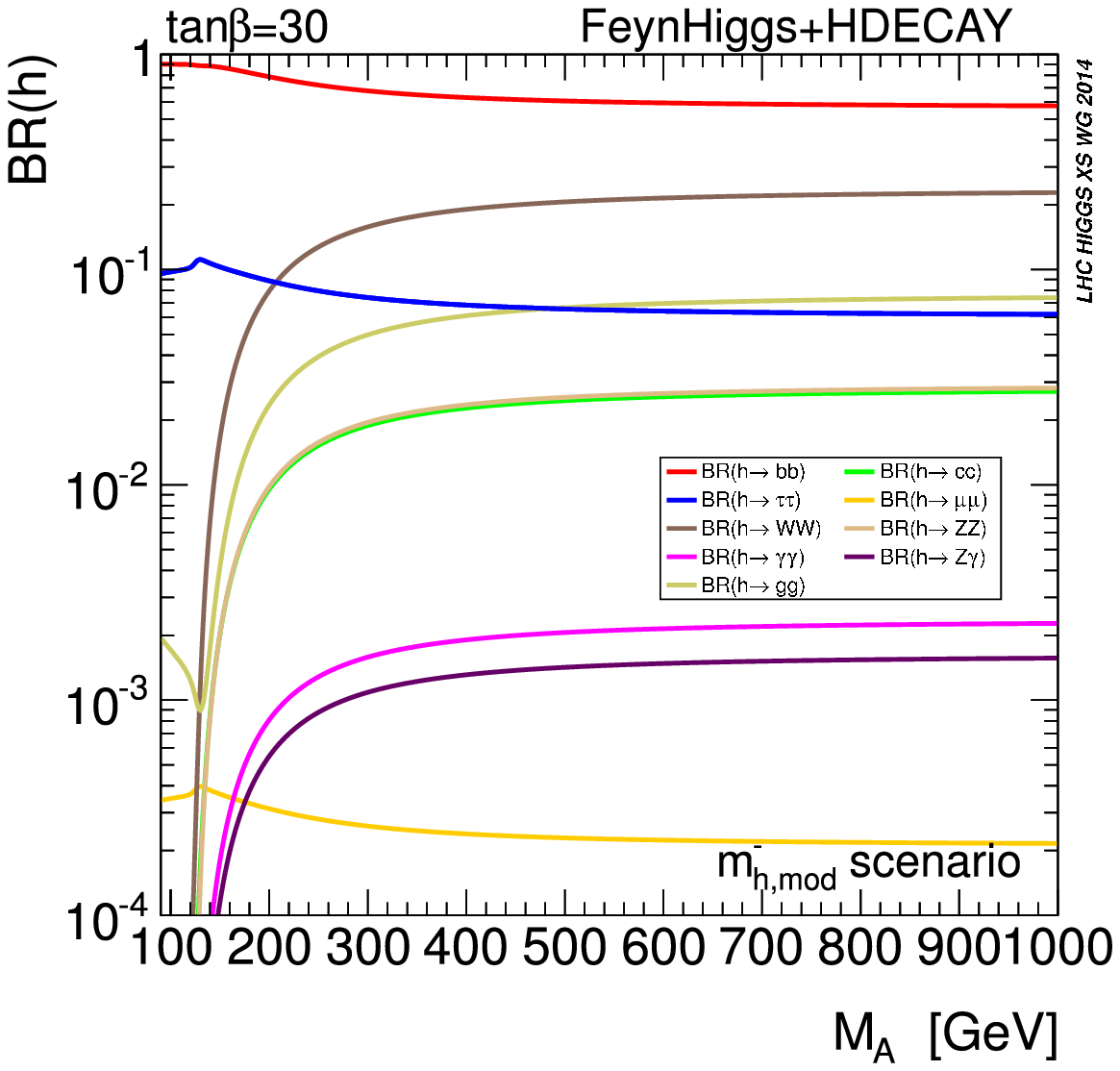}}
\put(-270,170.0){\includegraphics{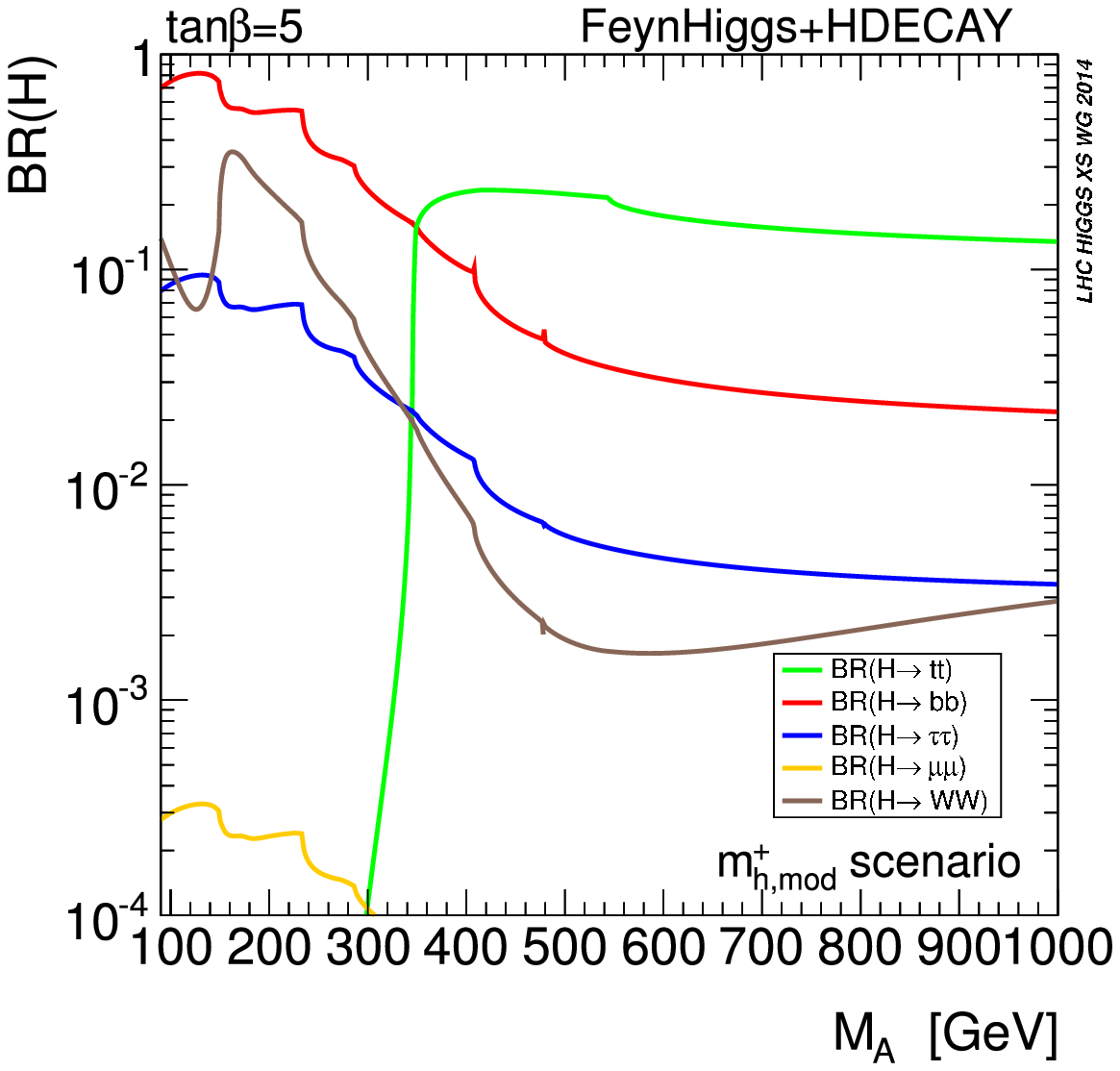}}
\put(-20,170.0){\includegraphics{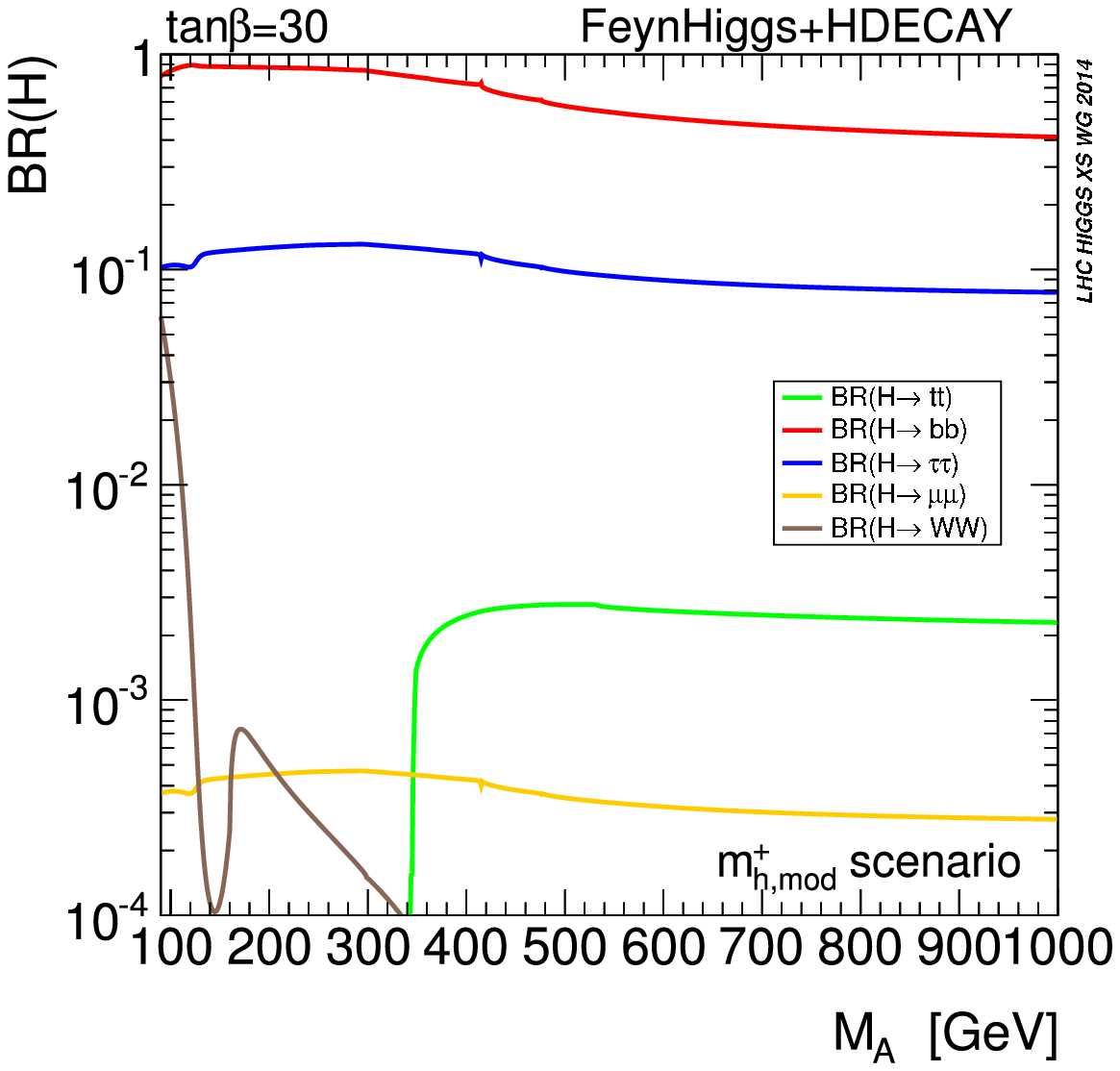}}
\put(-270,-30.0){\includegraphics{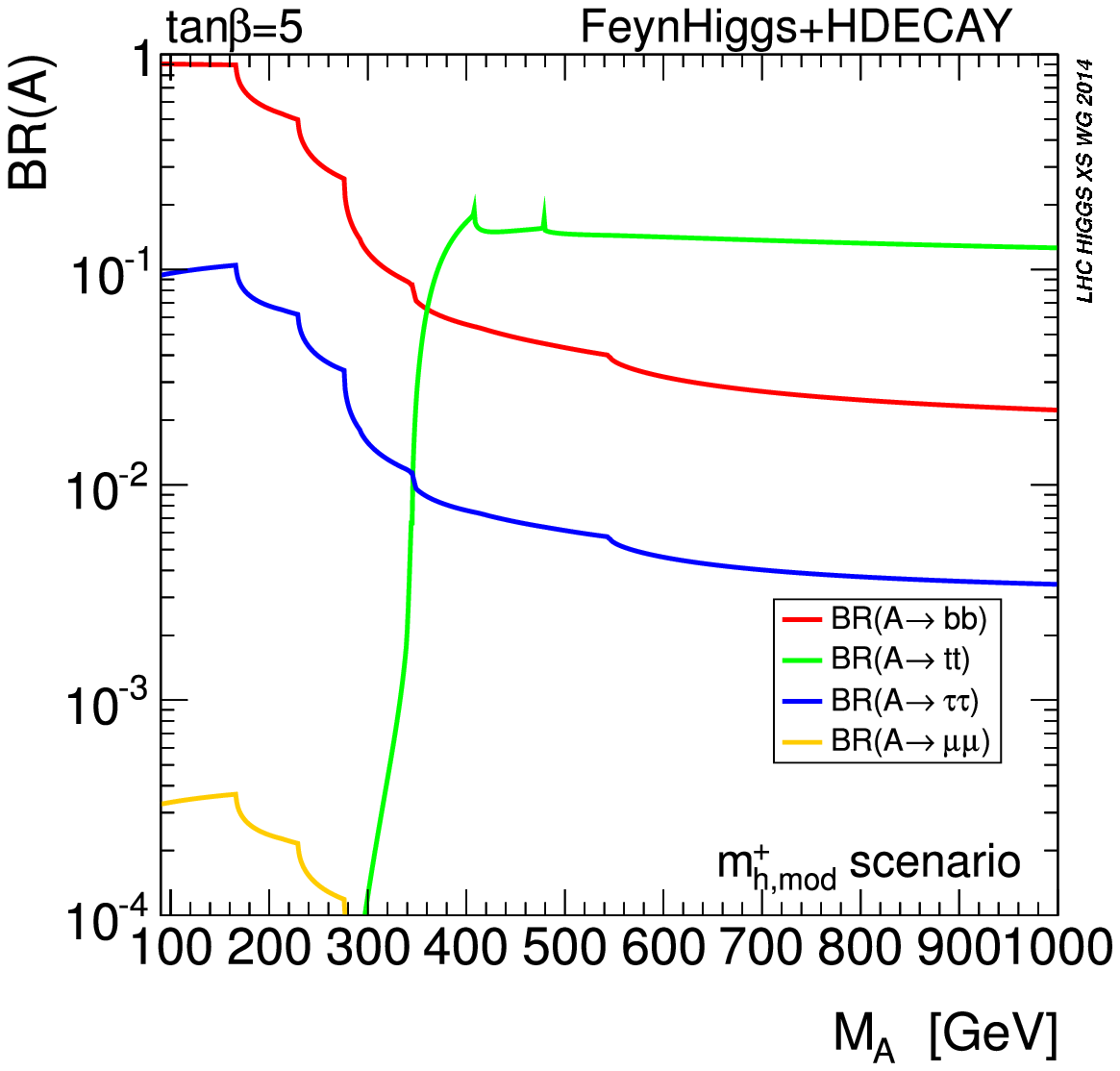}}
\put(-20,-30.0){\includegraphics{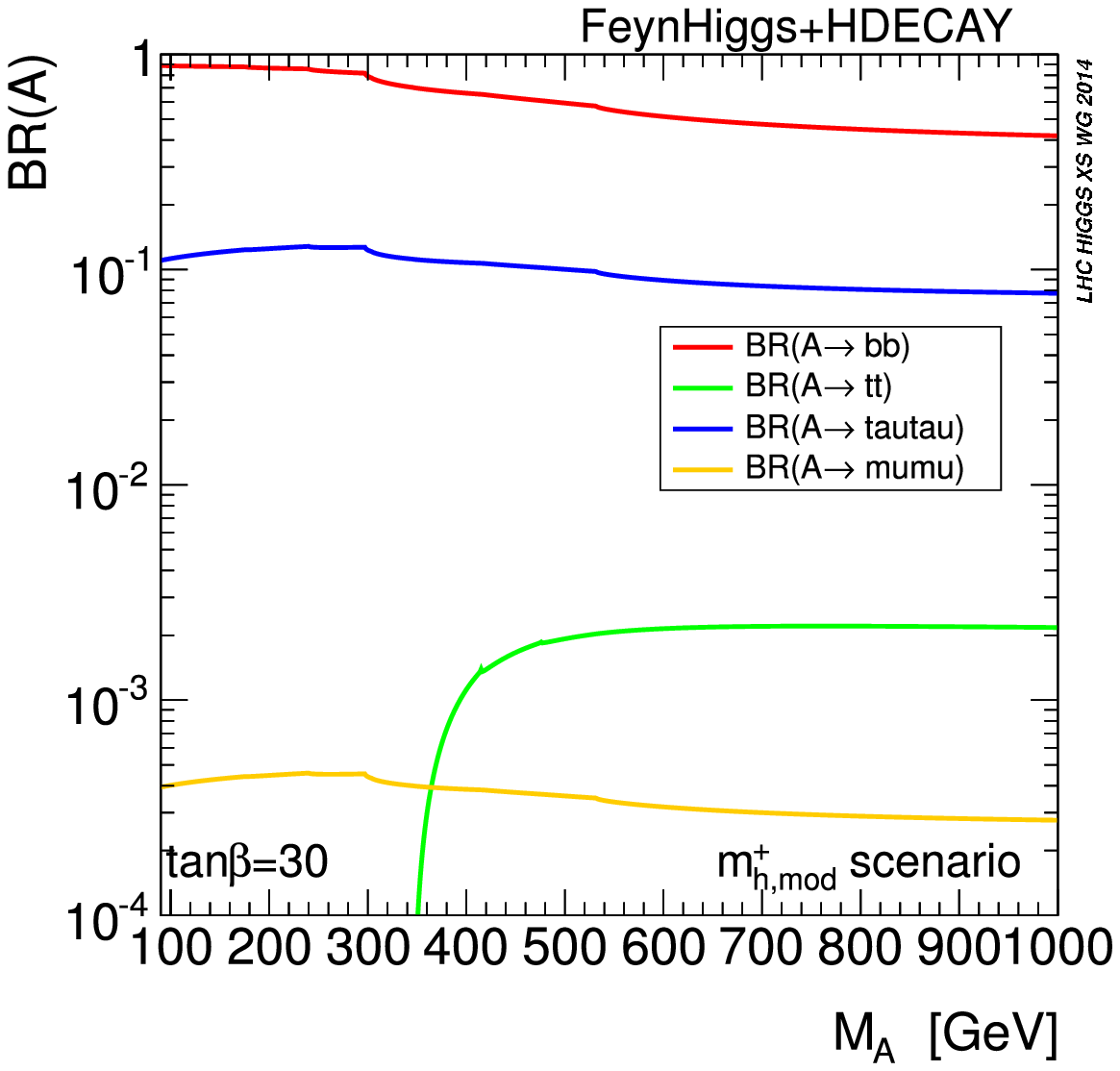}}
\end{picture}
\caption{\it Neutral MSSM Higgs branching ratios as functions of the
corresponding Higgs mass within the $m_{h}^{mod+}$ scenario
\cite{benchmark} for two values of $\mbox{tg}\beta$ obtained by a
combination of {\sc FeynHiggs} \cite{feynhiggs} and {\sc Hdecay}
\cite{hdecay}. From Ref.~\cite{yr4}.}
\label{fig:mssmnbr}
\end{center}
\end{figure}

The final results for the MSSM Higgs branching ratios have been obtained
with the public codes {\sc FeynHiggs} \cite{feynhiggs} and {\sc Hdecay}
\cite{hdecay} using the SM input parameters of the previous section.
Fig.~\ref{fig:mssmnbr} shows the neutral Higgs branching ratios for two
values of $\mbox{tg}\beta$ within the $m_{h}^{mod+}$ benchmark scenario
\cite{benchmark} that is defined as
\begin{eqnarray}
M_{SUSY} = 1000~\mbox{GeV},~\mu = M_2 = 200~\mbox{GeV},
A_b = A_\tau = A_t,~m_{\tilde g} = 1500~\mbox{GeV},~M_{\tilde{l}_3} =
1000~\mbox{GeV}, \nonumber \\
X_t^{OS} = 1.5\,M_{SUSY}~(\mbox{FD calculation}),~
X_t^{\overline{\rm MS}} = 1.6\,M_{SUSY}~(\mbox{RG calculation})
\hspace*{3.8cm}
\end{eqnarray}
The kinks visible in these plots are due to the opening of new decay
modes according to the SM and SUSY-particle masses of the final-state
particles. Fig.~\ref{fig:mssmcbr} displays the corresponding charged
Higgs branching ratios within the $m_{h}^{mod+}$ benchmark scenario
\cite{benchmark} for two values of $\mbox{tg}\beta$. The related
uncertainties are not shown in the plots.
\begin{figure}[hbtp]
\begin{center}
\begin{picture}(150,240)(0,0)
\put(-180,-20.0){\includegraphics{brc10.eps}}
\put(70,-20.0){\includegraphics{brc50.eps}}
\end{picture}
\caption{\it Charged MSSM Higgs branching ratios as functions of the charged
Higgs mass within the $m_{h}^{mod+}$ scenario \cite{benchmark} for two
values of $\mbox{tg}\beta$ obtained by a combination of {\sc FeynHiggs}
\cite{feynhiggs} and {\sc Hdecay} \cite{hdecay}. From Ref.~\cite{yr3}.}
\label{fig:mssmcbr}
\end{center}
\end{figure}

\section{Higgs-Boson Production}
\subsection{\it Gluon fusion: $gg\to H$}
\subsubsection{\it Standard Model}
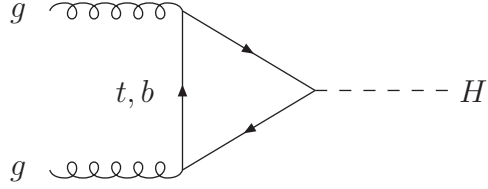
\begin{figure}[hbt]
\begin{center}
\setlength{\unitlength}{1pt}
\begin{picture}(180,100)(0,0)

\Gluon(0,20)(50,20){-3}{5}
\Gluon(0,80)(50,80){3}{5}
\ArrowLine(50,20)(50,80)
\ArrowLine(50,80)(100,50)
\ArrowLine(100,50)(50,20)
\DashLine(100,50)(150,50){5}
\put(155,46){$H$}
\put(25,46){$t,b$}
\put(-15,18){$g$}
\put(-15,78){$g$}

\end{picture}  \\
\setlength{\unitlength}{1pt}
\caption{\label{fg:gghlodia} \it Diagrams contributing to $gg\to H$
at lowest order.}
\end{center}
\end{figure}

The gluon-fusion mechanism \cite{glufus}
\begin{displaymath}
pp \to gg \to H
\end{displaymath}
dominates Higgs-boson production at the LHC in the entire relevant Higgs
mass range.  The gluon coupling to the Higgs boson in the SM is mediated
by triangular top- and bottom-quark loops, see Fig.~\ref{fg:gghlodia}.
Since the Yukawa coupling of the Higgs particle grows with the quark
mass, the form factor reaches a constant value for large loop quark
masses. If the masses of heavier quarks beyond the third generation are
fully generated by the Higgs mechanism, these particles would add the
same amount to the form factor as the top quark in the asymptotic heavy
quark limit. Thus gluon fusion can serve as a counter of the number of
heavy quarks, the masses of which are generated by the conventional
Higgs mechanism. On the other hand within the three-generation SM gluon
fusion will allow to measure the top quark Yukawa coupling. This,
however, requires a precise knowledge of the cross section within the SM
with three generations of quarks.

The partonic cross section can be derived from the gluonic width of the
Higgs boson at lowest order \cite{glufus},
\begin{eqnarray}
\hat\sigma_{LO} (gg\to H) & = & \sigma_0 \delta
(1 - z) \label{eq:gghlo} \\
\sigma_0 = \frac{\pi^2}{8M_H^3} \Gamma_{LO} (H\to gg)
& = & \frac{G_{F}\alpha_{s}^{2}(\mu_R)}{288 \sqrt{2}\pi} \
\left| \sum_{Q} A_Q^H (\tau_{Q}) \right|^{2}
\nonumber
\end{eqnarray}
with the scaling variables defined as $z=M_H^2/\hat s$, $\tau_Q=4M_Q^2/
M_H^2$. The variable $\hat{s}$ denotes the partonic c.m.~energy squared
and $\mu_R$ the renormalization scale. The amplitudes $A_Q^H(\tau_Q)$
are given in Eq.~(\ref{eq:ftau}).

In the narrow-width approximation the hadronic cross section can be cast into
the form \cite{glufus}
\begin{equation}
\sigma_{LO}(pp\to H) = \sigma_0 \tau_H \frac{d{\cal L}^{gg}}{d\tau_H}
\end{equation}
with the gluon luminosity
\begin{equation}
\frac{d{\cal L}^{gg}}{d\tau} = \int_\tau^1 \frac{dx}{x}~g(x,\mu_F^2)
g(\tau /x,\mu_F^2)
\label{eq:gglum}
\end{equation}
at the factorization scale $\mu_F$, and the scaling variable is defined,
in analogy to the Drell--Yan process, as $\tau_H = M^2_H/s$, with $s$
specifying the total hadronic c.m.~energy squared. The bottom-quark
contributions interfere destructively with the top loop and decrease the
cross section by about 10\% at LO.

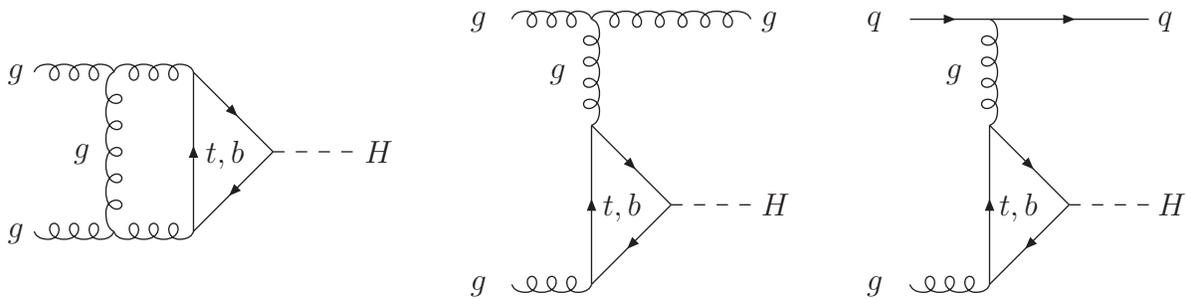
\begin{figure}[hbt]
\begin{center}
\setlength{\unitlength}{1pt}
\begin{picture}(450,100)(-10,0)

\Gluon(0,20)(30,20){-3}{3}
\Gluon(0,80)(30,80){3}{3}
\Gluon(30,20)(60,20){-3}{3}
\Gluon(30,80)(60,80){3}{3}
\Gluon(30,20)(30,80){3}{5}
\ArrowLine(60,20)(60,80)
\ArrowLine(60,80)(90,50)
\ArrowLine(90,50)(60,20)
\DashLine(90,50)(120,50){5}
\put(125,46){$H$}
\put(65,46){$t,b$}
\put(-10,18){$g$}
\put(-10,78){$g$}
\put(15,48){$g$}

\Gluon(180,100)(210,100){3}{3}
\Gluon(210,100)(270,100){3}{6}
\Gluon(180,0)(210,0){-3}{3}
\Gluon(210,100)(210,60){3}{4}
\ArrowLine(210,0)(210,60)
\ArrowLine(210,60)(240,30)
\ArrowLine(240,30)(210,0)
\DashLine(240,30)(270,30){5}
\put(275,26){$H$}
\put(215,26){$t,b$}
\put(165,-2){$g$}
\put(165,98){$g$}
\put(195,78){$g$}
\put(275,98){$g$}

\ArrowLine(330,100)(360,100)
\ArrowLine(360,100)(420,100)
\Gluon(330,0)(360,0){-3}{3}
\Gluon(360,100)(360,60){3}{4}
\ArrowLine(360,0)(360,60)
\ArrowLine(360,60)(390,30)
\ArrowLine(390,30)(360,0)
\DashLine(390,30)(420,30){5}
\put(425,26){$H$}
\put(365,26){$t,b$}
\put(315,-2){$g$}
\put(315,98){$q$}
\put(425,98){$q$}
\put(345,78){$g$}

\end{picture}  \\
\setlength{\unitlength}{1pt}
\caption{\label{fg:gghqcddia} \it Typical diagrams contributing to the
virtual and real QCD corrections to $gg\to H$ at NLO.}
\end{center}
\end{figure}
\paragraph{QCD corrections.}
In the past the (two-loop) NLO QCD corrections to the gluon-fusion cross
section, Fig.~\ref{fg:gghqcddia}, have been calculated including the
full mass dependences \cite{higgsqcd,hgaga1,hgg,gghsm0,gghsm,phd}. They
consist of virtual corrections to the basic $gg\to H$ process and real
corrections due to the associated production of the Higgs boson with
massless partons,
\begin{displaymath}
gg \rightarrow Hg \hspace{0.5cm} \mbox{and} \hspace{0.5cm}
gq \rightarrow Hq,~q\overline{q} \rightarrow Hg
\end{displaymath}
These subprocesses contribute to the Higgs production at ${\cal O}
(\alpha_s^3)$. The virtual corrections rescale the lowest-order fusion
cross section with a coefficient depending only on the ratios of the
Higgs and quark masses. Gluon radiation leads to two-parton final states
with invariant energy $\hat{s}\ge M_H^2$ in the $gg,gq$ and
$q\overline{q}$ channels at NLO. In general the hadronic cross section
can be split into seven parts \cite{higgsqcd,hgaga1,hgg,gghsm,phd},
\begin{equation}
\sigma(pp \rightarrow H+X) = \sigma_{0} \left[ 1+ C
\frac{\alpha_{s}}{\pi} \right] \tau_{H} \frac{d{\cal L}^{gg}}{d\tau_{H}} +
\Delta \sigma_{gg} + \Delta \sigma_{gq} + \Delta \sigma_{q\bar{q}} +
\Delta \sigma_{qq} + \Delta \sigma_{qq'}
\end{equation}
where the finite parts of virtual corrections $C$ and the real
corrections $\Delta \sigma_{gg}$, $\Delta \sigma_{gq}$ and $\Delta
\sigma_{q\bar{q}}$ (same-flavour quark-antiquark initial states) start
to contribute at NLO, while $\Delta \sigma_{qq}$ (same-flavour
quark-quark and antiquark-antiquark initial states) and $\Delta
\sigma_{qq'}$ (different-flavour quark and antiquark initial states)
appear for the first time at NNLO. The renormalization scale $\mu_R$ of
$\alpha_s$ and the factorization scale $\mu_F$ of the parton densities
are fixed properly, in general at $\mu_R=\mu_F=M_H/2$.  The quark-loop
mass has been identified with the pole mass $M_Q$, while the QCD
coupling $\alpha_s$ and the parton density functions are defined in the
$\overline{\rm MS}$ scheme with five active flavours.

We define the NLO $K$ factor as the ratio
\begin{eqnarray}
K_{NLO} = \frac{\sigma_{NLO}}{\sigma_{LO}}
\end{eqnarray}

\begin{figure}[hbt]
\vspace*{-1.5cm}

\hspace*{3.0cm}
\epsfxsize=10cm \epsfbox{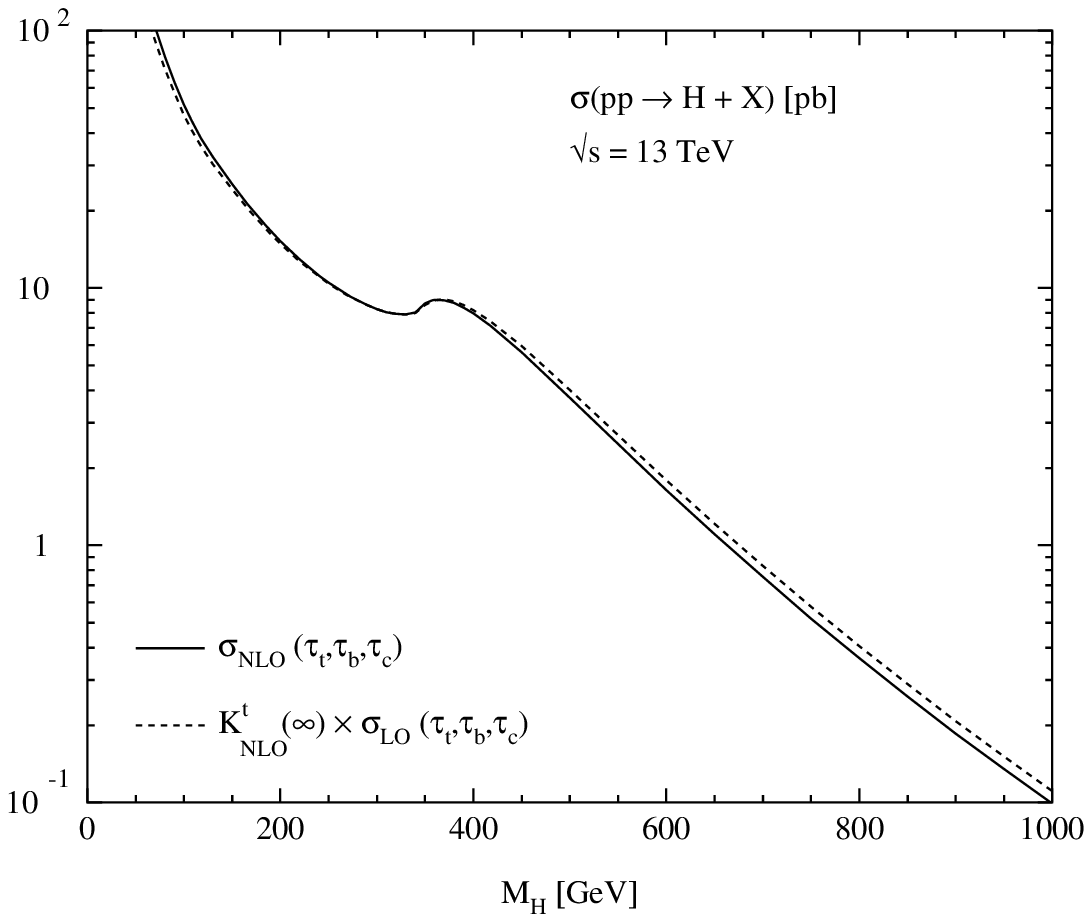}
\vspace*{-3.0cm}

\caption{\label{fg:gghapprox} \it Comparison of the exact and
approximate NLO cross section $\sigma(pp\to H+X)$ at the LHC with
c.m.~energy $\sqrt{s}=13$ TeV. The solid line shows the exact cross
section including the full $t,b,c$ quark mass dependence and the dashed
line corresponds to the approximation defined in
Eq.~(\ref{eq:gghapprox}). The renormalization and factorization scales
have been identified with half of the Higgs mass, $\mu_R=\mu_F=M_H/2$
and the PDF4LHC15 NLO parton densities \cite{pdf4lhc} with NLO strong
coupling ($\alpha_s(M_Z)=0.118$) have been adopted. The top mass has
been chosen as $M_t=172.5$ GeV, the bottom mass as $M_b=4.84$ GeV and
the charm mass as $M_c=1.43$ GeV.}
\end{figure}

The NLO corrections are positive and large, increasing the gluon-fusion
cross section at the LHC by about 60--90\%. The QCD corrections to
the bottom-quark contributions are significantly smaller if the bottom
mass is used in terms of the pole mass or the $\overline{\rm MS}$ mass
at the scale of the bottom mass. This choice is motivated by the
numerical cancellation of squared and single logarithms of the relative
QCD corrections at NLO \cite{higgsqcd}. This feature modifies the
destructive bottom-quark contribution to a reduction of the cross
section by about $6\%$ at NLO. Comparing the exact mass-dependent
results with the expressions in the heavy-quark limit, it turns out that
this asymptotic $K$ factor provides an excellent approximation even for
Higgs masses above the top-decay threshold\footnote{Large Higgs masses
are still relevant for off-shell Higgs bosons.}. We explicitly define the
approximation by
\begin{eqnarray}
\sigma_{app} & = & K_{NLO}^t (\infty) \times
\sigma_{LO}(\tau_t,\tau_b,\tau_c)
\label{eq:gghapprox} \\
K_{NLO}^t (\infty) & = & \lim_{M_t \to \infty} K_{NLO} \nonumber
\end{eqnarray}
where we neglect the $b$ quark contribution in $K_{NLO}^t (\infty)$,
while the leading order cross section $\sigma_{LO}$ includes the full
$t,b,c$ quark mass dependence. The comparison with the full massive NLO
result is presented in Fig.~\ref{fg:gghapprox}. The solid line
corresponds to the exact cross section and the broken line to the
approximate one. For Higgs masses below $\sim$ 1 TeV, the deviations of
the asymptotic approximation from the full NLO result are less than
15\%, whereas for $M_H=125$ GeV they amount to $\sim 5\%$, if the full
LO cross section is multiplied by the approximate K-factor. This
property of the NLO corrections suggests this to be true also at higher
orders, since it is a consequence of the dominating soft and collinear
gluon effects in the QCD corrections.

Within the heavy top-quark limit the NNLO \cite{gghnnlo} and N$^3$LO
\cite{gghn3lo1,gghn3lo} QCD corrections have been calculated. The NNLO
contributions increase the production cross section by about 20\% beyond
NLO, while the N$^3$LO corrections range at the level of a few per-cent.
These results indicate that the gluon-fusion cross section is under
theoretical control despite the large size of the NLO corrections. This
is further corroborated by the results obtained by a soft and collinear
gluon resummation on top of the N$^3$LO result. This approach resums the
dominant factorizing contributions from soft and collinear gluon effects
up to all perturbative orders. The soft corrections provide the leading
ones close to the production threshold, while collinear gluon effects
are of subleading order. Both can be treated systematically. The soft
and collinear contributions provide a reasonable approximation of the
full fixed-order results and thus a reliable estimate of missing
higher-order effects beyond the fixed-order corrections. This
resummation has been performed at the NNLL level \cite{gghnnll} and the
N$^3$LL level \cite{gghn3ll} in the heavy top-quark limit. The sizes of
the different logarithmic orders follow roughly the pattern of the
corresponding fixed-order corrections. Quite recently also finite
top-mass effects have been included in the resummation framework
\cite{gghnllmt} at the NLL order, where they are known exactly.
Resummation effects beyond N$^3$LO yield only a per-cent increase of the
cross section for the central scale choice. However, they provide an
approximation of effects beyond N$^3$LO and contribute to a
sophisticated estimate of the residual uncertainties by elaborating on
the uncertainties due to the matching to the fixed-order expression.

\paragraph{Electroweak corrections.}
The electroweak corrections to the gluon-fusion cross section have been
computed approximately first. The leading top mass corrections of ${\cal
O}(G_F M_t^2)$ coincide with the corrections to the gluonic decay mode
of Eq.~(\ref{eq:hggelw}) and are thus small \cite{hbb3,abdelgambino}. This
calculation has been refined by the determination of the NLO electroweak
corrections due to light-fermion loops \cite{hggelwf} and finally by the
full numerical integration of the exact NLO corrections to the top- and
$W,Z$-induced electroweak corrections \cite{hggelw}. The electroweak
corrections coincide with the ones to the $H\to gg$ decay. The NLO
electroweak corrections have been extended by a calculation of the mixed
QCD-electroweak corrections in the limit $M_H^2 \ll M_W^2$
\cite{gghqcdelw} which can be attributed to the corrections of the
effective $Hgg$ coupling.  However, it is unclear how reliable this
approximation is in practice.  Due to the dominance of soft and
collinear gluon effects the bulk of the electroweak corrections will
factorize from the pure QCD corrections. In the following the
electroweak corrections will be combined with the QCD corrections in
factorized form.

\begin{figure}[hbt]

\vspace*{0.5cm}
\hspace*{3.0cm}
\epsfxsize=12cm \epsfbox{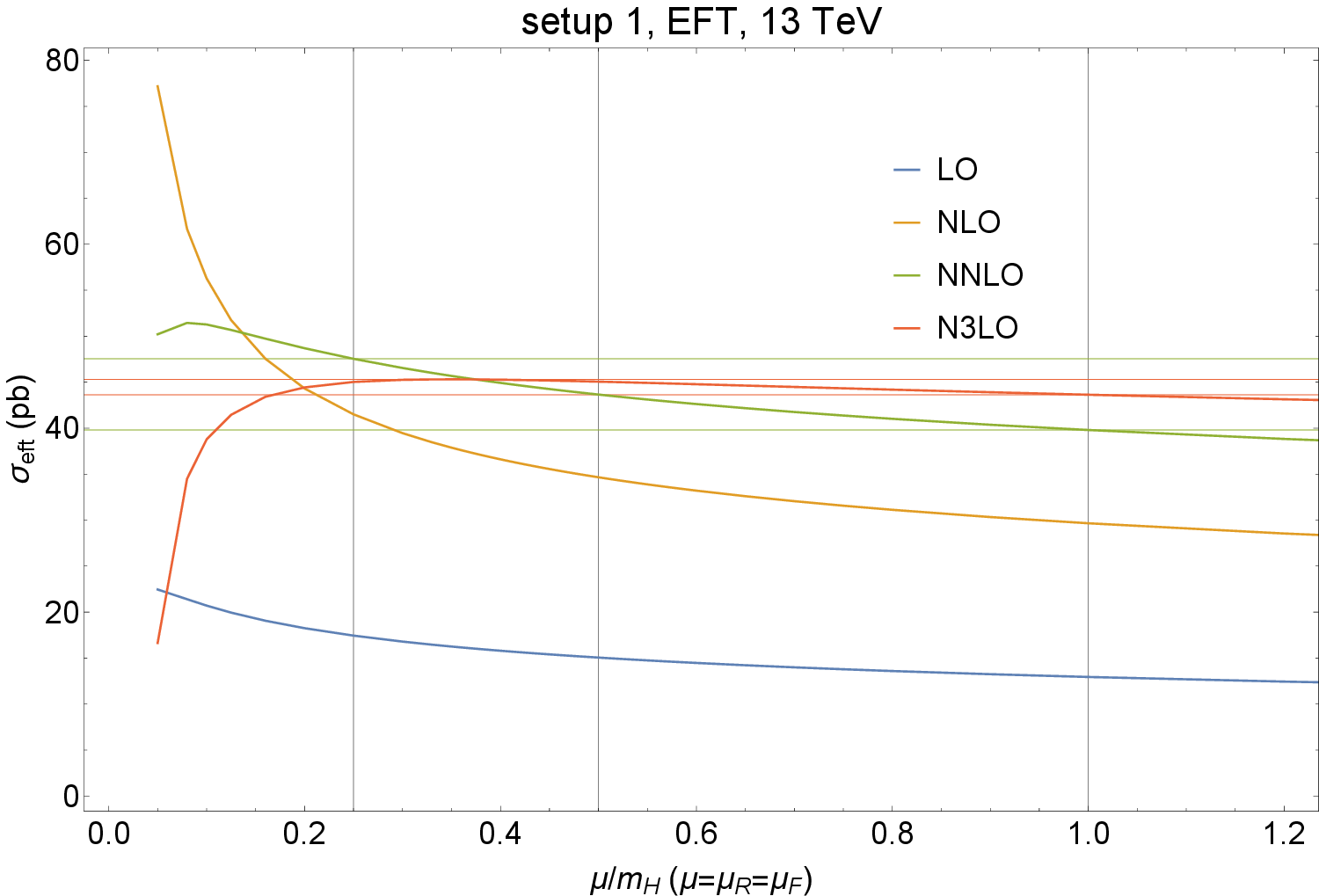}
\vspace*{0.0cm}

\caption{\label{fg:gghscale} \it Renormalization- and
factorization-scale dependence of the gluon fusion cross section as a
function of the identified scales in units of the Higgs boson mass at
different perturbative orders for a Higgs mass $M_H=125$ GeV at the LHC
with $\sqrt{s}=13$ TeV within the heavy-top-quark limit. From
Ref.~\cite{gghn3lo}.}
\end{figure}

\paragraph{Total cross section.}
Theoretical uncertainties in the prediction of the Higgs cross section
originate from three major sources, the dependence of the cross section
on the parton densities, the unknown corrections beyond N$^3$LO and the
parametric uncertainties originating from the input value of the strong
coupling $\alpha_s$ and to a lesser extent of the top and bottom quark
masses. The missing quark-mass effects beyond NLO have been estimated as
less than 1\% by an explicit large top-mass expansion of the NNLO
corrections beyond the heavy top-quark limit \cite{gghnnlomt}.  The
total uncertainty of the prediction for the total gluon-fusion cross
sections has been estimated as 4\% and can dominantly be traced back to
the renormalization- and factorization-scale dependence as well as the
PDF+$\alpha_s$ uncertainties \cite{yr4}. The renormalization- and
factorization-scale dependence is depicted in Fig.~\ref{fg:gghscale} for
the known perturbative orders in the heavy top-quark limit. A
significant reduction from $50-100\%$ to the few-per-cent level is
visible from LO to N$^3$LO.  The results for the cross section are
contained in the public codes {\sc Higlu} \cite{higlu} up to NNLO as
well as {\sc SusHi} \cite{sushi}, {\sc iHix} \cite{ihix} and {\sc
ggHiggs} \cite{gghiggs} up to N$^3$LO. The soft/collinear-gluon-resummed
results can be obtained with the code {\sc Troll} for corrections beyond
N$^3$LO \cite{troll}.

\paragraph{Transverse-momentum distribution.}
Apart from the total cross section also distributions are relevant for
analyzing the Higgs properties. Of particular interest in this context
is the transverse-momentum distribution of the Higgs boson that arises
due to the additional radiation of gluons. The LO contributions are part
of the real corrections to the gluon-fusion cross section at NLO. The
first results were obtained a long time ago \cite{gghptlo} that include
the full quark-mass dependence at LO. The NLO QCD corrections were
calculated in the heavy top-quark limit \cite{gghptnlo} that is expected
to be reliable up to $p_T$-values below the $t\bar t$ threshold as can
be inferred from a large top-mass expansion \cite{gghptnlomtexp}.
Recently a more rigorous inclusion of finite top-mass effects at NLO has
been studied by using exact results partially \cite{gghptnlomt}.  These
calculations are extended by the recent derivation of the NNLO QCD
corrections in the heavy-top limit \cite{gghptnnlo}. These predictions
have been matched to a soft-gluon resummed expression that provides a
regular prediction for small values of $p_T$ thus removing the spurious
singularities in $\log(M_H^2/p_T^2)$ in fixed-order expressions
\cite{gghptres,gghptnnll}. The basic structure of matched fixed-order
and resummed expression can be sketched as
\begin{equation}
\frac{d\sigma}{dp_T^2} = \frac{d\sigma^{(res)}}{dp_T^2} - \left[
\frac{d\sigma^{(res)}}{dp_T^2} \right]_{f.o.} +
\frac{d\sigma^{(f.o.)}}{dp_T^2}
\label{eq:ptres}
\end{equation}
where $d\sigma^{(res)}$ denotes the resummed and $d\sigma^{(f.o.)}$ the
fixed-order differential cross section. The resummed part is given by
\begin{equation}
\frac{d\sigma^{(res)}}{dp_T^2} = \sum_{ab = gg,gq,q\bar q} \int_0^1 dx_1
\int_0^1 dx_2 f_a(x_1,\mu_F^2) f_b(x_2,\mu_F^2)
\frac{d\hat\sigma_{ab}^{(res)}}{dp_T^2}
\end{equation}
with the partonic cross section kernel resummed in impact-parameter
space \cite{gghptres,gghptnnll},
\begin{equation}
\frac{d\hat\sigma^{(res)}}{dp_T^2} = z\int_0^\infty db \frac{b}{2}
J_0(bp_T) {\cal W}_{ab}(b,M_H,\hat s = M_H^2/z; \alpha_s(\mu_R^2),
\mu_R^2, \mu_F^2)
\end{equation}
where $\mu_R (\mu_F)$ denotes the renormalization (factorization) scale
and $J_0(x)$ the 0th-order Bessel function. Introducing Mellin moments
$f_N = \int_0^1 dz z^{N-1} f(z)$ the resummed factor ${\cal W}_{ab,N}$
factorizes into a (perturbative) hard part ${\cal H_N}$ and an exponential
factor,
\begin{equation}
{\cal W}_{ab,N} = {\cal H}_N \exp{{\cal G}_N}
\end{equation}
where the functional dependences have been suppressed. The exponent
${\cal G}_N$ can be expanded as
\begin{equation}
{\cal G}_N = L g^{(1)}(\alpha_s L) + g_N^{(2)}(\alpha_s L) +
\sum_{n=3}^\infty \left( \frac{\alpha_s}{\pi} \right)^{n-2}
g_N^{(n)}(\alpha_s L)
\end{equation}
Again functional dependences on the renormalization and resummation
scales have been suppressed. The basic ingredient of the resummation
formula is the large logarithmic expansion parameter
\begin{equation}
L = \log \frac{Q^2 b^2}{b_0^2}
\end{equation}
with $b_0=2 e^{-\gamma_E}$ ($\gamma_E = 0.5772\ldots$ is the Euler
number) and $Q$ denoting the resummation scale. In order to impose the
unitarity constraint that the integration over the transverse momentum
reproduces the known fixed-order cross section the logarithmic parameter
can be shifted to \cite{gghptnnll}
\begin{equation}
L \to \bar L = \log \left(\frac{Q^2 b^2}{b_0^2} + 1 \right)
\end{equation}
without modifying the limit of large $L$ (i.e.~large $b$) but
introducing a damping of effects for small impact parameters $b$ that
are not covered by the resummation. The expressions for the terms
$g_N^{(i)}$ can be found in \cite{gghptnnll}. The consistent matching
between the resummed and fixed order expressions requires the
subtraction of the fixed-order expansion of the resummed expression as
shown by the second term in Eq.~(\ref{eq:ptres}). This matching has been
performed at NNLL+NNLO level \cite{gghptnnll}\footnote{Here NNLO refers
to the corresponding total cross section, while it is NLO for the $p_T$
distribution at large $p_T$ values.}. In addition finite quark-mass
effects have been included in the resummed expressions at NLL+NLO, too
\cite{gghptnllmt, gghptcomp}\footnote{NLO refers to LO-mass effects on
the $p_T$ distribution at large $p_T$ values.}. The latter step requires
to use different matching scales $Q$ for the pure top-induced
contributions and those that involve bottom loops.  The resummation work
has been extended to the case of the $p_T$ distribution for fixed
rapidity at the NNLL+NNLO level \cite{gghptnnlly}.  During the last
years the full NLO expressions have been implemented in the {\sc Powheg}
box \cite{powheg} providing in this way a reliable NLO event generator
\cite{gghpow}. A second matching to parton showers has been performed by
combining {\sc SusHi} \cite{sushi} and the {\sc Mc@nlo} \cite{mg5amcnlo}
framework in the code {\sc aMCSusHi} \cite{amcsushi} at NLO accuracy,
too.

\subsubsection{\it Minimal supersymmetric extension}
\begin{figure}[hbt]
\begin{center}
\setlength{\unitlength}{1pt}
\begin{picture}(180,100)(0,0)

\Gluon(0,20)(50,20){-3}{5}
\Gluon(0,80)(50,80){3}{5}
\ArrowLine(50,20)(50,80)
\ArrowLine(50,80)(100,50)
\ArrowLine(100,50)(50,20)
\DashLine(100,50)(150,50){5}
\put(155,46){$\Phi$}
\put(20,46){$t,b,\tilde q$}
\put(-15,18){$g$}
\put(-15,78){$g$}

\end{picture}  \\
\setlength{\unitlength}{1pt}
\caption{\label{fg:mssmgghlodia} \it Typical diagram contributing to
$gg\to \Phi$ at lowest order.}
\end{center}
\end{figure}
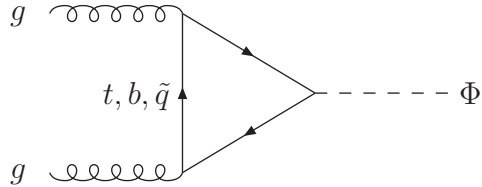

The gluon-fusion mechanism \cite{glufus}
\begin{displaymath}
pp \to gg \to \Phi
\end{displaymath}
dominates the neutral MSSM Higgs boson production at the LHC in the
phenomenologically relevant Higgs mass ranges for small and moderate
values of $\mbox{tg$\beta$}$. Only for large $\mbox{tg$\beta$}$ can the
associated $\Phi b\bar b$ production channel develop a larger cross
section due to the enhanced Higgs couplings to bottom quarks
\cite{htt,att}.  Analogous to the gluonic decay modes, the gluon
coupling to the neutral Higgs bosons in the MSSM is built up by loops
involving top and bottom quarks as well as squarks, see
Fig.~\ref{fg:mssmgghlodia}.

The partonic cross sections can be obtained from the gluonic widths of
the Higgs bosons at lowest order \cite{higgsqcd,SQCD}:
\begin{eqnarray}
\hat\sigma^\Phi_{LO} (gg\to \Phi) & = & \sigma^\Phi_0 \delta
(1 - z) \label{eq:mssmgghlo} \\
\sigma^\Phi_0 & = & \frac{\pi^2}{8M_\Phi^3}\Gamma_{LO}(\Phi\to gg)
\nonumber \\
\sigma^{h/H}_0 & = & \frac{G_{F}\alpha_{s}^{2}(\mu)}{288 \sqrt{2}\pi} \
\left| \sum_{Q} g_Q^{h/H} A_Q^{h/H} (\tau_{Q})
+ \sum_{\widetilde{Q}} g_{\widetilde{Q}}^{h/H} A_{\widetilde{Q}}^{h/H}
(\tau_{\widetilde{Q}}) \right|^{2} \nonumber \\
\sigma^A_0 & = & \frac{G_{F}\alpha_{s}^{2}(\mu)}{128 \sqrt{2}\pi} \
\left| \sum_{Q} g_Q^A A_Q^A (\tau_{Q}) \right|^{2} \nonumber
\end{eqnarray}
where the scaling variables are defined as $z=M_\Phi^2/\hat s$, $\tau_i
=4M_i^2/M_\Phi^2~~(i=Q,\widetilde{Q})$, and $\hat{s}$ denotes the
partonic c.m.~energy squared. The amplitudes $A_{Q,\widetilde{Q}}^\Phi
(\tau_{Q,\widetilde{Q}})$ are defined in Eqs.~(\ref{eq:mssmhgg},
\ref{eq:mssmagg}), and the MSSM couplings
$g_Q^\Phi,g^\Phi_{\widetilde{Q}}$ can be found in Tables \ref{tb:hcoup}
and \ref{tb:hsqcoup}.  In the narrow-width approximation the hadronic
cross sections are given by
\begin{equation}
\sigma_{LO}(pp\to \Phi) = \sigma^\Phi_0 \tau_\Phi \frac{d{\cal L}^{gg}}
{d\tau_\Phi}
\end{equation}
with the gluon luminosity defined in Eq.~(\ref{eq:gglum}) and the
scaling variables $\tau_\Phi = M^2_\Phi/s$ where $s$ specifies the total
hadronic c.m.~energy squared. For small $\mbox{tg$\beta$}$ the top loop
contribution is dominant, while for large values of $\mbox{tg$\beta$}$
the bottom quark contribution is strongly enhanced. If the squark masses
are less than $\sim 400$ GeV, their contribution is significant, and for
squark masses beyond $\sim 500$ GeV they can safely be neglected
\cite{SQCD}.

\paragraph{QCD corrections.}
In the past the full two-loop QCD corrections to the quark and squark
loops of the gluon-fusion cross section were calculated
\cite{higgsqcd,hgaga1,gghsqnlo,gghsusy}. In complete analogy to the SM case
they consist of virtual corrections to the basic $gg\to \Phi$ process
and real corrections due to the associated production of the Higgs
bosons with massless partons.  Thus the contributions to the final
result for the hadronic cross section can in complete analogy to the SM
case be classified as
\begin{equation}
\sigma(pp \rightarrow \Phi +X) = \sigma^\Phi_{0} \left[ 1+ C^\Phi
\frac{\alpha_{s}}{\pi} \right] \tau_\Phi \frac{d{\cal
L}^{gg}}{d\tau_\Phi} +
\Delta\sigma^\Phi_{gg} + \Delta\sigma^\Phi_{gq} +
\Delta\sigma^\Phi_{q\bar{q}} + \Delta\sigma^\Phi_{qq} +
\Delta\sigma^\Phi_{qq'}
\label{eq:mssmgghqcd5}
\end{equation}
where $\Delta\sigma^\Phi_{qq}$ and $\Delta\sigma^\Phi_{qq'}$ start to
contribute at NNLO. The analytic NLO expressions for arbitrary Higgs
boson and quark masses are rather involved and can be found in
\cite{higgsqcd,hgaga1}. As in the SM case the (s)quark-loop masses have been
identified with the pole masses $M_Q\, (M_{\widetilde{Q}})$, while the
QCD coupling and PDFs of the proton are defined in the $\overline{\rm
MS}$ scheme with five active flavours.  The axial $\gamma_5$ coupling
can be regularized in the 't~Hooft--Veltman scheme \cite{thoovel} or its
extension by Larin \cite{larin}, which preserves the chiral symmetry in
the massless quark limit by the addition of supplementary counter terms
and fulfills the non-renormalization theorem \cite{adlerbardeen} of the
ABJ anomaly \cite{ABJ} at vanishing momentum transfer. The same result
can also be obtained with the scheme of Ref.~\cite{kreimer} that gives up
the cyclicity of the traces involving Clifford matrices.

Similar to the SM the leading terms in the heavy-quark limit provide a
reliable approximation for small $\mbox{tg$\beta$}$ up to Higgs masses
of $\sim 1$ TeV with a maximal deviation of $\sim 25\%$ for
$\mbox{tg$\beta$}\lsim 5$ in the intermediate mass range.  The squark
contributions in the heavy-squark limit coincide with the heavy-quark
case apart from the mismatch related to the effective Lagrangians of
Eqs.~(\ref{eq:leff}, \ref{eq:leffsq}). The genuine SUSY--QCD corrections
are the same as for the gluonic decay width as discussed in Section
\ref{sc:hgg}. They are moderate for small values of $\mbox{tg}\beta$
\cite{hggsqcd0,hggsqcd}, but can be large for large $\mbox{tg}\beta$ due
to the contribution of the $\Delta_b$ terms and a sizeable remainder
beyond these approximate contributions \cite{hggsqcd}.

In the opposite limit, where the Higgs mass is much larger than the
quark mass, the analytic results coincide with the SM expressions for
both the scalar and pseudoscalar Higgs particles up to NLO at which
the results for small quark masses are known \cite{higgsqcd}. This
coincidence reflects the restoration of the chiral symmetry in the
massless quark limit. The leading logarithmic structure is the same as
in the SM case, see Eq.~(\ref{eq:hgglog}). The non-Abelian part of these
logarithms has not been resummed so that the uncertainties for large
values of $\mbox{tg}\beta$ are sizeable due to the dominance of the
bottom-quark loops involving the large logarithms.

The QCD corrections are positive and large, increasing the MSSM Higgs
production cross sections at the LHC by up to about 100\%.
The effect of the pure squark loops (without gluino exchange) on the scalar
Higgs $K$ factors is of moderate size and lead to a maximal modification
by about 10\%. Squark mass effects on top of the LO ones can amount to
$\sim 20\%$ for large Higgs masses \cite{gghsqnlo}. For the top-loop
contributions alone the N$^3$LO corrections can be used consistently.
SUSY-electroweak corrections are unknown so far.

\paragraph{Transverse-momentum distribution.}
The results of the {\sc Powheg} implementation of the full LO matrix
elements for the transverse-momentum distribution of Ref.~\cite{gghpow}
and {\sc aMCSusHi} \cite{amcsushi} can also be used for the MSSM, since
they also include variable top and bottom Yukawa couplings. The code of
Ref.~\cite{gghpow} includes squark-loop contributions in addition. A
rigorous comparison between the {\sc Powheg} implementation and
analytical resummation approaches \cite{gghptnllmt} as well as {\sc
aMCSusHi} has been performed \cite{gghptcomp, yr4}. This comparison
addressed in particular different approaches for the setting of the
resummation scale of the bottom-loop contributions. In scenarios where
the top loops provide the dominant contribution the differences between
the codes are in the range of up to about 50\%, while for scenarios with
bottom-loop dominance they are larger and can reach  100\%
\cite{gghptcomp, yr4}. These differences will only be reduced by
calculating the NLO corrections to the transverse-momentum distribution
including the full mass dependence.

\subsection{\it Vector-boson fusion: $qq\to qqV^*V^* \to qqH$}
\subsubsection{\it Standard Model}
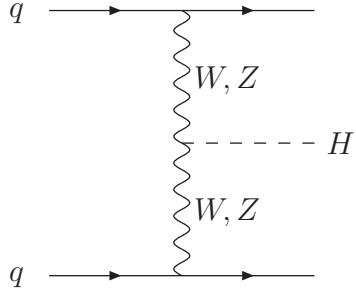
\begin{figure}[hbt]
\begin{center}
\setlength{\unitlength}{1pt}
\begin{picture}(120,110)(0,0)

\ArrowLine(0,0)(50,0)
\ArrowLine(50,0)(100,0)
\ArrowLine(0,100)(50,100)
\ArrowLine(50,100)(100,100)
\Photon(50,0)(50,50){3}{5}
\Photon(50,50)(50,100){3}{5}
\DashLine(50,50)(100,50){5}
\put(105,46){$H$}
\put(-15,-2){$q$}
\put(-15,98){$q$}
\put(55,21){$W,Z$}
\put(55,71){$W,Z$}

\end{picture}  \\
\setlength{\unitlength}{1pt}
\caption{\label{fg:vvhlodia} \it Diagram contributing to $qq \to qqV^*V^*
\to qqH$ at lowest order.}
\end{center}
\end{figure}
\noindent
At the LHC the second important Higgs production channel is the
vector-boson-fusion (VBF) mechanism (see Fig.~\ref{fg:vvhlodia})
\cite{vvh,vvhqcd}.  For intermediate Higgs masses the
vector-boson-fusion cross section is about one order of magnitude
smaller than the gluon-fusion one.  The cross section can be
approximated by the $t$-channel diagrams of the type shown in
Fig.~\ref{fg:vvhlodia} within $\sim 1\%$ accuracy, i.e.~without any
colour-cross talk between the quark lines, while interference effects
for identical quark flavours and $s$-channel contributions are at the
per-cent level after subtracting the corresponding Higgs-strahlung
component from the $s$-channel contributions \cite{vbfnlo}.  Within the
structure-function approach the leading order partonic
vector-boson-fusion cross section \cite{vvh} can be cast into the form
($V = W,Z$):
\begin{eqnarray}
d\sigma_{LO} & = & \frac{1}{4} \frac{\sqrt{2}G_F^3M_V^8 q_1^2 q_2^2}
{[q_1^2-M_V^2]^2 [q_2^2-M_V^2]^2} \nonumber \\
& & \left\{ F_1(x_1,\mu_F^2) F_1(x_2,\mu_F^2) \left[ 2+\frac{(q_1
q_2)^2}{q_1^2q_2^2} \right] \right. \nonumber \\
&&+\frac{F_1(x_1,\mu_F^2)F_2(x_2,\mu_F^2)}{P_2q_2}
\left[\frac{(P_2q_2)^2}{q_2^2}-
M_P^2+\frac{1}{q_1^2}\left(P_2q_1-\frac{P_2q_2}{q_2^2}q_1q_2\right)^2
\right] \nonumber \\
&&+\frac{F_2(x_1,\mu_F^2)F_1(x_2,\mu_F^2)}{P_1q_1}
\left[\frac{(P_1q_1)^2}{q_1^2}-
M_P^2+\frac{1}{q_2^2}\left(P_1q_2-\frac{P_1q_1}{q_1^2}q_1q_2\right)^2
\right] \nonumber \\
& & +\frac{F_2(x_1,\mu_F^2)F_2(x_2,\mu_F^2)}{(P_1q_1)(P_2q_2)}
\left[P_1P_2 - \frac{(P_1q_1)(P_2q_1)}{q_1^2} -
\frac{(P_2q_2)(P_1q_2)}{q_2^2} \right.  \nonumber \\
& & \left. \hspace*{4.5cm}+\frac{(P_1q_1)(P_2q_2)(q_1q_2)}{q_1^2q_2^2}
\right]^2 \nonumber \\
& &\left.  +\frac{F_3(x_1,\mu_F^2)F_3(x_2,\mu_F^2)}{2(P_1q_1)(P_2q_2)}
\left[ (P_1P_2)(q_1q_2) - (P_1q_2)(P_2q_1) \right] \right\} dx_1 dx_2
\frac{dP\!S_3}{\hat s}
\label{eq:vvhlo}
\end{eqnarray}
where $dP\!S_3$ denotes the three-particle phase space of the
final-state particles, $M_P$ the proton mass, $P_{1,2}$ the proton
momenta and $q_{1,2}$ the momenta of the virtual vector bosons $V^*$.
The functions $F_i(x,\mu_F^2)~(i=1,2,3)$ are the usual structure
functions from deep-inelastic scattering processes at the factorization
scale $\mu_F$:
\begin{eqnarray}
F_1(x,\mu_F^2) & = & \sum_q (v_q^2+a_q^2) [q(x,\mu_F^2) + \bar q(x,\mu_F^2)] \nonumber \\
F_2(x,\mu_F^2) & = & 2x \sum_q (v_q^2+a_q^2) [q(x,\mu_F^2) + \bar q(x,\mu_F^2)] \nonumber \\
F_3(x,\mu_F^2) & = & 4 \sum_q v_qa_q [-q(x,\mu_F^2) + \bar q(x,\mu_F^2)]
\label{eq:stfu}
\end{eqnarray}
where $v_q\, (a_q)$ are the (axial) vector couplings of the quarks $q$
to the vector bosons $V$: $v_q = a_q = \sqrt{2}$ for $V=W$ and $v_q = 2I_{3q}
- 4e_q \sin^2\theta_W$, $a_q = 2 I_{3q}$ for $V=Z$. $I_{3q}$ is the third weak
isospin component and $e_q$ the electric charge of the quark $q$.

\begin{figure}[ht]
\begin{center}
\begin{picture}(150,280)(0,0)
\put(-170,-10.0){\includegraphics{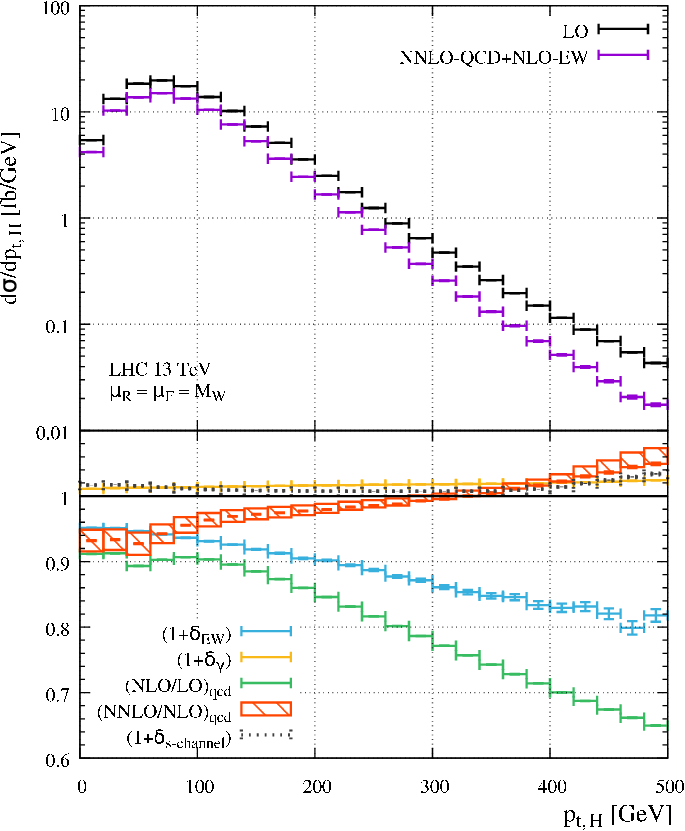}}
\put(100,-10.0){\includegraphics{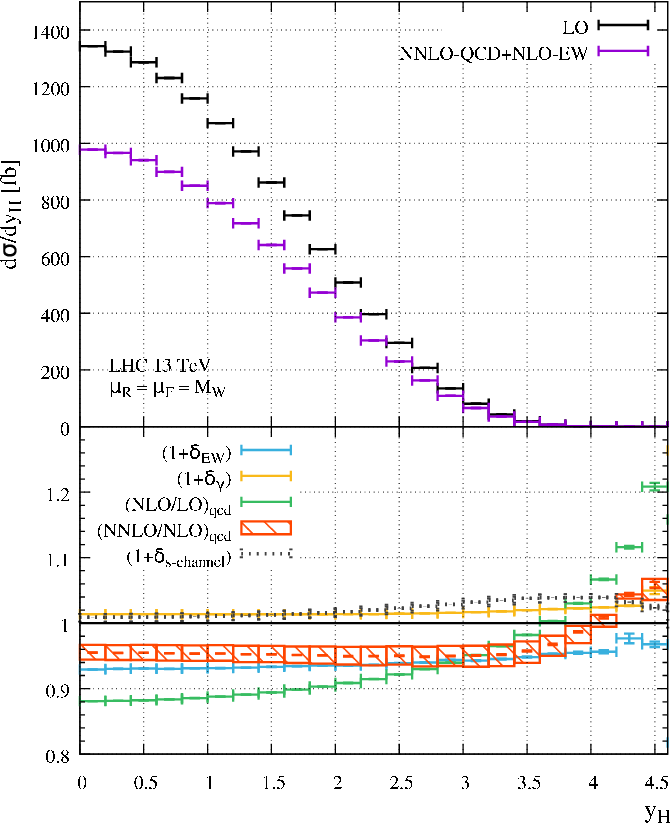}}
\end{picture}
\caption{\it Transverse-momentum and rapidity distributions in VBF at LO and
including the NNLO QCD and NLO electroweak corrections for $\sqrt{s}=13$
TeV and $M_H=125$ GeV. Upper plot: differential cross sections; lower
plot: individual corrections. From Ref.~\cite{yr4} (page 90).}
\label{fg:vbfcorr}
\end{center}
\end{figure}

In the past the NLO QCD corrections have been calculated first within
the structure function approach \cite{vvhqcd}. Since, at lowest order,
the proton remnants are color singlets, no color will be exchanged
between the first and the second incoming (outgoing) quark line and
hence the QCD corrections only consist of the well-known corrections to
the structure functions $F_i(x,\mu_F^2)~(i=1,2,3)$. The final result for
the NLO QCD-corrected cross section leads to the replacements
\begin{eqnarray}
F_i(x,\mu_F^2) & \to & F_i(x,\mu_F^2) + \Delta F_i(x,\mu_F^2,Q^2) \hspace*{1cm} (i=1,2,3)
\nonumber \\
\Delta F_1(x,\mu_F^2,Q^2) & = & \frac{\alpha_s(\mu_R)}{\pi}\sum_q (v_q^2+a_q^2)
\int_x^1 \frac{dy}{y} \left\{ \frac{2}{3} [q(y,\mu_F^2) + \bar q(y,\mu_F^2)]
\right. \nonumber \\
& &
\left[ -\frac{3}{4} P_{qq}(z) \log \frac{\mu_F^2z}{Q^2} + (1+z^2) {\cal D}_1(z)
- \frac{3}{2} {\cal D}_0(z) \right. \nonumber \\
& & \left. \hspace*{6cm} + 3 - \left(
\frac{9}{2} + \frac{\pi^2}{3} \right) \delta(1-z) \right]
\nonumber \\
& & \left. + \frac{1}{4} g(y,\mu_F^2) \left[ -2 P_{qg}(z) \log
\frac{\mu_F^2z}{Q^2(1-z)} + 4z(1-z)
- 1 \right] \right\} \\
\Delta F_2(x,\mu_F^2,Q^2) & = & 2x\frac{\alpha_s(\mu_R)}{\pi}\sum_q (v_q^2+a_q^2)
\int_x^1 \frac{dy}{y} \left\{ \frac{2}{3} [q(y,\mu_F^2) + \bar q(y,\mu_F^2)]
\right. \nonumber \\
& &
\left[ -\frac{3}{4} P_{qq}(z) \log \frac{\mu_F^2z}{Q^2} + (1+z^2) {\cal D}_1(z)
- \frac{3}{2} {\cal D}_0(z) \right. \nonumber \\
& & \left. \hspace*{3.0cm} + 3 + 2z - \left(
\frac{9}{2} + \frac{\pi^2}{3} \right) \delta(1-z) \right]
\nonumber \\
& & \left. + \frac{1}{4} g(y,\mu_F^2) \left[ -2P_{qg}(z) \log
\frac{\mu_F^2z}{Q^2(1-z)}
+ 8z(1-z) - 1 \right] \right\} \\
\Delta F_3(x,\mu_F^2,Q^2) & = & \frac{\alpha_s(\mu_R)}{\pi} \sum_q 4 v_q a_q
\int_x^1 \frac{dy}{y} \left\{ \frac{2}{3} [-q(y,\mu_F^2) + \bar
q(y,\mu_F^2)]
\right. \nonumber \\
& &
\left[ -\frac{3}{4} P_{qq}(z) \log \frac{\mu_F^2z}{Q^2} + (1+z^2) {\cal D}_1(z)
- \frac{3}{2} {\cal D}_0(z) \right. \nonumber \\
& & \left. \left. \hspace*{3cm} + 2 + z - \left(
\frac{9}{2} + \frac{\pi^2}{3} \right) \delta(1-z) \right] \right\}
\end{eqnarray}
where $z=x/y$ and the functions $P_{qq}, P_{qg}$ denote the
Altarelli--Parisi splitting functions, which are given by \cite{apsplit}
\begin{eqnarray}
P_{qq}(z) & = & \frac{4}{3} \left\{ 2{\cal D}_0(z)-1-z+\frac{3}{2}\delta(1-z)
\right\} \nonumber \\
P_{qg}(z) & = & \frac{1}{2} \left\{ z^2 + (1-z)^2 \right\}
\end{eqnarray}
The plus distributions are defined as
\begin{equation}
{\cal D}_i(z) = \left( \frac{\log^i(1-z)}{1-z} \right)_+
\end{equation}
The physical scale $Q$ is given by $Q^2 = -q_i^2$ for $x=x_i~(i=1,2)$.
These expressions have to be inserted in Eq.~(\ref{eq:vvhlo}) and the
full result expanded up to NLO. The typical renormalization and
factorization scales are fixed by the vector-boson momentum transfer
$\mu_R=\mu_F=Q$.  The size of the QCD corrections amounts to about 10\%
and is thus small \cite{vvhqcd} as can be inferred from
Fig.~\ref{fg:vbfcorr} that displays the individual corrections to the
cross section \cite{yr4}. These results have been extended by the
calculation of the full NLO QCD corrections including interference
contributions between the $s,t,u$-channels thus confirming the smallness
of the additional contributions at NLO, too \cite{vbfnlo}. The NLO
electroweak corrections reduce the cross section by about 10\%
\cite{vbfelw} so that there is a significant interplay between the
QCD and electroweak corrections, if the latter are defined in the
$G_F$-scheme, see Fig.~\ref{fg:vbfcorr}. For a reliable prediction of
the vector-boson-fusion cross section both corrections have to be taken
into account. The NLO QCD and electroweak corrections are also known
exclusively and have been implemented in the public Monte Carlo programs
HAWK \cite{hawk} and VBFNLO \cite{vbfnloprog}.  The full NLO results can
also be generated with the {\sc Mg5\_amc@nlo} framework \cite{mg5amcnlo}
and are available within the {\sc Powheg} box \cite{powheg} so that a
full matching to parton showers can be used. The NNLO QCD corrections
have been obtained in the structure-function approach giving rise to a
per-cent effect on the total cross section \cite{vbfnnloprog,vbfnnlo},
see Fig.~\ref{fg:vbfcorr}. These are implemented in the public codes
{\sc Vbf@nnlo} \cite{vbfnnloprog} and {\sc proVBF} \cite{vbfnnlo} with
the latter providing exclusive results. Up to NNLO non-factorizing
contributions have been shown to be small
\cite{vbfelw,vbfnnloprog,vbfnonfac}. The theoretical calculations have
been extended to N$^3$LO very recently \cite{vbfn3lo}. The N$^3$LO
corrections are tiny, but reduce the theoretical uncertainties
significantly.

\subsubsection{\it Minimal supersymmetric extension}
\begin{figure}[hbt]
\begin{center}
\setlength{\unitlength}{1pt}
\begin{picture}(120,120)(0,0)

\ArrowLine(0,0)(50,0)
\ArrowLine(50,0)(100,0)
\ArrowLine(0,100)(50,100)
\ArrowLine(50,100)(100,100)
\Photon(50,0)(50,50){3}{5}
\Photon(50,50)(50,100){3}{5}
\DashLine(50,50)(100,50){5}
\put(105,46){$h,H$}
\put(-15,-2){$q$}
\put(-15,98){$q$}
\put(55,21){$W,Z$}
\put(55,71){$W,Z$}

\end{picture}  \\
\setlength{\unitlength}{1pt}
\caption{\label{fg:mssmvvhlodia} \it Diagram contributing to $qq \to
qqV^*V^*
\to qqh/qqH$ at lowest order.}
\end{center}
\end{figure}
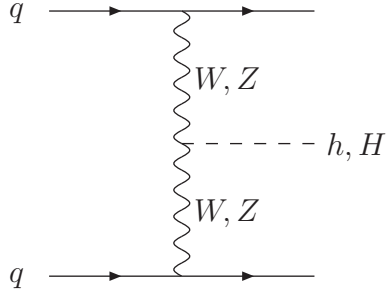
\noindent
Due to the absence of vector boson couplings to pseudoscalar Higgs
particles $A$, only the scalar Higgs bosons $h,H$ can be produced via
the vector-boson-fusion mechanism at tree level (see
Fig.~\ref{fg:mssmvvhlodia}).  However, these processes are suppressed
with respect to the SM cross section due to the MSSM couplings
($g_V^{h/H} = \sin (\alpha - \beta)/\cos (\alpha - \beta)$),
\begin{equation}
\sigma(pp\to qq\to qq+h/H) = \left(g_V^{h/H} \right)^2 \sigma
(pp\to qq\to qqH_{SM})
\end{equation}
It turns out that the vector-boson-fusion mechanism is less relevant in
the MSSM, because for large heavy scalar Higgs masses $M_H$, the MSSM
couplings $g_V^H$ are very small apart from the decoupling regime where
the light scalar Higgs boson becomes SM-like. The relative pure QCD
corrections are the same as for the SM Higgs particle, i.e.~$\sim 10\%$
\cite{vvhqcd}. The genuine SUSY--QCD corrections amount to less than a
per cent and are thus small \cite{vbfsqcd}. Some years ago the
SUSY--electroweak corrections have been determined \cite{vbfselw}. Being
moderate in size in general they can modify the VBF cross section by
$\sim 10\%$.

\subsection{\it Higgs-strahlung: $q\bar q\to V^* \to VH$}
\subsubsection{\it Standard Model}
\label{sc:higgsstrahlung}
\begin{figure}[hbt]
\begin{center}
\setlength{\unitlength}{1pt}
\begin{picture}(160,120)(0,-10)

\ArrowLine(0,100)(50,50)
\ArrowLine(50,50)(0,0)
\Photon(50,50)(100,50){3}{5}
\Photon(100,50)(150,100){3}{6}
\DashLine(100,50)(150,0){5}
\put(155,-4){$H$}
\put(-15,-2){$\bar q$}
\put(-15,98){$q$}
\put(65,65){$W,Z$}
\put(155,96){$W,Z$}

\end{picture}  \\
\setlength{\unitlength}{1pt}
\caption{\label{fg:vhvlodia} \it Diagram contributing to $q\bar q \to V^*
\to VH$ at lowest order.}
\end{center}
\end{figure}
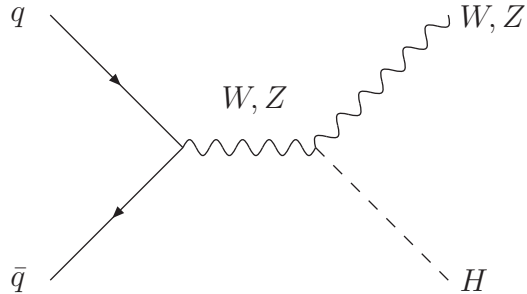
\noindent
The Higgs-strahlung mechanism $q\bar q\to V^* \to VH~(V=W,Z)$ (see
Fig.~\ref{fg:vhvlodia}) is important in the intermediate Higgs mass
range due to the possibility to tag the associated vector boson and to
reconstruct the $H\to b\bar b$ decay for strongly boosted Higgs bosons
by using jet-substructure techniques \cite{vhbb,substructure}. Its cross
section is about one to two orders of magnitude smaller than the
gluon-fusion cross section for the relevant Higgs mass range. The
partonic cross section can be expressed at lowest-order as \cite{vhv}
\begin{equation}
\hat\sigma_{LO}(q\bar q\to VH)=\frac{G_F^2 M_V^4}{288\pi Q^2}(v_q^2 + a_q^2)
\sqrt{\lambda(M_V^2,M_H^2;Q^2)} \frac{\lambda(M_V^2,M_H^2;Q^2) + 12 M_V^2
/ Q^2}{(1-M_V^2/Q^2)^2}
\label{eq:vhvpart}
\end{equation}
where $\lambda(x,y;z) = (1-x/z-y/z)^2-4xy/z^2$ denotes the usual
two-body phase-space factor and $v_q\, (a_q)$ are the (axial) vector
couplings of the quarks $q$ to the vector bosons $V$, which have been
defined after Eq.~(\ref{eq:stfu}).  At leading order the partonic
c.m.~energy squared $\hat s$ coincides with the invariant mass $Q^2 =
M^2_{VH}$ of the Higgs--vector-boson pair squared, $\hat s=Q^2$. The
hadronic cross section can be obtained from convolving
Eq.~(\ref{eq:vhvpart}) with the corresponding $q\bar q$-luminosity,
\begin{equation}
\sigma_{LO}(pp \to q\bar q\to VH) = \int_{\tau_0}^1 d\tau \sum_q
\frac{d{\cal L}^{q\bar q}}{d\tau} \hat\sigma_{LO}(Q^2=\tau s)
\end{equation}
with $\tau_0 = (M_H+M_V)^2/s$ and $s$ the total hadronic c.m.~energy
squared and
\begin{equation}
\frac{d{\cal L}^{q\bar q}}{d\tau} = \int_\tau^1 \frac{dx}{x}~\left[
q(x,\mu_F^2) \bar q(\tau /x,\mu_F^2) + \bar q(x,\mu_F^2) q(\tau
/x,\mu_F^2) \right]
\end{equation}

\begin{figure}[ht]
\begin{center}
\begin{picture}(150,280)(0,0)
\put(-170,-10.0){\includegraphics{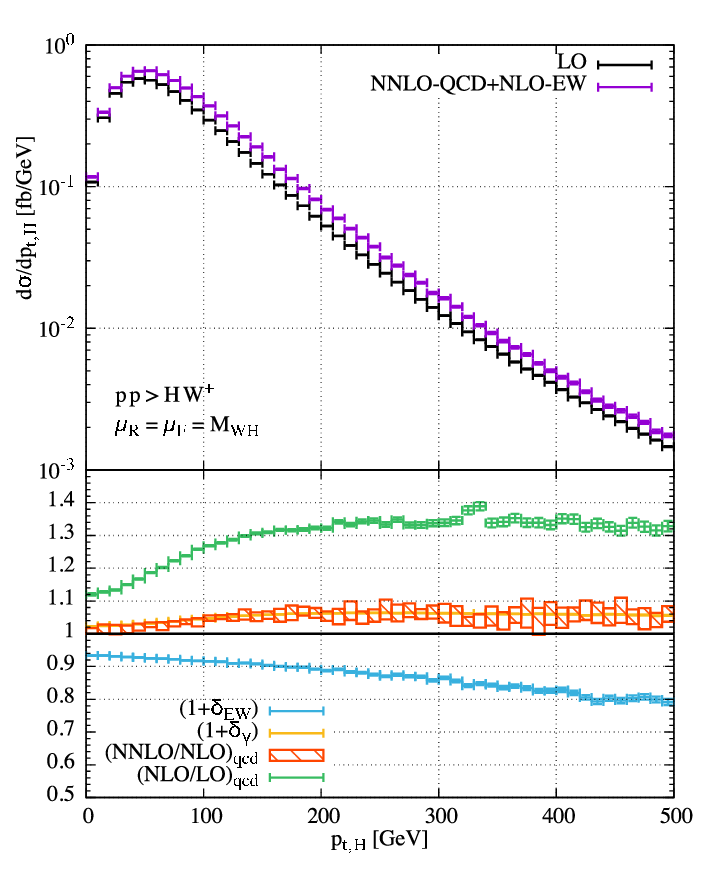}}
\put(100,-10.0){\includegraphics{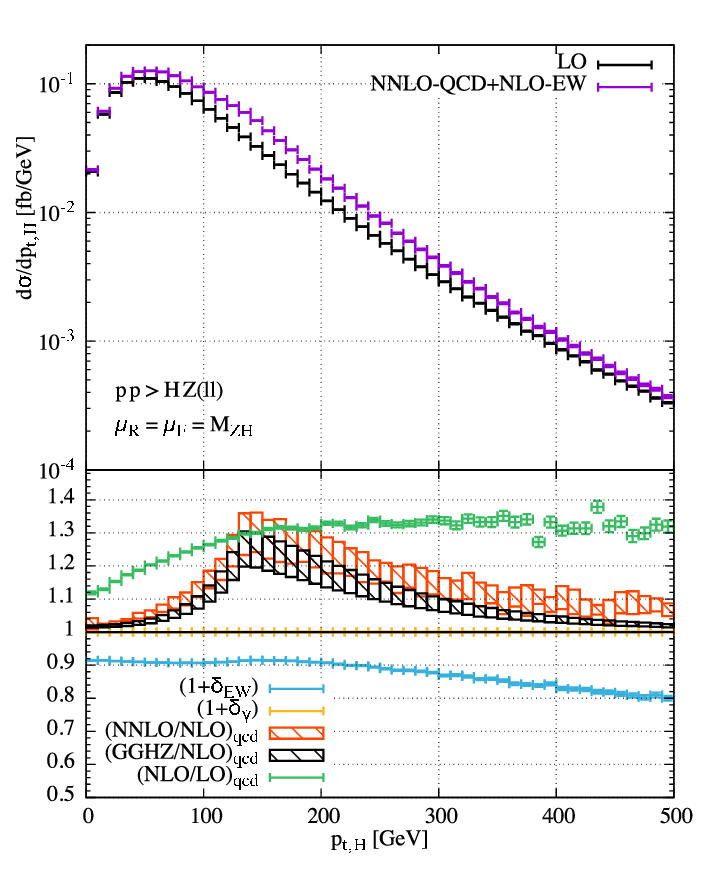}}
\end{picture}
\caption{\it Transverse-momentum distributions of $W^+H$ (left) and $ZH$
(right) production at LO and including the NNLO QCD and NLO electroweak
corrections for $\sqrt{s}=13$ TeV and $M_H=125$ GeV. Upper plot:
differential cross sections; lower plot: individual corrections. From
Ref.~\cite{yr4} (pages 100, 104).}
\label{fg:vhcorr}
\end{center}
\end{figure}

\begin{figure}[hbt]
\begin{center}
\setlength{\unitlength}{1pt}
\begin{picture}(180,90)(40,0)

\Gluon(0,20)(50,20){-3}{5}
\Gluon(0,80)(50,80){3}{5}
\ArrowLine(50,20)(50,80)
\ArrowLine(50,80)(100,50)
\ArrowLine(100,50)(50,20)
\Photon(100,50)(150,50){3}{5}
\Photon(150,50)(200,80){3}{5}
\DashLine(150,50)(200,20){5}
\put(205,76){$Z$}
\put(205,16){$H$}
\put(25,46){$t,b$}
\put(-15,18){$g$}
\put(-15,78){$g$}

\end{picture}
\begin{picture}(180,90)(-40,0)

\Gluon(0,20)(50,20){-3}{5}
\Gluon(0,80)(50,80){3}{5}
\ArrowLine(50,20)(50,80)
\ArrowLine(50,80)(100,80)
\ArrowLine(100,80)(100,20)
\ArrowLine(100,20)(50,20)
\Photon(100,80)(150,80){3}{5}
\DashLine(100,20)(150,20){5}
\put(155,76){$Z$}
\put(155,16){$H$}
\put(25,46){$t,b$}
\put(-15,18){$g$}
\put(-15,78){$g$}

\end{picture}  \\
\setlength{\unitlength}{1pt}
\caption{\label{fg:ggzhdia} \it Diagrams contributing to $gg\to ZH$
at NNLO.}
\end{center}
\end{figure}
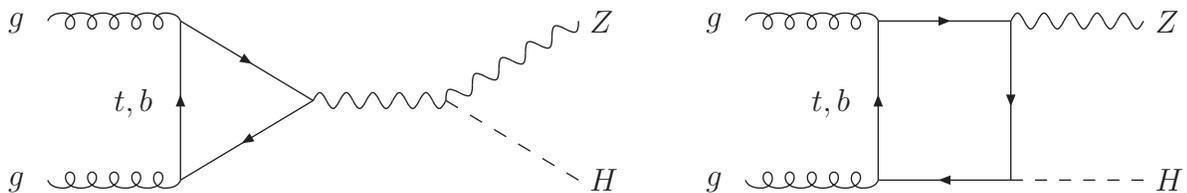

The NLO QCD corrections are identical to the corresponding corrections
to the Drell--Yan process \cite{vhvqcd}.  The natural scale of the
process is given by the invariant mass of the Higgs--vector-boson pair
in the final state, $\mu_R=\mu_F=Q$. The NLO QCD corrections increase
the total cross section by about 30\% and are thus moderate
\cite{vhvqcd}, see Fig.~\ref{fg:vhcorr}. Some time ago the NNLO QCD
corrections have been determined \cite{vhnnlo}. For the $ZH$ final state
there is a sizeable contribution from the loop-induced $gg\to ZH$
process, see Fig.~\ref{fg:ggzhdia}. It contributes $\sim 20\%$ to the
total cross section as can be inferred from Fig.~\ref{fg:vhcorr}, while
the rest of the NNLO corrections amounts to about 5\%.  Recently the NLO
QCD corrections to $gg\to ZH$ have been calculated in the
heavy-top-quark limit \cite{gg2zhnlo}. They increase this contribution
significantly. The calculation of the QCD corrections has been
supplemented by the full NLO electroweak corrections \cite{vhelw}.
Within the $G_F$-scheme they decrease the cross section by about 10\%
and are thus of the same relevance as the NNLO QCD corrections, see
Fig.~\ref{fg:vhcorr}. The NLO QCD and electroweak corrections are
implemented in the public program {\sc Hawk} \cite{hawk} including
leptonic decays of the final-state vector boson. The NNLO QCD
corrections are implemented in the public codes {\sc Hvnnlo}
\cite{hvnnlo} that includes final-state Higgs decays into leptons and
bottom quarks, {\sc Mcfm} \cite{mcfm}, {\sc Vhnnlo} \cite{vhnnlo}, {\sc
Vh@nnlo} \cite{vhannlo} and {\sc Nnlops} \cite{nnlops}. The NLO results
have been matched to parton showers within the {\sc Powheg} box
\cite{powheg} and the {\sc Mg5\_amc@nlo} framework \cite{mg5amcnlo}.

\subsubsection{\it Minimal supersymmetric extension}
\begin{figure}[hbt]
\begin{center}
\setlength{\unitlength}{1pt}
\begin{picture}(160,120)(0,-10)

\ArrowLine(0,100)(50,50)
\ArrowLine(50,50)(0,0)
\Photon(50,50)(100,50){3}{5}
\Photon(100,50)(150,100){3}{6}
\DashLine(100,50)(150,0){5}
\put(155,-4){$h,H$}
\put(-15,-2){$\bar q$}
\put(-15,98){$q$}
\put(65,65){$W,Z$}
\put(155,96){$W,Z$}

\end{picture}  \\ 
\setlength{\unitlength}{1pt}
\caption{\label{fg:mssmvhvlodia} \it Diagram contributing to $q\bar q
\to V^* \to h/H + V$ at lowest order.} 
\end{center}
\end{figure}
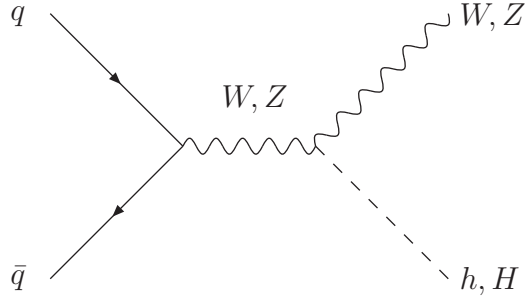
For the same reasons as in the vector-boson-fusion mechanism case, the
Higgs-strahlung off $W,Z$ bosons, $q\bar q\to V^* \to Vh/VH~(V=W,Z)$
(see Fig.~\ref{fg:mssmvhvlodia}), is smaller for the scalar MSSM Higgs
particles $h,H$ than for the SM Higgs boson apart from the decoupling
regime for the light scalar $h$. The cross sections can be easily
related to the SM cross sections,
\begin{equation}
\sigma(pp \to V+h/H) = \left(g_V^{h/H} \right)^2
\sigma(pp \to VH_{SM})
\end{equation}
which is true up to NNLO QCD apart from the loop-induced $gg\to Z+h/H$
process that involves the large top Yukawa coupling in addition to the
$HZZ$ coupling, see Fig.~\ref{fg:ggzhdia}.  Pseudoscalar couplings to
intermediate vector bosons are absent so that pseudoscalar Higgs
particles cannot be produced at tree level in this channel.  The
relative QCD corrections are the same as in the SM case and thus of
moderate size \cite{vhvqcd} in general up to NNLO. Only the
contamination from $gg\to Z+h/H$ will be weighted differently due to the
top Yukawa coupling factor. Its contribution ranges at about 20\%. The
genuine SUSY--QCD corrections are small \cite{vbfsqcd}, while the
genuine SUSY-electroweak corrections are unknown so far.

\subsection{\it Higgs bremsstrahlung off top and bottom quarks}
\subsubsection{\it Standard Model}
\begin{figure}[hbt]
\begin{center}
\setlength{\unitlength}{1pt}
\begin{picture}(360,120)(0,-10)

\ArrowLine(0,100)(50,50)
\ArrowLine(50,50)(0,0)
\Gluon(50,50)(100,50){3}{5}
\ArrowLine(100,50)(125,75)
\ArrowLine(125,75)(150,100)
\ArrowLine(150,0)(100,50)
\DashLine(125,75)(150,50){5}
\put(155,46){$H$}
\put(-15,98){$q$}
\put(-15,-2){$\bar q$}
\put(65,65){$g$}
\put(155,98){$t$}
\put(155,-2){$\bar t$}

\Gluon(250,0)(300,0){3}{5}
\Gluon(250,100)(300,100){3}{5}
\ArrowLine(350,0)(300,0)
\ArrowLine(300,0)(300,50)
\ArrowLine(300,50)(300,100)
\ArrowLine(300,100)(350,100)
\DashLine(300,50)(350,50){5}
\put(355,46){$H$}
\put(235,98){$g$}
\put(235,-2){$g$}
\put(355,98){$t$}
\put(355,-2){$\bar t$}

\end{picture}  \\
\setlength{\unitlength}{1pt}
\caption[ ]{\label{fg:httlodia} \it Typical diagrams contributing to
$q\bar q/gg \to t\bar tH$ at lowest order.}
\end{center}
\end{figure}

\paragraph{$t\bar tH$ production.}
In the intermediate Higgs mass range the cross section of the associated
production of the Higgs boson with a $t\bar t$ pair is smaller than
those of the Higgs-strahlung processes \cite{htt}, but of significant
size, since it allows to tag the additional $t\bar t$ pair and the Higgs
boson decays $H\to b\bar b$, $H\to WW$ as well as the rare photonic
decay mode $H\to \gamma\gamma$ on the long time run.  This process is
generated by gluon--gluon and $q\bar q$ initial states at leading order
(see Fig.~\ref{fg:httlodia}).  At the LHC the gluon--gluon channel is
dominant due to the enhanced gluon structure function analogous to the
gluon-fusion mechanism. The NLO QCD corrections to $t\bar tH$ production
have been calculated and result in a moderate increase of the cross
section by $\sim 20\%$ \cite{tthqcd0,tthqcd,tthqcd1}. The origin of this
moderate size is the strong phase-space suppression of the massive
three-particle threshold so that soft and collinear threshold effects
are strongly diminished. The main parts of the QCD corrections originate
from regions significantly above the production threshold and can be
approximated by a fragmentation approach involving first producing a
$t\bar t$ pair supplemented by the $t\to tH$ fragmentation in the
high-energy limit \cite{tthqcd0,tthfrag}.  Although this provides a bad
approximation for the magnitude of the cross section itself it leads to
a reasonable estimate of the relative QCD corrections \cite{tthqcd0}.
The full NLO results have recently been implemented in the {\sc Powheg}
box \cite{tthps}, matched to Sherpa \cite{tthsherpa} and generated
within the {\sc Mg5\_amc@nlo} framework \cite{tthqcd1} thus offering NLO
event generators matched to parton showers.  The NLO result has recently
been improved by a soft and collinear gluon resummation based on the
SCET approach starting from the boosted final-state particle
triplet\footnote{The recent alternative approach using conventional
threshold resummation techniques does not yield a sizeable contribution
beyond NLO \cite{tthnll0} due to the strong threshold suppression.}
\cite{tthnll} leading to a further increase of the cross section by
5-10\%. The residual scale dependence is reduced to the level of
$5-10\%$. Recently the electroweak corrections have been calculated for
$t\bar tH$ production \cite{tthelw}.  They range a the per-cent level
and are thus small.  Moreover, off-shell top-quark effects have been
determined at NLO in QCD \cite{tthoff} with leptonic top-quark decays
and turn out to be small for the inclusive $t\bar tH$ cross section.
However, they play a role in certain regions of phase space and are thus
of relevance for distributions.

\begin{figure}[hbt]
\begin{center}
\setlength{\unitlength}{1pt}
\begin{picture}(180,120)(40,-10)

\Gluon(0,0)(50,0){-3}{5}
\Gluon(0,80)(50,80){3}{5}
\Gluon(50,80)(80,80){3}{3}
\Gluon(50,80)(50,60){3}{2}
\ArrowLine(80,80)(100,100)
\ArrowLine(100,60)(80,80)
\ArrowLine(50,0)(50,60)
\ArrowLine(50,60)(100,30)
\ArrowLine(100,30)(50,0)
\DashLine(100,30)(150,30){5}
\put(155,26){$H$}
\put(105,96){$b$}
\put(105,56){$\bar b$}
\put(40,36){$t$}
\put(-15,-2){$g$}
\put(-15,78){$g$}

\end{picture}
\begin{picture}(180,120)(-40,-10)

\Gluon(0,20)(50,20){-3}{5}
\Gluon(0,80)(50,80){3}{5}
\ArrowLine(50,20)(50,80)
\ArrowLine(50,80)(100,80)
\ArrowLine(100,80)(100,20)
\ArrowLine(100,20)(50,20)
\Gluon(100,80)(130,80){3}{3}
\ArrowLine(130,80)(150,100)
\ArrowLine(150,60)(130,80)
\DashLine(100,20)(150,20){5}
\put(155,16){$H$}
\put(155,96){$b$}
\put(155,56){$\bar b$}
\put(40,46){$t$}
\put(-15,18){$g$}
\put(-15,78){$g$}

\end{picture}  \\
\setlength{\unitlength}{1pt}
\caption{\label{fg:bbhyt} \it Typical diagrams contributing to $gg\to
b\bar b H$ at NLO involving the top Yukawa coupling.}
\end{center}
\end{figure}
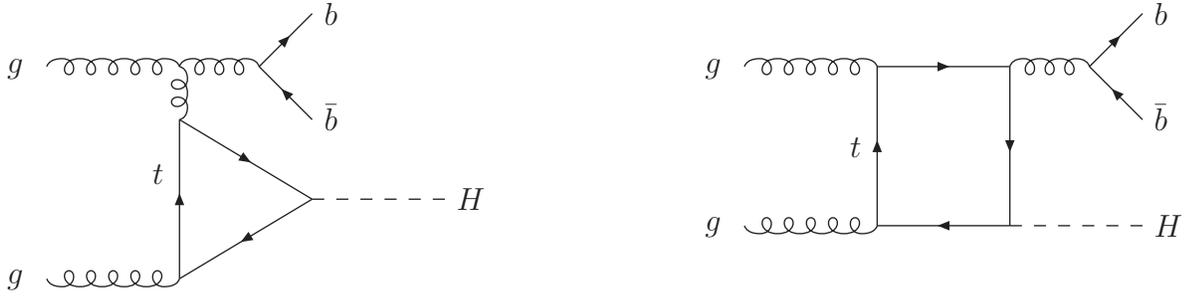

\paragraph{$b\bar bH$ production.}
Higgs bremsstrahlung off bottom quarks does not play a significant role
for the SM Higgs boson, but yields an important constraint on the bottom
Yukawa coupling. Its total cross section is of similar size as the
$t\bar tH$ production cross section. The results of $t\bar tH$
production can be taken over for $b\bar bH$ production. However, they
have to be transformed to the four-flavour-scheme (4FS) in order to
avoid artificial large logarithms initiated by the bottom mass in the
combination of the virtual and real corrections at NLO. In this way
finite bottom-mass effects can be taken into account consistently. The
NLO QCD corrections are positive and large. There is a decrease by about
10\% due to top-Yukawa induced contributions at NLO corresponding to the
diagrams of Fig.~\ref{fg:bbhyt} \cite{bbhqcd}.
However, the $p_T$-integration of the final-state bottom quarks
generates potentially large logarithmic contributions as part of the 4FS
result,
\begin{equation}
\hat\sigma(gg\to b\bar b H) = \int_{\tau_H}^1 dx_1 \int_{\tau_H/x_1}^1
dx_2~\hat b(x_1,Q^2) \hat b(x_2,Q^2) \hat\sigma(b\bar b\to H) +
\Delta\sigma
\label{eq:beff}
\end{equation}
where the effective bottom densities are given by
\begin{equation}
\hat b(x,Q^2) = P_{qg}(x) \log \frac{Q^2}{M_b^2}
\end{equation}
with a scale $Q$ significantly smaller than the Higgs boson mass as can
be inferred from the $p_T$ distribution of the bottom quark in the final
state \cite{bbhscale}. The logarithmic contributions contained in $\hat
b(x,Q^2)$ can be resummed by the DLGLAP evolution \cite{apsplit} of the
initial-state bottom PDFs. This requires to treat the bottom quark as a
massless active flavour, i.e.~to introduce the 5FS with bottom densities
of the proton.

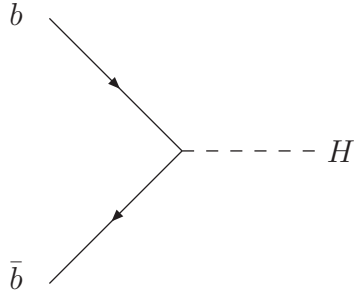
\begin{figure}[hbt]
\begin{center}
\setlength{\unitlength}{1pt}
\begin{picture}(100,110)(-20,0)

\ArrowLine(0,100)(50,50)
\ArrowLine(50,50)(0,0)
\DashLine(50,50)(100,50){5}
\put(105,46){$H$}
\put(-15,98){$b$}
\put(-15,-2){$\bar b$}

\end{picture}  \\
\setlength{\unitlength}{1pt}
\caption{\label{fg:bb2hdia} \it Diagram contributing to $b\bar b \to H$
within the 5FS at lowest order.}
\end{center}
\end{figure}
\begin{figure}[hbt]
\vspace*{-0.5cm}

\hspace*{2.0cm}
\epsfxsize=12cm \epsfbox{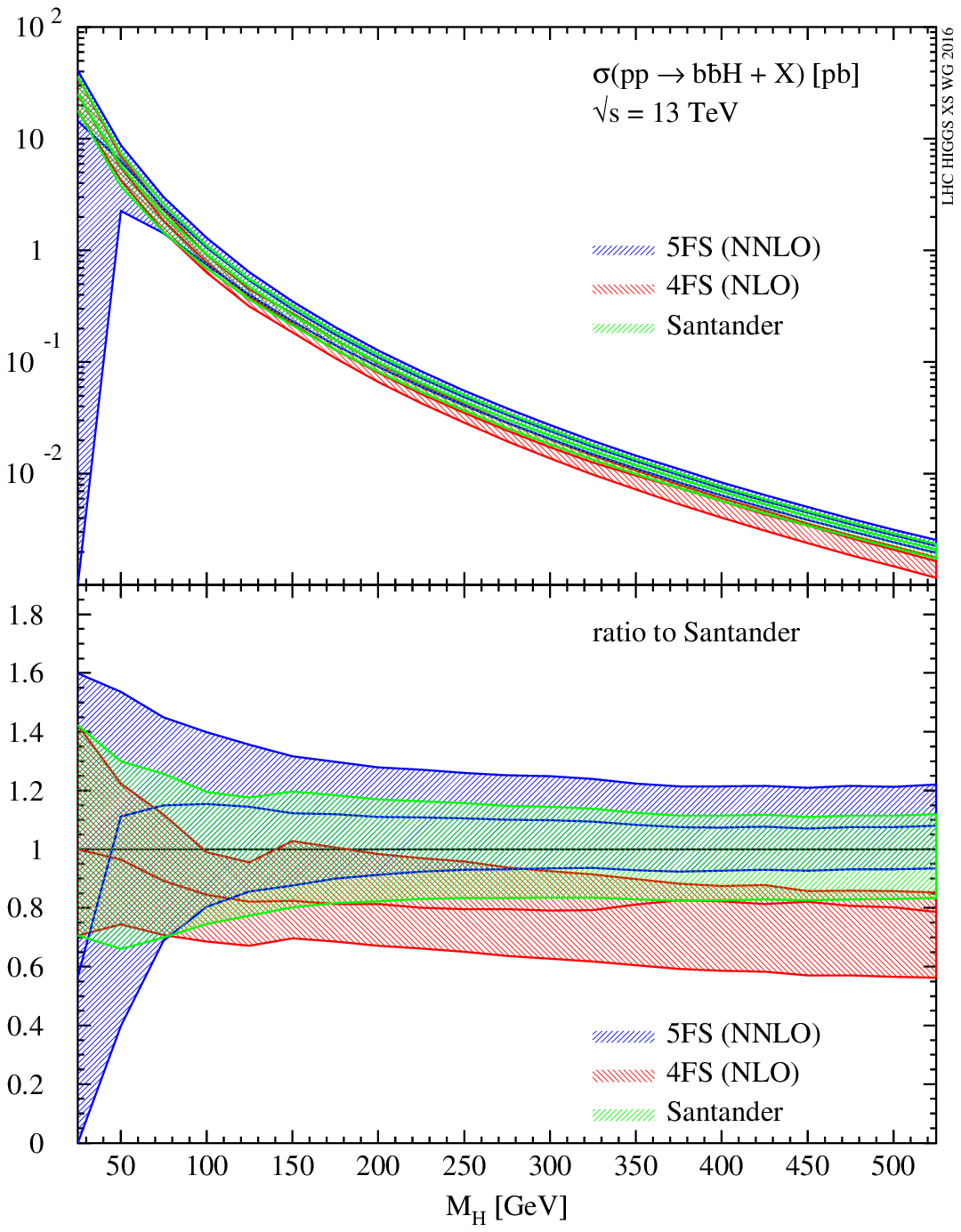}
\vspace*{-2.0cm}

\caption{\label{fg:bbhcomp} \it Comparison of the inclusive $b\bar bH$
production cross sections for the 4FS and 5FS as well as the result using
Santander matching. The lower plot displays the ratios to the central
Santander-matched prediction. The bands reflect the combined
uncertainties. From Ref.~\cite{yr4} (page 523).}
\end{figure}
The 5FS-calculation starts from the process $b\bar b\to H$ at LO, see
Fig.~\ref{fg:bb2hdia}, yielding the LO cross section
\begin{eqnarray}
\sigma_{LO} (pp\to H+X) & = & \sigma_0 \tau_H \frac{d{\cal L}^{b\bar
b}}{d\tau_H} \nonumber \\
\sigma_0 & = & \frac{\pi G_F \overline{m}_b^2(\mu_R)}{N_c \sqrt{2} M_H^2}
\end{eqnarray}
with the variable $\tau_H=M_H^2/s$ involving the hadronic c.m.~energy squared
$s$ and the parton luminosity
\begin{equation}
\frac{d{\cal L}^{b\bar b}}{d\tau} = \int_\tau^1 \frac{dx}{x}~\left[
b(x,\mu_F^2) \bar b(\tau /x,\mu_F^2) + \bar b(x,\mu_F^2) b(\tau
/x,\mu_F^2) \right]
\end{equation}
The 5FS relies on three approximations at LO, i.e.~neglecting
bottom-mass and off-shell effects and the transverse momenta of the
initial-state bottom quarks. The last two conditions are resolved by
adding the higher-order QCD corrections so that only the inclusion of
bottom mass effects is not possible in the 5FS framework. In the
small-mass limit the 4FS- and 5FS-calculations have to approach each
other at higher orders so that the comparison of both schemes yields an
estimate of missing higher-order corrections.  Using the running
$\overline{\rm MS}$ bottom mass $\overline{m}_b^2(\mu_R)$ the NLO QCD
corrections to $b\bar b\to H$ have been calculated \cite{bbhnlo},
\begin{eqnarray}
\sigma(b\bar b\to H) & = & \sigma_{LO} + \Delta\sigma_{b\bar b} +
\Delta\sigma_{bg} \nonumber \\
\Delta\sigma_{b\bar b} & = & \frac{\alpha_s(\mu_R)}{\pi} \int_{\tau_H}^1
d\tau \sum_q \frac{d{\cal L}^{b\bar b}}{d\tau} \sigma_{0}~z~\omega_{b\bar
b}(z) \nonumber \\
\Delta\sigma_{bg} & = & \frac{\alpha_s(\mu_R)}{\pi} \int_{\tau_H}^1 d\tau
\sum_{b,\bar b} \frac{d{\cal L}^{bg}}{d\tau} \sigma_{0}~z~\omega_{bg}(z)
\end{eqnarray}
with $z=\tau_H/\tau$, the coefficient functions
\begin{eqnarray}
\omega_{b\bar b}(z) & = & -P_{qq}(z) \log \frac{\mu_F^2}{\tau s}
+ \frac{4}{3}\left\{ \left(2\zeta_2-1 +
\frac{3}{2}\log\frac{\mu_R^2}{M_H^2} \right)\delta(1-z) + (1+z^2) \left[
2 {\cal D}_1(z) - \frac{\log z}{1-z} \right] + 1-z \right\} \nonumber \\
\omega_{bg}(z) & = & -\frac{1}{2} P_{qg}(z) \log \left(
\frac{\mu_F^2}{(1-z)^2 \tau s} \right) - \frac{1}{8}(1-z)(3-7z)
\end{eqnarray}
and the parton luminosity
\begin{equation}
\frac{d{\cal L}^{bg}}{d\tau} = \sum_{b,\bar b} \int_\tau^1
\frac{dx}{x}~\left[ b(x,\mu_F^2) g(\tau /x,\mu_F^2) + g(x,\mu_F^2)
b(\tau /x,\mu_F^2) \right]
\end{equation}
The calculation of the QCD corrections to $b\bar b\to H$ has been
extended to NNLO \cite{bbhnnlo}. The QCD corrections decrease the cross
section by about 30\% for the central scale choice equal to the Higgs
mass. At NNLO the $gg$-initiated diagrams of the 4FS (see
Fig.~\ref{fg:httlodia}) contribute for the first time so that the
kinematics of the spectator bottom quarks is restored. For a better
comparison of both schemes a general analysis of the logarithms related
to the $p_T$-integrated bottom quarks has been performed with the result
that the proper factorization scale to be chosen for the bottom
densities is significantly smaller than the Higgs boson mass within the
5FS \cite{bbhscale}. This analysis has been extended to all orders for
the logarithmic terms and confirms the smaller effective factorization
scale in the 5FS \cite{bbhscale2}. This decreases the 5FS result and
makes it agree better with the NLO 4FS result. In order to combine both
calculations within the 4FS and 5FS one has to avoid double counting of
common contributions according to Eq.~(\ref{eq:beff}). This has been
done empirically in the past by using the Santander matching
\cite{santander}
\begin{eqnarray}
\sigma & = & \frac{\sigma^{4FS}+w\sigma^{5FS}}{1+w} \nonumber \\
w & = & \log\frac{M_H}{m_b}-2
\label{eq:santander}
\end{eqnarray}
where the chosen weight $w$ pays attention to the fact that the common
terms of both calculations are logarithmic.  Fig.~\ref{fg:bbhcomp} shows
the comparison of the NLO 4FS result with the NNLO 5FS and
Santander-matched cross section as a function of the Higgs mass.

\begin{figure}[hbt]
\vspace*{-1.0cm}

\hspace*{2.0cm}
\epsfxsize=12cm \epsfbox{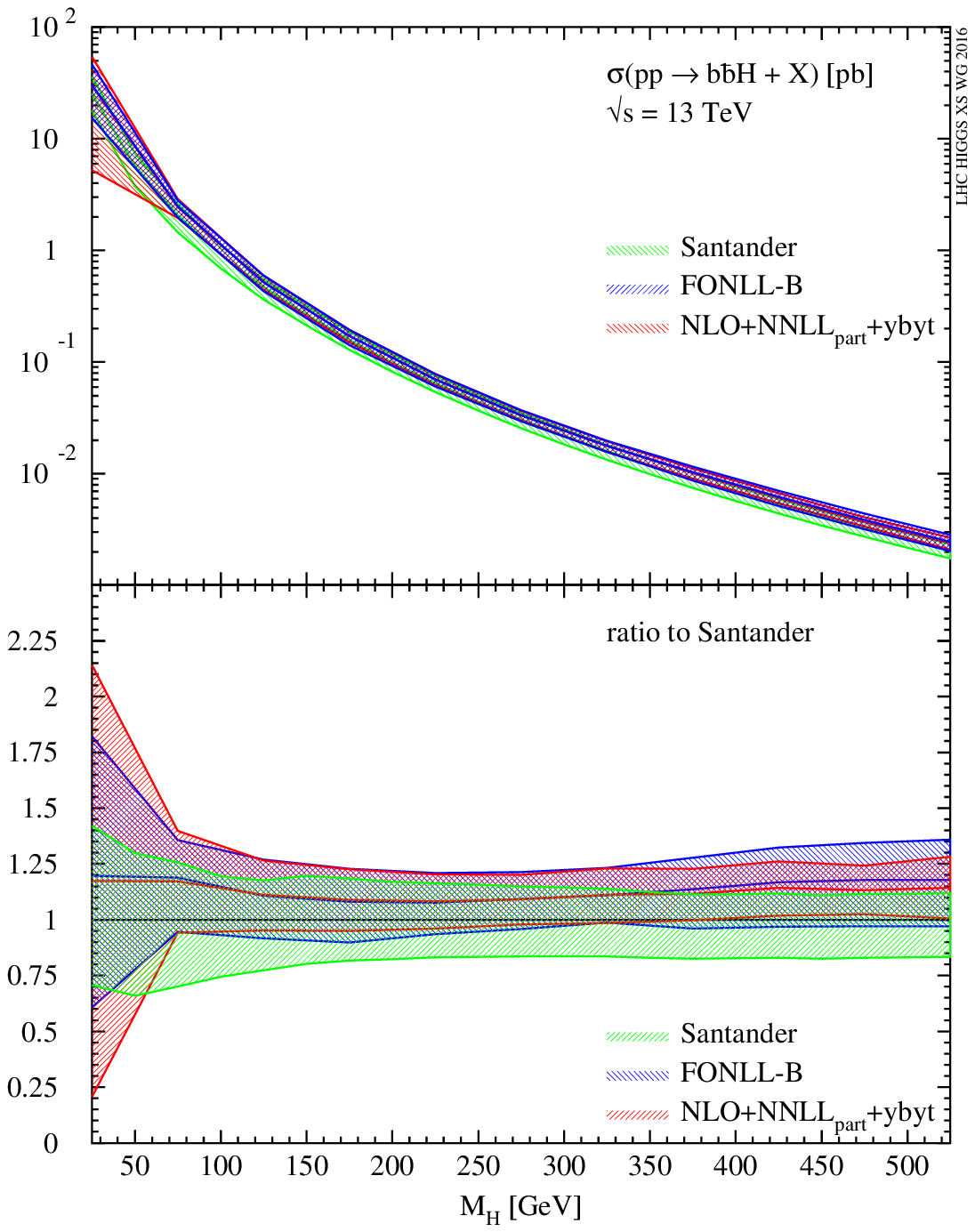}
\vspace*{-2.0cm}

\caption{\label{fg:bbhmatch} \it Comparison of the inclusive $b\bar bH$
production cross sections for the different matching procedures, i.e.
Santander matching, FONLL-B \cite{fonll} and NLO+NNLL$_{\rm
partial}$+ybyt \cite{bbhbonvini}. The lower plot displays the ratios to
the central Santander-matched prediction. The bands reflect the combined 
uncertainties. From Ref.~\cite{yr4} (page 524).}
\end{figure}
This empirically matched result has been improved by two different
procedures of a consistent matching of the 4FS and 5FS, one dubbed
'FONLL' \cite{fonll} and the other 'NLO+NNLL$_{\rm partial}$+ybyt
\cite{bbhbonvini}. Both approaches reexpress the quantities of one
scheme in terms of the other one and use suitably modified sets of the
PDF4LHC15 parton density functions that treat the particular region at
the input scale of the size of the bottom mass consistent with the
matching procedures \cite{yr4}. The perturbative counting within both
approaches is different so that a comparison of both approaches is
motivated to obtain an idea of their differences. Both results are shown
together with the Santander-matched result in Fig.~\ref{fg:bbhmatch} as
a function of the Higgs mass. Both consistent matching procedures agree
well within their respective uncertainties and come out slightly higher
than the cross sections obtained by Santander matching by about 20\%.
Both consistently matched results indicate a tendency towards the
original 5FS calculation for larger Higgs boson masses.

\paragraph{$tH$ production.}
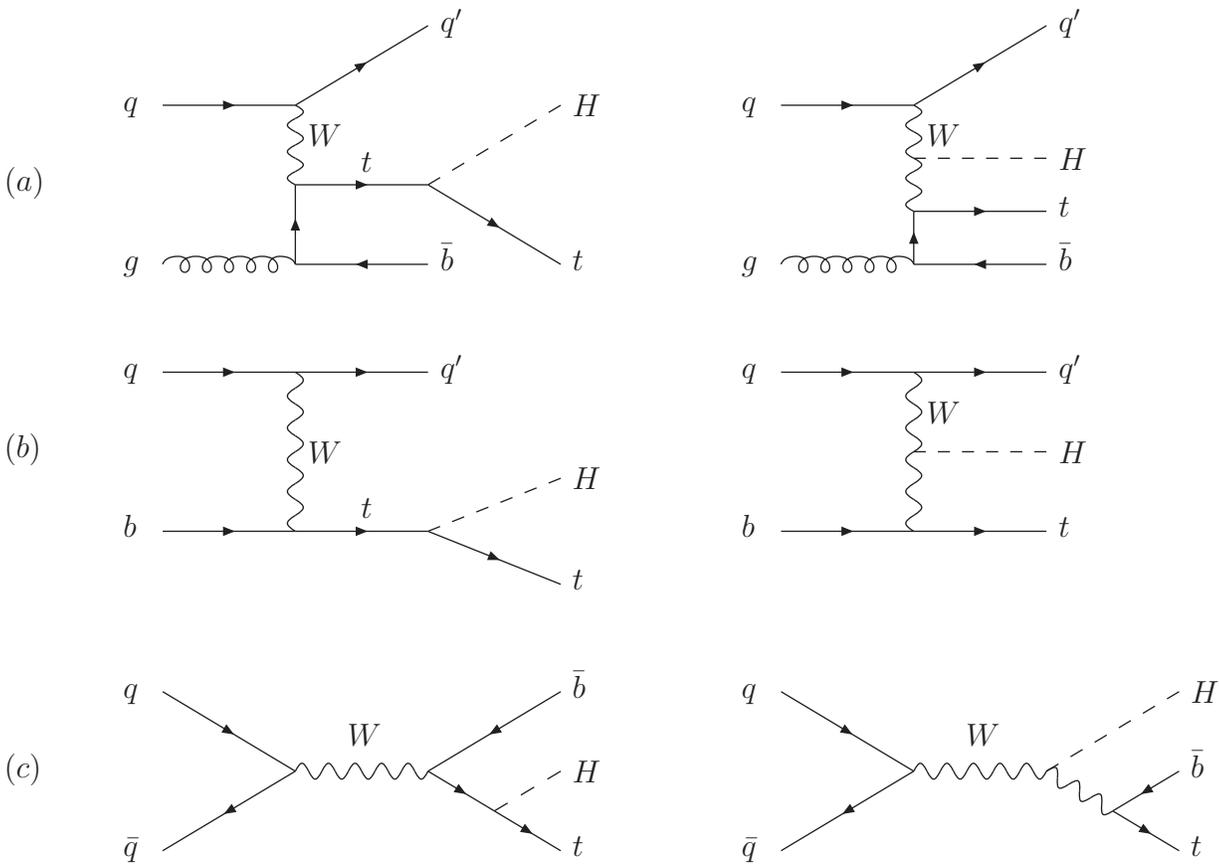
\begin{figure}[hbt]
\begin{center}
\setlength{\unitlength}{1pt}
\begin{picture}(100,100)(100,0)
\ArrowLine(0,80)(50,80)
\Gluon(0,20)(50,20){3}{5}
\ArrowLine(50,80)(100,110)
\Photon(50,80)(50,50){3}{3}
\ArrowLine(50,20)(50,50)
\ArrowLine(100,20)(50,20)
\ArrowLine(50,50)(100,50)
\ArrowLine(100,50)(150,20)
\DashLine(100,50)(150,80){5}
\put(-60,48){$(a)$}
\put(155,76){$H$}
\put(155,18){$t$}
\put(-15,78){$q$}
\put(-15,18){$g$}
\put(105,108){$q'$}
\put(55,65){$W$}
\put(75,55){$t$}
\put(105,18){$\bar b$}
\end{picture}
\begin{picture}(100,100)(-30,0)
\ArrowLine(0,80)(50,80)
\Gluon(0,20)(50,20){3}{5}
\ArrowLine(50,80)(100,110)
\Photon(50,80)(50,40){3}{4}
\ArrowLine(50,20)(50,40)
\ArrowLine(100,20)(50,20)
\ArrowLine(50,40)(100,40)
\DashLine(50,60)(100,60){5}
\put(105,56){$H$}
\put(105,38){$t$}
\put(-15,78){$q$}
\put(-15,18){$g$}
\put(105,108){$q'$}
\put(55,65){$W$}
\put(105,18){$\bar b$}
\end{picture} \\
\begin{picture}(100,100)(100,0)
\ArrowLine(0,80)(50,80)
\ArrowLine(0,20)(50,20)
\ArrowLine(50,80)(100,80)
\Photon(50,80)(50,20){3}{5}
\ArrowLine(50,20)(100,20)
\ArrowLine(100,20)(150,0)
\DashLine(100,20)(150,40){5}
\put(-60,48){$(b)$}
\put(155,36){$H$}
\put(155,-2){$t$}
\put(-15,78){$q$}
\put(-15,18){$b$}
\put(105,78){$q'$}
\put(55,46){$W$}
\put(75,25){$t$}
\end{picture}
\begin{picture}(100,100)(-30,0)
\ArrowLine(0,80)(50,80)
\ArrowLine(0,20)(50,20)
\ArrowLine(50,80)(100,80)
\Photon(50,80)(50,20){3}{5}
\ArrowLine(50,20)(100,20)
\DashLine(50,50)(100,50){5}
\put(105,46){$H$}
\put(105,18){$t$}
\put(-15,78){$q$}
\put(-15,18){$b$}
\put(105,78){$q'$}
\put(55,61){$W$}
\end{picture} \\
\begin{picture}(100,120)(100,0)
\ArrowLine(0,80)(50,50)
\ArrowLine(50,50)(0,20)
\Photon(50,50)(100,50){3}{5}
\ArrowLine(150,80)(100,50)
\ArrowLine(100,50)(125,35)
\ArrowLine(125,35)(150,20)
\DashLine(125,35)(150,50){5}
\put(-60,48){$(c)$}
\put(155,46){$H$}
\put(155,18){$t$}
\put(-15,78){$q$}
\put(-15,18){$\bar q$}
\put(155,78){$\bar b$}
\put(70,60){$W$}
\end{picture}
\begin{picture}(100,100)(-30,0)
\ArrowLine(0,80)(50,50)
\ArrowLine(50,50)(0,20)
\Photon(50,50)(100,50){3}{5}
\DashLine(100,50)(150,80){5}
\Photon(100,50)(125,35){3}{3}
\ArrowLine(125,35)(150,20)
\ArrowLine(150,50)(125,35)
\put(155,76){$H$}
\put(155,18){$t$}
\put(-15,78){$q$}
\put(-15,18){$\bar q$}
\put(155,48){$\bar b$}
\put(70,60){$W$}
\end{picture} \\
\setlength{\unitlength}{1pt}
\caption{\label{fg:thdia} \it Typical diagrams contributing to
$tH$ production at lowest order: (a) $t$-channel within the 4FS, (b)
$t$-channel within the 5FS, (c) $s$-channel.}
\end{center}
\end{figure}
Single-top quark production in association with a Higgs boson proceeds
in two different generic ways at LO, the dominant $t$-channel
contribution and the subleading $s$-channel one, see
Fig.~\ref{fg:thdia}. It allows to test the sign of the top Yukawa
coupling, since there are interference effects between the Higgs-boson
couplings to top quarks and $W$ bosons. The $t$-channel suffers from the
same problem as $b\bar bH$ production, i.e.~there is the alternative to
treat it within the 4FS or 5FS the latter starting with bottom
densities of the proton the DGLAP evolution of which resum potentially
large logarithms arising in the 4FS. Within the 5FS there is no
interference between the $t$- and $s$-channel contributions up to NLO.
The full NLO calculation has been performed by means of the {\sc
Mg5\_amc@nlo} framework \cite{thnlo}. The QCD corrections to the
$t$-channel contribution are small, i.e.~$\sim 10\%$, and of a similar
size as the difference between the 4FS and 5FS at NLO. In order to
reduce the difference between the 4FS and 5FS the central scale choice
has been reduced to $\mu_R=\mu_F=(M_H+M_t)/4$ for the $t$-channel
contributions, while the $s$-channel uses a central scale twice as large
\cite{yr4}. The theoretical uncertainties, estimated from the scale
dependence and the difference between both schemes, range at the level
of 5--15\% \cite{yr4}. Including the PDF+$\alpha_s$ uncertainties the
total uncertainty can be estimated to be in the same ball park.

\subsubsection{\it Minimal supersymmetric extension}
\paragraph{$t\bar t/b\bar b + \Phi$ production.}
\begin{figure}[hbt]
\begin{center}
\setlength{\unitlength}{1pt}
\begin{picture}(360,120)(0,-10)

\ArrowLine(0,100)(50,50)
\ArrowLine(50,50)(0,0)
\Gluon(50,50)(100,50){3}{5}
\ArrowLine(100,50)(125,75)
\ArrowLine(125,75)(150,100)
\ArrowLine(150,0)(100,50)
\DashLine(125,75)(150,50){5}
\put(155,46){$\Phi$}
\put(-15,98){$q$}
\put(-15,-2){$\bar q$}
\put(65,65){$g$}
\put(155,98){$t/b$}
\put(155,-2){$\bar t/\bar b$}

\Gluon(250,0)(300,0){3}{5}
\Gluon(250,100)(300,100){3}{5}
\ArrowLine(350,0)(300,0)
\ArrowLine(300,0)(300,50)
\ArrowLine(300,50)(300,100)
\ArrowLine(300,100)(350,100)
\DashLine(300,50)(350,50){5}
\put(355,46){$\Phi$}
\put(235,98){$g$}
\put(235,-2){$g$}
\put(355,98){$t/b$}
\put(355,-2){$\bar t/\bar b$}

\end{picture}  \\
\setlength{\unitlength}{1pt}
\caption{\label{fg:mssmhqqlodia} \it Typical diagrams contributing to
$q\bar q/gg \to \Phi Q\bar Q~~(Q=t,b)$ at lowest order.}
\end{center}
\end{figure}
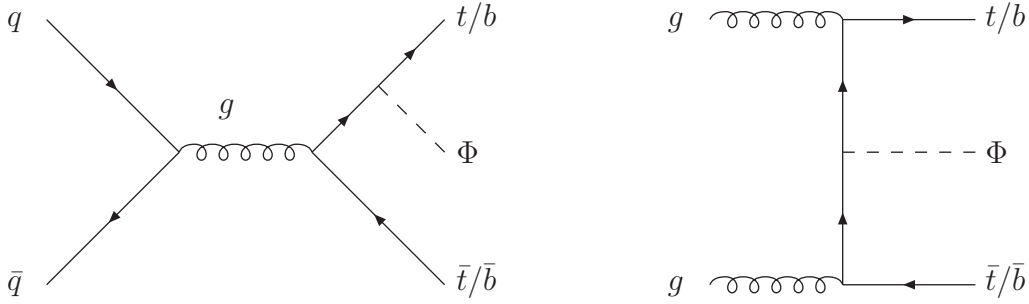
\noindent
The scalar Higgs cross sections for Higgs bremsstrahlung off heavy
quarks $Q$ can simply be related to the SM case at LO:
\begin{equation}
\sigma (pp\to h/H + Q\bar Q) = \left( g_Q^{h/H} \right)^2
\sigma (pp\to H_{SM}Q\bar Q)
\label{eq:tthmssm}
\end{equation}
This relation is also valid at NLO apart from diagrams involving closed
top/bottom loops coupling to the Higgs bosons, see Fig.~\ref{fg:bbhyt}.
The LO expressions for the pseudoscalar Higgs boson \cite{att} are
similarly involved as the scalar case and will not be presented here.

The top quark coupling to MSSM Higgs bosons is suppressed with respect
to the SM for $\mbox{tg$\beta$} >1$. Therefore Higgs bremsstrahlung off
top quarks $pp\to\Phi t\bar t$ is less important for MSSM Higgs
particles. On the other hand Higgs bremsstrahlung off bottom quarks $pp
\to Hb\bar b$ will be the dominant Higgs production channel for large
$\mbox{tg$\beta$}$ due to the strongly enhanced bottom quark Yukawa
couplings \cite{htt}. The QCD corrections for the scalar Higgs bosons
can be inferred from the SM cross section by Eq.~(\ref{eq:tthmssm}) but
with the properly rescaled diagrams of Fig.~\ref{fg:bbhyt}. The NLO QCD
corrections to pseudoscalar $t\bar tA$ production have been calculated
in Ref.~\cite{tthqcd1} yielding a slightly larger but moderate increase
of the cross section. The full SUSY-QCD corrections have been calculated
quite recently for $t\bar t\phi$ production \cite{tthsqcd, tthsqcd0} and
for $b\bar b\phi$ production in the 4FS \cite{tthsqcd} and the 5FS
\cite{bbhsqcd}. Being of moderate size for small values of
$\mbox{tg}\beta$, i.e.~similar to the pure QCD corrections, they can be
large for $b\bar b\phi$ production due to the $\Delta_b$ corrections to
the bottom Yukawa couplings.  However, the remainder after absorbing the
leading terms in the effective resummed Yukawa couplings turns out to be
small \cite{tthsqcd}. The SUSY-electroweak corrections are known for
$b\bar\phi$ production within the 5FS \cite{bbhselw}. The latter can be
well approximated by the radiatively improved Higgs masses and couplings
and the electroweak $\Delta_b$ terms for large values of
$\mbox{tg}\beta$.

\begin{figure}[hbt]
\begin{center}
\begin{picture}(150,180)(0,0)
\put(-300,-90.0){\includegraphics{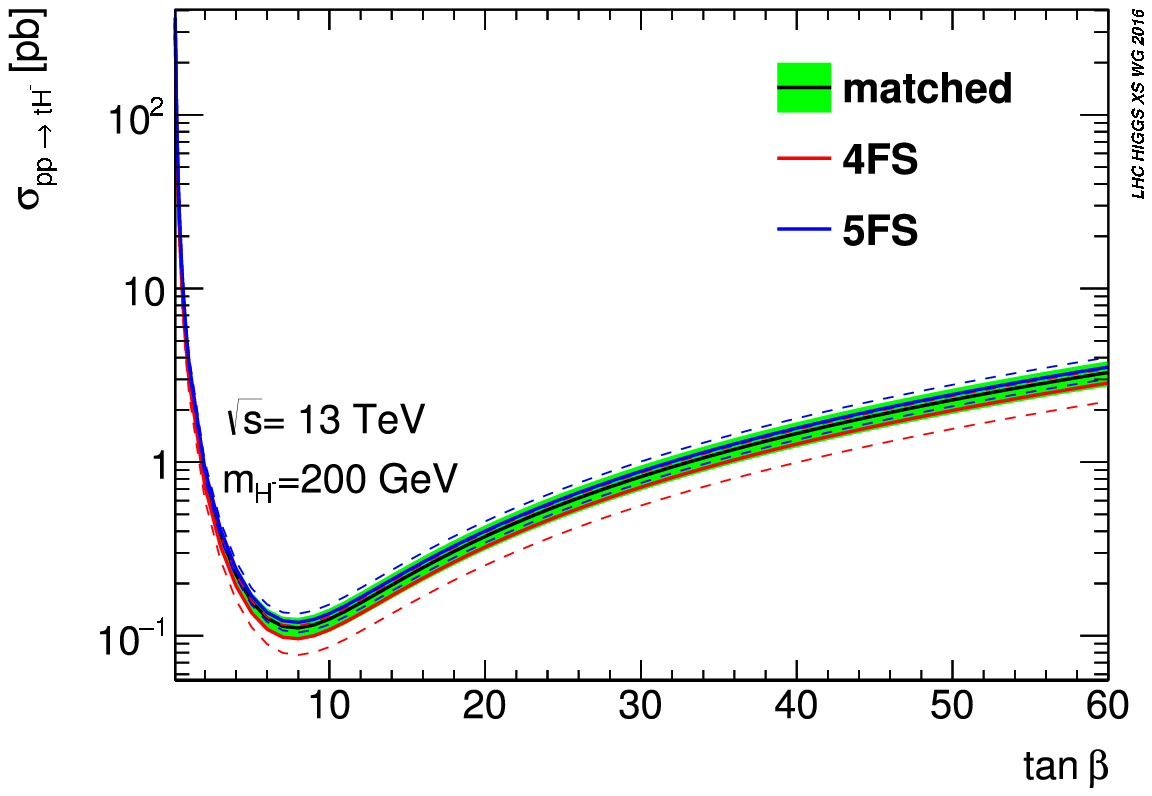}}
\put(-40,-90.0){\includegraphics{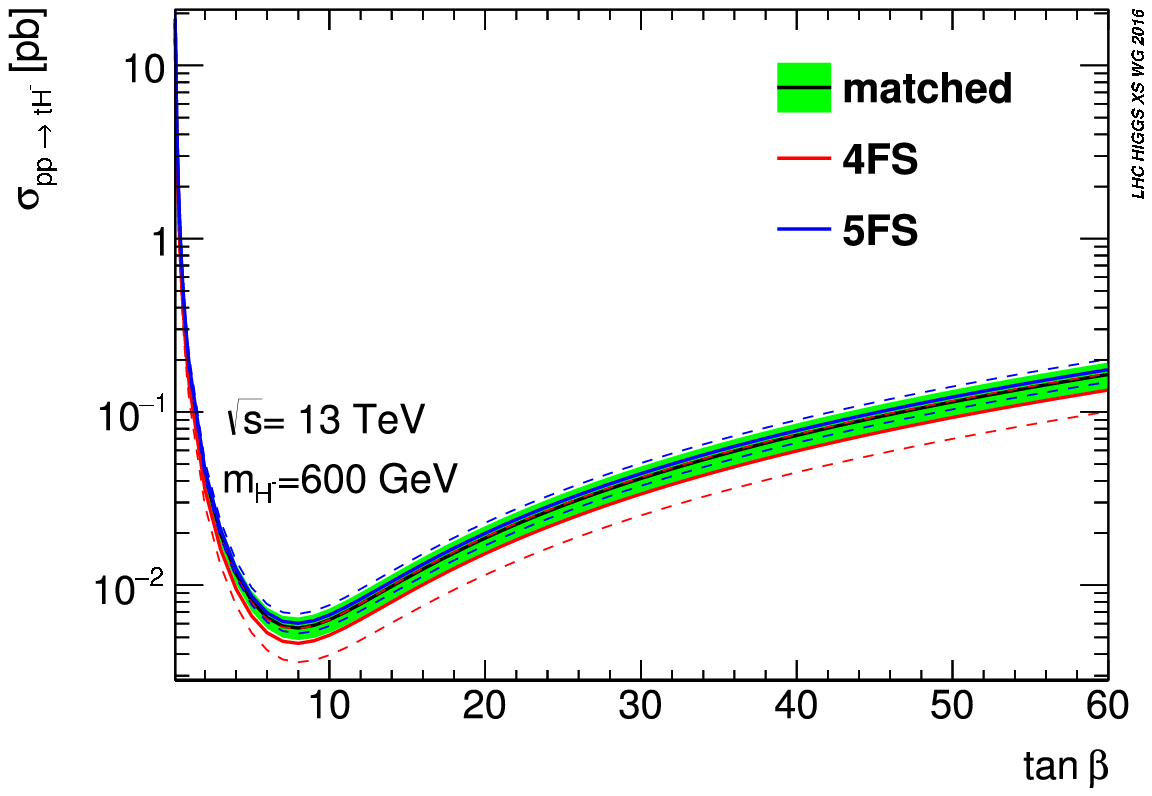}}
\end{picture}
\caption{\it Charged Higgs production cross section in association with top
quarks as functions of $\mbox{tg}\beta$ for two values of the charged
Higgs mass. The dashed lines show the uncertainty bands of the 4FS and
5FS, while the green bands are the combined results with
Santander-matching. From Ref.~\cite{yr4} (page 543).}
\label{fg:proch}
\end{center}
\end{figure}

\paragraph{Charged Higgs production.}
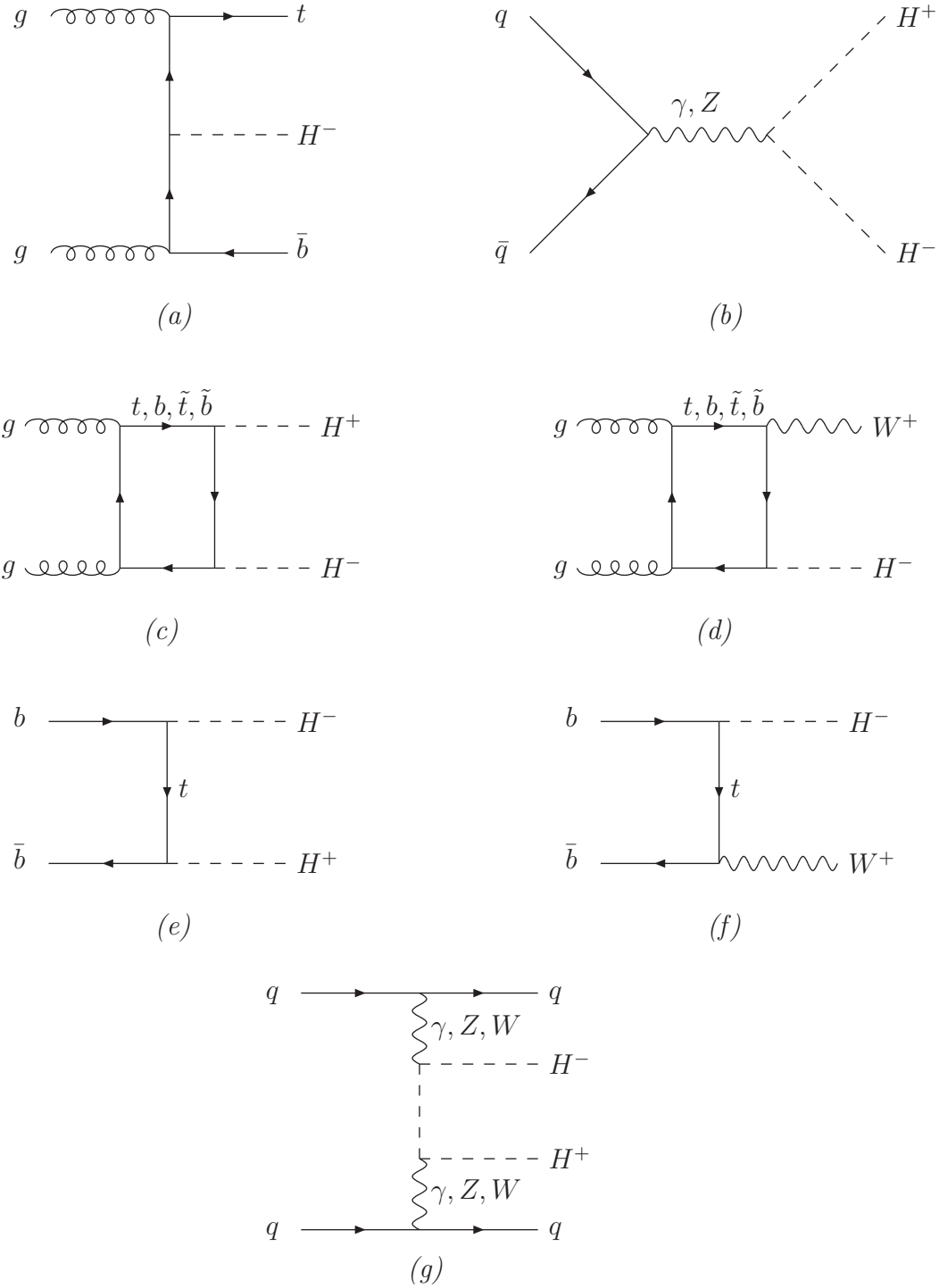
\begin{figure}[hbtp]
\begin{center}
\begin{picture}(130,110)(300,0)
\Gluon(250,0)(300,0){3}{5}
\Gluon(250,100)(300,100){3}{5}
\ArrowLine(350,0)(300,0)
\ArrowLine(300,0)(300,50)
\ArrowLine(300,50)(300,100)
\ArrowLine(300,100)(350,100)
\DashLine(300,50)(350,50){5}
\put(355,46){$H^-$}
\put(235,98){$g$}
\put(235,-2){$g$}
\put(355,98){$t$}
\put(355,-2){$\bar b$}
\put(295,-30){{\it (a)}}
\end{picture}
\begin{picture}(130,110)(-20,0)
\ArrowLine(0,100)(50,50)
\ArrowLine(50,50)(0,0)
\Photon(50,50)(100,50){3}{5}
\DashLine(100,50)(150,100){5}
\DashLine(100,50)(150,0){5}
\put(155,96){$H^+$}
\put(155,-4){$H^-$}
\put(-15,-2){$\bar q$}
\put(-15,98){$q$}
\put(60,60){$\gamma,Z$}
\put(75,-30){{\it (b)}}
\end{picture} \\[1.5cm]
\begin{picture}(130,110)(70,0)
\Gluon(10,20)(50,20){-3}{4}
\Gluon(10,80)(50,80){3}{4}
\ArrowLine(50,20)(50,80)
\ArrowLine(50,80)(90,80)
\ArrowLine(90,80)(90,20)
\ArrowLine(90,20)(50,20)
\DashLine(90,80)(130,80){5}
\DashLine(90,20)(130,20){5}
\put(0,18){$g$}
\put(0,78){$g$}
\put(55,85){$t,b,\tilde t,\tilde b$}
\put(135,76){$H^+$}
\put(135,16){$H^-$}
\put(60,-10){{\it (c)}}
\end{picture}
\begin{picture}(130,110)(-30,0)
\Gluon(10,20)(50,20){-3}{4}
\Gluon(10,80)(50,80){3}{4}
\ArrowLine(50,20)(50,80)
\ArrowLine(50,80)(90,80)
\ArrowLine(90,80)(90,20) 
\ArrowLine(90,20)(50,20)
\Photon(90,80)(130,80){3}{4}
\DashLine(90,20)(130,20){5}
\put(0,18){$g$}
\put(0,78){$g$}
\put(55,85){$t,b,\tilde t,\tilde b$}
\put(135,76){$W^+$}
\put(135,16){$H^-$}
\put(60,-10){{\it (d)}}
\end{picture} \\[0.5cm]
\begin{picture}(130,110)(50,0)
\ArrowLine(0,80)(50,80)
\ArrowLine(50,80)(50,20)
\ArrowLine(50,20)(0,20) 
\DashLine(50,80)(100,80){5}
\DashLine(50,20)(100,20){5}
\put(105,76){$H^-$}
\put(105,16){$H^+$}
\put(-15,78){$b$}
\put(-15,18){$\bar b$}
\put(55,48){$t$}
\put(45,-10){{\it (e)}}
\end{picture}
\begin{picture}(130,110)(-50,0)
\ArrowLine(0,80)(50,80)
\ArrowLine(50,80)(50,20)
\ArrowLine(50,20)(0,20)
\Photon(50,20)(100,20){3}{6}
\DashLine(50,80)(100,80){5}
\put(105,76){$H^-$}
\put(105,16){$W^+$}
\put(-15,78){$b$}
\put(-15,18){$\bar b$}
\put(55,48){$t$}
\put(45,-10){{\it (f)}}
\end{picture} \\[0.5cm]
\begin{picture}(130,120)(10,0)
\ArrowLine(0,100)(50,100)
\ArrowLine(50,100)(100,100)
\ArrowLine(0,0)(50,0)
\ArrowLine(50,0)(100,0)
\Photon(50,100)(50,70){3}{3}
\Photon(50,30)(50,0){3}{3}
\DashLine(50,70)(100,70){5}
\DashLine(50,30)(100,30){5}
\DashLine(50,70)(50,30){5}
\put(55,83){$\gamma,Z,W$}
\put(55,13){$\gamma,Z,W$}
\put(105,66){$H^-$}
\put(105,26){$H^+$}
\put(105,98){$q$}
\put(105,-2){$q$}
\put(-15,98){$q$}
\put(-15,-2){$q$}
\put(45,-20){{\it (g)}}
\end{picture} \\[1.0cm]
\caption{\label{fg:hcpro} \it Typical diagrams for charged Higgs boson
production mechanisms at leading order:
{\it (a)} $gg\to H^-t\bar b$, {\it (b)} $q\bar q\to H^+H^-$,
{\it (c)} $gg\to H^+H^-$, {\it (d)} $gg\to W^+H^-$, {\it (e)} $b\bar
b\to H^+H^-$, {\it (f)} $b\bar b\to W^+H^-$, (g) $qq\to qqH^+H^-$.}
\end{center}
\end{figure}
Charged Higgs bosons are dominantly produced at the LHC in association
with top quarks $gg,q\bar q\to t\bar bH^-$ and the charge-conjugated
process, see Fig.~\ref{fg:hcpro}a. For charged Higgs-boson masses
$M_{H^\pm} < m_t-m_b$ the dominant contribution to this process
factorizes into $t\bar t$ pair production with the subsequent (anti)top
decay into a charged Higgs and (anti)bottom quark. This region is then
suitably described by the QCD- and electroweak-corrected $t\bar t$
production cross section \cite{ttqcd, ttelw} multiplied with the
radiatively corrected branching ratio of the top decays \cite{tdecqcd,
tdecelw}. However, for larger charged Higgs masses the top decays are
kinematically closed and charged Higgs bosons are produced in terms of
the full process of Fig.~\ref{fg:hcpro}a.  For the inclusive rate the
final-state bottom quark gives rise to the options to calculate this
process in the 4FS or the 5FS, where the latter starts from the process
$gb\to tH^-$ and the charge-conjugated process and thus uses bottom PDFs
of the proton. Both calculations have been performed up to NLO QCD
resulting in moderate scale dependences \cite{ch4fs,ch5fs}. While the
5FS is available exclusively for all distributions involving the top
quark and charged Higgs boson, the 4FS allows to determine distributions
involving the bottom quark in the final state in addition. Both
calculations have been combined by means of the Santander matching of
Eq.~(\ref{eq:santander}) \cite{santander} with the weight $w=\log
M_{H^\pm}/m_b - 2$. For small values of $\mbox{tg}\beta$ the genuine
SUSY-QCD corrections are of moderate size \cite{ch4fs,ch5fs}, while for
large $\mbox{tg}\beta$ they are completely dominated by $\Delta_b$-terms
with a small remainder.  Thus for the compilation of the charged-Higgs
production cross section the effective bottom Yukawa coupling
\begin{equation}
\tilde{g}_b^{H^\pm} = \tilde{g}_b^A
\end{equation}
(combined with the top Yukawa coupling factor $g_t^{H^\pm} = g_t^A$) has
been used that can be translated into an effective $\mbox{tg}\beta$
value \cite{proch}. The result for the continuum production cross
section without SUSY-QCD corrections is shown in Fig.~\ref{fg:proch} for
two charged Higgs masses as a function of $\mbox{tg}\beta$.  The QCD
corrections enhance the 4FS cross section by about 60\% \cite{ch4fs},
while they are moderate for the 5FS \cite{ch5fs}.

\begin{figure}[hbt]
\begin{center}
\begin{picture}(150,350)(0,0)
\put(-90,-80.0){\includegraphics{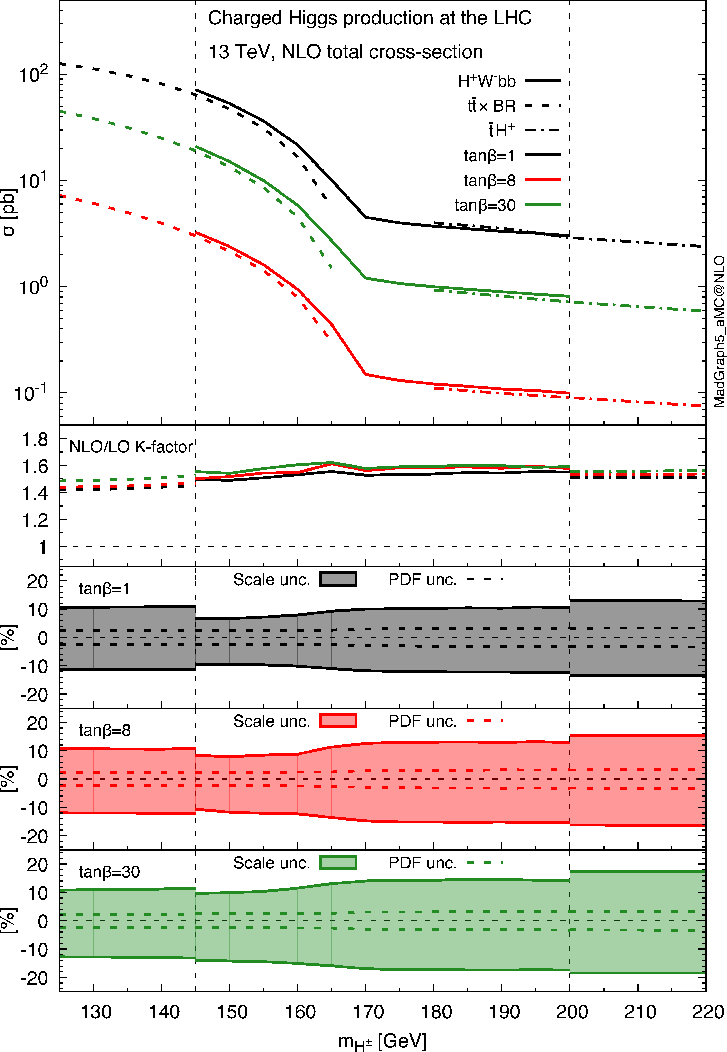}}
\end{picture}
\caption{\it Charged Higgs production cross section in association with top
quarks as a function of the charged Higgs mass in the low-, high- and
intermediate-mass regimes at the LHC for $\sqrt{s}=13$ TeV. The upper
plot shows the central predictions in all three regimes for three values
of $\mbox{tg}\beta$. The middle plot exhibits the NLO K factor defined
as the ratio between the NLO and LO cross sections. The lower plots
display the scale- and PDF-uncertainty bands for the three values of
$\mbox{tg}\beta$. From Ref.~\cite{ch4fi}.}
\label{fg:prochint}
\end{center}
\end{figure}

An open problem for a long time has been the treatment of the
intermediate region between resonant top quark decays and the continuum
contribution for large charged Higgs masses. This has been solved
recently by a full NLO calculation in the 4FS within the complex-mass
scheme for the intermediate top quarks \cite{ch4fi}. The NLO QCD
corrections turn out to be large in this mass regime, too, while the
results nicely interpolate between the low- and high-mass regimes as can
be inferred from Fig.~\ref{fg:prochint}.

The second important charged Higgs production process is charged Higgs
pair production in a Drell--Yan type process (see Fig.~\ref{fg:hcpro}b)
\cite{chdylo}
\begin{displaymath}
pp\to q\bar q\to H^+H^-
\end{displaymath}
which is mediated by $s$-channel photon and $Z$-boson exchange. The NLO
QCD corrections can be taken from the Drell--Yan process and are of
moderate size as in the case of the neutral Higgs-strahlung process
discussed before \cite{vbfsqcd}. The genuine SUSY--QCD corrections,
mediated by virtual gluino and squark exchange in the initial state, are
small \cite{vbfsqcd}.

Charged Higgs pairs can also be produced from $gg$ initial states by the
loop-mediated process \cite{gg2hc,bb2hc} (see Fig.~\ref{fg:hcpro}c)
\begin{displaymath}
pp\to gg\to H^+H^-
\end{displaymath}
where the dominant contributions emerge from top and bottom quark loops
as well as stop and sbottom loops, if the squark masses are light
enough. The NLO corrections to this process are unknown. This cross
section is of similar size as the bottom-initiated process \cite{bb2hc}
(see Fig.~\ref{fg:hcpro}e)
\begin{displaymath}
pp\to b\bar b\to H^+H^-
\end{displaymath}
which relies on the approximations required by the introduction of the
bottom densities as discussed before and is known at NLO
\cite{bb2hcnlo}. The SUSY--QCD corrections are of significant size due
to the $\Delta_b$ terms related to the bottom Yukawa coupling. The pure
QCD corrections and the genuine SUSY--QCD corrections can be of opposite
sign.

Charged Higgs bosons can be produced in association with a $W$
boson \cite{gg2hcw1,gg2hcw2} (see Fig.~\ref{fg:hcpro}d)
\begin{displaymath}
pp\to gg\to H^+W^- \qquad \mbox{and~c.c.}
\end{displaymath}
which is generated by top-bottom quark loops and stop-sbottom loops, if
the squark masses are small enough. This process is known at LO only.
The same final state also arises from the process \cite{gg2hcw1,bb2hcw}
(see Fig.~\ref{fg:hcpro}f)
\begin{displaymath}
pp\to b\bar b\to H^+W^- \qquad \mbox{and~c.c.}
\end{displaymath}
which is based on the approximations of the 5FS. The QCD corrections
have been calculated and turn out to be of moderate size
\cite{bb2hcwnlo}.

Finally, charged Higgs-boson pairs can be produced in vector-boson-fusion
\cite{vbfch} (see Fig.~\ref{fg:hcpro}g)
\begin{displaymath}
qq\to qq V^*V^* \to qq H^+H^-
\end{displaymath}
The LO cross section is independent of $\mbox{tg}\beta$ and can be
sizeable within the MSSM. However, the calculation of Ref.~\cite{vbfch}
is not consistent with the parton picture, since small quark masses have
been introduced for the accompanying quarks in order to regulate the
collinear divergences of photon-exchange contributions. For a more
reliable prediction QED-mass factorization has to be performed already at
LO.

\subsection{\it Summary of single-Higgs boson production cross sections}
\subsubsection{\it Standard Model}
\begin{figure}[hbt]
\vspace*{-0.5cm}

\hspace*{-4.0cm}
\begin{turn}{-90}%
\epsfxsize=9cm \epsfbox{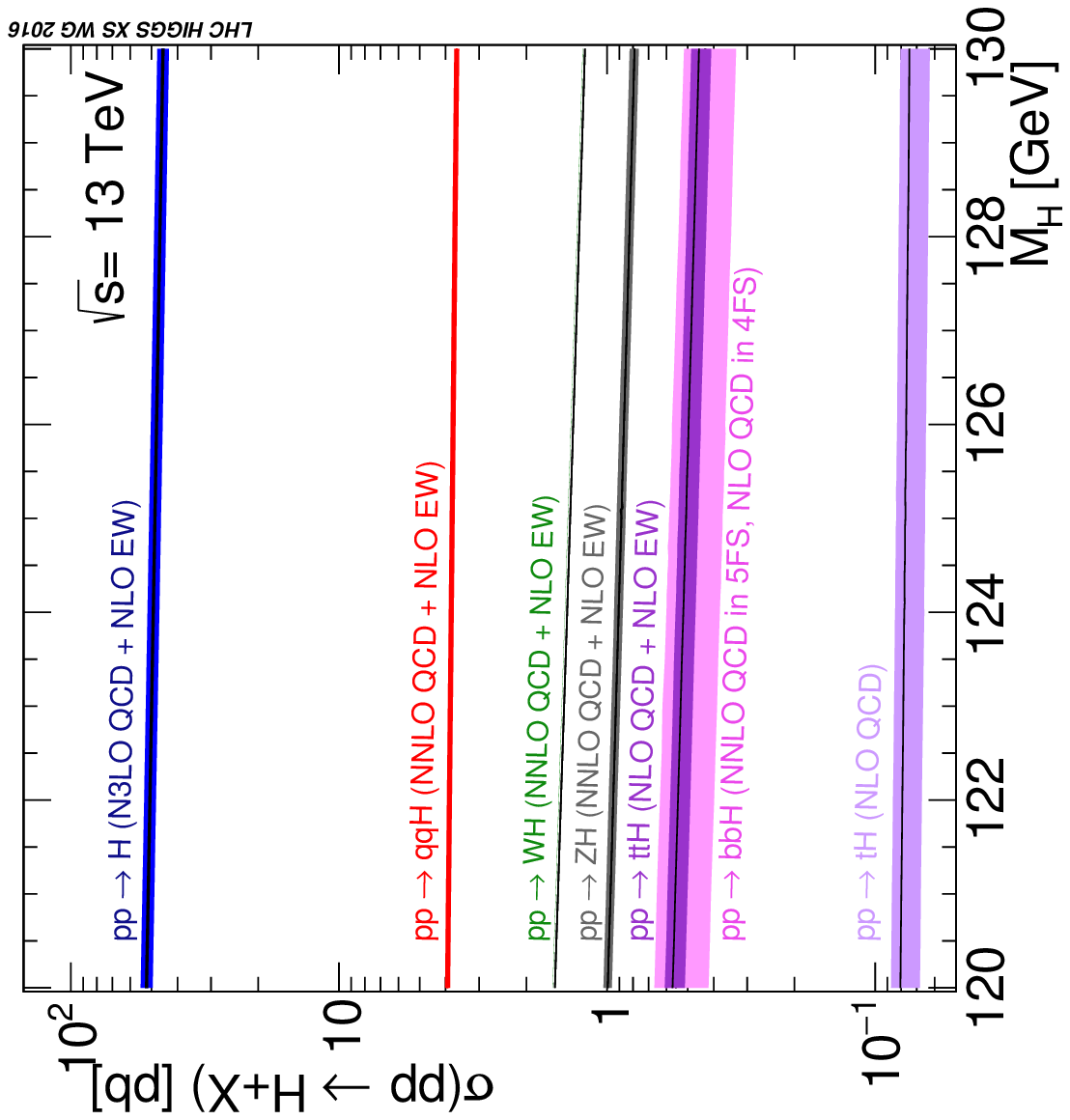}
\end{turn}
\hspace*{-4.0cm}
\begin{turn}{-90}%
\epsfxsize=9cm \epsfbox{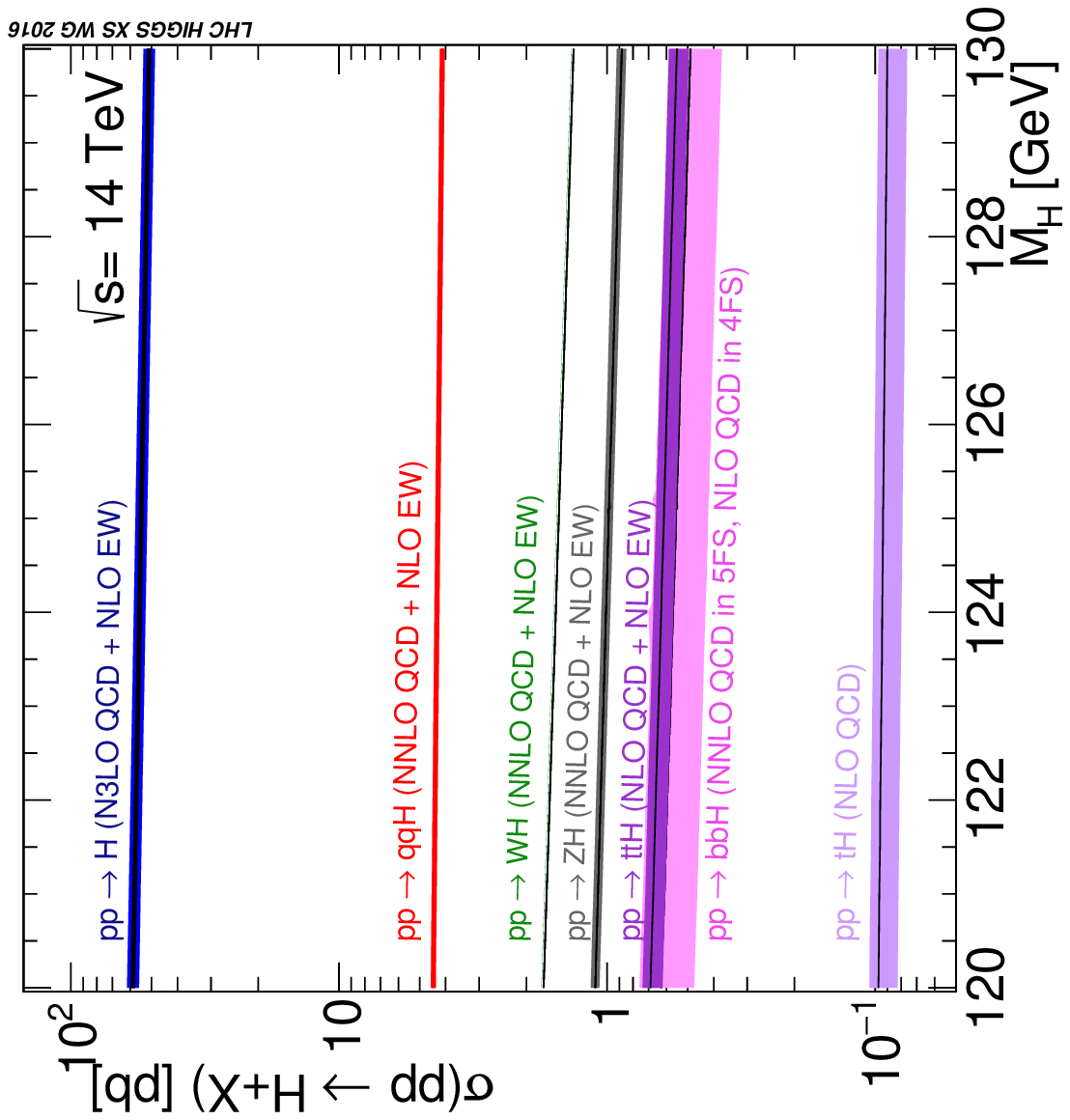}
\end{turn}
\vspace*{-0.5cm}

\caption{\label{fg:hprodcxn} \it Higgs boson production cross sections
as a function of the Higgs mass for $13$ and $14$ TeV c.m.~energy at the
LHC including the most up-to-date higher-order corrections as indicated
at the shown cross section bands. The size of the bands reflects the
total estimated theoretical uncertainties. From Ref.~\cite{yr4} (page
275).}
\end{figure}
\begin{figure}[hbt]
\vspace*{-0.0cm}

\hspace*{-0.0cm}
\begin{turn}{-90}%
\epsfxsize=9cm \epsfbox{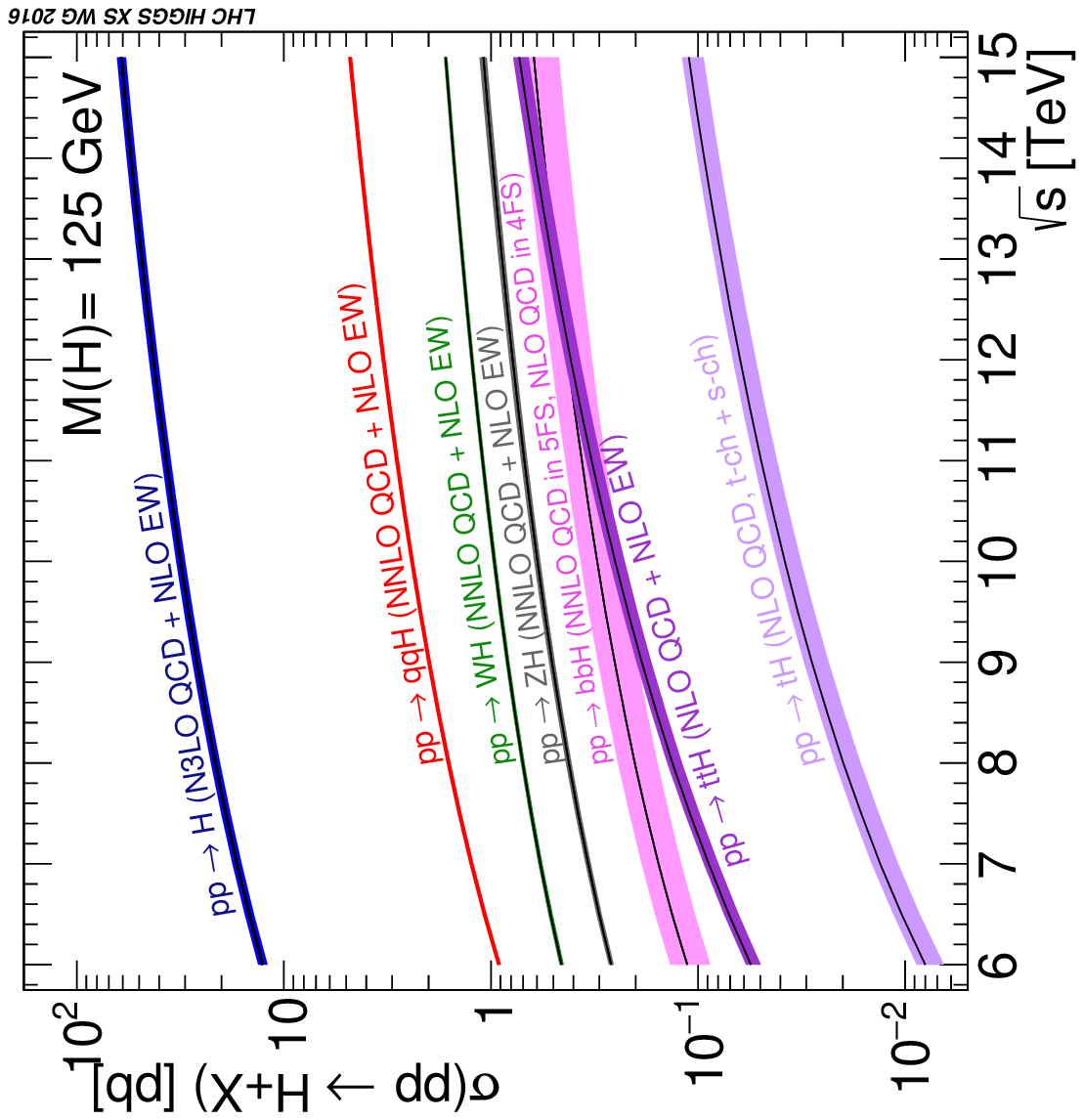}
\end{turn}
\vspace*{-1.0cm}

\caption{\label{fg:hcxnen} \it Higgs boson production cross sections as
a function of the c.m.~energy at the LHC for a Higgs mass $M_H=125$ GeV
including the most up-to-date higher-order corrections as indicated at
the shown cross section bands. The size of the bands reflects the total
estimated theoretical uncertainties. From Ref.~\cite{yr4} (page 276).}
\end{figure}

All Higgs boson production cross sections have been updated with the
known higher-order corrections and the most recent parton density
functions, i.e.~the PDF4LHC15 sets \cite{pdf4lhc}, where NLO densities
have been used consistently for NLO predictions and NNLO densities for
NNLO predictions. Using the same values of the input parameters as for
the branching ratios discussed in Section \ref{sc:br} and their
uncertainties a rigorous analysis has been performed to derive a
sophisticated prediction of the central cross section values and their
uncertainties. The results are shown in Fig.~\ref{fg:hprodcxn} as a
function of the Higgs mass for 13 and 14 TeV c.m.~energy at the LHC. The
size of the coloured bands represents the individual sums of the
theoretical and parametric uncertainties. All production cross sections
with results beyond NLO in QCD exhibit a small residual uncertainty in
the few-per-cent range. Only the cross sections for $t\bar tH$, $b\bar
bH$ and $tH$ production develop larger uncertainties due to the problems
discussed in the previous sections. The theoretical and parametric
uncertainties of each production process have been added in quadrature.
The gluon-fusion cross sections can be predicted with a total (Gaussian)
uncertainty of about 5\%, the vector-boson-fusion and $WH$
Higgs-strahlung channels with less than 3\% uncertainty and the $ZH$
Higgs-strahlung channel with about 4\% uncertainty due to the novel loop
contributions from $gg\to ZH$ as discussed in Section
\ref{sc:higgsstrahlung}. The uncertainties of $t\bar tH$ production
amount to about $10-15\%$, for $s$- and $t$-channel $tH$ production to about
$15-20\%$ and for $b\bar bH$ production to about $20-25\%$.
Fig.~\ref{fg:hcxnen} shows the energy dependence of the Higgs production
cross sections for a Higgs mass $M_H=125$ GeV. It is visible that all
cross sections develop a similar rising slope apart from $t\bar tH$ and
$tH$ that grow stronger due to the larger phase-space suppression for
smaller energies.

\subsubsection{\it Minimal supersymmetric extension}
\begin{figure}[hbtp]
\begin{center}
\begin{picture}(150,240)(0,0)
\put(-280,-20.0){\includegraphics{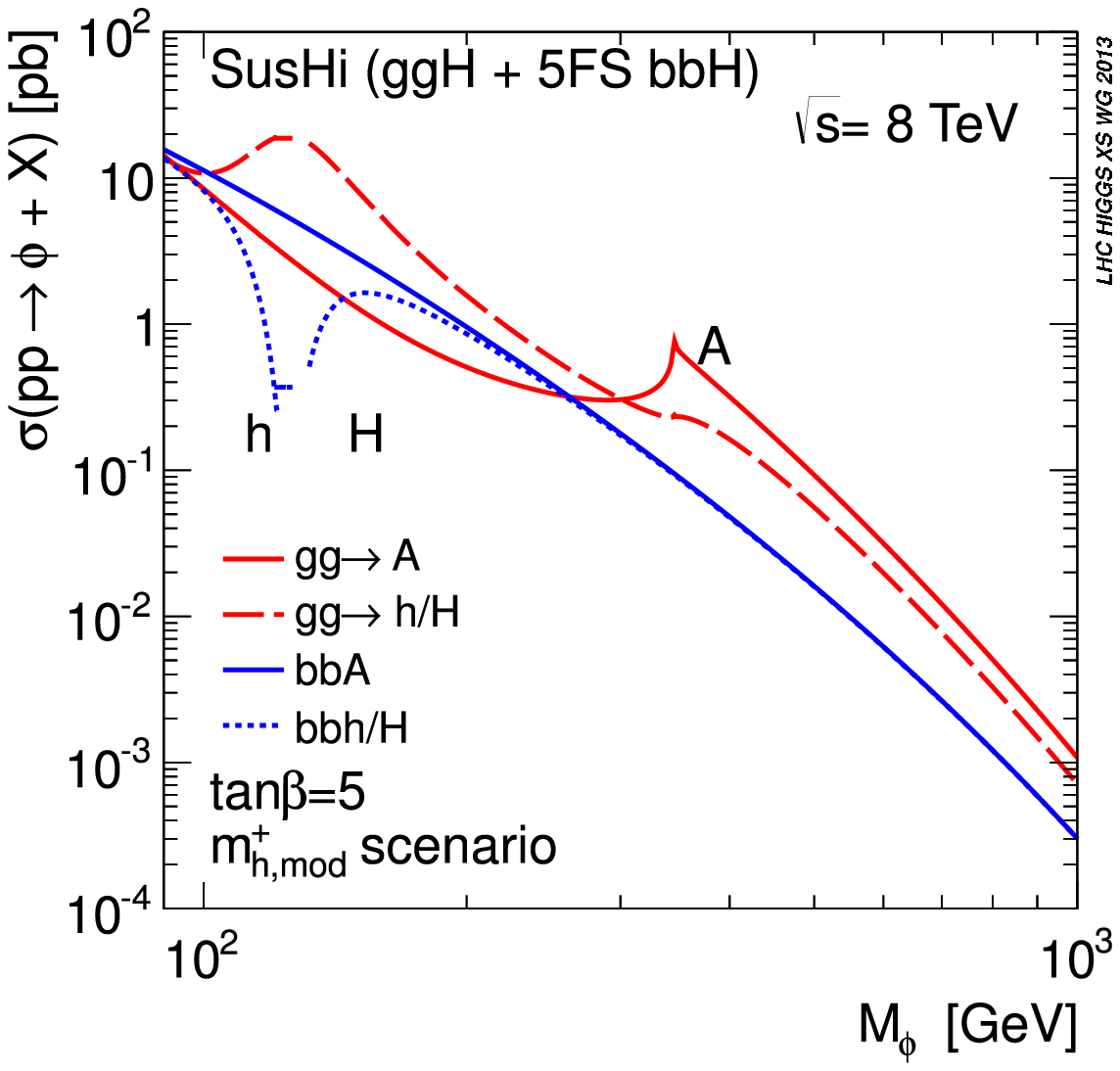}}
\put(-30,-20.0){\includegraphics{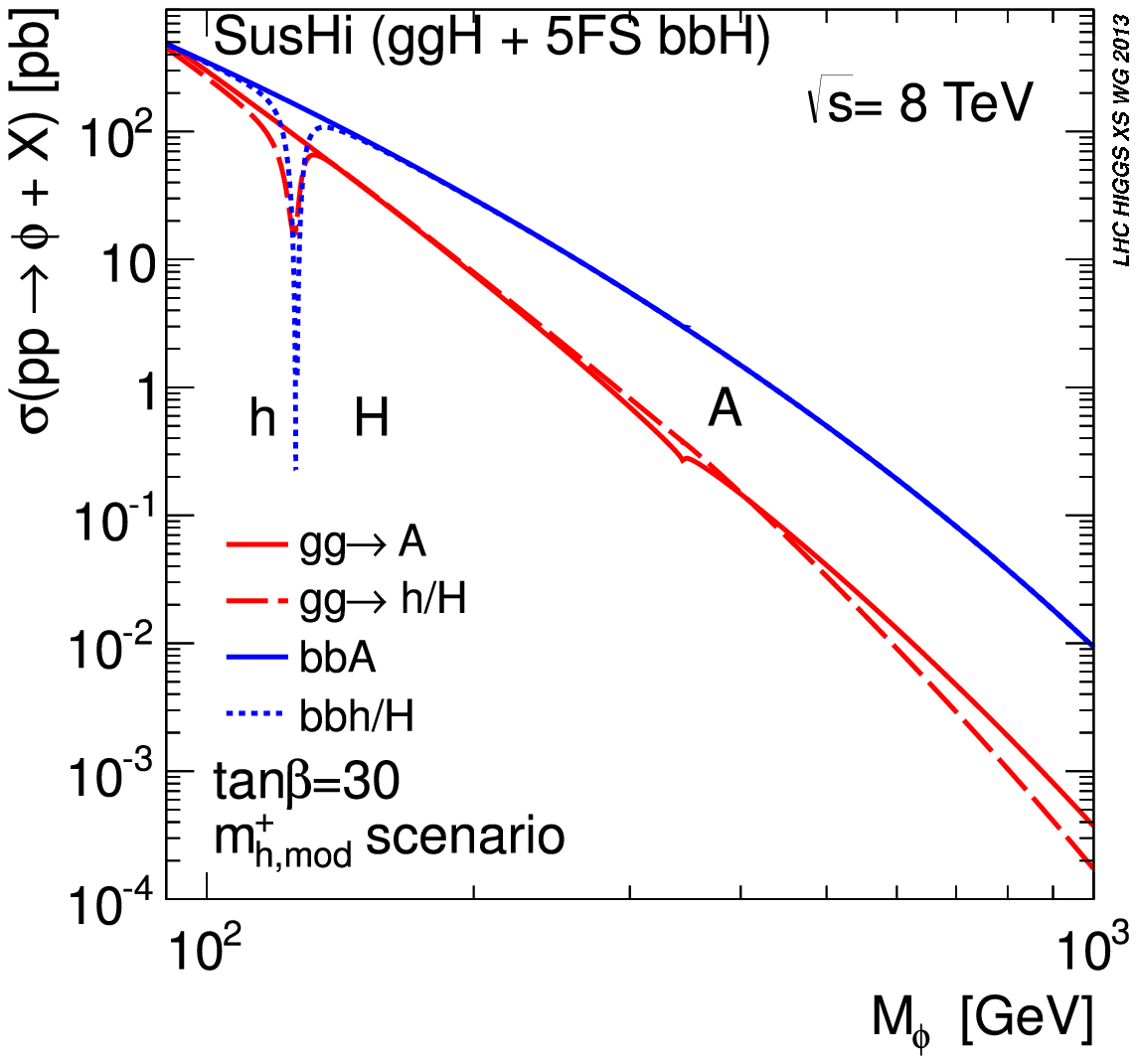}}
\end{picture}
\caption{\it Neutral MSSM Higgs production cross sections at the LHC for a
c.m.~energy $\sqrt{s}=8$ TeV as functions of the corresponding Higgs
mass within the $m_{h}^{mod+}$ scenario \cite{benchmark} for two values
of $\mbox{tg}\beta$ obtained by {\sc SusHi} \cite{sushi}. From
Ref.~\cite{yr3}.}
\label{fig:mssmcxn}
\end{center}
\end{figure}

An analogous update of the production cross sections as for the SM case
has been made also for the MSSM Higgs-boson production cross sections.
The public code {\sc Sushi} \cite{sushi} has been used as the preferred
choice since this includes the full NLO QCD corrections to the
gluon-fusion cross section and the NNLO QCD corrections in the heavy
top-quark limit for the top-loop contributions\footnote{Meanwhile also
the N$^3$LO QCD corrections have been included in {\sc Sushi}
\cite{sushi}.}. Moreover, electroweak corrections originating from
light-fermion loops \cite{hggelwf} are taken into account in this
program. For the $b\bar b\Phi$ production cross section {\sc Sushi}
contains the NNLO QCD-corrected 5FS-cross section that agrees with the
4FS results within about 20\% for the adopted scale choices \cite{yr4}.
SUSY-QCD corrections are included in the heavy SUSY-particle limit for
the gluon-fusion process and in the $\Delta_b$ approximation for $b\bar
b\phi$ production. The compiled cross sections within the $m_{h}^{mod+}$
scenario \cite{benchmark} are shown in Fig.~\ref{fig:mssmcxn} for two
values of $\mbox{tg}\beta$. The left plot exhibits the dominance of the
gluon-fusion process for smaller values of $\mbox{tg}\beta$ and the
right one the dominance of $b\bar b\phi$ production for large values of
$\mbox{tg}\beta$. Uncertainties have not been added to these results,
but have been analyzed in detail in Ref.~\cite{mssmunc}.

\subsection{\it Higgs boson pair production}
\subsubsection{\it Standard Model}
\begin{figure}[hbtp]
\begin{center}
\setlength{\unitlength}{1pt}
\begin{picture}(100,100)(100,0)
\Gluon(0,80)(50,80){3}{5}
\Gluon(0,20)(50,20){3}{5}
\ArrowLine(50,20)(50,80)
\ArrowLine(50,80)(100,80)
\ArrowLine(100,80)(100,20)
\ArrowLine(100,20)(50,20)
\DashLine(100,80)(150,80){5}
\DashLine(100,20)(150,20){5}
\put(-60,48){$(a)$}
\put(155,76){$H$}
\put(155,16){$H$}
\put(-15,78){$g$}
\put(-15,18){$g$}
\put(30,48){$t,b$}
\end{picture}
\begin{picture}(100,100)(-20,0)
\Gluon(0,80)(50,80){3}{5}
\Gluon(0,20)(50,20){3}{5}
\ArrowLine(50,20)(50,80)
\ArrowLine(50,80)(100,50)
\ArrowLine(100,50)(50,20)
\DashLine(100,50)(150,50){5}
\DashLine(150,50)(200,80){5}
\DashLine(150,50)(200,20){5}
\put(205,76){$H$}
\put(205,16){$H$}
\put(-15,78){$g$}
\put(-15,18){$g$}
\put(30,48){$t,b$}
\put(146,46){\Large \textcolor{red}{$\bullet$}}
\put(146,60){\textcolor{red}{$\lambda$}}
\end{picture}  \\
\begin{picture}(100,100)(100,0)
\ArrowLine(0,80)(50,80)
\ArrowLine(50,80)(100,80)
\ArrowLine(0,20)(50,20)
\ArrowLine(50,20)(100,20)
\Photon(50,80)(50,20){3}{5}
\DashLine(48,60)(100,60){5}
\DashLine(52,40)(100,40){5}
\put(-60,48){$(b)$}
\put(105,78){$q$}
\put(105,18){$q$}
\put(105,56){$H$}
\put(105,36){$H$}
\put(-15,78){$q$}
\put(-15,18){$q$}
\put(20,48){$W,Z$}
\end{picture}
\begin{picture}(100,100)(-20,0)
\ArrowLine(0,80)(50,80)
\ArrowLine(50,80)(100,80)
\ArrowLine(0,20)(50,20)
\ArrowLine(50,20)(100,20)
\Photon(50,80)(50,20){3}{5}
\DashLine(50,50)(100,50){5}
\DashLine(100,50)(150,80){5}
\DashLine(100,50)(150,20){5}
\put(105,78){$q$}
\put(105,18){$q$}
\put(155,76){$H$}
\put(155,16){$H$}
\put(-15,78){$q$}
\put(-15,18){$q$}
\put(20,63){$W,Z$}
\put(96,46){\Large \textcolor{red}{$\bullet$}}
\put(96,60){\textcolor{red}{$\lambda$}}
\end{picture}  \\
\begin{picture}(100,100)(100,0)
\ArrowLine(0,80)(50,50)
\ArrowLine(50,50)(0,20)
\Photon(50,50)(100,50){3}{5}
\Photon(100,50)(150,80){3}{5}
\DashLine(100,50)(150,20){5}
\DashLine(120,60)(150,42){5}
\put(-60,48){$(c)$}
\put(155,78){$W,Z$}
\put(155,38){$H$}
\put(155,16){$H$}
\put(-15,78){$q$}
\put(-15,18){$\bar q$}
\put(60,60){$W,Z$}
\end{picture}
\begin{picture}(100,100)(-20,0)
\ArrowLine(0,80)(50,50)
\ArrowLine(50,50)(0,20)
\Photon(50,50)(100,50){3}{5}
\Photon(100,50)(150,80){3}{5}
\DashLine(100,50)(125,35){5}
\DashLine(125,35)(150,20){5}
\DashLine(125,35)(150,50){5}
\put(155,78){$W,Z$}
\put(155,46){$H$}
\put(155,16){$H$}
\put(-15,78){$q$}
\put(-15,18){$\bar q$}
\put(60,60){$W,Z$}
\put(121,31){\Large \textcolor{red}{$\bullet$}}
\put(121,45){\textcolor{red}{$\lambda$}}
\end{picture}  \\
\begin{picture}(100,100)(100,0)
\Gluon(0,80)(50,80){3}{5}
\Gluon(0,20)(50,20){3}{5}
\ArrowLine(50,20)(50,80)
\ArrowLine(50,80)(100,80)
\ArrowLine(100,80)(100,50)
\ArrowLine(100,50)(100,20)
\ArrowLine(100,20)(50,20)
\Photon(100,80)(150,80){3}{5}
\DashLine(100,50)(150,50){5}
\DashLine(100,20)(150,20){5}
\put(155,76){$Z$}
\put(155,46){$H$}
\put(155,16){$H$}
\put(-15,78){$g$}
\put(-15,18){$g$}
\put(30,48){$t,b$}
\end{picture}
\begin{picture}(100,100)(-20,0)
\Gluon(0,80)(50,80){3}{5}
\Gluon(0,20)(50,20){3}{5}
\ArrowLine(50,20)(50,80)
\ArrowLine(50,80)(100,80)
\ArrowLine(100,80)(100,20)
\ArrowLine(100,20)(50,20)
\Photon(100,80)(150,80){3}{5}
\DashLine(100,20)(150,20){5}
\DashLine(125,20)(150,35){5}
\put(155,76){$Z$}
\put(155,31){$H$}
\put(155,16){$H$}
\put(-15,78){$g$}
\put(-15,18){$g$}
\put(30,48){$t,b$}
\put(121,16){\Large \textcolor{red}{$\bullet$}}
\put(121,30){\textcolor{red}{$\lambda$}}
\end{picture} \\
\begin{picture}(100,100)(100,0)
\Gluon(0,80)(50,80){3}{5}
\Gluon(0,20)(50,20){3}{5}
\ArrowLine(50,20)(50,40)
\ArrowLine(50,40)(50,60)
\ArrowLine(50,60)(50,80)
\ArrowLine(50,80)(100,80)
\ArrowLine(100,20)(50,20)
\DashLine(50,60)(100,60){5}
\DashLine(50,40)(100,40){5}
\put(-60,48){$(d)$}
\put(105,78){$t$}
\put(105,18){$\bar t$}
\put(105,56){$H$}
\put(105,36){$H$}
\put(-15,78){$g$}
\put(-15,18){$g$}
\end{picture}
\begin{picture}(100,100)(-20,0)
\Gluon(0,80)(50,80){3}{5}
\Gluon(0,20)(50,20){3}{5}
\ArrowLine(50,20)(50,50)
\ArrowLine(50,50)(50,80)
\ArrowLine(50,80)(100,80)
\ArrowLine(100,20)(50,20)
\DashLine(50,50)(100,50){5}
\DashLine(100,50)(150,80){5}
\DashLine(100,50)(150,20){5}
\put(105,78){$t$}
\put(105,18){$\bar t$}
\put(155,76){$H$}
\put(155,16){$H$}
\put(-15,78){$g$}
\put(-15,18){$g$}
\put(96,46){\Large \textcolor{red}{$\bullet$}}
\put(96,60){\textcolor{red}{$\lambda$}}
\end{picture}  \\
\setlength{\unitlength}{1pt}
\caption{\label{fg:hhdia} \it Diagrams contributing to Higgs-boson pair
production: (a) gluon fusion, (b) vector-boson fusion, (c) double
Higgs-strahlung and (d) double Higgs bremsstrahlung off top quarks. The
contribution of the trilinear Higgs coupling is marked in red.}
\end{center}
\end{figure}
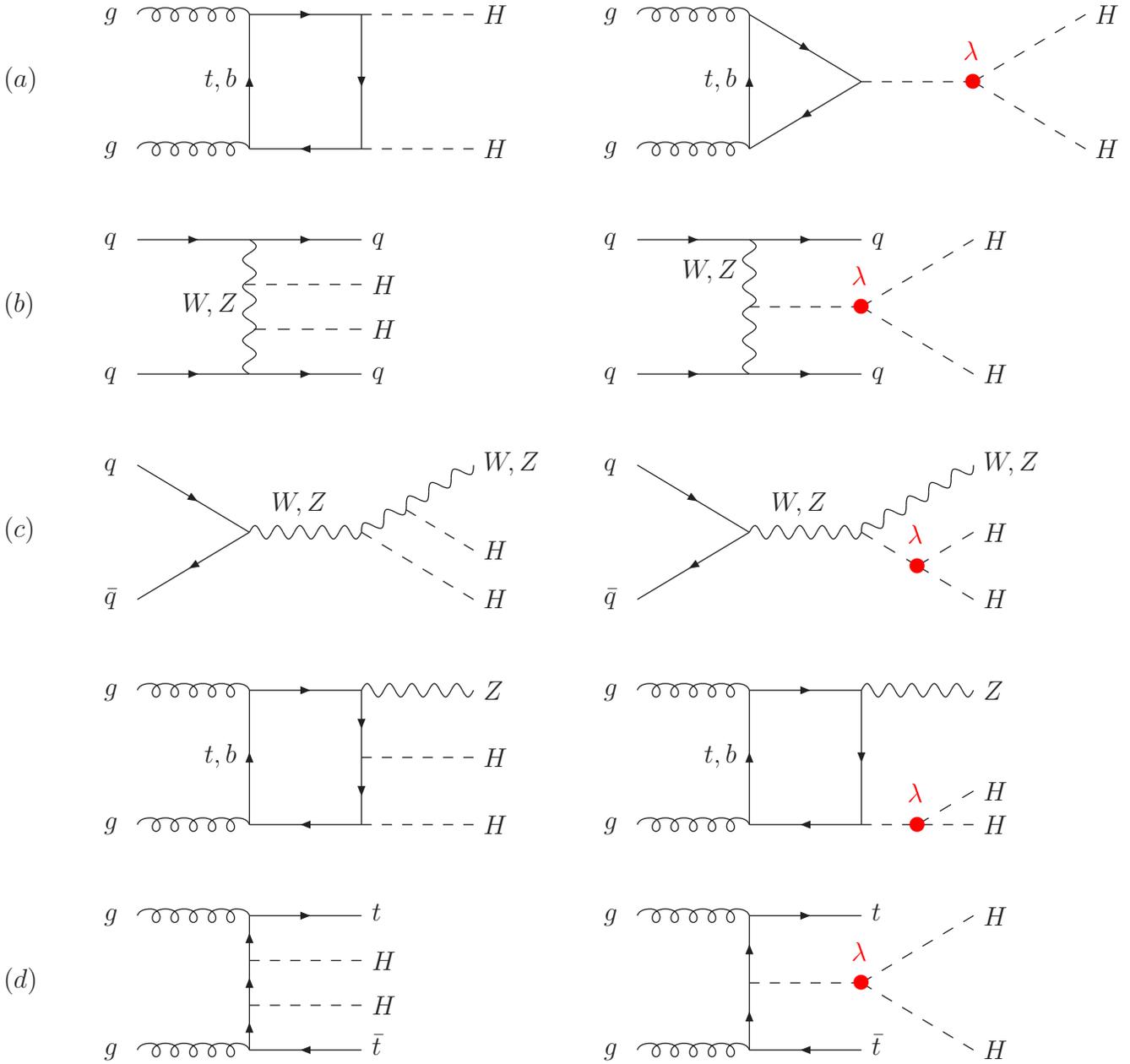
\begin{figure}[hbt]
\vspace*{0.5cm}

\hspace*{-0.0cm}
\epsfxsize=8cm \epsfbox{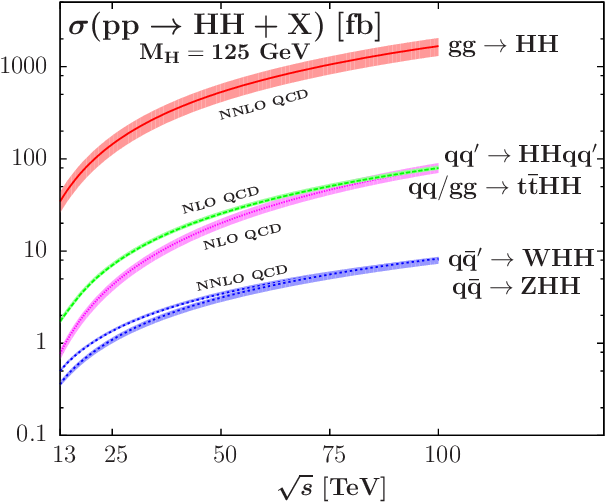}
\hspace*{1.5cm}
\epsfxsize=8cm \epsfbox{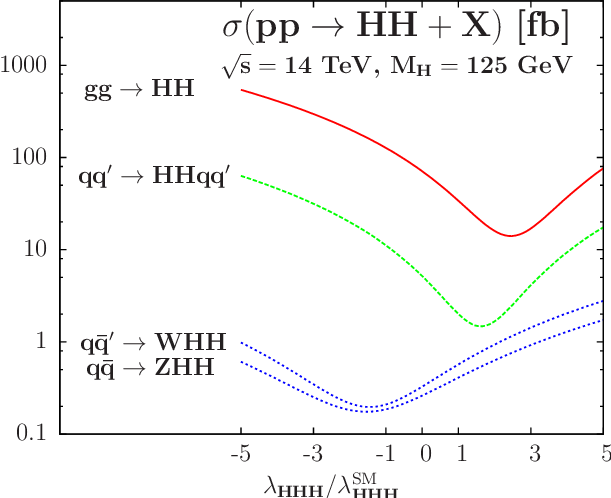}
\vspace*{-0.0cm}

\caption{\label{fg:hhcxns} \it Left: Higgs-boson pair production cross
sections as a function of the c.m.~energy at the LHC for a Higgs mass
$M_H=125$ GeV. The bands represent the theoretical uncertainties. From
Ref.~\cite{hh100}. Right: Dependence of the production cross sections on the
trilinear Higgs coupling $\lambda_{HHH}$. From Ref.~\cite{hhvbfnlo}.}
\end{figure}

Higgs-boson pair production is the first process that allows to obtain
direct access to the trilinear self-coupling of the Higgs boson and
paves the way to the Higgs potential at the origin of electroweak
symmetry breaking\footnote{Additional indirect sensitivity to the
trilinear Higgs coupling is provided by the single-Higgs production
processes due to its contribution to the electroweak corrections
\cite{lambdaind}.}. Higgs boson pairs are dominantly produced in the
gluon-fusion process and to a lesser extent in vector-boson fusion,
double Higgs-strahlung and double Higgs bremsstrahlung off top quarks,
see Fig.~\ref{fg:hhdia}. The production cross sections are shown as a
function of the collider energy for a Higgs mass $M_H=125$ GeV in
Fig.~\ref{fg:hhcxns}. Gluon fusion dominates the production of
Higgs-boson pairs by more than an order of magnitude, while the
production modes roughly follow the pattern of single-Higgs boson
production. Since the diagrams involving the trilinear Higgs coupling
are only a subset for each process the sensitivity to the trilinear
Higgs coupling is reduced compared to the size of the individual cross
sections. The sensitivities of the individual production cross sections
to the trilinear Higgs coupling are shown in Fig.~\ref{fg:hhcxns}.
The locations of the minima of the cross sections in terms of the value
of $\lambda$ are different for the production mechanisms \cite{hhvbfnlo}.

\paragraph{Gluon Fusion.}
The gluon-fusion process is mediated by top- and to a lesser extent
bottom-quark loops. There are box- and triangle contributions at LO, see
Fig.~\ref{fg:hhdia}a. The box contributions are dominant with a
destructive interference to the triangle diagrams \cite{gghhlo}. The
sensitivity to the trilinear Higgs coupling follows the rough behaviour
$\Delta \sigma/\sigma \sim -\Delta\lambda/\lambda$ so that the
uncertainties of the production cross section are immediately translated
to the uncertainties of the extraction of the trilinear self-coupling.

In the past the NLO QCD corrections have been determined in the limit of
heavy top quarks \cite{gghhnlo}. Similar to the single Higgs case they
increase the cross section by up to 100\%. Due to the fact that the
invariant mass of the final-state Higgs-boson pair is much larger than
in the single-Higgs case the heavy top-quark limit is expected to work
less reliable in the Higgs-pair case. The first attempt to estimate
finite top-mass effects at NLO was by means of a systematic heavy-top
expansion of the inclusive cross section \cite{gghhnloexp} that gave an
estimate of about 10\% for the finite top-mass effects beyond LO. The
second attempt kept the virtual corrections in the heavy-top limit but
included the real corrections exactly \cite{gghhnloreal}. This resulted
in a 10\%-decrease of the NLO cross section. Very recently the full NLO
calculation has been completed by numerical methods implying a decrease
of the total cross section by about 14\% at the LHC for a c.m.~energy of
13 TeV \cite{gghhnlomt} so that the heavy-top limit works still
reasonably well for the total cross section. For large invariant
Higgs-pair masses the finite mass effects can reach a level of --20\%.
The NLO result has been extended by the NNLO QCD corrections in the
heavy top-quark limit that lead to a rise of the total cross section by
about 20\% \cite{gghhnnlo}. Within a heavy top-quark expansion NNLO top
mass effects have been estimated to be at the 5\%-level
\cite{gghhnnloexp}. Finally a NNLL soft and collinear gluon resummation
has been added which adds a contribution of 5-10\% beyond NNLO
\cite{gghhnnll}. Including the most up-to-date theoretical status for
the prediction of the total cross section the theoretical uncertainties
due to the scale dependence is reduced to about 5\%. Together with the
PDF+$\alpha_s$ uncertainties one obtains a total uncertainty of about
10\% for the total cross section at the LHC \cite{yr4}. Recently a
public code has been constructed for fully exclusive Higgs-boson pair
production via gluon fusion including NNLO QCD corrections at parton
level \cite{gghhnnloexcl}.

\paragraph{Vector-boson Fusion.}
Higgs-boson pair production via vector-boson fusion is dominated by
$t$-channel contributions as in the single-Higgs case, see
Fig.~\ref{fg:hhdia}b. The NLO QCD corrections can be taken from deep
inelastic lepton-nucleon scattering analogous to single-Higgs production
\cite{hhvbfnlo, gghhnloreal}. They increase the cross section by about
10\%. Within the same approach the NNLO QCD corrections have been
obtained \cite{hhvbfnnlo}. They range at the per-cent level. The
residual theoretical and parametric uncertainties amount to about 3-4\%
at the LHC \cite{yr4}.

\paragraph{Double Higgs-strahlung.}
Double Higgs-strahlung proceeds along the same lines as
single-Higgs-strahlung, i.e.~the Higgs boson pair is produced in
association with a $W$ or $Z$ boson, see Fig.~\ref{fg:hhdia}c. The NLO
and NNLO QCD corrections can be taken over from the corresponding
calculation for the Drell--Yan process, since the final state is only
weakly interacting \cite{hhvbfnlo,gghhnloreal}. The only difference to
the Drell--Yan process emerges from the additional loop-mediated $gg\to
ZHH$ process, see Fig.~\ref{fg:hhdia}c. This has been added to the NNLO
QCD corrections. The QCD corrections increase the production cross
sections by about 30\%, while the $gg\to ZHH$ adds another 20--30\% to
$ZHH$ production \cite{hhvbfnlo}. The residual theoretical uncertainties
range at the 3\%-level for $WHH$ production and at the level of 4\% for
$ZHH$ production \cite{yr4}. A fully differential calculation for $WHH$
final states has recently been completed \cite{whhnlo}.

\paragraph{Double Higgs bremsstrahlung off top quarks.}
Higgs-boson pair production in association with top quarks is generated
by analogous diagrams to the single-Higgs case, see Fig.~\ref{fg:hhdia}d
\cite{hhvbfnlo,tthhlo}. The NLO QCD corrections have been calculated recently
with the {\sc Mg5\_amc@nlo} tool \cite{gghhnloreal}. They modify the total
cross section by about 20\% and reduce the scale dependence to a level
of less than 5\%. The total theoretical and parametric uncertainties for
this production process amount to about $5-6\%$ at the LHC with
c.m.~energies of 13 and 14 TeV \cite{yr4}.

\subsubsection{\it Minimal supersymmetric extension}
\begin{figure}[hbt]
\begin{picture}(100,100)(-30,-5)
\SetScale{0.8}
\Gluon(0,100)(50,100){3}{6}
\Gluon(0,0)(50,0){3}{6}
\ArrowLine(50,100)(100,50)
\ArrowLine(100,50)(50,0)
\ArrowLine(50,0)(50,100)
\DashLine(100,50)(150,50){5}
\DashLine(150,50)(200,100){5}
\DashLine(150,50)(200,0){5}
\put(165,76){$\phi_1$}
\put(165,-5){$\phi_2$}
\put(90,50){$\phi,Z$}
\put(-15,-3){$g$}
\put(-15,78){$g$}
\put(20,38){$t,b$}
\SetScale{1}
\end{picture}
\begin{picture}(100,100)(-150,-5)
\SetScale{0.8}
\Gluon(0,100)(50,100){3}{6}
\Gluon(0,0)(50,0){3}{6}
\ArrowLine(50,100)(140,100)
\ArrowLine(140,100)(140,0)
\ArrowLine(140,0)(50,0)
\ArrowLine(50,0)(50,100)
\DashLine(140,100)(190,100){5}
\DashLine(140,0)(190,0){5}
\put(165,76){$\phi_1$}
\put(165,-5){$\phi_2$}
\put(-15,-3){$g$}
\put(-15,78){$g$}
\put(20,38){$t,b$}
\SetScale{1}
\end{picture}
\caption{\it \label{fg:ggha} Generic diagrams describing neutral
Higgs-boson pair production in gluon--gluon collisions ($\phi,\phi_i =
h,H,A$).}
\end{figure}
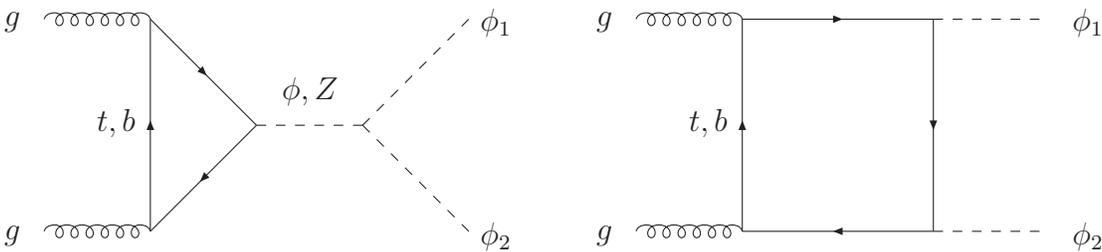
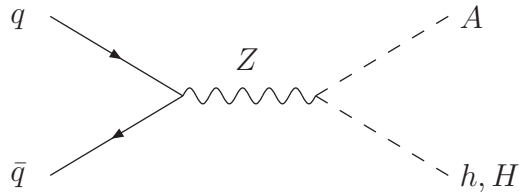
\begin{figure}[hbt]
\begin{center}
\begin{picture}(160,100)(0,0)
\ArrowLine(0,80)(50,50)
\ArrowLine(50,50)(0,20)
\Photon(50,50)(100,50){3}{5}
\DashLine(100,50)(150,80){5}
\DashLine(100,50)(150,20){5}
\put(155,16){$h,H$}
\put(-15,18){$\bar q$}
\put(-15,78){$q$}
\put(70,60){$Z$}
\put(155,76){$A$}
\end{picture}  \\
\setlength{\unitlength}{1pt}
\caption{\label{fg:hady} \it Diagram contributing to $q\bar q \to
Ah,AH$ at lowest order.}
\end{center}
\end{figure}

In the MSSM there are several possible neutral Higgs-pair final states.
The gluon-fusion mechanism yields $hh, hH, HH, hA, HA, AA$ final states
that emerge from analogous diagrams as in the SM-Higgs case. In the
mixed scalar-pseudoscalar channel, i.e.~$gg\to h/H + A$, off-shell
$Z$-boson exchange contributes to the $s$-channel diagrams, too, see
Fig.~\ref{fg:ggha}. However, these final states are dominated by
Drell-Yan production mechanisms, see Fig.~\ref{fg:hady}. Since the QCD
corrections to MSSM Higgs pair production via gluon fusion are only
known in the heavy-top-quark limit so far, reliable NLO predictions of
the corresponding cross sections are only possible for small values of
$\mbox{tg}\beta$ where the top loops provide the dominant contributions
and for not too large external Higgs masses. The QCD corrections are
large and positive, increasing the cross sections by up to 100\%
\cite{gghhnlo}. The SUSY--QCD corrections are known in the limit of
large SUSY-particle and top masses \cite{gghhsqcd} which is also
reliable for smaller values of $\mbox{tg}\beta$ where the bottom-loop
contributions are suppressed. The impact of SUSY-contributions starts to
be sizeable for squark mass below about 1 TeV. The MSSM cross sections
for the gluon-fusion processes range below 10 fb in all regions where
none of the Higgs bosons involved in the $s$-channel becomes resonant.
However, for small values of $\mbox{tg}\beta$ and below the $t\bar
t$-threshold there are sizeable regions where the heavy scalar Higgs
boson can become resonant and decays into $hh$ final states, $gg\to H\to
hh$. In these regions the cross sections become large, since the
dominant piece factorizes into the single heavy-scalar Higgs production
cross section and the branching ratio of the $H\to hh$ decay
\cite{gghhnlo}.

Mixed scalar-pseudoscalar Higgs boson production is dominated by the
Drell--Yan process $q\bar q\to h/H+A$. The QCD corrections can be
translated from the corresponding Drell--Yan process and are of moderate
size \cite{gghhnlo}. The cross sections for these processes can reach
the level of about 100 fb.

\section{Summary}
In this review the decay widths and branching ratios of SM and MSSM
Higgs bosons have been discussed. All relevant higher order corrections,
which are dominated by QCD corrections, have been summarized according
to the present state of the art.  At the LHC the SM Higgs particle is
produced predominantly by gluon fusion $gg\to H$, followed by
vector-boson fusion $VV\to H$ ($V=W,Z$) and, to a lesser extent,
Higgs-strahlung off vector bosons, $V^*\to VH$, and top quarks,
$gg/q\bar q\to t\bar t H$. The cross sections of these production
channels have been described including all known QCD and electroweak
corrections, which are important in particular for the dominant
gluon-fusion mechanism where they increase the production cross section
by about a factor of two. The QCD corrections to the subleading
production processes are moderate in case of the electroweak
vector-boson fusion and Higgs-strahlung but also for $t\bar tH$
production, since in the latter process the threshold region is strongly
phase-space suppressed. For vector-boson fusion and Higgs-strahlung the
electroweak corrections are of similar size as the QCD ones. The partial
decay widths are all known with per-cent or sub-per-cent precision. 

In the MSSM the neutral Higgs bosons will mainly be produced via gluon
fusion $gg\to \Phi$. However, through the enhanced $b$ quark couplings,
Higgs bremsstrahlung off $b$ quarks, $gg/q\bar q\to b\bar b \Phi$, will
dominate for large $\mbox{tg}\beta$.  All other Higgs production
mechanisms, i.e.\ vector-boson fusion and Higgs-strahlung off vector
bosons or $t\bar t$ pairs, will be less important than in the SM. The
potentially large SUSY--QCD and -electroweak corrections to the bottom
Yukawa coupling constitute a major ingredient for accurate predictions of
production and decay processes involving the bottom Yukawa coupling. For
configurations with on-shell bottom quarks the leading $\Delta_b$ terms
introduced in Section \ref{sec:mssmintro} provide an excellent
approximation to the full NLO results, while cases with far off-shell
bottom quarks as e.g.~the gluon-fusion processes $gg\to h/H/A$ or the
reverse gluonic Higgs decays develop sizeable corrections beyond the
$\Delta_b$-approximation for the genuine supersymmetric part of the
radiative corrections.

\vspace*{1cm}

\noindent
{\large \bf Acknowledgements} \\
I am grateful to J.~Baglio, S.~Dittmaier and M.~M.~M\"uhlleitner for
reading the manuscript and valuable comments and discussions.

\end{document}